\documentclass[11pt,a4paper]{mybook}

\usepackage{psfrag}
\usepackage{natbib}
\usepackage{fancyheadings}
\renewcommand{\chaptermark}[1]{\markboth{#1}{}}

\usepackage{amsmath}  
\usepackage{amsfonts}
\usepackage{amssymb}
\usepackage{vmargin} 
\usepackage{epsfig}
\usepackage{subfigure} 

\usepackage{doublespace} 
\newcommand{\clearemptydoublepage}{\newpage{\pagestyle{empty}\cleardoublepage}}

\newcommand{\bflangle}{\Bigg\langle\!\!\!\Bigg\langle\!}
\newcommand{\bfrangle}{\!\Bigg\rangle\!\!\!\Bigg\rangle}
\newcommand{\flangle}{\Big\langle\hspace{-1.3mm}\Big\langle\!}
\newcommand{\frangle}{\!\Big\rangle\hspace{-1.3mm}\Big\rangle}
\newcommand{\ds}{\displaystyle}
\newcommand{\be}{\begin{equation}}
\newcommand{\ee}{\end{equation}}
\newcommand{\Real}{{\rm Re}}
\newcommand{\Tr}{{\rm Tr}}
\newcommand{\bea}{\begin{eqnarray}}
\newcommand{\eea}{\end{eqnarray}}
\newcommand{\bml}{\begin{multline}}
\newcommand{\eml}{\end{multline}}
\newcommand{\pa}[1]{\partial_{#1}}

\newcommand{\LE}{{\cal E}}
\newcommand{\LA}{{\cal A}}
\newcommand{\LK}{{\cal K}}
\newcommand{\LV}{{\cal V}}
\newcommand{\LH}{{\cal H}}

\newcommand{\nn}{\nonumber}
\newcommand{\bm}[1]{\mbox{\boldmath $#1$}}
\newcommand{\bms}[1]{\mbox{\boldmath ${\scriptstyle #1}$}}
\newcommand{\hbm}[1]{\hat{\mbox{\boldmath $#1$}}}
\newcommand{\hbms}[1]{\hat{\mbox{\boldmath ${\scriptstyle #1}$}}}

\newcommand{\ord}{{\cal O}}
\newcommand{\graphd}{\begin{array}{l}\begin{picture}(20,25)

        \put (0,0) {\vector(1,0){13}}
        \put (0,0) {\line(1,0){20}}
        \put (4,4) {\vector(1,0){9}}
        \put (4,4) {\line(1,0){12}}
        
        \put (16,4){\line(0,1){12}}
        \put (16,4){\vector(0,1){9}}
        \put (16,16){\line(-1,0){12}}
        \put (16,16){\vector(-1,0){9}}
        \put (4,16){\line(0,-1){12}}
        \put (4,16){\vector(0,-1){9}}

        \put (20,0) {\vector(0,1){13}}
        \put (20,0) {\line(0,1){20}}
        \put (20,20) {\vector(-1,0){13}}
        \put (20,20) {\line(-1,0){20}}
        \put (0,20) {\vector(0,-1){13}}
        \put (0,20) {\line(0,-1){20}}

\end{picture}\end{array}}

\newcommand{\grapha}{\begin{array}{l}\begin{picture}(20,25)

        \put (0,0) {\line(4,1){16}}
        \put (4,4) {\line(4,-1){16}}
        \put (16,4){\line(0,1){12}}
        \put (16,4){\vector(0,1){9}}
        \put (16,16){\line(-1,0){12}}
        \put (16,16){\vector(-1,0){9}}
        \put (4,16){\line(0,-1){12}}
        \put (4,16){\vector(0,-1){9}}

        \put (20,0) {\vector(0,1){13}}
        \put (20,0) {\line(0,1){20}}
        \put (20,20) {\vector(-1,0){13}}
        \put (20,20) {\line(-1,0){20}}
        \put (0,20) {\vector(0,-1){13}}
        \put (0,20) {\line(0,-1){20}}

\end{picture}\end{array}}

\newcommand{\graphdd}{\begin{array}{l}\begin{picture}(20,25)

        \put (0,0) {\vector(1,0){13}}
        \put (0,0) {\line(1,0){20}}
        \put (4,4) {\vector(0,1){9}}
        \put (4,4) {\line(0,1){12}}
        
        \put (4,16){\line(1,0){12}}
        \put (4,16){\vector(1,0){9}}
        \put (16,16){\line(0,-1){12}}
        \put (16,16){\vector(0,-1){9}}
        \put (16,4){\line(-1,0){12}}
        \put (16,4){\vector(-1,0){9}}

        \put (20,0) {\vector(0,1){13}}
        \put (20,0) {\line(0,1){20}}
        \put (20,20) {\vector(-1,0){13}}
        \put (20,20) {\line(-1,0){20}}
        \put (0,20) {\vector(0,-1){13}}
        \put (0,20) {\line(0,-1){20}}

\end{picture}\end{array}}

\newcommand{\plaquette}{\begin{array}{l}\begin{picture}(20,25)

        \put (0,0) {\vector(1,0){13}}
        \put (0,0) {\line(1,0){20}}
        \put (20,0) {\vector(0,1){13}}
        \put (20,0) {\line(0,1){20}}
        \put (20,20) {\vector(-1,0){13}}
        \put (20,20) {\line(-1,0){20}}
        \put (0,20) {\vector(0,-1){13}}
        \put (0,20) {\line(0,-1){20}}

\end{picture}\end{array}}
\newcommand{\rectangleA}{\begin{array}{l}\begin{picture}(40,25)

        \put (0,0) {\line(1,0){20}}
        \put (0,0) {\vector(1,0){13}}
        \put (20,0) {\line(1,0){20}}
        \put (20,0) {\vector(1,0){13}}  
        \put (40,0) {\vector(0,1){13}}
        \put (40,0) {\line(0,1){20}}
        \put (40,20) {\vector(-1,0){13}}
        \put (40,20) {\line(-1,0){20}}
        \put (20,20) {\vector(-1,0){13}}
        \put (20,20) {\line(-1,0){20}}
        \put (0,20) {\vector(0,-1){13}}
        \put (0,20) {\line(0,-1){20}}

\end{picture}\end{array}}

\newcommand{\rectanglelong}{\begin{array}{l}\begin{picture}(60,25)

        \put (0,0) {\line(1,0){20}}
        \put (0,0) {\vector(1,0){13}}
        \put (40,0) {\line(1,0){20}}
        \put (40,0) {\vector(1,0){13}}  
        \put (24,0) {\circle*{1}}
        \put (28,0) {\circle*{1}}
        \put (27,-7) {$m$}
        \put (32,0) {\circle*{1}}
        \put (36,0) {\circle*{1}}
        \put (60,0) {\vector(0,1){13}}
        \put (60,0) {\line(0,1){20}}
        \put (60,20) {\vector(-1,0){13}}
        \put (60,20) {\line(-1,0){20}}
        \put (24,20) {\circle*{1}}
        \put (28,20) {\circle*{1}}
        \put (32,20) {\circle*{1}}
        \put (36,20) {\circle*{1}}
        \put (20,20) {\vector(-1,0){13}}
        \put (20,20) {\line(-1,0){20}}
        \put (0,20) {\vector(0,-1){13}}
        \put (0,20) {\line(0,-1){20}}

\end{picture}\end{array}}

\newcommand{\boldx}{\boldsymbol{x}}
\newcommand{\eqn}[1]{Eq.~(\ref{#1})}

\newcommand{\eqns}[2]{Eqs.~(\ref{#1}) and (\ref{#2})}
\newcommand{\eqnss}[2]{Eqs.~(\ref{#1}---\ref{#2})}
\newcommand{\fig}[1]{Fig.~\ref{#1}}
\newcommand{\figs}[2]{Figs.~\ref{#1} and~\ref{#2}}

\newcommand{\figsss}[3]{Figs.~\ref{#1},~\ref{#2} and~\ref{#3}}
\newcommand{\tab}[1]{Table~\ref{#1}}

\newcommand{\tabss}[2]{Tables~\ref{#1}---\ref{#2}}
\newcommand{\sect}[1]{Section~\ref{#1}}
\newcommand{\sects}[2]{Sections~\ref{#1} and~\ref{#2}}

\newcommand{\rcite}[1]{Ref.~\cite{#1}}
\newcommand{\rcites}[2]{Refs.~\cite{#1} and~\cite{#2}}

\newcommand{\chap}[1]{Chapter~\ref{#1}}
\newcommand{\chaps}[2]{Chapters~\ref{#1} and~\ref{#2}}



\begin{document}

%
%

\title{{\Huge \bf Improvement and Analytic Techniques 
in Hamiltonian Lattice Gauge Theory}\\}
\vspace{2cm}

\author{{\Large Jesse Carlsson}\\\\
Submitted in total fulfilment of the requirements\\
of the degree of Doctor of Philosophy\\
\\School of Physics\\
The University of Melbourne}
\date{April 2003\\
\vspace{3cm}
{\it Produced on acid-free paper}}

\maketitle
\pagestyle{plain}
\cleardoublepage
\pagestyle{fancyplain}
\pagenumbering{roman}

\chapter*{Abstract}

This thesis is concerned with two topics in Hamiltonian lattice
gauge theory (LGT): improvement and the application of analytic techniques. 
On the topic of improvement, we develop a direct method for
improving lattice Hamiltonians for gluons, 
in which linear combinations of gauge
invariant lattice operators are chosen to cancel the lowest order
discretisation errors. A number of improved Hamiltonians
are derived and the level of improvement tested for the 
simple cases of U(1) in 3+1 dimensions and SU(2) in 2+1
dimensions. In each case the improved calculations are closer to the
continuum limit, when working at a given lattice spacing, than their
unimproved counterparts.\\

On the topic of analytic methods, we extend the techniques that have
been used in 2+1 dimensional SU(2) variational calculations
for many years, to the general case of SU($N$). For this purpose a
number of group integrals are calculated in terms of Toeplitz
determinants. As generating functions these group integrals allow the
calculation of all matrix elements appearing in variational glueball mass
calculations in 2+1 dimensions. Making use of these analytic
techniques, the lowest five glueball masses, or massgaps, in various symmetry
sectors are calculated, with improved and unimproved Hamiltonians, for
$2\le N \le 5$. For
simplicity, a minimisation basis containing only rectangular states
is used. Large scaling regions are obtained for each of the five
lowest mass states in the $J^{PC}=0^{++}$ and $0^{--}$ sectors. 
The results are comparable to competing calculations. However,
the enumeration of the states obtained requires further work. \\

The large $N$ limit of the glueball mass spectrum is also
explored. Making use of analytic variational techniques, glueball mass
calculations with $N$ as large as 25 are carried out on a desktop
computer and the results extrapolated accurately to the
$N\rightarrow\infty$ limit. Evidence of leading order
$1/N^2$ finite $N$ corrections is obtained for a number of excited
states. An interesting empirical observation is made which
demonstrates a close similarity between the glueball mass spectrum
 and the simple harmonic oscillator spectrum in 2+1 dimensions.\\

We finish with a feasibility study of applying analytic Hamiltonian 
techniques in 3+1 dimensional calculations. After developing analytic
techniques specific to 3+1 dimensions, we study the variational 
glueball mass spectrum for SU($N$) on a single cube,
with a minimisation basis containing only two states. Memory limitations do not allow
calculations with $N \ge 8$. We observe
promising signs of an approach to asymptotic scaling in the $1^{+-}$
sector as $N$ is increased, warranting larger volume studies with
additional states in the minimisation basis.

\pagestyle{plain}\cleardoublepage\pagestyle{fancyplain}

\chapter*{Declaration}
\hrule
\vspace{1cm}
This is to certify that
\begin{itemize}
\item[(i)] this thesis comprises only my original work,
\item[(ii)] due acknowledgement has been made in the text to all other material used,
\item[(iii)] this thesis is less than 100,000 words in length, 
exclusive of table, maps, bibliographies, appendices and footnotes.
\end{itemize}
\vspace{2cm}
\hfill .......................................\\

\vspace{0.5cm}
\hfill .......................................\\
\vspace{1cm}
\hrule
\pagestyle{plain}\cleardoublepage\pagestyle{fancyplain}

\chapter*{Acknowledgements}

I would like to thank my supervisor, Professor Bruce H. J. McKellar, for
providing me with an enjoyable three years of study. 
His knowledge and speed were inspirational and his care in supervision 
was appreciated. I would also like to thank Dr. John A. L. MacIntosh and
Dr. Lloyd C. L. Hollenberg for interesting and useful 
discussions along the way. Finally, I would like to thank Elaine 
Fallshaw and D-Mo for their help in proofreading. 
   
\pagestyle{plain}\cleardoublepage\pagestyle{fancyplain}

\tableofcontents
\listoffigures
\listoftables
\clearemptydoublepage

\pagenumbering{arabic}
\chapter{Introduction}

\section{Historical Background}
\label{historicalbackground-1}

Quantum chromodynamics (QCD) is the accepted theory of high energy
strong interactions. It is the gauge theory describing the
interactions of coloured particles, quarks and gluons, which are
commonly accepted to be the fundamental degrees of freedom of hadronic
matter. \\

The non-abelian nature of QCD results in a collection of non-trivial
complications when compared to the archetypal gauge theory quantum
electrodynamics (QED). Most serious is the phenomenon of asymptotic freedom
which dictates that the QCD coupling constant, $g^2$, is not in fact
constant. Rather it depends on the energy scale of the
process in question. The calculation of the one-loop $\beta $
function~\cite{Gross:1973id,Politzer:1973fx}
demonstrated that the QCD coupling constant is small only for high
energy processes; a fact observed in experiment. This discovery was
instrumental in the establishment of QCD as a theory of strong interactions.\\

Asymptotic scaling poses a serious problem to quantitative tests of
QCD. The traditional 
perturbative techniques used in high precision tests of 
 QED rely on expansions about a small coupling constant. 
Consequently such techniques are only applicable to high energy QCD
processes. Perturbative QCD has had a  great deal of success
explaining such processes. \\

The first quantitative demonstration of the validity of perturbative
QCD was in the violation of Bjorken scaling in deep-inelastic
lepton-hadron scattering (for example see
\rcite{Gribov:1972ri}). More recently high order QCD calculations of
jet rates in hadron colliders have shown impressive agreement with
experimental 
data. Less striking tests of perturbative QCD have been made in
the analysis of $\tau $ lepton decays, heavy-quarkonium decay and
also in $e^+ e^-$ collisions. For a recent review see \rcite{PDBook}.\\

For low energy processes the QCD coupling constant is
large. Consequently traditional perturbative techniques can not be
used to test the validity of QCD in this regime. A great deal of
interesting physics appears in this nonperturbative regime. One would
like to test whether the same theory that successfully explains a
broad range of high energy
processes also gives rise to quark confinement, provides a mechanism
for chiral symmetry breaking, gives rise to the correct
values for the hadronic mass spectrum and explains other low-energy
phenomena. Without
the use of non-perturbative techniques the situation is dire. One
cannot calculate such mental quantities as the 
masses of mesons and baryons even if given the coupling constant and
the constituent quark masses. The need to probe the low-energy regime of QCD has given rise
to many QCD motivated models~\cite{Chodos:1974je,Cornwall:1983zn,Isgur:1985bm,Johnson:2000qz}. However, only one first
principles approach to nonperturbative QCD is currently known and that
is lattice gauge theory (LGT).  \\

LGT was originally developed in the Lagrangian formulation by Wilson
in 1975~\cite{Wilson:1974sk}. 
A year later the corresponding Hamiltonian formulation
was published by Kogut and Susskind~\cite{Kogut:1975ag}. 
The basic approximation of LGT is
the replacement of continuous and infinite space-time by a discrete
lattice of sites and links. The essential difference between the
Hamiltonian and Lagrangian formulations is that in the former time
is left continuous. Using LGT one can presumably
calculate the full range of hadronic properties from mass spectra and
wave functions to decay constants and matrix elements. The hadronic
matrix elements appearing in the weak decays of hadrons are of
particular interest. Phenomenologists are waiting for lattice
estimates to such matrix elements in order to improve the theoretical 
understanding of the CKM matrix~\cite{Beneke:2002ks,Ryan:2001ej}. Another area of recent interest has been
in the study of QCD in extreme conditions. A knowledge of the phase
diagram of QCD is relevant to the understanding of high density quark
systems such as the early universe, neutron stars and heavy ion
collisions. The scope of LGT is
summarised in \rcite{Gupta:1997nd}.\\

In practice LGT cannot, at this stage, predict the complete range of
hadronic phenomena. The major stumbling block is the requirement of
time on the largest supercomputers in existence to calculate basic
quantities with accuracy. In the early days of LGT strong coupling
expansions in the Hamiltonian approach provided compelling evidence
for the confinement of quarks by showing that the potential energy
between two infinitely heavy quarks grows linearly with their
separation~\cite{Kogut:1975ag,Kogut:1980sg}.  
This was without the complication of dynamical
quarks and in the strong coupling limit which is far from the real
world physics of the continuum. 
The inclusion of two flavours of quarks 
and the extension to physical couplings came
much later and required the numerical techniques of Lagrangian
LGT. The qualitative result of the strong coupling expansions did not
however change (for example see \rcite{Bali:2000vr} and references within).\\

Most LGT calculations to date have been performed using Monte Carlo
techniques in the Lagrangian approach. Until recently most simulations
used the so called quenched approximation in which the vacuum polarization
effects of the QCD vacuum are explicitly removed. The calculation of
certain hadron masses are within 10\% of the
experimentally observed results using the quenched
approximation~\cite{Aoki:1999yr}. Calculations utilising full QCD with two flavours of quarks
have been performed and show significant improvements over the
quenched approximation~\cite{Yoshie:2002ru}. At this stage the advancement of LGT in the Lagrangian
approach is restricted by the available computational
resources.\\

There are some areas of research where progress in the Lagrangian
formulation has been significantly slow. Some examples are in 
calculations involving excited states and the determination of the QCD phase
diagram. Such slow progress and the need for significant computational
resources in Lagrangian LGT suggests that alternative methods should
be pursued in parallel. A viable alternative is Hamiltonian LGT, which
may have a number of advantages. It has been suggested that Hamiltonian LGT
could more readily handle finite density QCD~\cite{Gregory:1999pm}. 
Indeed, the problems
encountered in finite density QCD in the Lagrangian formulation has
prompted a return to the strong coupling expansions of early
Hamiltonian LGT~\cite{Bringoltz:2002ug}. 
Additional areas in which Hamiltonian LGT may
have an advantage over the Lagrangian approach are in calculations 
involving excited sites, time dependent quantities and wave functions. 
An appealing aspect of the Hamiltonian approach is that it reduces LGT
to a many body problem. As such a host of analytic techniques
from many body theory are applicable. Techniques such as the coupled
cluster
method~\cite{Coester:1958,Coester:1960,Ligterink:2000ug,McKellar:2000zk},
plaquette
expansion~\cite{Hollenberg:1993bp,Hollenberg:1994pv,McIntosh:2001uk,McIntosh:2001fm} 
and variational methods~\cite{Arisue:1983tt,Arisue:1990wv}
have had success in recent years in calculations of glueball masses.
 These calculations however have been performed without the
inclusion of quarks, often in less dimensions that the physical 3+1
and often using lower dimensional gauge groups than the
physical case of SU(3). \\ 

The self interactions of gluons in QCD allow the existence of massive states
containing only gluon degrees of
freedom~\cite{Fritzsch:1972,Fritzsch:1973pi}. Although QCD provides a theory of strong interactions we have
little understanding of the physical states of the theory. The
constituent quark model provides an excellent phenomenological
description of meson and baryon spectroscopy. QCD inspired models
however predict the existence of other hadrons. These hadrons include
not only the glueballs, which are often described as bound states of
two and three gluons, but other more exotic states as well. So called
hybrid particles, which have both quark and excited gluon degrees of
freedom, have been predicted to exist.  Until the properties of the
physical states of QCD can be reliably calculated, and these
calculations stand up to experimental tests, we cannot claim to understand the
nonperturbative aspects of QCD. From both theoretical and
experimental sides there have been significant developments in the
search for glueballs in recent years.\\

On the theoretical side calculations of the glueball mass spectrum have
been dominated by LGT calculations in the Lagrangian
approach. The calculations of the last decade are nearing agreement at 
least for the lowest mass glueballs. The emerging picture is that the
scalar glueball with quantum numbers $J^{PC} = 0^{++}$ is the
lightest with a mass in the range 1.4--1.8 GeV. Interestingly the
next lowest mass states in the $PC=++$ sector have spin ordering 2, 3,
1, whereas the ordering in the $PC=-+$ sector is 1, 3, 2, 0. The status
of LGT glueball calculations in the Lagrangian approach is reviewed in
\rcites{Bali:2001nc}{Morningstar:2001nu}.\\

On the experimental side progress has been hampered by numerous
technical difficulties. Perhaps most significant is the fact
that glueballs can mix strongly with nearby $q\bar{q}$ resonances. 
To search for glueballs one needs to probe glue rich processes. Recent
searches have taken place in radiative $J/\psi$ decay, $p\bar{p}$
annihilations and central production in $pp$ collisions. A review of
search techniques is presented in
\rcite{Godfrey:1998pd}.
Perhaps the most promising glueball
candidates are the $f_0(1500)$ and $f_0(1710)$ resonances (the
subscript denotes the total angular momentum and the argument the mass
in MeV). The interpretation of these resonances has been the subject
of much debate, particularly with regard to mixing with $q\bar{q}$
states. Some authors argue that the $f_0(1500)$ has a dominant glueball
contribution while the glueball content of $f_0(1710)$ is
minimal~\cite{Amsler:2002ey,Amsler:1996td}. Other authors argue the
opposite case~\cite{Lee:1999kv,Sexton:1995kd}. Improved data from
non-appearance searches in $\gamma \gamma$ collisions 
could possibly resolve this issue~\cite{Acciarri:2000ex}. The basis for these studies lies in
the fact that the glueball to two photon widths are small compared to
those of mesons since gluons do not couple directly with photons. Such
studies will also be important in the identification of other glueball
candidates such as $f_J(2220)$~\cite{Benslama:2002pa}.\\

The focus of most of the lattice community, as far as glueballs are
concerned, has been the physically interesting case of SU(3) glueballs
in 3+1 dimensions. In recent years the study of glueballs belonging to higher
dimensional gauge groups in both 2+1 and 3+1 dimensions 
has received intense interest from the string theory community. The
reason for this is that the SU($N$) glueball mass spectrum in the
large $N$ limit provides an
ideal laboratory for quantitative tests of Maldacena's conjecture and
its later extensions. We defer the discussion of this topic to
\sect{largenbackground}.\\

While accurate studies of the SU(3) glueball spectrum in 3+1
dimensions may be some way off in Hamiltonian LGT, studies of the SU($N$)
glueball spectrum in 2+1 dimensions are definitely feasible. In fact
it appears that the Hamiltonian approach has a distinct advantage over
the Lagrangian approach in that excited states are easily
accessed. Current Lagrangian calculations have extracted the three
lowest mass states in some sectors in the large $N$ limit. 
With Hamiltonian techniques it is
feasible to even triple this number. The accurate simulation of large
mass excited states is of immediate interest to QCD inspired 
models of glueballs such as the bag model~\cite{Chodos:1974je}, potential models~\cite{Cornwall:1983zn}
and the flux-tube model~\cite{Isgur:1985bm} and its
extensions~\cite{Johnson:2000qz}. The last of these in fact shows remarkable
agreement with the lowest excited state masses calculated on the
lattice in 2+1 dimensions for 
all but the $0^{-+}$ state~\cite{Johnson:2000qz}.\\
 
The results so far obtained in Lagrangian LGT have been made possible
by an extensive improvement programme initiated by Symanzik nearly
twenty years ago~\cite{Symanzik:1983dc,Symanzik:1983gh}. 
The formulation of QCD on a lattice introduces
errors which can be expressed in powers of the lattice spacing $a$ and the
QCD coupling and start at $\ord(a^2)$. When performing a lattice 
calculation in the Lagrangian
approach the most important factor in determining the computational
cost is the lattice spacing. In fact it has been predicted that the
cost of a full QCD simulation varies as $a^{-7.5}$~\cite{Jegerlehner:2000xt}.
Thus it is very important to be able to work on coarse lattices. This
of course is a move away from the $a\rightarrow 0$ limit in which the
real world physics of the continuum is retrieved. This realisation has
motivated the development of 
improved actions (for a review see \rcite{Lepage:1996jw}). 
These actions aim to systematically remove the
errors caused by placing QCD on a lattice. In practice these 
actions allow one be closer to the
continuum limit when working with a given lattice spacing. The additional 
computational cost of working with the more complicated improved 
actions is by far 
outweighed by the benefits of working on coarse lattices. The
development of improved actions has been a crucial factor in the progress of
LGT in the last 15 years. It has allowed routine calculations to be
performed on work stations and has brought the numerical
solution of full QCD tantalisingly close. \\

In contrast the improvement of Hamiltonian LGT has hardly begun. 
Perhaps the most comprehensive study into the derivation of improved
Hamiltonians to date is due to Luo, Guo, Kr{\"o}ger and
Sch{\"u}tte~\cite{Luo:1998dx}. Their study started with an improved
Lagrangian and derived improved Hamiltonians with the standard
transfer matrix and Legendre transformation techniques. Using these
techniques it became clear that the 
construction of an improved kinetic Hamiltonian with a finite number
of terms is a non-trivial problem.  Using a more direct method
improved Hamiltonians can be constructed with relative ease. In this
thesis we propose a method for the direct improvement of Hamiltonian
LGT 
which is in the spirit of the original Kogut-Susskind derivation. This
topic is the focus of \chap{constructingandimproving}.\\

We now move on to a description of the outline of the thesis.
  
\section{Outline of the Thesis}

As a form of introduction, in \chap{gaugetheoryonalattice}, the fundamental approximations of LGT are detailed. The basic steps involved in a Hamiltonian lattice calculation are also discussed as well as the important topic of how the real world physics of the continuum is retrieved.\\

\chaps{constructingandimproving}{testingimprovement} are concerned
primarily with the topic of improvement. In
\chap{constructingandimproving} we develop a technique which allows
the straightforward construction of lattice Hamiltonians for gluons in $d+1$ dimensions. A variety of improved Hamiltonians are derived using this method. These improved 
Hamiltonians are devoid of the lowest order discretisation errors 
present in the unimproved Kogut-Susskind Hamiltonian. In \chap{testingimprovement} we present some
simple checks on improvement for the cases of U(1) in 3+1 dimensions
and SU(2) in 2+1 dimensions. In this way we test the level of
improvement achieved by a selection of the Hamiltonians derived in \chap{constructingandimproving}.\\

In \chap{analytictechniques} we move on to the topic of analytic
methods. Here we introduce techniques which allow analytic lattice
calculations of SU($N$) glueball masses in 2+1 dimensions for general
$N$. Such techniques extend the SU(2) methods of
Arisue~\cite{Arisue:1983tt} to the general case of SU($N$). These
techniques are applied in \chap{sunmassgaps} in the calculation of
glueball masses in 2+1 dimensions with $2\le N\le5$. Kogut-Susskind, improved and tadpole improved Hamiltonians, all of which are derived in \chap{constructingandimproving}, are used and their results compared.\\

In \chap{largenphysics} we move on to an analytic study of large $N$ physics 
on the lattice.  Here we 
study the large $N$ glueball mass spectrum in 2+1 dimensions using the Kogut-Susskind 
Hamiltonian. Using the analytic techniques of
\chap{analytictechniques}, calculations with $N$ as large as 25 are
possible on a desktop computer, allowing an accurate extrapolation to
the $N\rightarrow \infty$ limit. \chaps{sunmassgaps}{largenphysics} contain the key results of this thesis.\\

In \chap{3+1dimensions} we move on to explore the viability of adapting
the analytic techniques, used with success in
\chaps{sunmassgaps}{largenphysics}, to the physically interesting case
of 3+1 dimensions. Here, the difficulties of a move to 3+1 dimensions
are discussed. Calculations of glueball masses within a simplistic,
one cube, model are presented as a first step in testing the worth of analytic variational techniques in 3+1 dimensions.\\

\chap{thesisconclusion} contains a summary of our findings and our conclusions.

\chapter{Gauge Theory on a Lattice}
\label{gaugetheoryonalattice}

\section{Introduction}

In this chapter we describe the basic steps involved in Hamiltonian
LGT calculations. Our intention is to introduce the fundamental 
approximations of LGT and the topics of importance to
recovering the real world physics of the continuum. We start with an explanation of how quarks and gluons are put on a lattice in
\sect{thelattice}. In this section the important difference in
the building blocks of LGT and continuum gauge theory are discussed. We
then move on to explain the steps involved in performing a
calculation in Hamiltonian LGT. We finish in
\sect{extractingcontinuumphysics} with a discussion of the
important topic of how the real world physics of the continuum is
extracted from a calculation on the lattice.   

\section{The Lattice}
\label{thelattice}

The fundamental simplification of LGT is the
replacement of continuous and infinite space-time with a finite and
discrete lattice of sites and links. A label, $l=(\bm{x},i)$, is
attached to each link. Here $\bm{x}$ is the starting point
of link $l$ and $i$  its direction. 
 Quark fields are essentially
treated in the same way on the lattice as they are in continuum QCD. In both
theories they are represented by Grassmann variables. However, 
in LGT the quark fields are restricted to lattice sites. As in the
continuum, the fermion fields can be integrated out of the partition function
exactly giving a highly non-local fermion determinant, $\det M$. The
calculation of this determinant is the key factor in the computational
expense of full QCD simulations in Lagrangian LGT. The
quenched approximation, in which the vacuum polarisation effects of the
QCD vacuum are explicitly removed, gives
$\det M$ a constant value, reducing both the computational expense and the
connection to real world physics of LGT. It is in this approximation that
the bulk of Lagrangian LGT calculations have been carried out. \\

\begin{figure}
\centering

\psfrag{x+ai}{$\bm{x}+a \hbm{i}$}
\psfrag{U}[][]{$U_i(\bm{x})$}
\psfrag{x}{$\bm{x}$}
\psfrag{i}{$i$}
\psfrag{j}{$j$}
\includegraphics[width=7.0cm]{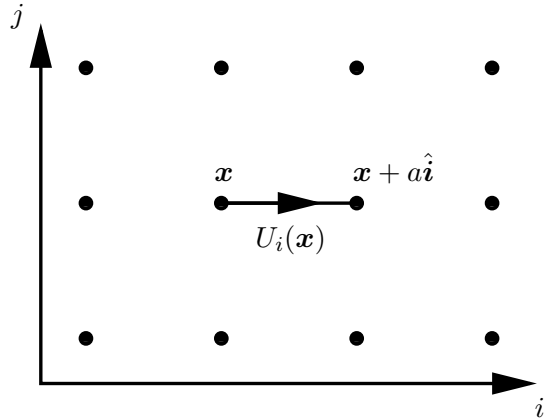}

\caption{The link operator.}
\label{link}
\end{figure}

\begin{figure}
\centering
\hspace{-1cm}
\subfigure[Continuum QCD] 
                     {
                         \label{notlattice}
                \psfrag{quark}{quark}
                \psfrag{gluon}{gluon}         
                \includegraphics[width=9.0cm]{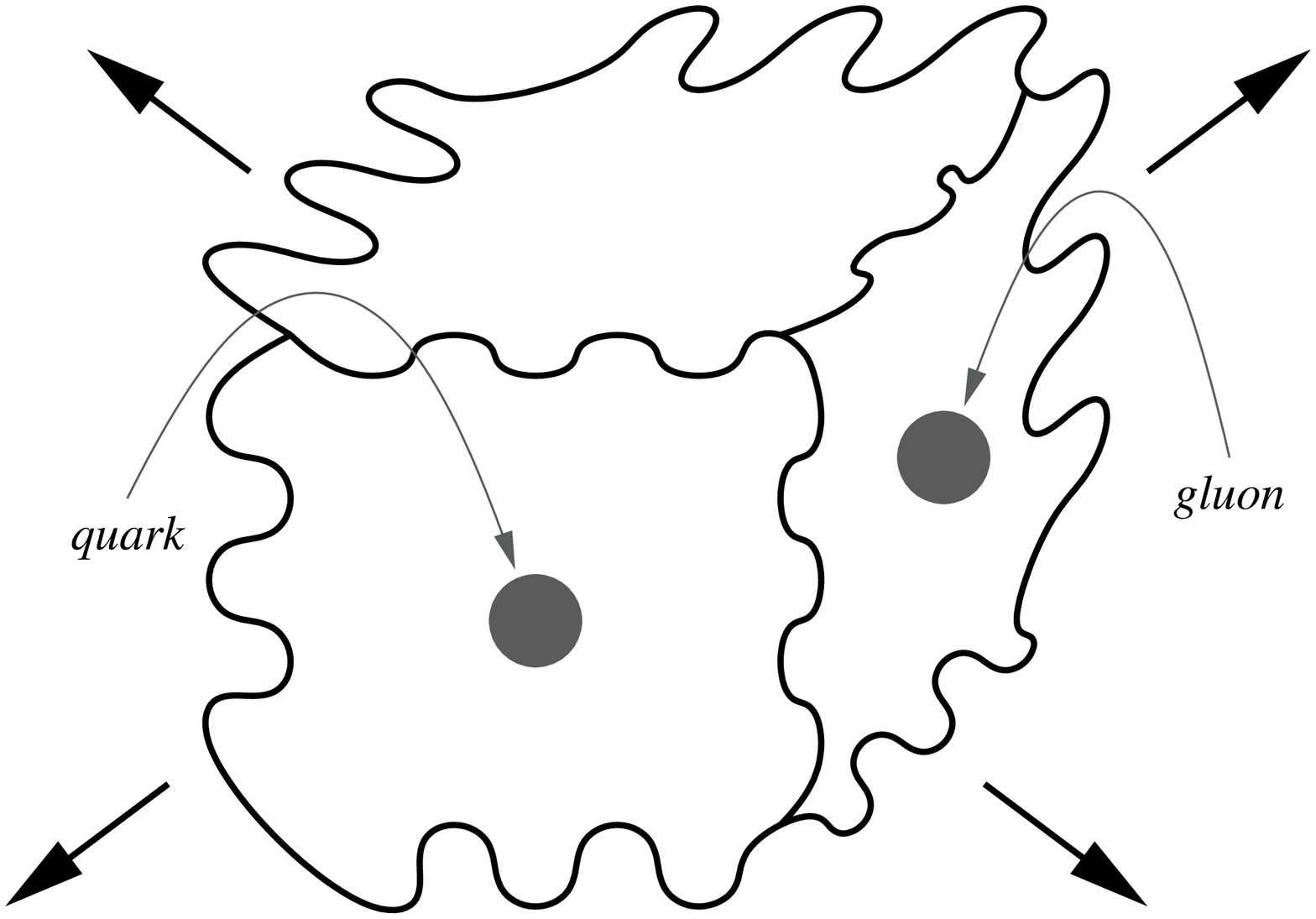}
                     }                   \hspace{-1cm}
 \subfigure[Lattice QCD] 
                     {
                         \label{lattice}
                        \psfrag{quark}{quark}
                        \psfrag{gluon}{gluon}
                        \psfrag{a}{$a$}
                        \psfrag{L}{$L$}       
                         \includegraphics[width=7.0cm]{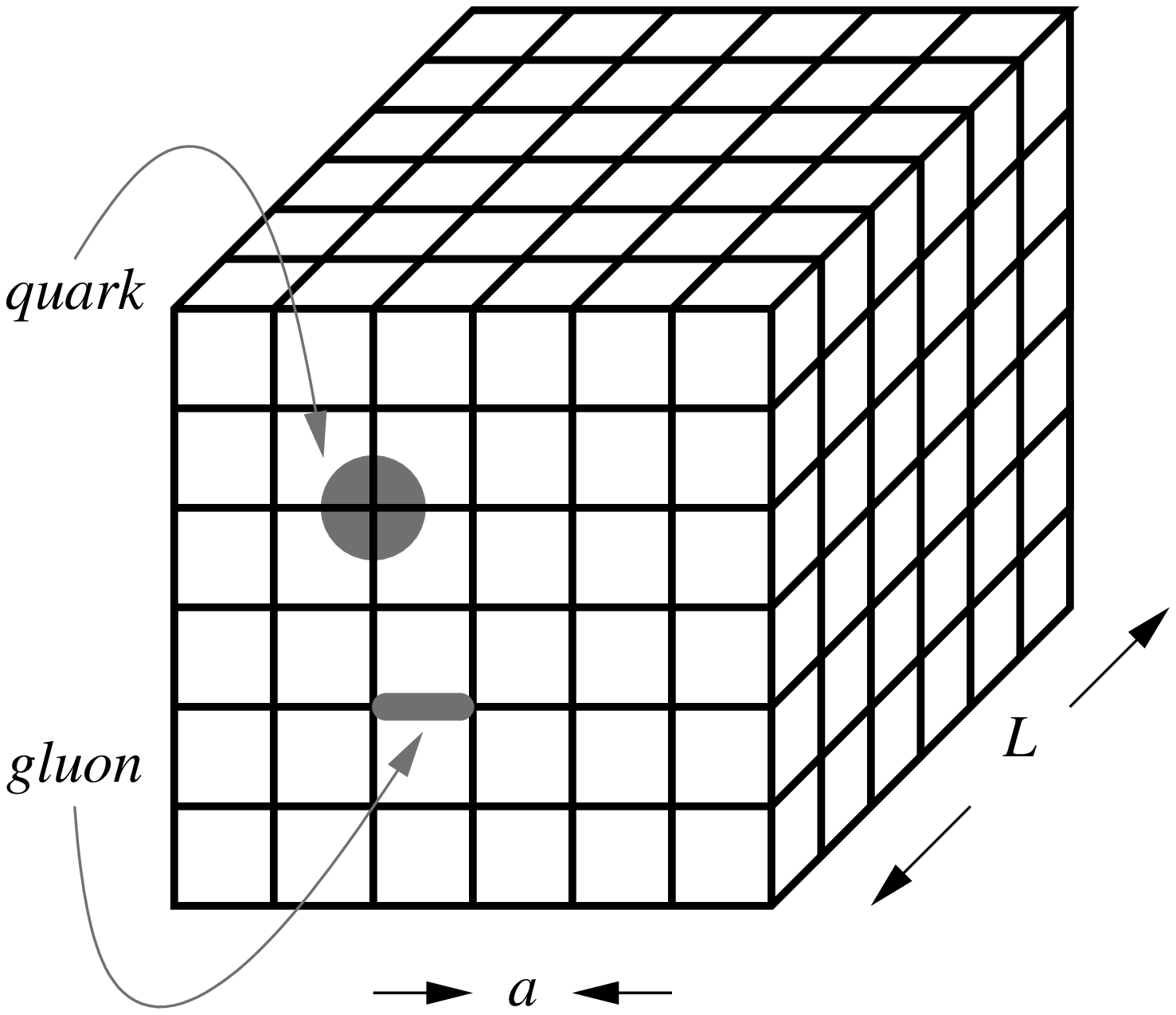}
                     }
\caption{A cartoon comparing continuum QCD with lattice QCD.}
\label{cartoon}
\end{figure}

In LGT gluons are incorporated differently to continuum gauge theory in order to maintain manifest gauge invariance on
the lattice. In continuum SU($N$) gauge theory gluons are represented
by the SU($N$) valued fields $A_\mu(x)$. In LGT gluons are represented by Wilson lines (or link
operators) along the links joining lattice sites. We denote the link
operator on link $l$ by $U_l$. The link operator on the same link but
taken in the opposite direction is given by $U^\dagger_l$. At times it will be convenient to use
the more detailed notation, $U_i(\bm{x})$, for the link operator on link 
$l=(\bm{x},i)$. It is conventional 
to represent the link operator by a directed line segment joining the
sites $\bm{x}$ and $\bm{x}+a\hbm{i}$, as shown in
\fig{link}. Here $\hbm{i}$ is a unit vector in the $i$ direction. The details of this construction and precisely how it
relates to the continuum theory will be introduced in
\sect{gaugeinvariance}. The fundamental differences between
continuum gauge theory and LGT are caricatured in \fig{cartoon}. As in
\fig{cartoon} it is conventional to denote the lattice spacing 
by $a$.  \\

Having discussed the fundamental approximation of LGT and briefly introduced
its building blocks we now move on to a description of the steps
involved in a complete Hamiltonian LGT calculation. The first step is
choosing an appropriate lattice.

\section{Defining a Lattice}
\label{definingalattice}
In Hamiltonian LGT one works on a
spatial lattice. The time coordinate remains continuous. Contrastingly, 
the Lagrangian
formulation discretises all dimensions and works in Euclidean
space. An obvious advantage of the Hamiltonian approach is that one
works with a lattice with one less dimension. The cartoon in
\fig{cartoon} illustrates a cubic lattice in three
dimensions. LGT 
is not, however, restricted to 3 dimensions and the lattice itself is
not restricted to a cubic arrangement.\\

 For reasons of simplicity, it has been conventional to
study LGTs built on cubic lattices. Not a great deal of effort 
has been devoted to
exploring the effect of using lattices with different
structures~\cite{WichmannPhD}. 
One could however imagine that working on a non-cubic lattice
could be advantageous in some situations. For example, if the degrees of freedom were
arranged into products of link operators (which is often convenient) around the elementary faces of the lattice
(for example squares of side length $a$ in the cubic case), working on
non-cubic lattices could be advantageous when the number of faces and number of links
need to be related in a precise way to incorporate constraints such as
Gauss' law~\cite{Ligterink:2000ug}. In this thesis we work on a $d$
dimensional spatial lattice with $d=2$ and 3.\\

When performing numerical LGT calculations the primary restriction
is computer memory. The value of every link operator on the lattice
needs to be stored in memory at a given time. Since
the number of links on the lattice increases quickly as the volume
increases, one is in practice limited to lattices with small
volumes. The majority of work in LGT to date has involved small volume
numerical simulations in the Lagrangian approach. Typical calculations on supercomputers with speeds of the
order of hundreds of gigaflops\footnote{1 flop = 1 floating point
operation per second} with two flavours of
dynamical quarks are performed on $24^3\times 48$ lattices. It has
been estimated that with access to a 10 teraflop machine $36^3\times
96$ lattices will be able to be used for equivalent calculations 
with similar effort~\cite{Jegerlehner:2000xt}. \\
 
Variational studies in 
Hamiltonian LGT in 3+1 dimensions (which go beyond the
strong coupling limit), to our knowledge, have
only dealt with small lattices; typically with sizes smaller than
$6^3$~\cite{Chin:1988at, Long:1988qe}. We discuss the viability of
using analytic Hamiltonian techniques in 3+1 dimensions 
in \chap{3+1dimensions}.\\

Having defined a lattice we now move on to the important step of 
constructing a Hamiltonian.

\section{Constructing a Hamiltonian}
\label{constructingahamiltonian}
When constructing a Hamiltonian on the lattice there are three
important points to consider. Firstly, the lattice Hamiltonian must reduce to
its continuum counterpart in the limit of vanishing lattice
spacing. From here on we refer to this limit as the continuum limit. 
Secondly, we must incorporate as many symmetries of the
continuum theory as possible. Finally, the lattice Hamiltonian must not 
be prohibitively expensive to use. 
These points are closely related and often an attempt to address one
point is detrimental to another. For example, if one attempts to
build a Hamiltonian that is closer to its continuum  counterpart for a
given lattice spacing, one with a weak dependence on $a$, one is left 
unavoidably with a more computationally intensive Hamiltonian. 
Another example relates to restoring rotational symmetry. 
On the lattice, continuous rotational
symmetry is broken down to a discrete analogue which depends on the
particular lattice in use. One can attempt to construct Hamiltonians with
improved rotational symmetry but again the resulting Hamiltonian is
more computationally expensive than the simplest versions. \\

One symmetry that is essential to incorporate in the lattice
Hamiltonian is local gauge invariance. One may ask why it is important to
uphold local gauge invariance when other symmetries such as rotational
invariance and Lorentz invariance are explicitly broken. The reason
lies in the fact that with local gauge invariance upheld the bare gluon mass is
zero and the couplings appearing at multiple gluon and gluon-quark
vertices are all equal. Without local gauge invariance these couplings
would need
to be tuned independently in order to retrieve real world
physics. This is a complication that would become prohibitively expensive
when processes with loop
diagrams were taking into consideration. \\

In building lattice Hamiltonians we have a great deal of
freedom. Our only restrictions are gauge invariance and the need to
obtain the correct continuum limit. 
Our approach is to start with gauge invariant lattice operators and
form linear combinations such that the correct continuum limit is
obtained. In this way Hamiltonians, which couple links with various
separations, can be easily constructed. This process is discussed in
detail in \sects{thekogutsusskindhamiltonian}{classicalimprovement}.\\

Having constructed a Hamiltonian the next step in a Hamiltonian LGT
calculation is to choose an appropriate wave function.

\section{Choosing a Wave Function}
\label{choosingawavefunction}

In this section we discuss the issues involved in choosing a
wave function. We start with a discussion of lattice symmetries and
finish by mentioning some practical concerns.\\

When performing a LGT calculation one is usually interested in studying a
particle with certain symmetry properties. For instance in this thesis
we study the masses of glueballs in 2+1 and 3+1
dimensions. These glueballs can be classified according to their total
angular momentum, $J$, charge conjugation, $C$, and parity, $P$ eigenvalues. 
In order to study these particles we need to 
construct suitable wave functions. The
restrictions on the wave function are firstly, that it be gauge invariant, 
and secondly, possess the symmetry properties of the continuum state we
wish to study. As we will see in \sect{gaugeinvariance-new}, the former forces us to construct our state from Wilson
loops, the latter requires us to choose particular combinations of 
Wilson loops with certain symmetry properties.\\

 Suppose for example we work with a Hamiltonian symmetric under parity
transformations\footnote{We use the notation, $\hat{X}$, to represent
an operator and, $X$ to represent an eigenvalue.}, $\hat{P}$, charge
conjugation, $\hat{C}$, and rotations by
$\pi/2$. Then the Hamiltonian commutes with $\hat{C}$, $\hat{P}$ and the lattice
spin operator $\hat{J}$. Consequently the eigenstates of the Hamiltonian can be labelled
by their respective $\hat{C}$, $\hat{P}$, and $\hat{J}$ eigenvalues. $\hat{C}$ has the
effect of replacing the lattice coupling, which we denote by $e$, by
$-e$. This is equivalent to the replacement $U_l \rightarrow U_l^\dagger
$. To define the parity operator and explore rotations on the lattice we
need to be more careful. We start with a description of parity.\\ 

With three spatial dimensions, the effect of $\hat{P}$ is
to replace lattice sites $\bm{x}$ by $-\bm{x}$. However in two spatial
dimensions, with one less axis to rotate about, the same prescription (of course with the $z$
component taken to be zero) is equivalent to a pure rotation. In 2+1 
dimensions we can define $\hat{P}$ as a reflection in one axis. For
example, in this thesis we will take
\bea
\hat{P} | \psi(x,y) \rangle = | \psi(x,-y) \rangle.
\eea
Next we discuss the more difficult topic of rotational symmetry.\\

The rotational symmetry properties of lattice operators are 
restricted to the particular rotational symmetries of the lattice. 
In the 3+1 dimensional continuum 
rotational symmetry is continuous and the spin of a
state is characterised by the irreducible representations of the
rotation group
SO(3). The situation on the lattice is quite different. On a cubic
lattice the SO(3) rotational symmetry of the continuum is broken down
to the cubic or octahedral group O. Since the lattice Hamiltonian is
symmetric under cubic rotations its eigenstates must be classified in
terms of the irreducible representations of O, not of SO(3). The
classification of lattice states in terms of the irreducible
representations of O was developed by Berg and
Billoire~\cite{Berg:1983kp}.
An analogous situation occurs in 2+1 dimensions. 
As we will see in \sect{classificationofstates} the result is an ambiguity in spin 
identification on the lattice. For instance on a 2 dimensional square 
lattice one can only determine the continuum spin of a state modulo 4 with
standard  techniques~\cite{Teper:1998te}. One can
determine the spins of the lowest mass states when using smeared
links within the Lagrangian approach~\cite{Teper:1998te}. This
approach, however, does not allow the spins of excited states to be
determined. Techniques are available to
counter this problem~\cite{Johnson:1998ev,Johnson:2002qt} but currently their use is at an exploratory stage.\\

In 3+1 dimensions the situation is
less problematic. Since there is an additional axis to rotate about, one can
count the multiplicity of states to help determine continuum
spins. Additional methods, specific to 3+1 dimensions, have been
developed to distinguish spin 0 and spin 4 states with a careful
choice of lattice operators~\cite{Liu:2001wq}.\\

A further consideration in the construction of appropriate wave
functions is translational invariance. On the lattice, states with
particular momenta are constructed in the usual way via Fourier
transforms. In this thesis we will be concerned with the calculation of
masses, for which translationally invariant zero momenta states
suffice.\\

The last issue to discuss is one of a practical nature.
A benefit of the Hamiltonian approach is
that analytic techniques are available for use. This is only the case if one makes use of wave functions which result in expectation values
for which analytic results are available. If the expectation values
are not able to be
handled analytically one must resort to Monte Carlo techniques for
their calculation~\cite{Long:1988qe}. Unsurprisingly wave functions which allow
analytic calculations have less appealing continuum limits than those
which must be handled numerically~\cite{Long:1988qe}. \\

Having discussed the issues involved in choosing a trial state we now
move on to discuss the important topic of how continuum physics is
extracted from a lattice calculation.

\section{Extracting Continuum Physics}
\label{extractingcontinuumphysics}

In this section we explain how continuum results are extracted from a
lattice calculation in 2+1 and 3+1 dimensions. This will serve as an
introduction to the form in which the results of later chapters are
presented. In the context of this thesis the primary difference
between 2+1 and 3+1 dimensions is the dimension of the coupling constant. As
we shall see in \sect{TheKineticHamiltonian} the coupling
constant, $e^2$, is dimensionless in 3+1 dimensions but has the units of
mass in 2+1 dimensions. We start with a general discussion of how
continuum results are extracted from a LGT calculation. We then
specialise to the cases of 3+1 and 2+1 dimensions. \\

When performing a calculation on the lattice one is always interested
in eventually extracting continuum physics. 
The process of unmasking continuum physics however is highly nontrivial. 
Naively one would expect continuum results to be
obtained simply in the limit as $a\rightarrow 0$. However, 
if this naive limit is
taken in a lattice calculation of a given physical quantity either zero or
infinity is obtained depending on the dimension of the
quantity under consideration. To achieve physical results one must
tune the coupling constant in such a way that a finite result is
obtained for {\it all} physical quantities. \\

To illustrate the process consider the calculation of a physical
quantity, $Q$, on the lattice with dimension in units of
the lattice spacing $d(Q)$. As an example $Q$ could be the mass
of a physical state, in which case $d(Q) = -1$. In Hamiltonian LGT one
typically calculates $Q$
for a number of couplings, arriving at
\bea
 Q = a^{d(Q)} f(e).
\label{scale}
\eea
Here $f$ is a dimensionless function which may depend on the coupling,
$e$, either
directly or via the parameters of the wave functions used in the
calculation of $Q$. It is clear from
\eqn{scale} that in the naive limit, $a\rightarrow 0$ , $Q$ approaches either
infinity or zero (depending on the sign of $d(Q)$). It is clear that for
a finite result one must let
$e$ vary with $a$ in such a way that a sensible result is obtained for
$Q$ as $a\rightarrow 0$. It is only in this limit that the real
world physics  of the continuum is recovered. 
The nontrivial aspect of this limit
is immediately clear. The dependence of $e$ on $a$ must be chosen so
that {\it all} physical observables have a well defined $a\rightarrow 0$
limit. The complexity lies in the fact that for each physical
observable $Q$ there is a different $f(e)$ in \eqn{scale}.  
The coupling must be tuned in such a way that
as it approaches its $a\rightarrow 0$ limit all
physical observables tend to constant values. \\

The lattice is a regularisation scheme which reduces the continuum
theory to a theory with a finite number of
degrees of freedom. The process of tuning the coupling constant described above is the renormalisation process
which precisely removes the dependence of the calculation on the
regularisation scheme.
Precisely how the coupling depends on $a$ is dictated by the
Callan-Symanzik $\beta$ function
\bea
\beta(e) = a\frac{d e}{da}.
\label{betadef}
\eea 
One can make specific predictions about the form of $\beta$ in 
weakly coupled QCD. Including up to two loop contributions it has been
shown~\cite{Caswell:1974gg,Jones:1974mm} that for 3+1 dimensional 
SU($N$) gauge theory with $n_f$ flavours of quarks, 
\bea
\frac{\beta(e)}{e} = \frac{e^2}{16 \pi^2}\left(\frac{11 N}{3}-\frac{2
n_f}{3}\right)
+ \left(\frac{e^2}{16 \pi^2}\right)^2 \left[\frac{34 N^2}{3}- \frac{10 N
n_f}{3}- \frac{n_f(N^2-1)}{N} \right] + \ord(e^6).
\label{betaperturbative}
\eea
A four loop
calculation of $\beta(e)$ is also available~\cite{vanRitbergen:1997va}. 
In this thesis we will always be concerned with the quarkless case 
$n_f=0$. Making use of \eqn{betaperturbative} one can integrate the
one loop part of 
\eqn{betadef} directly to obtain for the quarkless case
\bea
\Lambda_L = \frac{1}{a}\left(\frac{48 \pi^2}{11} \xi \right)^{51/121}
e^{-\frac{24 \pi^2}{11} \xi}, 
\label{latticescale}
\eea 
where $\xi$ is the inverse 't Hooft coupling $\xi = (N e^2)^{-1}$.
The integration constant, $\Lambda_L$, with dimensions of mass is commonly called the lattice
scale parameter. It can be related perturbatively, at the one~\cite{Dashen:1981vm,Hasenfratz:1981tw}
 and two loop level~\cite{Luscher:1995np}, to the scale parameters
obtained from other regularisation
schemes.
Such results are important when comparing results from LGT to results obtained with other regularisation schemes. It is important to
note that by inverting \eqn{latticescale} it is clear that we must have
$e\rightarrow 0$ as $a\rightarrow 0$.
Since they are different
regularisation schemes, the scale
parameters of Hamiltonian and Lagrangian LGT also differ. Results for their
ratios are available~\cite{Hasenfratz:1981tw,Hamer:1996ub} and are 
necessary for the comparison  of results
obtained in the two formulations in the continuum limit.\\

When attempting to extract continuum physics
from a Hamiltonian LGT calculation one generally looks for a range of 
couplings for 
which \eqn{latticescale} applies. If such a range of couplings,
called a scaling window, exists all observables with dimensions of
mass must scale in the same way as $\Lambda_L$; all masses must
be proportional to $\Lambda_L$ in a scaling window. The window does
not, in practice, extend all the way to $e=0$. Generally the
approximations used to perform the calculation break down before
then. One generally assumes that if the approximations
used in the calculations were improved the scaling window 
would extend closer to $e=0$.\\ 

To demonstrate how continuum physics is extracted in practice consider
the calculation of a mass on the lattice in pure SU($N$) gauge
theory. We calculate the mass, $M(e)$ for a range of couplings. To
observe asymptotic scaling we require 
\bea
a M(e) = M^\ast a \Lambda_L 
\label{asymptoticscaling}
\eea
for some range of
couplings. Here $M^\ast$ is the continuum limit mass in units of
$\Lambda_L$. Making use of \eqn{latticescale} we see that 
\bea
 \log( a M(e)\xi^{-51/121})  = \log( M^\ast) +
 \frac{51}{121}\log\left(\frac{48 \pi^2}{11}\right) -\frac{24
 \pi^2}{11} \xi .
\eea 
Thus in a scaling window a plot of  $\log( a M(e)\xi^{-51/121})$
versus $\xi $ yields a straight line with gradient $-24
 \pi^2 /11$. The observation of such behaviour allows the constant of
proportionality, $M^\ast$, to be extracted. We refer to the type of
scaling where all masses scale as \eqn{latticescale} as ``asymptotic
scaling''. If such behaviour is observed one can be confident that the
lattice spacing is small enough that continuum physics is being
revealed and an accurate value for the continuum limit mass is obtained.\\

A less general type of scaling can be seen in the ratios of masses. 
Since all masses scale as \eqn{asymptoticscaling} for small enough
lattice spacing, the ratios of masses become constant in a scaling
window. However, a ratio of masses may happen to be constant on a
much larger scaling window, one where the lattice scale parameter does not
obey \eqn{latticescale} for its entire range of couplings. In such a
region the non-zero errors, which depend on $a$
and $e$, cancel to some extent and the ratio reproduces the
correct continuum ratio. We refer to this more general situation as
``scaling'' which is not to be confused with asymptotic scaling.\\ 

The final step in obtaining a result in familiar units is to fix the
value of $\Lambda_L$ in physical units (MeV for example). This is most commonly done by
by comparing the string tension $\sqrt{\sigma}$ calculated on the
lattice in units of $\Lambda_L$ with the physical value in units of 
MeV obtained from the spectroscopy of heavy quarkonia. 
The string tension is the coefficient of the linearly rising potential
energy as a  quark-antiquark pair are separated. We introduce the string tension in \sect{staticqqbarpotential}. A commonly used
value for SU(3) in 3+1 dimensions 
is $\sqrt{\sigma}=440 \pm 38$ MeV~\cite{Teper:1997am}.\\

The above discussion treats the case of 3+1 dimensions. To renormalise
the results of an LGT calculation on a lattice with different numbers
of dimensions requires a
different treatment. For the case of 2+1 dimensions, with which we are
concerned for the bulk of this thesis, the situation is particularly simple.
For this case the coupling constant, $e^2$, has dimensions of mass, in contrast
to the 3+1 dimensional case for which it is dimensionless. The coupling
constant can therefore be used to explicitly set the scale of the lattice
theory. This has become standard practice in 2+1 dimensional LGT, both
in the Lagrangian and Hamiltonian approaches. 
Consequently, in the limit of vanishing lattice spacing all
masses calculated on the lattice must approach a constant value when
measured in units of the coupling,
\bea
M(e) = M^\ast e^2.
\eea 
Here $M^\ast $ is the continuum limit mass in units of $e^2$. 
In 2+1 dimensional LGT, the Hamiltonian and Lagrangian formulations
again 
present different regularisations of the continuum gauge theory. To 
compare the results of the two formulations one needs to acknowledge
that the coupling constants of the two formulations differ. Their
ratio can be calculated perturbatively~\cite{Hamer:1996ub} for small
couplings in pure SU($N$) LGT and is close to unity.

\section{Conclusion}

In this chapter we have introduced the basic approximations made in
LGT. We have discussed briefly the steps involved in a Hamiltonian LGT
calculation, introducing in particular the important steps of choosing
wave functions and extracting continuum physics. Our intention was to
provide a method-independent introduction to the process of a
Hamiltonian LGT calculation in order to introduce the notation in which the results of
later chapters are presented. In the next chapter
we consider in detail the process of constructing and improving
lattice 
Hamiltonians.

\chapter{Constructing and Improving Lattice Hamiltonians}
\label{constructingandimproving}

\section{Introduction}

In this chapter we develop a technique which allows the
straightforward construction of Hamiltonians
for gluons in $d+1$ dimensions. We start with a historical background
in \sect{historicalbackground}
to motivate the need for additional Hamiltonians beyond the original
Kogut-Susskind Hamiltonian. In \sect{thekogutsusskindhamiltonian} we derive the simplest
Hamiltonian for gluons on the lattice, the 
Kogut-Susskind Hamiltonian, focussing on the
errors occurring when it is compared to its continuum counterpart. These
errors can be divided into two classes, classical and quantum. The
origin of each class of error is discussed in
\sect{gaugeinvariance}. 
We then turn our attention to the removal of classical errors in
\sect{classicalimprovement}, deriving the simplest classically
improved lattice Hamiltonian
by adding appropriately weighted gauge invariant lattice operators. We
refer to this technique as direct improvement. In
 \sect{tadpoleimprovement} we discuss the topic of tadpole
improvement, a technique for correcting a particular type of quantum error, which has revolutionised Lagrangian LGT in the last 15
years. In particular we
discuss its implementation in the Hamiltonian formulation. We finish
the chapter in \sect{additionalimprovedhamiltonians} with 
a
derivation of an improved lattice Hamiltonian which couples distant lattice
sites. This highlights the ease with which improved Hamiltonians can
be constructed with the direct method of improvement.

\section{Historical Background}
\label{historicalbackground}
The Hamiltonian formulation of LGT was developed by Kogut and
Susskind~\cite{Kogut:1975ag}. Their Hamiltonian was derived by
demanding the correct continuum limit be obtained in the limit of
vanishing lattice spacing. This was the style in which the 
Wilson action was derived a year earlier. Creutz showed that the
Kogut-Susskind Hamiltonian could be derived from the Wilson action
using the transfer matrix method~\cite{Creutz:1984m}. Later Kogut
demonstrated that the same could be done by taking the continuous time
limit of the Wilson action and performing a canonical Legendre
transformation~\cite{Kogut:1983ds}.\\

When building a gauge theory on a lattice, the infinite number of degrees of
freedom of continuum theory is reduced to a large but manageable
number. The price to pay, however, is that errors are made. In order
to obtain an accurate extrapolation to continuum physics these errors
need careful attention.\\  


As discussed in \sect{historicalbackground-1} much work in the last fifteen years has been devoted to improving
lattice actions. 
The motivation for improvement in the action
formulation is computational cost. Since the cost of a Lagrangian LGT
calculation varies as $a^{-7.25}$~\cite{Jegerlehner:2000xt}, it is by far more efficient to build
improved theories than to work on finer lattices. The improvement
programme, in particular tadpole improvement, has allowed accurate 
calculations to be performed on
relatively coarse lattices on desktop computers and brought the
simulations full QCD within the reach of today's most powerful
computers.\\

In contrast, the improvement of lattice Hamiltonians has only recently
begun. With most computational techniques in the Hamiltonian
formulation, as one moves closer to the continuum limit calculations 
require the symbolic manipulation of an increasing number of
increasingly complicated lattice operators. The motivation for
improvement 
in the Hamiltonian approach is that when working at a given lattice
spacing (or coupling) one is closer to the continuum limit when using
an improved Hamiltonian. In this
way the effort required to reach the continuum limit is reduced,
 provided the improved Hamiltonian is not prohibitively difficult to
work with.   
Perhaps the most extensive treatment to date is due to Luo,
Guo, Kr{\"o}ger and Sch{\"u}tte who discussed the classical and
tadpole improvement of
Hamiltonian LGT for gluons~\cite{Luo:1998dx}. 
In their study it was
discovered that deriving an improved Hamiltonian from a classically
improved action, whether by transfer matrix or Legendre transformation
methods, results in a highly non-local kinetic term with an
infinite number of terms. To derive a kinetic Hamiltonian with only
nearest neighbour terms it was found necessary to start with an
improved action with an infinite number of terms, coupling arbitrarily
distant lattice sites. \\

With the technique of Luo et al the order $a^2$ errors are removed from the
Kogut-Susskind Hamiltonian. However,
generating Hamiltonians with further improvement would seem exceedingly
difficult. This is because one would need to start from a
L\"uscher-Weisz improved action with non-planar terms~\cite{Luscher:1985zq}.
For this reason we propose a move to the direct approach.
That is, to construct improved Hamiltonians directly by adding appropriate
gauge invariant terms and fixing their coefficients so that errors are
cancelled. \\

Tadpole improvement has been studied briefly in the Hamiltonian
approach~\cite{Luo:1998dx,Fang:2000vm}. 
However there has been some disagreement between
possible implementations. Other forms of quantum improvement have 
not yet been considered in detail in Hamiltonian LGT.\\     

In this chapter we derive the Kogut-Susskind Hamiltonian for gluons 
and develop a
technique for constructing improved Hamiltonians directly. This
approach is in the spirit of the original Kogut-Susskind derivation and
parallels the strategies used for constructing improved actions. For
simplicity we will not consider quarks at this stage.\\

\section{The Kogut-Susskind Hamiltonian}
\label{thekogutsusskindhamiltonian}

\subsection{Preliminaries}

In this section we derive the Kogut-Susskind Hamiltonian for pure
SU($N$) gauge theory in $d$+1 dimensions. We maintain a
continuous time coordinate and work on a $d$ dimensional spatial lattice.\\

We start with the formal
continuum Hamiltonian operator expressed in terms of the
chromoelectric and chromomagnetic fields:
\bea
H = \int d^d x \Tr \left( \bm{E}^2 + \bm{B}^2 \right).   
\label{continuumham}
\eea
The space-like components of the continuum chromomagnetic field are given in
terms of the space-like components of the field strength tensor, $F_{ij}$,  by
\bea
B_i = \frac{1}{2} \varepsilon_{ijk} F_{jk},
\label{Bdef}
\eea
where $i=1,2,\ldots,d$ is the Dirac index labelling the space-like
directions and  summation over repeated indices is understood. $\varepsilon_{i_1\ldots i_n}$ is the totally antisymmetric
Levi-Civita tensor defined to be 1 if $\{i_1,\ldots,i_n\}$ is an even
permutation of $\{1,2,\ldots,n\}$, $-1$ if it is an odd permutation
and 0 otherwise (i.e.~if an index is repeated). The field strength tensor is given in terms of the
continuum gluon field, $A_\mu$, by 
\bea
F_{\mu \nu} = \pa{\mu} A_\nu - \pa{\nu} A_\mu -i e \left[A_\mu,A_\nu \right],
\eea
where $e$ is the QCD coupling.

The chromoelectric and chromomagnetic fields can each be expressed in
terms of their colour components as follows,
\bea
E_i = \lambda^a E_i^a \qquad B_i   = \lambda^a B_i^a.
\label{Edef}
\eea
Here $\{\lambda^a : 1\le a\le N^2-1\}$ is a basis for SU($N$). It is
common to use the Gell-Mann basis, in which case $\lambda^a = G^a/2$,
where $\{G^a : 1\le a\le N^2-1\}$ is the set of traceless $N\times N$ Gell-Mann
matrices. It is conventional to
normalise the basis as follows,
\bea
\Tr \left(\lambda^a \lambda^b\right) &=& \frac{1}{2} \delta_{a b}, 
\label{lambdanormalisation}
\eea
so that \eqn{Edef} can be inverted to write the colour components of
the chromoelectric and chromomagnetic fields as,
\bea
E^a_i = 2 \Tr \left(E_i \lambda^a\right) \quad {\rm and}\quad  B^a_i = 2 \Tr \left(B_i
\lambda^a\right) .
\eea
A final relation that is of use in this chapter is the commutation
relation,
\bea
\left[ \lambda^a,\lambda^b \right] &=& if^{abc}\lambda^c,
\label{structureconsts}
\eea
where the $f^{abc}$ are the totally antisymmetric, real structure
constants for SU($N$).

\subsection{Errors in Lattice Gauge Theory}
\label{gaugeinvariance}

Before discussing Hamiltonian improvement we must first understand
how deviations between LGT and its continuum
counterpart arise. The deviations can be separated
into two classes, classical and quantum errors, which will be described
in what follows. We start with classical errors.\\

Rather than being
constructed from gluon fields, pure gauge theory on
the lattice is built from link operators 
\bea
U_l = e^{i e a \LA_l}.
\label{linkoperator} 
\eea
Here $e$ is the dimensionful QCD coupling, $l$ labels the link in question and
${\cal A}_l$ is the lattice gluon field on the link $l$. In what
follows we will 
adopt the convention of writing the lattice version of a continuum
quantity in a calligraphic typeface.
We define the lattice gluon field on link $l$ 
to be the average continuum gluon field $A_\mu$ along link $l$,
\bea
\LA_l = \frac{1}{a} \int_l d\bm{x}\cdot \bm{A}.
\label{Adef}
\eea  
The definitions of \eqns{linkoperator}{Adef} lead to the diagrammatic
representation of the link operator shown in \fig{link}. \\

In practice the lattice gluon field is defined at only 
one point on or nearby a link. This leads to
interpolation errors in the integral in \eqn{Adef}. For example,
by choosing to evaluate the gluon field at $\bm{x}$, the midpoint of
the link, we have,
\bea
\LA_{l} &=& \frac{1}{a} \int_{-a/2}^{a/2} dt A_i(\bm{x}+t\hbm{i}),
\eea
where $\hbm{i}$ is a unit vector in the direction of link
$l$. The interpolation errors can be obtained by expanding in a Taylor
series as follows:
\bea
\LA_{l} &=& \frac{1}{a} \int_{-a/2}^{a/2} dt \left[ A_i(\bm{x}) + t \pa{i}
A_i(\bm{x}) + \frac{1}{2} t^2 \pa{i}^2
A_i(\bm{x})+ \ord(a^3)\right] \nn\\
&=& A_i(\bm{x}) + \frac{a^2}{24} \pa{i}^2 A_i(\bm{x}) +\frac{a^4}{1920} \pa{i}^2 A_i(\bm{x}) \ldots
\label{LAexpand}
\eea
We see that the lattice gluon field reduces to its continuum
counterpart in the $a\rightarrow 0$ limit, but that  they 
differ by interpolation 
errors of order $a^{2}$. From \eqn{LAexpand} we build the sequence of approximations to the lattice gluon field: 
\bea
\LA^{(0)}_i(\bm{x}) &=& A_i(\bm{x}) \nn\\
\LA^{(1)}_i(\bm{x}) &=& A_i(\bm{x}) + \frac{1}{24}a^2 \partial^2_i A_i(\bm{x}) 
\label{gapprox}\\
\LA^{(2)}_i(\bm{x}) &=& A_i(\bm{x}) + \frac{1}{24}a^2 \partial^2_i A_i(\bm{x}) +
\frac{1}{1920}a^4 \partial^4_i A_i(\bm{x})\nn\\
             &\vdots& \nn
\eea

Having discussed classical errors we now move on to the more difficult
topic of quantum errors in LGT. 
Quantum errors arise in LGT in two different contexts. Firstly, the
lattice acts as an ultraviolet regulator 
allowing the simulation
of only those 
states with momenta less than $\pi/a$. The absence of high
momentum states results in a deviation between the lattice and
continuum theories. Secondly, non-physical
interactions arise due to the use of the link operator in constructing
the lattice theory. 
To demonstrate this we expand the link operator in powers of $e$, 
\be
 U_\mu(x) = 1 + i e a A_\mu(x) -\frac{e^2 a^2}{2!} A_\mu(x)  A_\mu(x) + \dots, 
\ee 
and note that the interaction of {\em any} number of gluons is
allowed. Naively, the unphysical interactions are suppressed by
powers of $a$. However, when contracted, products of pairs of gluon fields
produce ultraviolet divergences ($\propto 1/a^2$) which {\em exactly}
cancel the $a$ dependence of the expansion. These terms can be
uncomfortably large and result in what are known as tadpole errors.\\

In the last decade the improvement programme has led to a good
understanding of both classical and quantum errors in quark and
gluon actions
 (See \rcite{Lepage:1996jw} and references within). 
In contrast, only the lowest order
classical errors have been corrected in the Kogut-Susskind
Hamiltonian. Conjectures have been made about the structure of a
quantum improved Hamiltonian~\cite{Luo:1998dx}, 
but a perturbative study, which would be needed to fix its precise
form,  has not yet been
carried out.\\

We describe how tadpole improved lattice Hamiltonians can be constructed 
in \sect{tadpoleimprovement}.

\subsection{Gauge Invariance on the Lattice}
\label{gaugeinvariance-new}
Under a local gauge transformation, $\Lambda(\bm{x})$, 
the link operator, $U_i(\bm{x})$, transforms as follows,
\bea
U_i(\bm{x}) \rightarrow  \Lambda(\bm{x})  U_i(\bm{x})\Lambda^\dagger(\bm{x}+a\hbm{i}).
\label{gaugetransform}
\eea 
Consider the closed path, $C$, on the lattice consisting of links $(\bm{x}_1,i_1)$,
$(\bm{x}_1+a \hbm{i}_1,i_2),\ldots,(\bm{x}_1-a \hbm{i}_m,i_m)$. The
ordered product of link operators along $C$ is given by
\bea
P^C_{i}(\bm{x}) &=& U_{i_1}(\bm{x}_1)U_{i_2}(\bm{x}_1 +a \hbm{i}_2) \ldots U_{i_m}(\bm{x}_1-a \hbm{i}_m).
\eea 
Making use of \eqn{gaugetransform}, under a local gauge transformation $\Gamma_i(\bm{x})$ we have
\bea
P^C_{i}(\bm{x}) &\rightarrow&   U_{i_1}(\bm{x}_1) P^C_{i}(\bm{x}) U^\dagger_{i_1}(\bm{x}_1).
\eea
Since the trace of a matrix does not depend on the choice of
representation (i.e.~$\Tr (A U A^\dagger) = \Tr U $), by taking the
trace of $P^C_{i}(\bm{x})$ we obtain a gauge invariant operator. The
traces of closed loops are known as Wilson loops and are the building
blocks of LGT. The simplest Wilson loop on the
lattice is called the plaquette. It is the trace of the ordered
product of link operators around a square of side length $a$. Using
different combinations of Wilson loops one can construct lattice
theories which have different approaches to the continuum limit.

\subsection{Commutation Relations}

In continuum gauge theory, when employing canonical quantisation the
following equal time commutation relations are postulated,
\bea
\left[E_i^a(\bm{x}),A_j^b(\bm{y})\right] &=& i \delta_{ij}\delta_{ab}
\delta(\bm{x} - \bm{y}) \nn\\
\left[A_i^a(\bm{x}),A_j^b(\bm{y})\right] &=&
\left[E_i^a(\bm{x}),E_j^b(\bm{y})\right] = 0.
\label{contcom}
\eea
When constructing a lattice Hamiltonian the building blocks should be
related in such a way that the correct continuum commutation
relations are restored in the limit of zero lattice spacing. 
To construct the corresponding commutation relations on the
lattice we need to use the
lattice chromoelectric field, $\LE$, and the link operator, $U_l$. While
we can determine the continuum behavior of the link operator with a
simple Taylor expansion about $a$, the
requirement that the correct continuum commutation relations be
restored as $a\rightarrow 0$ defines the relationship between the
lattice  and continuum chromoelectric fields. \\

Let us now consider the commutation relations between the lattice
chromoelectric field and the link operator,
\bea
\left[ \LE_l^a, U_{m} \right] &=& \left[ \LE_l^a, e^{iga \LA_m} \right].
\eea
Making use of the Campbell-Baker-Hausdorff formula for $N\times N$
matrices $A$ and $B$,
\bea
e^{-A} B e^{A} = B+ \left[B,A\right]+\frac{1}{2!}\left[\left[B,A\right],A\right]+\frac{1}{3!}\left[\left[\left[B,A\right],A\right],A\right]+\dots,
\label{cbh}
\eea
 and \eqn{gapprox} gives to lowest order in $a$
\bea
\left[ \LE_l^{(0)}{}^a, U_{m} \right] &=&iea \left[ \LE_l^{(0)}{}^a,\LA_m^{(0)} \right]U_m
 \nn\\ 
&=& iea \left[ \LE_l^{(0)}{}^a,A_j^b(\bm{y}) \right]\lambda^b U_m + \ord(a^2).
\label{ecoma}
\eea
Here $m$ is the link $(\bm{y},j)$.
In order to produce the correct continuum limit commutation relations, 
the lattice electric field  must be 
proportional its continuum counterpart to lowest order in $a$;
\bea
\LE_l^{(0)}{}^a = \alpha \left[E^a_i(\bm{x}) + \ord(a^2)\right].
\label{LEdef1}
\eea
Here $l$ is the link $(\bm{x},i)$ and $\alpha$ is a constant 
which we now determine. Substituting
\eqn{LEdef1} in \eqn{ecoma} and using the continuum commutation
relations \eqn{contcom} gives
\bea
\left[ \LE_l^{(0)}{}^a, U_{m} \right] &=& iea \alpha \left[
E_i^a(\bm{x}),A_j^b(\bm{y}) \right]\lambda^b U_m + \ord(a^2) \nn\\
&=&   - e a \alpha \delta_{ij} \delta(\bm{x}-\bm{y})\lambda^b U_m +
\ord(a^2) \nn\\ 
&=&  - \alpha e a^{1-d} \delta_{ij} \delta_{ab}\delta_{\bms{x}\bms{y}}
\lambda^b U_m + \ord(a^2).
\eea   
In the last line we have related the discrete and continuous
delta functions by $\delta_{\bms{x}\bms{y}}
\approx a^d \delta(\bm{x}-\bm{y})$ in the small $a$ limit. We set
$\alpha = -a^{d-1}/e$ so that the
commutation relations between the lattice chromoelectric field and the
link operator do not depend on either the coupling or the lattice
spacing. This leads to, 
\bea
\LE_l^{(0)}{}^a  = -\frac{a^{d-1}}{e}\left[ E_i^a(\bm{x}) + \ord(a^2)\right],
\label{Edef2}
\eea  
and the conventional commutation relation,
\bea
\left[\LE_l^{(0)}{}^a ,U_m\right]  = \delta_{lm} \lambda^a U_l+ \ord(a^2).
\label{comeu}
\eea
Proceeding exactly as above, or simply by pre and post-multiplying
both sides of \eqn{comeu} by $U_m^\dagger$, we can show that the
analogous commutation relation for $U_m^\dagger$ is given by
\bea
\left[\LE_l^{(0)}{}^a ,U_m^\dagger\right]  = -\delta_{lm} U_l^\dagger \lambda^a + \ord(a^2).
\label{comeudag}
\eea

\subsection{The Kinetic Hamiltonian}
\label{TheKineticHamiltonian}
Using \eqn{Edef2} we see that the kinetic part of a lattice
Hamiltonian with order $a^2$ errors can be defined as follows
\bea
\LK^{(0)} &=& \frac{e^2 a^{3-d}}{2a} \sum_{l} \LE_l^a \LE_l^a .
\label{3.23}
\eea
 By comparison with the kinetic part of the continuum Hamiltonian of
 \eqn{continuumham}, it is clear that \eqn{3.23} has the correct form
 in $a\rightarrow 0$ limit. To show this we substitute
 \eqn{Edef2} in \eqn{3.23} which leads to, 
\bea
\LK^{(0)} &=& \frac{e^2 a^{3-d}}{2a} \sum_{l} \LE_l^a \LE_l^a  = \frac{ a^d}{2}\sum_{\bms{x},i} \left[ 
E_i^a(\bm{x}) E_i^a(\bm{x})  +\ord(a^2) \right] \nn\\
&\approx & \int d^d \bm{x} \Tr \left(\bm{E}^2 \right) +\ord(a^2).
\eea
Here, in the small $a$ limit, we have replaced the sum over $\bm{x}$
by a $d$ dimensional integral. 
This is the correct continuum limit kinetic Hamiltonian up to order
$a^2$ corrections.
Since the Hamiltonian has dimensions of mass, from \eqn{3.23}
we see that in $d+1$ dimensions the coupling constant, $e^2$, has units
of $({\rm length})^{d-3}$. Thus, in 2+1 dimensions the coupling constant has
dimensions of mass and in 3+1 dimensions it is dimensionless. 
We can define a dimensionless coupling by 
\bea
g^2 = e^2 a^{3-d}.
\label{couplings-relation}
\eea
In terms of the dimensionless coupling the kinetic
Hamiltonian is 
\bea
\LK^{(0)} &=& \frac{g^2}{2a}  \sum_{l} \LE_l^a \LE_l^a .
\label{kindef}
\eea

\subsection{The Potential Term}
\label{pottermsect}
We have seen in \sect{gaugeinvariance-new} 
that gauge invariant operators on the
lattice are built from the traces of products of link operators around
closed loops. The simplest loop, the square of side length $a$ is
called the plaquette. Consider a plaquette which joins the lattice
sites $\bm{x}$, $\bm{x} + a \hbm{i}$,  $\bm{x} + a (\hbm{i}+\hbm{j})$
and  $\bm{x} + a \hbm{j}$. We define the plaquette operator on such a
plaquette by
\bea
P_{ij}(\bm{x}) = 1 - \frac{1}{N} \Real \Tr \left[ U_i(\bm{x})
U_j(\bm{x}+a\hbm{i}) U^\dagger_i(\bm{x}+a \hbm{j}) U^\dagger_j(\bm{x})\right].
\eea
We will often use the equivalent pictorial expression to
condense notation:
\bea
P_{ij}(\bm{x}) =  1 - \frac{1}{N} \Real \plaquette. 
\label{plaqop-chap2}
\eea
The directed square denotes the traced ordered product of link 
operators, starting with $U_i(\bm{x})$, around an elementary square, or plaquette, of the lattice,
\bea
\plaquette := \Tr \left[U_i(\bm{x})U_j(\bm{x}+\bm{i}a)
U^\dagger_i(\bm{x}+\bm{j}a) U^\dagger_j(\bm{x})\right].
\eea
The reason for choosing the particular form of \eqn{plaqop-chap2} 
for the plaquette
operator, $P_{ij}$, becomes obvious once we expand it in powers of $a$.  Making
use of the definition of the link operator (\eqn{linkoperator}) and Stokes'
theorem we have,
\bea
P_{ij}(\bf{x}) &=& 1 - \frac{1}{N}\Real \Tr \exp\left( i e \oint_\square
d\bm{x}\cdot \bm{A} \right)
\nn\\
&=&
1 - \frac{1}{N}\Real \Tr \exp\left[ i e \int_{-a/2}^{a/2}ds \int_{-a/2}^{a/2}
dt F_{ij}(\bm{x} + s\hbm{i}+ t\hbm{j}) + \ord(A^2)\right].
\eea
Expanding the exponential in the above and expanding $F_{ij}$ in a
Taylor series in $a$ leads to 
\bea
P_{ij}(\bf{x}) &=& \frac{e^2 a^4}{2N} \Tr \left[ F_{ij}(\bm{x})^2
\right] + \frac{e^2 a^6}{24 N} \Tr \left[ F_{ij}(\bm{x}) (D_i^2 +
D_j^2) F_{ij}(\bm{x})\right] + \ord(a^8).
\label{plaqexp}
\eea
Here $D_i = \partial_i - i e A_i$ is the gauge covariant derivative.
The potential term of a lattice Hamiltonian with order $a^2$
corrections is thus given by
\bea
\LV^{(0)} &=& \frac{2 N}{e^2 a} \frac{1}{a^{3-d}}  \sum_{\bms{x},i<j}
P_{ij}(\bm{x}) =  \frac{2 N}{g^2 a} \sum_{\bms{x},i<j}
P_{ij}(\bm{x}) . 
\label{potdef}
\eea 
To see this we make use of \eqn{plaqexp} to give
\bea
\LV^{(0)} &=&  a^{d}  \sum_{\bms{x},i<j} \Tr \left[F_{ij}(\bm{x})^2 \right] + \ord(a^2) \nn\\
&\approx &   \sum_{i<j} \int d^d \bm{x} \Tr \left[F_{ij}(\bm{x})^2
\right] + \ord(a^2) \nn\\
&=& \int d^d \bm{x} \Tr \left( \bm{B}^2
\right) + \ord(a^2).
\eea
The last line follows simply from the definition of the chromomagnetic
field in \eqn{Bdef}.\\

Combining the kinetic and potential parts (\eqns{kindef}{potdef}
respectively) leads to the Kogut-Susskind Hamiltonian,
\bea
\LH^{(0)} &=&  \LK^{(0)} +  \LV^{(0)} \nn\\
          &=&  \frac{g^2}{2a}  \sum_{l} \LE_l^a \LE_l^a +
          \frac{2N}{g^2 a}  \sum_{\bms{x},i<j}
P_{ij}(\bm{x}), 
\label{KSdef}
\eea
which has the correct continuum form up to order $a^2$ corrections.

\section{Classically Improved Hamiltonians}
\label{classicalimprovement}

\subsection{Introduction}

In this section we derive a classically improved Hamiltonian
directly by adding additional gauge invariant 
terms and fixing their coefficients in
order to cancel errors. As a first step we aim to correct the
classical order $a^2$ errors arising in the Kogut-Susskind Hamiltonian
of \eqn{KSdef}. We treat the kinetic and potential terms of the
Hamiltonian separately. We start with the potential term.

\subsection{Improving the Potential Term}

 To improve the potential term of the Kogut-Susskind Hamiltonian, we
follow the process of improving the Wilson action. We introduce the rectangle
operator  $R_{ij}(\bm{x})$ in the $i$-$j$ plane (with 
the long side in the $i$ direction),
\bea
R_{ij}(\bm{x}) = 1- \frac{1}{N} \Real\rectangleA .
\eea
Here the Wilson loop joins the lattice sites $\bm{x}$,
$\bm{x}+2a\hbm{i}$, $\bm{x}+a(2\hbm{i} +\hbm{j})$ and
$\bm{x}+a\hbm{j}$. Following the procedure of the previous section
we can expand the rectangle operator in powers of the lattice spacing
as follows
\bea
R_{ij}(\bm{x}) =  \frac{4e^2a^4}{2N}\Tr \left[F_{ij}(\bm{x})^2\right] +
\frac{4e^2a^6}{24N}\Tr \left[
F_{ij}(\bm{x})(4D_i^2+D_j^2)F_{ij}(\bm{x})\right]+\ord(
a^8).
\label{rectexp}
\eea
We see that the expansion of $R_{ij}(\bm{x})$ gives precisely the same
terms as the expansion of $P_{ij}(\bm{x})$ in \eqn{plaqexp}, but
weighted differently. This allows us to take a linear combination of 
$R_{ij}(\bm{x})$ and $P_{ij}(\bm{x})$ and fix the coefficients so
that the order $a^2$ errors cancel. We write the improved potential
term as the linear combination
\bea
\LV^{(1)} = \frac{2N}{a g^2}\sum_{\bms{x},i<j} \left\{X P_{ij}(\bm{x}) +
\frac{Y}{2}\left[R_{ij}(\bm{x})+R_{ji}(\bm{x})\right]\right\} ,\label{v1def}
\eea
Substituting \eqns{plaqexp}{rectexp} into \eqn{v1def} we have
\bea
\LV^{(1)} &=& a^d \sum_{\bms{x},i<j} \Big\{\left(X+4Y\right)\Tr \left[ F_{ij}(\bm{x})^2
\right]  \nn\\
&&  + \frac{a^2}{12}\left(X+ 10 Y\right) \Tr \left[ F_{ij}(\bm{x}) (D_i^2
+D_j^2) F_{ij}(\bm{x})\right] \Big\}+ \ord(a^4).
\eea
In order to obtain the correct continuum limit we must have $X +4Y =
1$ and to cancel the order $a^2$ error we require $X+10 Y = 0$.  Solving
simultaneously gives $X = 5/3$ and $Y=-1/6$. 
This leads to the improved potential term
\bea
\LV^{(1)} = \frac{2N}{ag^2}\sum_{\bms{x},i<j} \left\{\frac{5}{3}
P_{ij}(\bm{x})
-\frac{1}{12}\left[R_{ij}(\bm{x})+R_{ji}(\bm{x})\right]\right\}. \label{v1def1}
\eea
In principle, the next lowest order classical errors
 could be corrected by
including additional, more complicated Wilson loops in the potential
term. This has not been done because many additional diagrams are
required to cancel the large number of order $a^4$ error terms. 
Since these errors are overwhelmed by order $a^2 g^2$ quantum
errors in the Lagrangian approach, addressing quantum corrections in the
Hamiltonian approach would seem to be of more immediate importance.

\subsection{Improving the Kinetic Term}
 
Constructing an improved  kinetic Hamiltonian with a finite number of terms has
proven to be a nontrivial exercise. Luo, Guo, Kr\"oger and Sch\"utte
demonstrated an interesting trade off when using either the transfer
matrix or Legendre transformation methods to derive an improved
Hamiltonian~\cite{Luo:1998dx}. Both techniques
require the starting point to be an improved action. 
When one starts from an improved action incorporating rectangular
terms the resulting Hamiltonian has infinitely many terms and couples links 
which are arbitrarily far apart. To produce a Hamiltonian which
couples only nearest neighbour links, it was found necessary to start
from a carefully constructed highly non-local improved action.\\

Here we demonstrate an alternative approach, similar in nature to the
classical improvement of the potential term in the previous section. 
One only needs
to include additional gauge invariant terms with appropriate continuum
behaviour in the kinetic Hamiltonian.
The coefficients of the additional terms are chosen 
so that the order $a^2$ errors vanish. \\

Perhaps the most important property of the electric field is that it
generates group transformations. Mathematically, this translates
to the lattice electric fields and link operators satisfying the commutation relations
given by \eqns{comeu}{comeudag}.
It is desirable for these to hold on the lattice for any degree of
approximation. Let us consider what happens to these commutation
relations for the approximation labelled by the
superscript (1) in \eqn{gapprox}. Making use of the 
Campbell-Baker-Hausdorff formula  (\eqn{cbh}) we have,
\bea
[\LE^{(1)a}_i(\bm{x}), U_j(\bm{y})] &=&
iea [\LE^{(1)a}_i(\bm{x}),
\LA^{(1)}_j(\bm{y})] U_j(\bm{y}) \nn\\
&=& iea [\LE^{(1)a}_i(\bm{x}),
A^{b}_j(\bm{y})+  \frac{a^2}{24}\partial_j^2 A^b_j(\bm{y})]
\lambda^b U_j(\bm{y}). 
\eea
We observe that if
the lattice electric field is taken to be related to the continuum electric
field via \eqn{Edef2}, an order $a^2$ error arises in the commutation
relation. To cancel this error we set
\bea
\LE^{(1)\alpha}_i(\bm{x}) = -\frac{a^{d-1}}{e}\left[E^\alpha_i(\bm{x}) - \frac{a^2}{24}\partial_i^2
E^\alpha_i(\bm{x})\right]. \label{e1}
\eea
We can take this to order $a^4$ by setting
\bea
\LE^{(2)\alpha}_i(\bm{x}) =  -\frac{a^{d-1}}{e}\left[ E^\alpha_i(\bm{x}) - \frac{a^2}{24}\partial_i^2
E^\alpha_i(\bm{x}) 
+ \frac{7a^4}{5760}\partial_i^4 E^\alpha_i(\bm{x})\right]. \label{e2}
\eea
In this way a sequence of approximations to the {\em lattice}
electric field in terms of the continuum electric field can be constructed.\\

Making use of these approximations we can analyse the   
classical errors arising in the kinetic Hamiltonian. 
To cancel these errors we take the approach of adding new terms and
fixing their coefficients in order to cancel the order $a^2$ error. 
We have a great deal of freedom in choosing additional terms. They are
restricted only by gauge invariance and the need for an appropriate continuum 
limit.\\

To understand the construction of gauge invariant kinetic terms
involving the lattice electric field, it is
important to recall that the lattice electric field transforms as follows under a local gauge transformation, $\Lambda(\bm{x})$:  
\bea
\LE_i(\bm{x}) &\rightarrow & \Lambda(\bm{x}) \LE_i(\bm{x})
\Lambda^\dagger(\bm{x}).
\eea
The response of the link operator to a gauge transformation is given
in \eqn{gaugetransform}.
Consequently, the next most
complicated gauge invariant kinetic term we can construct (after $\Tr
\left[\LE_i(\bm{x})^2\right] $) couples nearest neighbour electric fields:
\bea
\Tr \left[\LE_i(\bm{x}) U_i(\bm{x}) \LE_i(\bm{x}+a\hbm{i}) U_i^\dagger(\bm{x})\right].
\eea 
More complicated gauge invariant terms are easily constructed. One
only needs to couple electric fields on different links anywhere
around a closed loop. Consequently, generating Hamiltonians with
higher degrees of improvement
would be seem to be more readily achieved within this approach.

\subsection{Continuum Limits}
\label{chap2-contlimits}
In this section we examine the small $a$ limit of gauge invariant
lattice operators involving electric fields for possible use in the
kinetic term of the Hamiltonian.\\

We start with the simplest possible gauge invariant operator
containing lattice electric fields, the kinetic term of the
Kogut-Susskind Hamiltonian,
\bea
\frac{2 a}{e^2 a^{3-d}}\LK^{(0)}&=& \sum_{\bms{x},i} \Tr \left[ \LE_i(\bm{x})\LE_i(\bm{x})
\right] = 
\frac{1}{2}\sum_{\bms{x},i} \LE^\alpha_i(\bm{x})\LE^\alpha_i(\bm{x}).
\eea
Making use of \eqn{e1} to express the lattice electric fields in terms
of their continuum counterparts gives
\bea
\frac{2 a}{e^2 a^{3-d}}\LK^{(0)}&=& 
\frac{a^{2d-2}}{2 e^2}\sum_{\bms{x},i} \left[E^\alpha_i(\bm{x})E^\alpha_i(\bm{x}) - \frac{a^2}{12}E^\alpha_i(\bm{x})\partial_i^2
E^\alpha_i(\bm{x})+\ord(a^4)\right].
\label{kssmalla}
\eea
In the small $a$ limit we can replace the sum over $\bm{x}$ 
with a $d$ dimensional integral;
\bea\label{kserror}
\LK^{(0)}&\!\!\!=\!\!\!& \frac{1}{2} \int d^d \bm{x} \sum_i \left[ 
E^\alpha_i(\bm{x})E^\alpha_i(\bm{x}) -
\frac{a^2}{12}E^\alpha_i(\bm{x})\partial_i^2 E^\alpha_i(\bm{x})+\ord(a^4)
\right].
\eea
\eqn{kserror} quantifies the order $a^2$ discrepancy between the
Kogut-Susskind kinetic Hamiltonian and its continuum counterpart. \\

We now move on to the next most complicated term; one with nearest
neighbour correlations. Let us consider the small $a$ limit of the term 
\bea
\Delta \LK &=& \sum_{\bms{x},i} \Tr \left[ \LE_i(\bm{x}) U_i(\bm{x})
\LE_i(\bm{x}+a\hbm{i}) U_i^\dagger(\bm{x})\right] \nn\\
&=& \sum_{\bms{x},i}
\LE_i^\alpha(\bm{x})\LE_i^\beta(\bm{x}+a\hbm{i})\Tr\left[
\lambda^\alpha   U_i(\bm{x}) \lambda^\beta U_i^\dagger(\bm{x})\right].
\eea
Assuming an infinite lattice we can shift the summation variable
$\bm{x}$ by any finite number of lattice sites without changing the result:
\bea
\Delta \LK &=& \frac{1}{2}\sum_{\bms{x},i} \Bigg\{
\LE_i^\alpha(\bm{x})\LE_i^\beta(\bm{x}+a\hbm{i})\Tr\left[
\lambda^\alpha   U_i(\bm{x}) \lambda^\beta U_i^\dagger(\bm{x})\right]
\nn\\
&& +\LE_i^\alpha(\bm{x}) \LE_i^\beta(\bm{x}-a\hbm{i})\Tr\left[
\lambda^\beta   U_i(\bm{x}-a\hbm{i}) \lambda^\alpha
U_i^\dagger(\bm{x}-a\hbm{i})\right] \Bigg\} \nn\\
&=&\frac{1}{2} \sum_{\bms{x},i} \Bigg\{
\LE_i^\alpha(\bm{x})\LE_i^\beta(\bm{x}+a\hbm{i})\Tr\left[
\lambda^\alpha \lambda^\beta +iea \left[\lambda^\beta,\lambda^\alpha
\right] \LA_i(\bm{x}) 
\right]
\nn\\
&& + \LE_i^\alpha(\bm{x})\LE_i^\beta(\bm{x}-a\hbm{i})\Tr\left[
\lambda^\beta\lambda^\alpha
- iea \left[\lambda^\beta,\lambda^\alpha
\right] \LA_i(\bm{x}-a\hbm{i})
\right] \Bigg\} + \ord(e^2 a^2).
\eea
Here we have used the definition of $U_l$ (\eqn{linkoperator}) and the
Campbell-Baker-Hausdorff identity (\eqn{cbh}).
Making use of the totally antisymmetric SU($N$) structure constants,
$f^{\alpha \beta \gamma}$, defined in \eqn{structureconsts},
and the relation  $\LA_i(\bm{x}) = A_i(\bm{x}+a\hbm{i}/2) + \ord(a^2)$, we
expand $\Delta \LK$
in $a$ keeping the lowest order terms to give
\bea
\Delta\LK &=&\frac{1}{2} \sum_{\bms{x},i} \Bigg\{ \frac{1}{2}
\LE_i^\alpha(\bm{x})\left[\LE_i^\beta(\bm{x}+a\hbm{i})+\LE_i^\beta(\bm{x}-a\hbm{i})
\right] 
\nn\\
&& + \Tr \left[ e a f^{\alpha \beta \gamma} \lambda^\gamma A_i(\bm{x})+\ord(e a^3)\right]\LE_i^\alpha(\bm{x})\left[\LE_i^\beta(\bm{x}+a\hbm{i})-\LE_i^\beta(\bm{x}-a\hbm{i})
\right]. 
\eea
Here we have made use of the antisymmetry of the structure
constants. We now expand the lattice chromoelectric fields in powers
of $a$ to obtain
\bea
\Delta\LK &=&\frac{1}{2} \sum_{\bms{x},i} \Bigg[
\LE_i^\alpha(\bm{x})\LE_i^\alpha(\bm{x})+\frac{1}{2}a^2\LE_i^\alpha(\bm{x})
\pa{i}^2\LE_i^\alpha(\bm{x}) \nn\\
&& + e a^2 f^{\alpha \beta \gamma}A^\gamma_i(\bm{x})
\LE_i^\alpha(\bm{x}) \pa{i} \LE_i^\beta(\bm{x}) + \ord(e a^3, a^4)\Bigg].
\label{impcont}
\eea
The final step is to express the lattice electric fields in terms of
their continuum counterparts. Let us work to order $a^2$ so that $\LE
= \LE^{(2)}+\ord(a^4) $. Substituting \eqn{e1} in \eqn{impcont} gives
\bea
\Delta\LK &=&\frac{a^{2d-2}}{2 e^2} \sum_{\bms{x},i} \Bigg[
 E_i^\alpha(\bm{x}) E_i^\alpha(\bm{x})+\frac{5}{12}a^2 E_i^\alpha(\bm{x})
\pa{i}^2 E_i^\alpha(\bm{x})+\ord(e a^2,a^4)\Bigg].
\label{impexp}
\eea 
It should be pointed out that while the Kogut-Susskind and nearest 
neighbour kinetic terms 
both produce the correct continuum limit with order $a^2$ classical 
corrections, the nearest neighbour term introduces a \emph{new}
quantum error of order $e a^2$. We briefly discuss quantum errors in \sect{quantumerrors}.

\subsection{The Improved Kinetic Term}

Having derived the small $a$ expansions of the two simplest gauge
invariant kinetic terms we can form a linear combination of them 
and choose the coefficients so that the correct continuum limit is
obtained and the order $a^2$ classical error vanishes. 
Incorporating nearest neighbour interactions leads to the 
simplest improved kinetic Hamiltonian: 
\bea
\LK^{(1)} = \frac{g^2}{a} \sum_{\bms{x},i} \Tr\left[ X
\LE_i(\bm{x})\LE_i(\bm{x}) + Y
\LE_i(\bm{x}) U_i(\bm{x}) \LE_i(\bm{x}+a\hbm{i}) U_i^\dagger(\bm{x})\right].
\label{k1def}
\eea
Substituting the small $a$ expansions from 
\eqns{kserror}{impexp} into this linear combination we obtain
\bea
\LK^{(1)} &\!\!\!=\!\!\!& a^d  \sum_{\bms{x},i} \Tr\left[ \left(\frac{X+Y}{2}\right)
E^\alpha_i(\bm{x})^2 + \left(\frac{5Y-X}{12}\right)
E_i(\bm{x})\pa{i}^2 E_i(\bm{x})+\ord(e a^2,a^4) \right] \nn\\
&\!\!\!=\!\!\!& \sum_{i}\int\! d^d\bm{x} \left[ \left(\!\frac{X\!+\!Y}{2}\!\right) E^\alpha_i(\bm{x})^2 +\left(\!\frac{5Y\!-\!X}{12}\!\right)
E_i(\bm{x})\pa{i}^2 E_i(\bm{x})+\ord(e a^2,a^4) \right].
\eea
To obtain the correct continuum limit we must set $X+Y=1$. To cancel
 the order $a^2$ error we require $5Y - X =0$. Solving these equations
 simultaneously gives $X = 5/6$ and $Y=1/6$. Substituting back into
 \eqn{k1def} results in the order $a^2$
improved kinetic Hamiltonian,
\bea
\LK^{(1)} = \frac{g^2}{a} \sum_{\bms{x},i} \Tr\left[ \frac{5}{6}
\LE_i(\bm{x})\LE_i(\bm{x}) + \frac{1}{6}
\LE_i(\bm{x}) U_i(\bm{x}) \LE_i(\bm{x}+a\hbm{i}) U_i^\dagger(\bm{x})\right]. \label{kimp1}
\eea
This is the result of Luo, Guo, Kr\"oger and Sch\"utte~\cite{Luo:1998dx}.
We can take the this to order $a^4$ by including next nearest neighbour
interactions. A similar calculation using $\LE \approx \LE^{(2)}$ from
\eqn{e2} gives 
\bea
\LK^{(2)} &=& \frac{g^2}{a}\sum_{\bms{x},i}\Tr\left[ \frac{97}{120}\LE_i(\bm{x})\LE_i(\bm{x}) 
+ \frac{1}{5}\LE_i(\bm{x})U_i(\bm{x})\LE_i(\bm{x}+a\hbm{i})U^\dagger_i(\bm{x}) \right. \nn\\
&& \left. \hspace{1cm}- \frac{1}{120}
\LE_i(\bm{x})U_i(\bm{x})U_i(\bm{x}+a\hbm{i})\LE_i(\bm{x}+2a\hbm{i})U^\dagger_i(\bm{x}+a\hbm{i})U^\dagger_i(\bm{x})\right]. 
\label{kimp2}
\eea

Combining \eqns{kimp1}{v1def1} leads to the simplest classically
improved lattice Hamiltonian devoid of $\ord(a^2)$ errors,
\bea
\LH^{(1)} &=& \LK^{(1)} + \LV^{(1)} \nn\\
&=& \frac{g^2}{a} \sum_{\bms{x},i} \Tr\left[ \frac{5}{6}
\LE_i(\bm{x})\LE_i(\bm{x}) + \frac{1}{6}
\LE_i(\bm{x}) U_i(\bm{x}) \LE_i(\bm{x}+a\hbm{i})
U_i^\dagger(\bm{x})\right] \nn\\
&&+ \frac{2N}{ag^2}\sum_{\bms{x},i<j} \left\{\frac{5}{3}
P_{ij}(\bm{x})
-\frac{1}{12}\left[R_{ij}(\bm{x})+R_{ji}(\bm{x})\right]\right\}.
\label{classimpham}
\eea

\section{Tadpole Improvement}
\label{tadpoleimprovement}

Tadpole improvement is an important step in producing more
continuum-like actions in LGT. Its implementation
for pure glue on both isotropic and anisotropic lattices is well known
in the Lagrangian approach. When tadpole improvement is included
agreement between Monte Carlo and perturbative calculations on the
lattice is achieved~\cite{Lepage:1996jw}. \\

Tadpole
improvement in the Lagrangian formulation is carried out at tree-level
simply by replacing all link operators, $U$, 
by the tadpole improved link operator, $U/u_0$, where $u_0$ is the mean
link operator. The scheme was developed a decade ago by Lepage and
Mackenzie ~\cite{Lepage:1993xa}. \\

In the Hamiltonian formulation the implementation of
tadpole improvement in the potential term follows the procedure in the
action approach; all links are divided by the mean link. However for
the kinetic term two conflicting implementations have been suggested. The earliest
 starts from a tadpole improved action and carries factors of $u_0$ into the
Hamiltonian~\cite{Luo:1998dx}. More recently it was suggested that no tadpole
improvement was necessary in the kinetic term of the improved
Hamiltonian~\cite{Fang:2000vm}. Here we present our own views on the 
correct implementation.\\
 
In the Hamiltonian approach
the question of whether the lattice electric field should be rescaled
under tadpole improvement
arises. This question is easily answered by considering the
commutation relations between the link operator and electric field:
\be
[\LE^\alpha_i(\bm{x}), U_j(\bm{y})] = \delta_{ij}\delta_{\bms{x}\bms{y}}\lambda^\alpha U_i(\bm{x}).
\ee 
We see that if we divide all link operators by $u_0$ we have
\be
[\LE^\alpha_i(\bm{x}), \frac{1}{u_0}U_j(\bm{y})] =
\delta_{ij}\delta_{\bms{x}\bms{y}}\lambda^\alpha \frac{1}{u_0}U_i(\bm{x}). 
\ee
We observe that the lattice electric field cannot be rescaled and still maintain the
correct commutation relations. Thus under tadpole improvement the
electric field cannot change. We must, however, divide the second of
the kinetic terms by a factor of $u_0^2$. Tadpoles arise in this term
because the chromoelectric and gluon fields do not commute.\\

Including tadpole improvement in \eqns{v1def1}{kimp1}  
leads to the simplest order $a^2$ tadpole improved Hamiltonian: 
\bea
\LH^{(1)} &=& \LK^{(1)} +\LV^{(1)}\nn\\
 &=& \frac{g^2}{a} \sum_{\bm{x},i} \Tr\left[\frac{5}{6}
\LE_i(\bm{x})\LE_i(\bm{x}) + \frac{1}{6u_0^2}
\LE_i(\bm{x}) U_i(\bm{x})\LE_i(\bm{x}+a\hbm{i}) U_i^\dagger(\bm{x})\right] \nn\\
&&+\frac{2N}{a g^2}\sum_{\bms{x},i<j} \left[\frac{5}{3} P_{ij}(\bm{x}) -
\frac{1}{12}\left(R_{ij}(\bm{x})+R_{ji}(\bm{x})\right)\right].
\label{tadimp}
\eea
Here, the tadpole improved plaquette and rectangle operators are defined as follows
\bea
P_{ij}(\bm{x}) &=& 1- \frac{1}{u_0^4 N} \Real
\plaquette \nn\\
R_{ij}(\bm{x}) 
&=& 1- \frac{1}{u_0^6 N} \Real \rectangleA .
\label{tadplaqrect}
\eea

\section{Additional Improved Hamiltonians}
\label{additionalimprovedhamiltonians}

In this section we use the direct method to calculate improved 
Hamiltonians coupling arbitrarily 
distant lattice sites in both the kinetic and potential terms. Such
Hamiltonians may be of use in calculations where vacuum wave functions
with long distance correlations are used. Such wave functions are used
in coupled cluster calculations of glueball masses. It is possible that
more accurate calculations of high mass excited states may be obtained
using such extended Hamiltonians. Again we consider the kinetic and
potential terms separately. We start with the potential term.\\ 

For the potential term we will couple lattice sites that are separated
by, at most, $m$ links. 
Consider the extended rectangle operator,
\bea
R_{m, ij}(\bm{x}) &=& 1 -
\frac{1}{N}\Real \rectanglelong .
\eea 
Here the rectangular Wilson loop is a $1\times m$ rectangle joining the
lattice sites $\bm{x}$, $\bm{x}+ma\hbm{i}$, $\bm{x}+a(m \hbm{i} +
\hbm{j})$ and   $\bm{x}+a\hbm{j}$. We can follow the procedure of the
last section to calculate the small $a$ expansion of
$R_{m,ij}(\bm{x})$. After some algebra, the procedure of
\sect{pottermsect} leads to 
\bea
R_{m, ij}(\bm{x}) &=& \frac{e^2 m^2 a^4}{2N}\Tr(F_{i j}^2) 
+\frac{e^2 m^2 a^6}{24N}\Tr\left[F_{ij}
(m^2 D^2_i+ D^2_j)F_{ij}\right] +\ord(a^8) .\label{Rmexp}
\eea
We can build an improved potential term as in the previous section by
taking an appropriate linear combination of plaquette and extended
rectangle operators. Let such an improved potential term be defined by
\bea
\LV^{(1)}_m &=& \frac{2N}{a g^2} \sum_{\bms{x},i<j} \left\{ X
P_{ij}(\bm{x}) + \frac{Y}{2}\left[R_{m,ij}(\bm{x})+R_{m,ji}(\bm{x})\right]\right\}.
\label{V1extdef}
\eea 
Substituting the small $a$ expansions from \eqns{Rmexp}{plaqexp} gives
\bea
\LV^{(1)}_m &=& a^d \sum_{\bms{x},i<j} \Bigg\{
(X+m^2 Y) \Tr(F_{i j}^2) \nn\\
&&
+\frac{e^2a^2}{12}\left[X+\frac{m^2(m^2+1)}{2} Y\right]\Tr\left[F_{ij}
(D^2_i+ D^2_j)F_{ij}\right] +\ord(a^4) 
\Bigg\}.
\label{3.66-LV}
\eea 
We see that in order to obtain the correct continuum limit and cancel
the order $a^2$ corrections we must have
\bea
X+m^2 Y &=& 1 \nn\\
X+\frac{m^2(m^2+1)}{2} Y &=& 0 . \nn
\eea
Solving these equations simultaneously results in
\bea
X = \frac{m^2+1}{m^2-1} \quad {\rm and} \quad Y = -\frac{2}{m^2(m^2-1)}.
\eea
Substituting these values in \eqn{V1extdef} leads to the order $a^2$
improved potential term
\bea
\LV^{(1)}_m &=&\frac{2N}{a g^2} \sum_{\bms{x},i<j}\left\{ \frac{m^2+1}{m^2-1}
P_{ij}(\bm{x})
- \frac{2}{m^2(m^2-1)} \left[R_{m,ij}(\bm{x})+R_{m,ji}(\bm{x})\right]\right\}.
\label{vterm-dist}
\eea
We note that the result for the simplest improved potential term given
by \eqn{v1def1} is obtained by setting $m=2$.

We can construct an extended kinetic term in a similar fashion. In 
the kinetic term we couple lattice sites that are separated by, at
most, $n$ links. We start by considering the small $a$ behaviour of the 
extended kinetic operator,
\bea
\Delta \LK^{(1)}_n &=&\sum_{\bms{x},i} \Tr\left[ \LE_i(\bm{x})U_{\bms{x}\rightarrow \bms{x}+na\hbms{i}}
\LE_i(\bm{x}+na\hbm{i})U^\dagger_{\bms{x}\rightarrow \bms{x}+na\hbms{i}}\right].
\eea
Here we have introduced the following notation for the path ordered product of link operators
joining sites $\bm{x}$ and $\bm{x}+na\hbm{i}$,
\bea
U_{\bms{x}\rightarrow \bms{x}+na\hbms{i}} =
\prod_{l=0}^{n-1}U_i(\bm{x}+la\hbm{i}).
\eea
The small $a$ expansion of $\Delta \LK^{(1)}_n$ can be determined
using the procedure of \sect{chap2-contlimits}. After some algebra we
arrive at
\bea
\Delta \LK^{(1)}_n &=& \frac{1}{2} \sum_{\bms{x},i} \left[
E^\alpha_i(\bm{x})^2 + \frac{6 n^2 -1}{12} a^2
E^\alpha_i(\bm{x})\pa{i} E^\alpha_i(\bm{x})
+\ord(a^4,e a^2)
\right].
\label{dlk1def}
\eea
We define the extended improved kinetic Hamiltonian by taking the
following linear 
combination of the standard Kogut-Susskind kinetic term and the
extended kinetic term $\Delta \LK^{(1)}_n$:
\bea
\LK^{(1)}_n &=&  \frac{g^2}{a} \sum_{\bms{x},i} \Tr\left[ 
X \LE_i(\bm{x}) \LE_i(\bm{x})+ Y
\LE_i(\bm{x})U_{\bms{x}\rightarrow \bms{x}+na\hbms{i}}
\LE_i(\bm{x}+na\hbm{i})U^\dagger_{\bms{x}\rightarrow \bms{x}+na\hbms{i}}\right].
\label{LK1extdef}
\eea
Again we fix the constants $X$ and $Y$ to produce the correct
continuum result and cancel the order $a^2$ correction
term. Substituting the small $a$ expansions from \eqns{kssmalla}{dlk1def}
into \eqn{LK1extdef} gives the following:
\bea
\LK^{(1)}_n &=&  \frac{a^d}{2} \sum_{\bms{x},i} \left\{ 
(X+Y) E^\alpha_i(\bm{x})^2 + \left[\frac{(6n^2-1)Y-X}{12}\right] a^2 
E^\alpha_i(\bm{x})\pa{i}^2 E^\alpha_i(\bm{x})\right\} .
\eea 
To obtain the correct continuum limit and cancel the order $a^2$ error
term we set
\bea
 X = 1-\frac{1}{6 n^2} \quad {\rm and} \quad Y = \frac{1}{6 n^2}.
\eea
Substituting in \eqn{LK1extdef} we obtain the following extended improved
kinetic term
\bea
\LK^{(1)}_n &=&  \frac{g^2}{a} \sum_{\bms{x},i} \Tr\Bigg[ 
\left(1-\frac{1}{6 n^2}\right) \LE_i(\bm{x}) \LE_i(\bm{x}) \nn\\
&&+
\frac{1}{6 n^2}
\LE_i(\bm{x})U_{\bms{x}\rightarrow \bms{x}+na\hbms{i}}
\LE_i(\bm{x}+na\hbm{i})U^\dagger_{\bms{x}\rightarrow \bms{x}+na\hbms{i}}\Bigg].
\label{kterm-dist}
\eea
We notice the result for the simplest improved kinetic term is
recovered by setting $n=1$.\\

Combining \eqns{kterm-dist}{vterm-dist} leads to the classically
improved Hamiltonian coupling lattice sites, at most, $n$-links apart in the
kinetic term and, at most, $m$-links apart in the potential term,
\bea
\LH^{(1)}_{n,m} &=&  \frac{g^2}{a} \sum_{\bms{x},i} \Tr\Bigg[ 
\left(1-\frac{1}{6 n^2}\right) \LE_i(\bm{x}) \LE_i(\bm{x}) \nn\\
&&+
\frac{1}{6 n^2}
\LE_i(\bm{x})U_{\bms{x}\rightarrow \bms{x}+na\hbms{i}}
\LE_i(\bm{x}+na\hbm{i})U^\dagger_{\bms{x}\rightarrow
\bms{x}+na\hbms{i}}\Bigg]\nn\\
&&+
\sum_{\bms{x},i<j}\left\{ \frac{m^2+1}{m^2-1}
P_{ij}(\bm{x})
- \frac{2}{m^2(m^2-1)} \left[R_{m,ij}(\bm{x})+R_{m,ji}(\bm{x})\right]\right\}.
\label{classimphamdistant}
\eea
It should be pointed out here that Luo, Guo, Kr{\"o}ger and
Sch{\"u}tte derived an improved Hamiltonian~\cite{Luo:1998dx} 
with a kinetic term which
is a sum over $n$ of $\Delta \LK^{(1)}_n$.\\

Having calculated some examples of classically improved Hamiltonians
we now move on to briefly discuss the more difficult topic of quantum
improvement.

\section{Quantum Errors}
\label{quantumerrors}

The correction of quantum errors has been instrumental in the progress
of Lagrangian LGT. While classically improved actions are built from
linear combinations of traced Wilson loops with the constant
coefficients chosen to cancel leading order $a^2$ errors, quantum
improved actions determine the coefficients to one-loop order giving
them a $g^2$ dependence. A method for determining these coefficients
in weak coupling perturbation theory was developed by L{\"u}scher and
Weisz~\cite{Luscher:1985zq}. Unable to find sufficient conditions in the physical
theory, L{\"u}scher and Weisz considered the situation where two
dimensions are compactified and the gauge field obeys twisted periodic
boundary conditions. This mechanism, familiar from Kaluza-Klein
theories, gives the gluons mass and provides many simple on-shell
quantities for the study of quantum improved actions. For the purpose
of calculating the coefficients of a quantum improved action at the
one-loop level, L{\"u}scher and Weisz used the masses of the asymptotic
gluon states and some simple scattering amplitudes. The result of
their SU(3) calculation in 3+1 dimensions, with errors starting at
$\ord
(g^4 a^4)$, 
is termed the L{\"u}scher-Weisz improved action, 
\bea
{\cal S}_{{\rm LW}} = \frac{6}{g^2}\left[ 
\left(\frac{5}{3}+0.2370 g^2 \right) {\cal L}_1 
- \left(\frac{1}{12}+0.02521 g^2 \right) {\cal L}_2
- 0.00441 g^2 {\cal L}_3 \right], 
\label{LW-action}
\eea     
with
\bea
{\cal L}_1  &=& \sum_{\bms{x},i}\left(1 - \frac{1}{N} \Real \plaquette
\right) \nn\\
{\cal L}_2  &=& \sum_{\bms{x},i<j}\left( 1- \frac{1}{N}
\Real\rectangleA \right) \\
{\cal L}_2  &=& \sum_{\bms{x},i<j<k}\left( 1- \frac{1}{N}
\Real \begin{array}{c}\includegraphics[width=1.125cm]{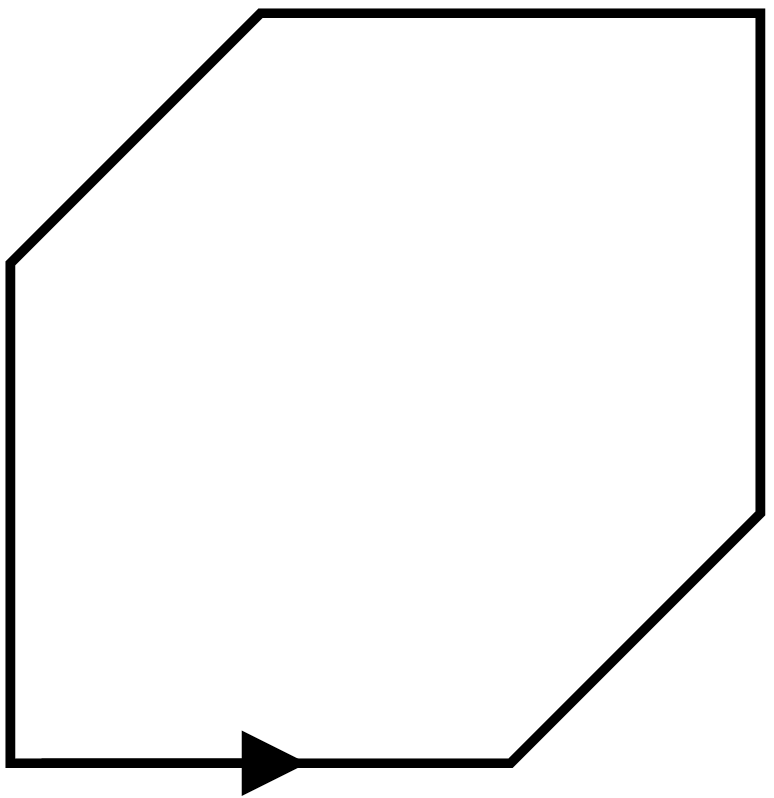}\end{array}\right). \nn
\eea
The classically improved action is recovered from \eqn{LW-action} by
modifying the coefficients so that only the order $g^0$ terms remain.
It is important to note that since the coefficient of ${\cal L}_3$ is
small, a quantum improved action is obtained, to good approximation,
by keeping only plaquette and rectangular terms. The tadpole
improvement of the L{\"u}scher-Weisz action at the one-loop level,
which we do not discuss here, is
straightforward~\cite{Alford:1995ui}. A convincing demonstration of
improvement using the tadpole improved L{\"u}scher-Weisz action 
was first shown in the restoration of rotational symmetry in the
$q\bar{q}$ potential~\cite{Alford:1995hw}. \\

In the Hamiltonian formulation quantum errors are
yet to be considered in detail. It would seem that their treatment
would require considerable effort. Luo, Guo, Kr{\"o}ger and
Sch{\"u}tte have suggested a form for a quantum improved Hamiltonian
by taking the time-like lattice spacing to zero in a quantum improved
L{\"u}scher-Weisz action, and then performing a Legendre 
transformation~\cite{Luo:1998dx}. To extend
their suggestion to a precise form would require new perturbative
calculations in the Lagrangian approach in the  $a_t\ll a_s$
limit\footnote{Here $a_t$ is the time-like and $a_s$ the space-like
lattice spacing.}, possibly following the techniques of L{\"u}scher and
Weisz described above. The remaining step would be to relate the
Lagrangian and Hamiltonian lattice couplings. This could be done using
the background field method, extending the work of Hamer for the unimproved
case~\cite{Hamer:1996ub}, to the improved case.




\section{Conclusion}

In this chapter we have introduced a direct method for the
construction of improved lattice Hamiltonians. This method differs
from Legendre transform and transfer matrix techniques in that one
does not need to start with a lattice Lagrangian. One simply needs to
construct suitable linear combinations of gauge invariant terms and
fix their coefficients so that the correct continuum limit is obtained
to the desired level of approximation. The advantage of the direct
approach is that it can be easily applied in the construction of
various improved Hamiltonians, as was demonstrated in
\sects{classicalimprovement}{additionalimprovedhamiltonians}. 
We have also discussed the
implementation of tadpole improvement within the Hamiltonian
formulation and how the construction of quantum improved Hamiltonians
may be possible.\\

The extension of the direct method to the construction of lattice
Hamiltonians for quarks and gluons is straightforward. One simply needs to
construct appropriate gauge invariant quark operators and form linear
combinations of them such that the correct continuum Hamiltonian is
obtained. The construction of quantum improved Hamiltonians is of more
immediate interest but their calculation seems to require significant
additional work.\\

Having constructed a variety of lattice Hamiltonians we now move on
testing the levels of improvement that they offer. In the next chapter
we introduce some simple tests of improvement. In
\chap{sunmassgaps} we continue the discussion of improvement by 
employing classically improved and tadpole improved Hamiltonians in the
calculation of glueball masses.

\chapter{Testing Improvement}
\label{testingimprovement}

\section{Introduction}
 
When performing a calculation on the lattice we are interested in
the continuum limit in which the lattice spacing is taken to
zero. It is only in this limit that physical results are obtained.
For an improved theory one would expect the presence of lattice
artifacts to be less pronounced, taking the lattice theory closer to 
continuum physics for a given lattice spacing. 
Whether or not this is the case for the improved Hamiltonians
generated in \chap{constructingandimproving}
will become evident in specific calculations.\\

In this chapter we perform some relatively simple checks on
improvement. The first check presented is for the simple case of U(1)
gauge theory. In this case we perform a perturbative calculation of the static
quark-antiquark potential in the strong coupling regime  
for both the Kogut-Susskind and improved Hamiltonians. One would
expect an improved Hamiltonian to exhibit a break from the strong
coupling limit at a larger coupling, corresponding to a larger lattice
spacing, than the unimproved case. We demonstrate in
\sect{staticqqbarpotential} that this is indeed the case.\\

The second check presented is a variational calculation of the
lattice specific heat for the case of SU(2) in 2+1
dimensions. 
For this case analytic results for the integrals involved are known
and have been used in
variational calculations for nearly 20 years. The peak in the lattice 
specific heat has been shown to correspond to the transition region 
from strong to weak coupling~\cite{Horn:1985ax}. One would expect 
the peak of the improved lattice specific heat to be appear at a
larger coupling (corresponding to a larger lattice spacing) than its unimproved
counterpart. We demonstrate in \sect{SU(2)SpecificHeat} 
that this is in fact the case.

\section{The U(1) Static Quark-Antiquark Potential}
\label{staticqqbarpotential}
 
Without resorting to a detailed computation, a relatively
straightforward check on improvement can be made in a perturbative 
calculation of the
static quark-antiquark potential in the strong coupling regime. 
We adopt the simplistic model of Kogut and
Susskind~\cite{Kogut:1975ag} and consider a source and a sink of
colour flux, each
infinitely heavy, separated by $r$ lattice sites. The potential
existing between the quark and antiquark is described in terms of
a string of excited links joining the particles. \\

Consider the case of U(1) on a 3 dimensional spatial lattice with $L$ links. 
The Kogut-Susskind Hamiltonian is given by
\bea
\LH = \frac{g^2}{2a}\sum_{\bms{x},i} \LE_i(\bm{x})^2 +
\frac{1}{a g^2}\sum_{\bms{x},i<j} P_{ij}(x). 
\eea
In the strong coupling limit ($g\gg 1$) only the
kinetic term survives. In this limit the potential term can be treated 
as a perturbation whose effect can be 
handled with standard Rayleigh-Schr\"odinger perturbation
theory. Before proceeding with the calculation, we first introduce the
Fock space of link excitations in which we intend to work.\\

We define the lattice vacuum by the direct product of all strong
coupling link vacuua:
\be
|0\rangle = |0\rangle_1\otimes |0\rangle_2 \otimes\cdots\otimes
|0\rangle_L . \label{strongvacuum}
\ee   
Here $|0\rangle_i$ denotes the strong coupling vacuum of the $i$-th
link upon which the
action of the lattice electric field gives $\LE_l |0\rangle_l = 0$. 
An excited link is defined by acting a link variable on the
vacuum. For example:
\be
|1\rangle_l = U_l |0\rangle_l, \quad |2\rangle_l = U_l U_l
|0\rangle_l, \quad\mbox{etc.}
\ee
Here the index $l$ is equivalent to the pair of indices $\bm{x}$ (starting
point) and $i$ (direction) which label the link in the notation $U_i(\bm{x})$.
Employing the commutation relations between 
$\LE $ and $U$ which, for the gauge group U(1), are given by
\be
[\LE_l,U_m]= \delta_{lm}U_l,
\ee
we see that excited link states are eigenstates
of the strong coupling Hamiltonian. In particular the energy
associated with a singly excited link is $g^2/2a$, and with a chain of
$r$ links $r g ^2/2a$.\\

We now return to the calculation of the static quark-antiquark
potential. The lowest
energy state is given by the direct line of flux joining the source
and sink with all links singly excited. Its energy is given by 
\be
        V^{(0)}_0 = \frac{g^2}{2a}r.
\ee  
Making use of Rayleigh-Schr\"odinger perturbation theory to include
the effects of the potential term, the lowest
order shifts in energy start at order $g^{-6}$. Taking into account
all possible contributions yields the static quark-antiquark potential
in 3+1 dimensions (relative to the perturbed vacuum),
\be
        V^{(0)}_{3+1} = \frac{r g^2}{2a}\left(1-\frac{2}{3 g^8}\right).
\ee
In 2+1 dimensions the perturbation is half the 3+1 dimensional
result. This is a result of there being only two possible orientations for a
plaquette sharing a link with a line of flux in 2+1 dimensions,
whereas in 3+1 dimensions there are four possible orientations. The
2+1 dimensional result is then
\be
V^{(0)}_{2+1} = \frac{rg^2}{2a}\left(1-\frac{1}{3g^8}\right). \label{vv1}
\ee

We can perform the same calculation with the 
classically improved Hamiltonian, which for U(1) is given by
\bea
H^{(1)} &=&
\frac{g^2}{2a}\sum_{\bms{x},i}\left[\frac{5}{6}\LE_i(\bm{x})^2+
\frac{1}{6}\LE_i(\bm{x}) \LE_i(\bm{x}+a\hbm{i})\right] \nn\\
&&\hspace{1cm} -\frac{1}{ a
g^2}\sum_{\bms{x},i<j}\left\{\frac{5}{3}P_{ij}(\bm{x})- \frac{1}{12}\left[R_{ij}(\bm{x}) + R_{ji}(\bm{x})\right]
\right\}.    
\eea 
The improved strong coupling 
ground state energy of the chain of $N$ excited links is
given by:
\be
V^{(1)}_0 = \frac{g^2}{2a}\left[ \frac{5}{6}r +
\frac{1}{6}(r-1)\right] = \frac{g^2}{2a}\left(r-\frac{1}{6}\right) .
\ee
The difference between this and the unimproved result arises because
the improvement term,
$\sum_{i,\bms{x}}\LE_i(\bm{x})\LE_i(\bm{x}+a\hbm{i})$,  
gives non-zero contributions only when acting on pairs of 
adjacent excited links. Consequently, in a straight 
chain of $r$ links there are only $r-1$ contributions from this 
term. This difference vanishes however in the limit of large separation.\\

The calculation of the lowest order perturbation in the improved case
is considerably more complicated than the unimproved case. 
Including rectangular Wilson loops in the potential term results in a large
number of contributions to the order $g^{-6}$ shift. Collecting all
contributions 
leads to the improved static quark-antiquark potential (relative
to the perturbed vacuum),
\be
V^{(1)}_{3+1} \approx \frac{r g^2}{2a}\left(1-\frac{3.7651}{g^8}\right) +
\frac{1.1844}{a g^6}. \label{vv2}
\ee
The 2+1 dimensional result can be derived as above:
\be
V^{(1)}_{2+1} \approx \frac{r g^2}{2a}\left(1-\frac{1.8825}{g^8}\right) +
\frac{0.59219}{a g^6}.
\ee
We notice a large change in the magnitude of the first order
perturbation in going from the Kogut-Susskind to the improved lattice
Hamiltonian. In the limit of large separations we ignore the trailing order
$g^{-6}$ terms which are independent of $r$.\\

\begin{figure}
\begin{center}
\includegraphics{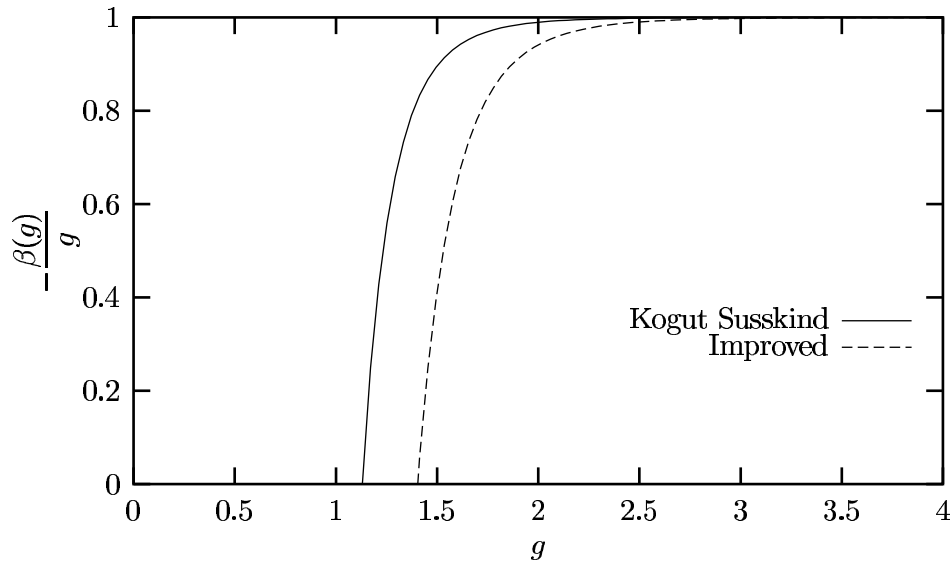}
\caption{
The Callan-Symanzik $\beta$ function versus coupling for U(1) in 3+1 dimensions.}
\end{center}
\label{beta}
\end{figure}

The string tension, $\sigma$, is defined as the coefficient of the
linear part of 
the quark-antiquark potential. It is a physical quantity and is 
measurable, in principle, in deep inelastic scattering experiments.
From \eqns{vv1}{vv2} we obtain the Kogut-Susskind and 
classically improved U(1) string tensions in 3+1 dimensions respectively,
\bea
\sigma^{(0)} &=& \frac{g^2}{2a^2}\left(1- \frac{2}{3g^8}\right)\label{ss1}\\
\sigma^{(1)} &\approx & 
\frac{g^2}{2a^2}\left(1- \frac{3.765}{g^8}\right).\label{ss2}
\eea
Since it is a physical quantity, the string tension must have a finite
continuum limit. For this to be the case
the coupling, $g$, must depend on the lattice
spacing. Precisely how is determined by removing the dependence of the
string tension on $a$ by setting 
\bea
\frac{d}{da}\sigma(a,g(a)) = 0 = \frac{\partial \sigma}{\partial a}
+ \frac{dg}{da}\frac{\partial \sigma}{\partial g}.
\eea 
This is known as the renormalisation group equation. 
It is usual to describe the dependence of the coupling on the lattice
spacing with the Callan-Symanzik $\beta$ function, which, for Abelian
gauge theories, we define by
\be
\beta(g) = -a\frac{dg}{da}.
\ee
The renormalisation group equation then allows us to write the $\beta$ 
function in terms of the string tension as follows
\be
\beta(g) = -a \frac{\partial \sigma}{\partial a}
\Big/\frac{\partial \sigma}{\partial g}.
\ee
With this result we can use \eqns{ss1}{ss2} to obtain the
Kogut-Susskind and improved $\beta$ functions shown in \fig{beta},
\bea
-\frac{\beta^{(0)}}{g} &=& 1- \frac{8}{3g^8} \label{bb1}\\
-\frac{\beta^{(1)}}{g} &\approx & 1- \frac{15.0604}{g^8}.\label{bb2}
\eea 
We can use this result to check the level of improvement offered by
the improved 
Hamiltonian. To do this we examine the break
away from the strong coupling limit, $-\beta/g = 1$. One would expect 
a calculation using an
 improved Hamiltonian to break
away from the strong coupling limit at a larger lattice spacing
(or equivalently a larger coupling) than one using an unimproved Hamiltonian.
We see from \fig{beta} that this is indeed the case. \\

To be more precise, suppose the break from the strong coupling limit occurs at a coupling 
$g_\ast^{(0)}$ for the Kogut-Susskind Hamiltonian and $g_\ast^{(1)}$
for the improved Hamiltonian. A relationship between
$g_\ast^{(0)}$ and $g_\ast^{(1)}$ can be deduced using \eqns{bb1}{bb2}:
\bea
g_\ast^{(1)} \approx \left(\frac{3\times 15.0604 }{8 }
\right)^{\frac{1}{8}}g_\ast^{(0)} \approx 1.24 g_\ast^{(0)}. \label{pert}
\eea 
The same result is obtained, to this order, in 2+1 dimensions.\\

Having observed improvement in a simple strong coupling calculation
for pure U(1) gauge theory we now move on to a more detailed check on
improvement away from the strong coupling limit.


  
\section{SU(2) Lattice Specific Heat in 2+1 Dimensions}
\label{SU(2)SpecificHeat}

\subsection{Introduction}

In this section we calculate lattice specific heats using various improved
SU(2) Hamiltonians in 2+1 dimensions. We start with the classically
improved Hamiltonian given in \eqn{classimphamdistant},
which couples links, at most, $n$ lattice sites apart in the kinetic term, and $m$ sites apart in the potential term. 
Following that we consider tadpole
improvement, discussing its practical implementation in the
Hamiltonian approach. Finally, we calculate the lattice specific heat with the 
order $a^4$ classically improved kinetic term. A discussion of the
results follows.

\subsection{Preliminaries}
\label{prelims-3}
Before proceeding with the calculation we need to make some
preliminary definitions. Consider for the time being SU($N$) 
pure gauge theory on a lattice with $N_p$ plaquettes.
The lattice specific heat is defined by
\bea
C_V = -\frac{\partial^2 \epsilon_0}{\partial \beta ^2},
\label{latticespecificheat}
\eea
where $\beta = N/g^2$ is the inverse coupling, not to be confused with
the Callan-Symanzik $\beta$ function of \sect{staticqqbarpotential}. 
From here on $\beta$ will always denote the inverse coupling. 
The vacuum energy density, $\epsilon_0$, is given by
\bea
\epsilon_0 = \frac{a}{N_p} \frac{\langle \phi_0 | \LH |
\phi_0 \rangle}{\langle \phi_0 | \phi_0 \rangle}.
\label{ed}
\eea
We compute the vacuum energy density variationally, using the
one-plaquette trial state,
\bea
|\phi_0\rangle = \exp\left( c \sum_{\bms{x},i<j} \Real
 \plaquette \right)|0\rangle. \label{oneplaquette}
\eea
Here $|0\rangle$ is the SU($N$) analogue of the strong coupling 
vacuum defined in
\eqn{strongvacuum}.
At this stage \eqn{ed} defines the vacuum energy density as a function
of the coupling and the variational parameter, $c$. 
The variational parameter is fixed as a function of the coupling  
by minimising the vacuum energy density at each coupling. 
Using the commutation relations between $\LE_l^a$ and $U_m$ in \eqns{comeu}{comeudag} the vacuum energy
density can be expressed in terms of the expectation values of
the Wilson loops that appear in the 
Hamiltonian under consideration. 
This can be done easily for all $N$ and any
number of dimensions. The difficulty lies in the calculation of the
expectation values themselves. There is no difficulty however 
for the special case of SU(2) in 2+1 dimensions for which analytic
results are available.

With the necessary preliminary definitions made, we now proceed to the
calculation of lattice specific heats using a selection of
the Hamiltonians derived in \chap{constructingandimproving}.

\subsection{The Extended Hamiltonian}
\label{theextendedhamiltonian}

In this section we calculate lattice specific heats for SU(2) LGT in 2+1
dimensions using the extended improved Hamiltonian $\LH^{(1)}_{n,m}$
of \eqn{classimphamdistant}. Let us denote the lattice specific heat
calculated with $\LH^{(1)}_{n,m}$ by
\bea
        C^{(n,m)}_V = -\frac{\partial^2 \epsilon^{(n,m)}_0}{\partial \beta ^2},
\label{CV-1}
\eea
where the vacuum energy density is given by
\bea
\epsilon^{(n,m)}_0 = \frac{a}{n_p} \frac{\langle \phi_0 | \LH^{(1)}_{n,m} |
\phi_0 \rangle}{\langle \phi_0 | \phi_0 \rangle}.
\eea
Here $| \phi_0 \rangle$ is the one-plaquette trial state of
\eqn{oneplaquette}. Making use of \eqn{simplify}, after some algebra,
$\epsilon^{(n,m)}_0$ can be expressed in terms of the expectation
values of plaquettes and extended rectangles as follows:
\bea
\epsilon^{(n,m)}_0 \!\!\!&=&\!\!\!
\left[\left( 1 -\frac{1}{6n^2}\right)
\frac{N^2-1}{2\beta}c
-\frac{2\beta}{N}\frac{m^2+1}{m^2-1}\right]\left\langle \plaquette
\right\rangle \nn\\
&&\hspace{1.1cm}
+
\frac{4\beta}{N}\frac{1}{m^2(m^2-1)}\left\langle\rectanglelong
\right\rangle +2\beta \left(1+\frac{2}{m^2}\right). \label{thisone-1} 
\eea  
For the special case of SU(2) in 2+1 dimensions, the expectation
values appearing in \eqn{thisone-1} can be evaluated analytically~\cite{Arisue:1983tt,Arisue:1990wv}, with
the result,
\bea
\left\langle\plaquette\right\rangle &\equiv & 
\frac{\langle\phi_0 | \plaquette |\phi_0\rangle}{\langle
\phi_0|\phi_0 \rangle} =
2\frac{I_2(4c)}{I_1(4c)}\nn\\
\left\langle\rectanglelong\right\rangle &=& 
\frac{1}{2^{m-1}} \left\langle \plaquette \right\rangle ^m
= 2 \left(\frac{I_2(4c)}{I_1(4c)}\right)^m .
\label{expvalssu2}
\eea
Here $I_n$ is the $n$-th order modified Bessel function of the first
kind defined for integers $n$ by
\bea
I_n(2x) = \sum_{k=0}^\infty \frac{x^{2k+n}}{k!(k+n)!}.
\eea 
Making use of \eqn{expvalssu2} in \eqn{thisone-1} gives the SU(2) improved
vacuum energy density in 2+1 dimensions,
\bea
\epsilon^{(n,m)}_0  \!\!\!&=&\!\!\! 
\left[\left( 1 -\frac{1}{6n^2}\right)
\frac{3}{\beta}c
-2\beta\frac{m^2+1}{m^2-1}\right]\frac{I_2(4c)}{I_1(4c)}\nn\\
&& +\frac{4\beta}{m^2(m^2-1)}\left(\frac{I_2(4c)}{I_1(4c)}\right)^m 
+2\beta\left(1+\frac{2}{m^2}\right).
\label{e0}
\eea 
The result of \eqn{e0} should be compared with the unimproved
result of Arisue, Kato and Fujiwara~\cite{Arisue:1983tt}, derived from
the Kogut-Susskind Hamiltonian:
\bea
\epsilon_0 =
\left(\frac{3c}{\beta}-2\beta\right)\frac{I_2(4c)}{I_1(4c)}
 + 2\beta.
\eea

With an analytic expression for the energy density as a function of
$\beta$ in hand, the lattice specific
heat may be calculated using \eqn{CV-1}. We now consider the effect of varying the
extent of the correlations in the 
kinetic term and potential term separately. We start
with the kinetic term. \\

\begin{figure}[t]
\begin{center}
\includegraphics{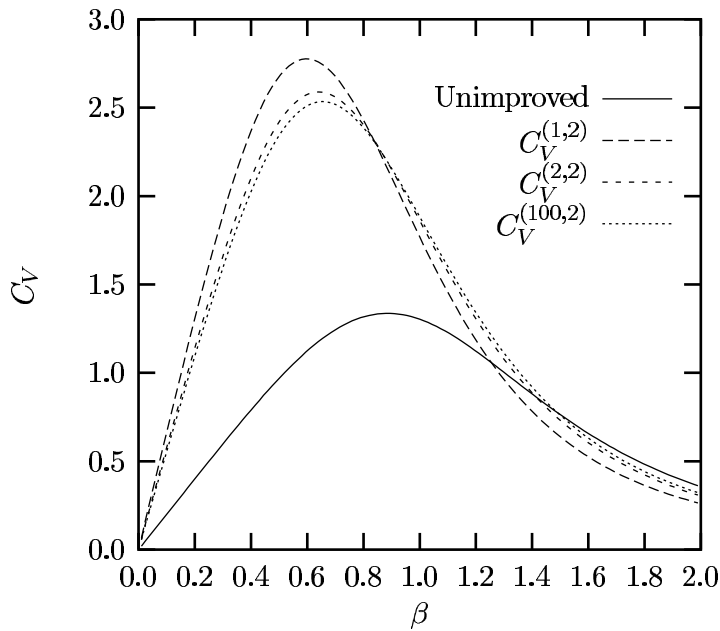}
\end{center}
\caption{The 2+1 dimensional SU(2) lattice specific heat calculated 
with various Hamiltonians. $C_V^{(1,2)}$, $C_V^{(2,2)}$ and
$C_V^{(100,2)}$ are improved results derived from Hamiltonians with varying kinetic terms. The unimproved result refers to the
calculation of Arisue, Kato and Fujiwara~\cite{Arisue:1983tt}.} 
\label{cvplot}
\end{figure}

\fig{cvplot} shows a plot of the lattice specific heats $C_V^{(1,2)}$,
$C_V^{(2,2)}$ and $C_V^{(100,2)}$ (those with 2$\times$1 rectangles in
the potential term), as well as the unimproved result. 
The location of the peak, $\beta^\ast$, indicates the crossover region
between the strong and weak coupling regimes~\cite{Horn:1985ax}. 
The graphs exhibit the behaviour expected from an improved
Hamiltonian. It is clear that for the improved Hamiltonians 
the transition from strong to weak coupling
occurs at a larger coupling (smaller $\beta$), and hence a larger
lattice spacing, than the unimproved case. The approximate values of 
$\beta^\ast$ are: 0.8850 for the unimproved case, 0.5955 ($n=1$), 0.6391
($n=2$) and  0.6533 ($n=100$). This implies that when
using an improved Hamiltonian one is closer to continuum physics when
working at a given lattice spacing. It should be noted that
the degree of improvement, does not vary greatly with $n$. This is 
expected, since distant correlations in the kinetic term will not
contribute to the energy density unless equally distant correlations are 
included in the trial state. For such complicated trial states, 
Monte Carlo techniques are necessary to calculate the required
expectation values. Such calculations will not be presented here. They
are made difficult by very shallow minima in the vacuum energy 
density~\cite{Long:1988qe}. \\

\begin{figure}[t]
\begin{center}
\includegraphics{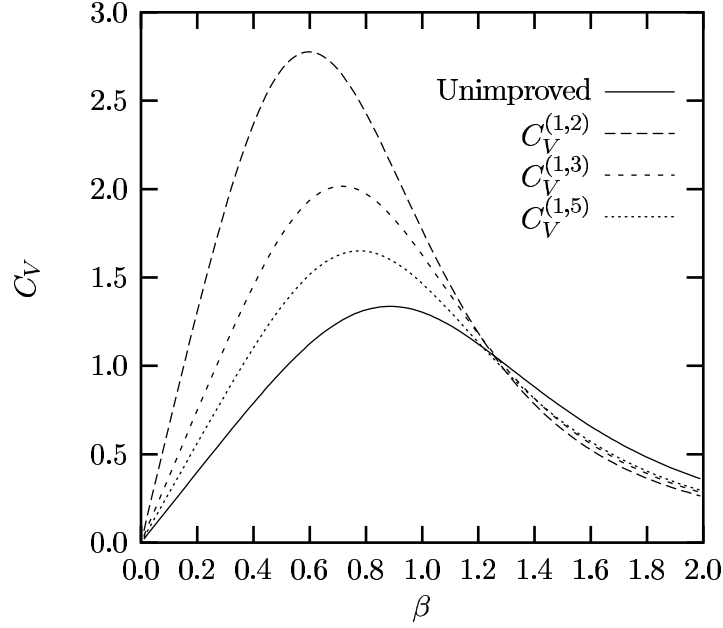}
\end{center}
\caption{
The 2+1 dimensional SU(2) lattice specific heat calculated with
various 
Hamiltonians. 
$C_V^{(1,2)}$, $C_V^{(1,3)}$ and $C_V^{(1,5)}$ are improved results
derived from Hamiltonians with varying potential terms.}
\label{cvplotpotential}
\end{figure}

We now examine the dependence of improvement on the potential term. 
\fig{cvplotpotential} shows a plot of the lattice specific heats $C_V^{(1,2)}$,
$C_V^{(1,3)}$ and $C_V^{(1,5)}$, as well as the unimproved result. 
Again, it is clear that for the improved Hamiltonians, 
the transition from strong to weak coupling
occurs at a larger coupling (smaller $\beta$), and hence a larger
lattice spacing, than for the unimproved case. 
The values of $\beta^\ast$ are: 0.5955 ($m=2$), 0.7133 ($m=3$) and
0.7768 ($m=5$). The degree of 
improvement is best for $m=2$ and becomes worse as $m$
 increases. This is to be expected since
the errors present in \eqn{3.66-LV} are $\ord(m^4 a^4)$,
which is large for large $m$. This suggests that there may be 
limits on the utility of improved Hamiltonians which couple distant
lattice sites. 
It is clear that the degree of improvement is, by far, more
sensitive to varying the potential term than varying the kinetic
term with the chosen trial state.  \\

Having examined classical improvement in the context of improved
Hamiltonians coupling distant lattice sites, we now move on to consider
the effect of tadpole improvement on the location of the lattice 
specific heat peak.

\subsection{Tadpole Improvement}
\label{tadpoleimprovement-1}

In this section we extend the work of \sect{theextendedhamiltonian} to
consider tadpole improved SU(2) LGT in 2+1 dimensions.
As discussed in \sect{tadpoleimprovement}, tadpole improvement is an
important step in the improvement of LGT. Tadpole errors arise when 
expectation values involving products of link operators are taken. When contracted, products 
of gluon fields, $A_i(x)$, produce ultraviolet divergences. Such
divergences spoil small $a$ perturbation series, leading to inaccurate 
results in the continuum limit. Lepage and Mackenzie demonstrated that
tadpole errors are largely cancelled by dividing each link operator 
by the mean link $u_0$~\cite{Lepage:1993xa}.\\

In this section we calculate the lattice specific heat using the simplest
tadpole improved Hamiltonian given by ~\eqn{theextendedhamiltonian}. 
We define the SU($N$) mean link in terms of the mean plaquette as follows:
\bea
 u_0^4 &=& \frac{1}{N} \left\langle \plaquette \right\rangle. 
\label{meanlink}
\eea
In the Lagrangian formulation, the more difficult choice of fixing to the
Landau gauge and calculating the mean link directly is often used. We
choose to define the mean link in terms of the mean plaquette
for simplicity. In this case one does not need to perform a gauge
fixing, which would be difficult in the Hamiltonian approach.\\

To incorporate tadpole improvement in $\LH^{(1)}_{n,m}$ is
straightforward. We follow the prescription discussed in
\sect{tadpoleimprovement} in which all links are divided by the
mean link and the lattice electric field operators remain unchanged. Making
use of this prescription leads to the tadpole improved version of $\LH^{(1)}_{n,m}$,
\bea
\LH^{(1)}_{n,m} &\!\!\!=\!\!\!&
\frac{g^2}{2a}\sum_{x,i}\Tr\left[\left(1-\frac{1}{6n^2}\right)
\LE_i(x)\LE_i(x) + \right. \nn\\
&&\left.
\frac{1}{6n^2 u_0^{2n}}\LE_i(x)U_{x\rightarrow x+nai}
\LE_i(x+nai)U^\dagger_{x\rightarrow x+nai} \right] \nn\\
&&+ \frac{2N}{ag^2}\sum_{x,i<j}\left\{
\frac{m^2+1}{m^2-1} P_{ij}(x) - \frac{1}{m^2(m^2-1)}\left[R_{ij}(x) 
+ R_{ji}(x)\right]\right\},
\eea
where the tadpole improved plaquette operator is given by
\eqn{tadplaqrect} and the tadpole improved extended rectangle operator is
defined by,
\bea
R_{ij}(x) &=& 1 - \frac{1}{Nu_0^{2m+2}}\Real \rectanglelong .
\eea
Here the mean link is given by \eqn{meanlink}.
Proceeding as in \sect{theextendedhamiltonian} we obtain the tadpole improved energy
density, 
\bea
\epsilon^{(n,m)}_0  \!\!\!&=&\!\!\! 
\left[\left( 1 -\frac{1}{6n^2}\right)
\frac{3}{\beta}c
-\frac{2\beta}{u_0^4}\frac{m^2+1}{m^2-1}\right]\frac{I_2(4 c)}{I_1(4 c)}\nn\\
&& +\frac{4\beta}{u_0^{2m+2}m^2(m^2-1)}\left(\frac{I_2(4 c)}{I_1(4
c)}\right)^m +2\beta\left(1+\frac{2}{m^2}\right).
\label{e0tadpole}
\eea  

\begin{figure}[t]
\begin{center}
\includegraphics{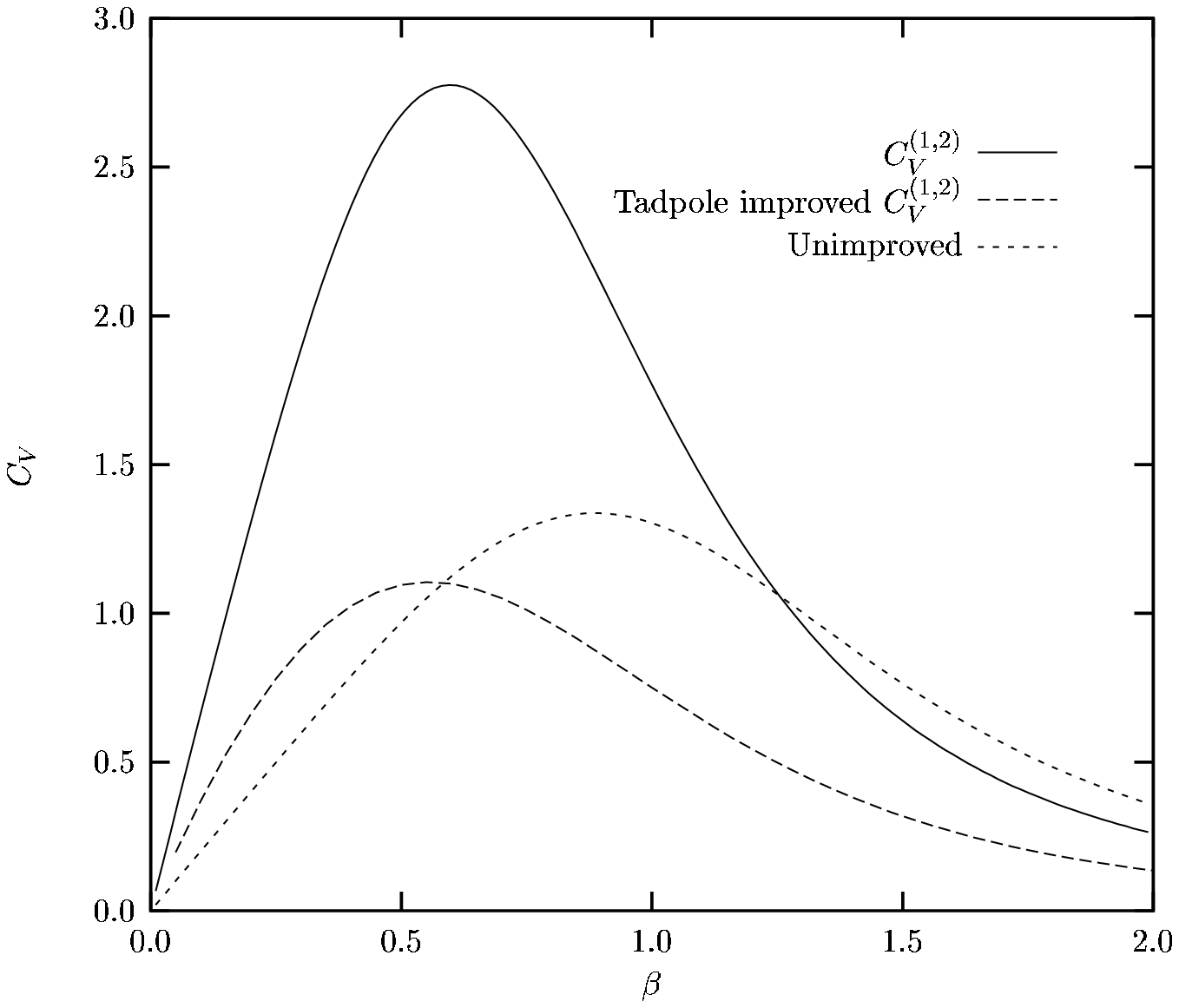}
\end{center}
\caption{
The tadpole improved 2+1 dimensional SU(2) lattice specific heat
compared with the standard improved and unimproved results. }
\label{cvtad}
\end{figure}

Incorporating tadpole improvement in a variational calculation is not
straightforward. This is because the mean plaquette depends
on the variational state, which is determined by minimising the energy
density. The energy density however depends on the mean plaquette. 
Such interdependence suggests the use of an iterative procedure for
the calculation of the energy density. The approach we adopt is as
follows. For a given $\beta$
and starting value of $u_0$ we minimise the energy density of
\eqn{e0tadpole} to fix the
variational state, $|\phi_0\rangle$. We then calculate a new mean
plaquette value using this trial state and substitute it in
\eqn{e0tadpole} to obtain a new 
expression for the energy density which is then minimised. 
This process is iterated until 
convergence is achieved, typically between five and ten iterations. The result of the procedure for the simple
case of $n=1$, $m=2$ is shown in \fig{cvtad}.  For
the purpose of comparison we also plot the unimproved lattice specific heat
and the most improved result of \sect{theextendedhamiltonian}. 
We observe a
greater level of improvement for the tadpole improved case ($\beta^\ast =
0.5565$) than the order $a^2$ classically improved case ($\beta^\ast =
0.5955$).\\

We now move on to consider the effect of incorporating order $a^4$
classical improvement in the kinetic Hamiltonian.

\subsection[Classical Order $a^4$ Improvement]{Classical Order $\bm{a^4}$ Improvement}

In this section we calculate the lattice specific heat using a Hamiltonian
with an order $a^4$ classically improved kinetic term with next nearest
neighbour interactions given in \eqn{kimp2}. 
For the potential term we use the standard order
$a^2$ improved result of \eqn{v1def1}. Ideally we would use an order
$a^4$ improved potential term so that all terms were improved
consistently to the same order. In principle, such a potential term
could be constructed by adding appropriately chosen additional, more
complicated operators to the order $a^2$ potential term of
\eqn{v1def1}. This will not be done here because many
additional operators are needed to cancel the large number of
contributions to the order $a^4$ error. Consequently in this section
we are not able to compare an order $a^4$ classically improved
calculation with its order $a^2$ equivalent. Instead we examine the
effect of incorporating easily calculated addtitional improvements in
the kinetic term of the Kogut-Susskind Hamiltonian.\\

\begin{figure}[t]
\begin{center}
\includegraphics{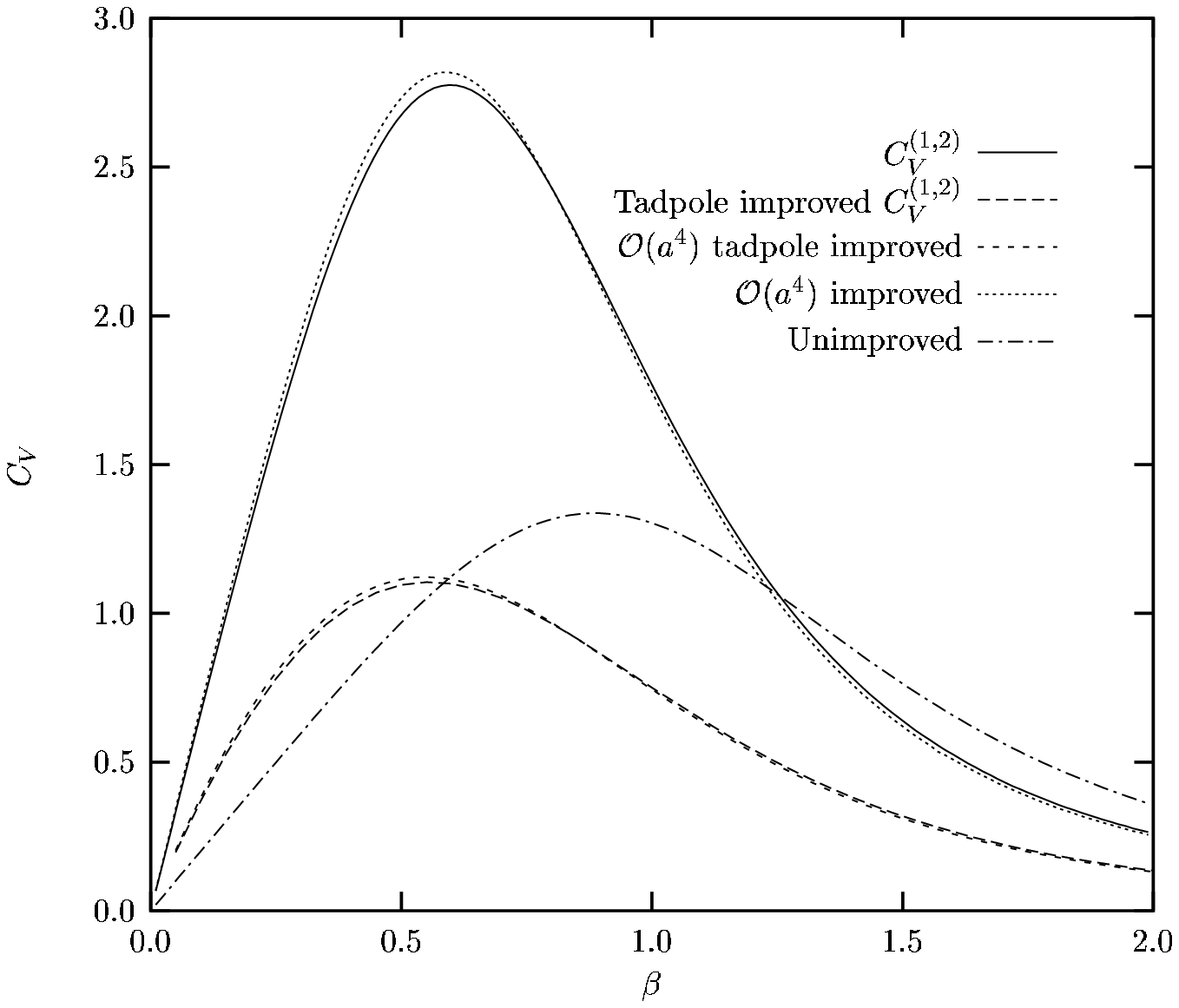}
\end{center}
\caption{
$\ord(a^4)$ classical and tadpole improved 2+1 dimensional SU(2)
lattice specific heats compared with the standard improved and 
unimproved results. }
\label{cvnextorder}
\end{figure}

Tadpole improvement is easily
incorporated in the Hamiltonian described in the preceding paragraph by dividing all links
by the mean link and leaving the lattice electric field operators unchanged as described in
\sect{tadpoleimprovement}. Making use of this prescription
leads to the Hamiltonian
\bea
\LH^{(2,1)} &=& \frac{g^2}{a}\sum_{\bms{x},i}\Tr\left[ \frac{97}{120}\LE_i(\bm{x})\LE_i(\bm{x}) 
+ \frac{1}{5 u_0^2}\LE_i(\bm{x})U_i(\bm{x})\LE_i(\bm{x}+a\hbm{i})U^\dagger_i(\bm{x}) \right. \nn\\
&& \left. \hspace{1cm}- \frac{1}{120u_0^4}
\LE_i(\bm{x})U_i(\bm{x})U_i(\bm{x}+a\hbm{i})\LE_i(\bm{x}+2a\hbm{i})U^\dagger_i(\bm{x}+a\hbm{i})U^\dagger_i(\bm{x})\right]
\nn\\
&& + \frac{2N}{ag^2}\sum_{\bms{x},i<j} \left\{\frac{5}{3}
P_{ij}(\bm{x})
-\frac{1}{12}\left[R_{ij}(\bm{x})+R_{ji}(\bm{x})\right]\right\},
\eea 
where the tadpole improved plaquette and rectangle operators are given
by \eqn{tadplaqrect}. With this Hamiltonian we obtain the following
tadpole improved SU(2) energy density in 2+1 dimensions:
\bea
\epsilon_0 \!\!\!&=&\!\!\! 
\left(\frac{97 c}{40\beta}
-\frac{10\beta}{3u_0^4}\right)\frac{I_2(4 c)}{I_1(4 c)}
+\frac{\beta}{3u_0^6}\left(\frac{I_2(4 c)}{I_1(4 c)}\right)^2 + 3\beta.
\eea
The improved and tadpole improved lattice specific heats for this case
are shown in \fig{cvnextorder} along with the results of
\sect{tadpoleimprovement-1} for comparison. Without tadpole
improvement we observe that the case of order $a^4$ improvement
in the kinetic term ($\beta^\ast = 0.5866$) displays marginal
improvement over the order $a^2$ improved case ($\beta^\ast =
0.5955$). Similarly small levels of improvement are achieved when tadpole
improvement is included ($\beta^\ast = 0.5475$) compared to the order
$a^2$ classically improved result of \sect{theextendedhamiltonian} ($\beta^\ast
= 0.5565$).  
As mentioned in \sect{theextendedhamiltonian}, such
marginal improvement is expected, since
the one-plaquette trial state is not particularly sensitive to distant
correlations in the kinetic term. A trial state more sensitive to distant
correlations in the kinetic term would require larger Wilson loops in
its exponent.

\subsection{Conclusion}

We have shown that the level of
improvement offered by an improved Hamiltonian does not depend significantly on the degree of separation
of the sites coupled in the kinetic term when using the one-plaquette
trial state of \eqn{oneplaquette}. The onset of weak 
coupling has been found to occur at $\beta \approx 0.6 $ for each
improved case, compared with $\beta \approx 1 $ for the unimproved
case. The tadpole improved Hamiltonian with order $a^4$ improvement in
the kinetic term produced a lattice specific heat peak at the largest
coupling which suggests that this Hamiltonian exhibits the largest
degree of improvement of those considered in this chapter. \\

The weak dependence on distant correlations in the kinetic
Hamiltonian is a result of only including plaquettes in the trial 
state. For such a choice the calculation of all matrix elements
encountered is analytic but the
improvement term does not explicitly contribute. This is because the
improvement term must act on an excitation of two neighbouring links
to produce a nonzero result. Such a configuration does not occur with
the plaquette trial state. If extended Wilson loops were included in the
trial state exponent one would expect significantly different results for the
dependence on separation in the kinetic term. One would expect the dependence
of the lattice specific heat peak on the separation in the kinetic term to be
significant until it matched the degree of  separation in the
trial state. Beyond that one would again expect a weak dependence.  \\

To extend the calculation to more complicated trial states requires a
change in technique. The required matrix elements must be calculated
using Monte Carlo techniques as no analytic results are available.

\section{Summary}

In this chapter we have performed some simple calculations designed to
test the level of improvement offered by the improved Hamiltonians
constructed in \chap{constructingandimproving}. In both a
strong coupling calculation of the U(1) string tension in 3+1 
dimensions and a
variational calculation of the SU(2) lattice specific heat in 2+1 dimensions 
we have observed that improved Hamiltonians do in fact provide
improvement. By improvement we mean that when working at a given
lattice spacing (or coupling) we are closer to the continuum limit
when working with an improved Hamiltonian.\\

In \chap{sunmassgaps} we return to the discussion of
improvement. There we make use of the simplest classical and tadpole
improved Hamiltonians, as well as the original Kogut-Susskind Hamiltonian, 
in calculations of SU($N$) glueball masses in 2+1 dimensions. Before
presenting these calculations it is necessary to introduce the
analytic techniques upon which the calculations rely. These techniques
do nothing more than allow the calculation of expectation values of
Wilson loops in 2+1 dimensions for general SU($N$).

\chapter{Analytic Techniques in Hamiltonian Lattice Gauge Theory}
\label{analytictechniques}
\section{Introduction}

In \chap{testingimprovement} it was shown that when working
with the one-plaquette trial state of \eqn{oneplaquette} the vacuum
energy density can be expressed in terms of expectation values of
plaquettes and rectangles. In principle, this allows the variational 
parameter, $c$, to be fixed as a function of $\beta$ in any number 
of dimensions and
for any gauge group. The difficulty lies in the
calculation of the plaquette and  rectangle expectation values. Monte
Carlo simulations can be used to calculate these expectation values in
any number of dimensions and for any gauge group. However, in 2+1 dimensions it
is possible to
calculate such expectation values analytically.\\

In this chapter we develop
techniques which allow the analytic calculation of  various
expectation values in 2+1 dimensions for SU($N$) gauge theory,
extending the long known analytic results for SU(2) to SU($N$). 
Analytic expressions are obtainable in 2+1 dimensions because the change of variables from links to plaquettes has unit
Jacobian~\cite{Batrouni:1984rb}. Consequently, the plaquettes on a 2 
dimensional spatial lattice are independent variables. For the case of 3+1 dimensions the
same change of variables produces a complicated Jacobian, which seems
only manageable on special lattices. The problems faced in
3+1 dimensional Hamiltonian LGT are discussed in \chap{3+1dimensions}.\\

Our motivation for considering SU($N$) expectation values is in their use for
calculating general SU($N$) glueball masses and in particular examining their
large $N$ limit. This limit is of current interest due to a proposed correspondence
between certain string theories and
large $N$ gauge theory. This correspondence was first conjectured by
Maldacena for supersymmetric gauge theories~\cite{Maldacena:1998re} and later extended by
Witten to pure SU($N$) gauge theories~\cite{Witten:1998qj}. The
Maldacena conjecture and the role of large $N$ glueball masses are
discussed further in \sect{largenbackground}.\\

The outline of this chapter is as follows.
In \sect{su3integrals} we consider the special cases of SU(2),
for which analytic results have been known and used for many years,
and SU(3), for which we rediscover
a useful generating function. In \sect{AnalyticresultsforSU} we
calculate three generating functions for general 
SU($N$) which provide analytic results for 
the expectation values that appear in calculations of glueball masses.   
We finish the chapter in \sect{analyticsummary} with a summary of the 
results obtained.

\section{Analytic Techniques for SU(2) and SU(3)}
\label{su3integrals}
In this section we discuss analytic techniques for use in the
calculation of group integrals for the special cases of SU(2) and
SU(3). Before deriving the generating functions which are useful in 
the calculation of
SU(2) and SU(3) expectation values, the concept of
group integration needs to be introduced. 
To do this we  
consider the expectation value of a plaquette with respect to the one
plaquette trial state of \eqn{oneplaquette}. We can write this
expectation value as an SU($N$) group integral as follows:
\bea
\left\langle \plaquette \right\rangle &=& \frac{
\int_{{\rm SU}(N)}\prod_{l}dU_l Z_1(0) \prod_p
e^{c\left[Z_1(p)  +\bar{Z_1}(p)\right]}} {\int_{{\rm SU}(N)}\prod_{l'}dU_{l'} \prod_{p'}
e^{c\left[Z_1(p')  +\bar{Z_1}(p')\right]}}.
\label{groupintegralintro}
\eea
Here the products over $l$ and $l'$ extend over all links on the
lattice, while the products over $p$ and $p'$ extend over all
plaquettes on the lattice. We introduce the notation $Z_1(p)$ to denote
the trace of plaquette $p$. For each integral in \eqn{groupintegralintro} the integration measure is 
given by the Haar measure (also called the invariant measure and less
commonly the Hurwitz measure)~\cite{Montvay:1994cy,Creutz:1984m}. 
For any compact group $G$, the Haar
measure is the unique measure $dU$ on $G$ which is left and right
invariant, 
\bea
\int_G dU f(U) = \int_G dU f(V U) = \int_G dU f(U V) \quad \forall V
\in G
\label{LRinvariance}
\eea
and normalised,
\bea
\int_G dU =1. \label{normalisation}
\eea
In \eqn{LRinvariance} $f$ is an arbitrary function over $G$. 

In 2+1 dimensions the variables in \eqn{groupintegralintro} can be
changed from links to plaquettes with unit
Jacobian~\cite{Batrouni:1984rb}. The plaquettes then become
independent variables allowing the cancellation of all but one  
group integral in \eqn{groupintegralintro}. All that remains is
\bea
\left\langle \plaquette \right\rangle &=&
\frac{\int_{{\rm SU}(N)} d V \Tr V e^{c{\rm Tr}(  V +  V^\dagger)}}
{\int_{{\rm SU}(N)} d V e^{c{\rm Tr}( V +V^\dagger)}}. 
\eea
Here, $V$ is a plaquette variable with $\Tr V = Z_1(0)$.   
For the case of SU(2), analytic expressions for the 
plaquette expectation value in terms of modified Bessel 
functions (\eqn{expvalssu2}) have been used in variational 
calculations for almost 20 years. \\

Integrals over SU(2) are easily handled using the parameterisation of
SU(2)~\cite{Montvay:1994cy},
\bea
U = \left(\begin{array}{cc} \ds\cos \frac{\rho}{2} + i \sin\frac{\rho}{2}
\cos \theta &\ds i \sin \frac{\rho}{2} \sin\theta e^{-i \phi} \\\ds
\ds i \sin \frac{\rho}{2} \sin\theta e^{i \phi} &\ds \cos \frac{\rho}{2} - i
\sin\frac{\rho}{2} \cos \theta \end{array}\right) \quad \forall U\in {\rm SU}(2),
\eea 
where the parameters $\rho$, $\theta$  and $\phi$ lie in the ranges
$0\le \theta \le \pi$ and $0\le \rho, \phi \le 2 \pi$. The Haar
measure corresponding to this parameterisation is
\bea
 d U = \frac{1}{4\pi^2}\sin^2\frac{\rho}{2}\sin\theta  d\rho d\phi d\theta.
\eea
With this information it is straightforward to show,
\bea
\int_{{\rm SU}(2)} dU e^{c \Tr U} = \frac{1}{c} I_1(2 c).
\label{su2gen}
\eea
All SU(2) integrals of interest to us can be obtained from
\eqn{su2gen} by differentiating appropriately with respect to $c$. For
example
\bea
\int_{{\rm SU}(2)} dU\Tr U e^{c \Tr U} &=& \frac{2}{c} I_2(2 c),
\label{su2plaq} \\
\int_{{\rm SU}(2)} dU(\Tr U)^2 e^{c \Tr U} &=& \frac{4}{c} I_1(2
c) - \frac{6}{c^2}I_2(2c) ,\,{\rm etc}\ldots
\eea
Here we have used the following property of the modified Bessel
function under differentiation:
\bea
\frac{d I_n}{dx}(x) = I_{n\pm1}\pm \frac{n}{x}I_n(x).
\eea

The corresponding SU(3) results, which have not been used, to our
knowledge, in Hamiltonian LGT, follow simply from a paper of
Eriksson, Svartholm and Skagerstam~\cite{Eriksson:1981rq}, in which the SU(3) integral,
\bea
\int_{{\rm SU}(3)} dU e^{{\rm Tr}(U M^\dagger +U^\dagger M)},
\eea
is calculated  for arbitrary $3\times 3$ matrices $M$.
Following their 
analysis, but treating $M$ and $M^\dagger$ as independent variables, 
with $M = c  1\hspace{-0.9mm}{\rm l}$, $M^\dagger = d
1\hspace{-0.9mm}{\rm l}$, and 
$1\hspace{-0.9mm}{\rm l}$ the $3\times 3$ unit matrix, leads to an 
expression for the SU(3) generating function:
\bea
{\cal Y}(c,d) &=&  \int_{{\rm SU}(3)} dU e^{{\rm Tr}(cU + dU^\dagger)} \nn\\
&=& \frac{i}{\pi} \oint_\Gamma dz \frac{e^{-(c^3+d^3)/z}}{z (z-cd)^{3/2}} 
J_1\left(\frac{2}{z}(z-cd)^{3/2}\right). 
\label{genexp}
\eea 
Here $\Gamma$ is a closed contour in the complex plane including the
pole at $z=0$ but excluding the pole at $z=cd$ and $J_n$ is the $n$-th
order Bessel function of the first kind defined, for integers $n$, by
\bea
J_n(2 x) = \sum_{k=0}^\infty \frac{(-1)^k x^{2k+n}}{k!(k+n)!}.
\eea
 To evaluate the contour integral in \eqn{genexp} we expand the
 integrand in power series about the pole at $z=0$ and use Cauchy's
 integral theorem to eventually obtain the following convergent series:
\bea
{\cal Y}(c,d) &=& 2 \sum_{k=0}^\infty \frac{1}{(k+1)!(k+2)!}
\sum_{l=0}^k\left(\!\begin{array}{c} 3k+3 \\ k-l \end{array}\!\right)
\frac{1}{l!}(cd)^{k-l}(c^3+d^3)^l . \label{Y}
\eea
This generating functional is extremely useful in the context of
Hamiltonian LGT. In principle, it permits an analytic investigation of
2+1 dimensional pure SU(3) gauge theory. The calculation of various matrix elements 
for all couplings, reduces to a exercise in differentiation. For example: 
\bea
\langle Z_1(p) \rangle &\equiv& \left\langle \plaquette \right\rangle 
= \left[\frac{1}{{\cal Y}}\frac{\partial{\cal Y} }{\partial c}\right]_{d=c}, \nn\\
\langle Z_1(p) \bar{Z_1}(p) \rangle
&\equiv& \left\langle\graphdd\right\rangle = \left[\frac{1}{{\cal Y}}\frac{\partial^2{\cal Y}}
{\partial c \partial d}\right]_{d=c} \mbox{, and}
\nn\\
\langle \left[Z_1(p)\right]^m \left[\bar{Z_1}(p)\right]^n \rangle 
&=& \left[\frac{1}{{\cal Y}}\frac{\partial^{m+n}{\cal Y} }
{\partial c^m \partial 
d^n}\right]_{d=c}.
\eea
As an application of \eqn{Y}
we can calculate the strong coupling limit ($c=0$) of $\langle \left[Z_1(p)\right]^m \left[\bar{Z_1}(p)\right]^n \rangle$. These integrals arise in coupled cluster
calculations~\cite{ConradPhD}. By differentiating~\eqn{Y}
appropriately, we observe
that non-zero strong coupling matrix elements occur only when $n+2m \equiv 0 \,{\rm mod}\, 3$. For this case we have,
\bea
\hspace{-1cm}\left\langle \left[Z_1(p)\right]^m \left[\bar{Z_1}(p)\right]^n 
\right\rangle_{c=0} 
&\!\!\!=\!\!\!&
\sum_{k} 
\left(\!\!\! 
\begin{array}{c}3k\!+\!3 \\ n\!+\!m\!+\!3 \end{array}
\!\!\!\right)
\frac{2 n! m! }{(k+1)!(k+2)!\left(\frac{n+2m}{3}\!-\!k\right)!
\left(\frac{m+2n}{3}\!-\!k\right)!},
\label{sc}
\eea
where the sum runs over all integers $\frac{n+m}{3}\le k \le {\rm min}\left(\frac{n+2m}{3},\frac{m+2n}{3}\right)$.
This strong coupling result has an equivalent combinatorial interpretation as 
the number of times the singlet representation appears in the direct
product of $m$ {\bf 3} and $n$ ${\bf\bar{3}}$
representations~\cite{Creutz:1984m}. Although formulas for the case of $m=n$
exist \cite{Gessel:1990,Stanley:1999}, we
have not seen a general expression for $m\ne n$ published elsewhere. 
These numbers are well known in combinatorics and are related to the Kostka
numbers \cite{Sagan:2001}. 

\section[Analytic Results for SU($N$) Integrals]{Analytic Results for SU($\bm{N}$) Integrals}
\label{AnalyticresultsforSU}

\subsection{Introduction}

Much work has been carried out on the topic of integration over the classical
compact groups. The
subject has been studied in great depth in the context of random
matrices and combinatorics. Many analytic results in terms of determinants are
available for integrals of various functions over unitary, orthogonal 
and symplectic groups~\cite{Baik:2001}. Unfortunately similar results
for SU($N$) are not to our knowledge available. 
The primary use of these integrals has been in the study of Ulam's problem concerning
the distribution of the length of 
the longest increasing subsequence in permutation
groups~\cite{Rains:1998,Widom:2001}. Connections between random
permutations and Young tableaux~\cite{Regev:1981} allow an interesting
approach to combinatorial problems involving Young tableaux. A
problem of particular interest is the counting of Young tableaux
of bounded height~\cite{Gessel:1990} which is closely related to the
problem of counting singlets in product representations mentioned in \sect{su3integrals}. 
Group integrals similar to those needed in
this thesis have also appeared in studies of the distributions of the
eigenvalues of random matrices~\cite{Diaconis:1994,Widom:1999}. \\

In the context of LGT not much has work been done in the last 20 years on the
subject of group integration. The last
significant development was due to Creutz who
developed a diagrammatic technique for calculating specific SU($N$)
integrals~\cite{Creutz:1978ub} using link variables. 
This technique allows strong coupling
matrix elements to be calculated for SU($N$)~\cite{ConradPhD}
and has more recently been used in the loop formulation of quantum
gravity where spin networks are of interest~\cite{DePietri:1997pj,Rovelli:1995ac,Ezawa:1997bv}.\\
  
In sections~\ref{simpleintegral},~\ref{anotherintegral} and~\ref{morecomplicated} we extend the results of \sect{su3integrals} to
calculate a number of SU($N$) integrals. As generating functions
these integrals allow the evaluation of all expectation
values appearing in variational calculations of SU($N$) glueball masses in
2+1 dimensions. To calculate these generating functions we work with
plaquette variables and make use of
techniques which have become standard practice in the fields of random matrices
and combinatorics. In
\sect{simpleintegral} we derive a generating function which
allows the calculation of integrals of the form
\bea
\int_{{\rm SU}(N)} dU ({\rm Tr} U)^m \overline{({\rm Tr} U)}^n e^{c ({\rm Tr} U + {\rm Tr}U^\dagger)}.
\eea
The work in \sect{morecomplicated} generalises the generating
function of \sect{simpleintegral} allowing the calculation of
more complicated integrals of the form
\bea
\int_{{\rm SU}(N)} dU \left[{\rm Tr}(U^l)\right]^m  e^{c ({\rm Tr} U + {\rm Tr}U^\dagger)}.
\eea
The third integral, which we calculate in \sect{anotherintegral}, while not directly relevant to the calculation of
massgaps here, is of importance in the study of combinatorics. It allows the
calculation of integrals of the form
\bea
\int_{{\rm SU}(N)} dU \left[{\rm Tr}(U^l)\right]^m
\left[\overline{{\rm Tr}(U^l)}\right]^n .
\eea

For each integral considered the approach is the same and proceeds as
follows. We
start with a calculation of a  U($N$) integral. For example in
\sect{simpleintegral} we calculate 
\bea
G_{{\rm U}(N)}(c,d) &=& \int_{{\rm U}(N)} dU e^{c {\rm Tr} U + d {\rm Tr} U^\dagger}.
\label{simplegen}
\eea
This is a generalisation of $G_{{\rm U}(N)}(c,c)$, an integral first
calculated by Kogut, Snow and
Stone~\cite{Kogut:1982ez}. We then make use of a result of Brower, Rossi and
Tan~\cite{Brower:1981vt} to extend the U($N$) integral to SU($N$) by
building the restriction, $\det U = 1$ for all $U \in$ SU($N$), into
the integration measure. In this way SU($N$)
generating functions can be obtained  as sums of determinants whose
entries are modified Bessel functions of the first kind.

\subsection{A Simple Integral}
\label{simpleintegral} 
In this section we introduce a useful technique for performing SU($N$)
integrals. We start with the U($N$) integral of \eqn{simplegen} and
calculate it using a technique which has become standard practice in the
study of random matrices and combinatorics.\\

Since the Haar measure is left and right invariant (see
\eqn{LRinvariance}) we can diagonalise $U$ inside the integral as 
\bea
 U = V \left(\begin{array}{cccc} e^{i\phi_0} & 0 &\cdots & 0\\
                                 0 & e^{i \phi_1}&       & \vdots \\
                                \vdots & & \ddots &   \\
                                0 & \cdots & & e^{i\phi_N}
                                 \end{array}\right) 
V^\dagger.
\eea
In terms of the set of variables, $\{\phi_k\}_{k=1}^{N}$, the Haar
measure factors as $dU = d\mu(\phi) d V$~\cite{Kogut:1982ez}. Since
the integrand is independent of $V$, the $V$ integral can be carried
out trivially using the normalisation of the Haar measure given by 
\eqn{normalisation}.\\

Making use of the Weyl parameterisation for U($N$)~\cite{Weyl:1946},
\bea
d\mu(\phi) &=& \prod_{i=1}^{N} \frac{d\phi_i}{2\pi} |\Delta(\phi)|^2 ,
\label{unparam}
\eea
where $\Delta(\phi)$ is the Vandermonde determinant, with implicit
sums over repeated indices understood,
\bea
\Delta(\phi) &=& \frac{1}{\sqrt{N!}} \varepsilon_{i_1 i_2\cdots i_N} e^{i
\phi_1(N-i_1)}e^{i
\phi_2(N-i_2)} \cdots  e^{i
\phi_N(N-i_N)},
\label{vandermonde}
\eea
 we can express the
U($N$) generating function as follows
\bea
G_{{\rm U}(N)}(c,d) &=& \int_0^{2
\pi}\frac{d\phi_1}{2\pi}\cdots\int_0^{2 \pi}\frac{d\phi_N}{2\pi}
\exp\left[\sum_{i=1}^{N}(c e^{i\phi_i}+d e^{-i\phi_i})\right]|\Delta(\phi)|^2.\label{Weyl}
\eea
Substituting \eqn{vandermonde} gives,
\bea
G_{{\rm U}(N)}(c,d) &=& 
\frac{1}{N!} 
\varepsilon_{i_1 i_2\ldots i_N}\varepsilon_{j_1 j_2\ldots j_N} \nn\\
&&\times \prod_{k=1}^N \int_0^{2
\pi}\frac{d\phi_k}{2\pi}\exp\left[i(j_k-i_k)\phi_k+ c e^{i \phi_k} +d
e^{-i \phi_k} \right]. \label{U(N)integral}
\eea
To simplify this further we need an expression for the integral,
\bea
g_n(c,d) &=& \int_0^{2\pi} \frac{dx}{2\pi} \exp(i n x + c e^{i x} + d
e^{-i x}),
\eea
which is easily handled by expanding the integrand in Taylor series in
$c$ and $d$,
\bea
g_n(c,d) &=&
\sum_{k=0}^\infty \sum_{l=0}^\infty \frac{c^k d^l}{k! l!}\int
\frac{dx}{2\pi} e^{i x\left(k-l + n \right)} \nn\\
&=& \sum_{k=0}^\infty  \frac{c^k d^{k+n}}{k! (k+n)!} \nn\\
&=& \left(\frac{d}{c}\right)^{n/2} I_n\left(2 \sqrt{c d}\right).
\label{neededint}
\eea
Making use of \eqn{neededint} in \eqn{U(N)integral} gives an
expression for $G_{{\rm U}(N)}(c,d)$ as a Toeplitz
determinant\footnote{A Toeplitz determinant is defined as a determinant of a
matrix whose $(i,j)$-th entry depends only on $j-i$.},
\bea
G_{{\rm U}(N)}(c,d) &=& 
\frac{1}{N!} 
\varepsilon_{i_1 i_2\ldots i_N}\varepsilon_{j_1 j_2\ldots j_N}
\prod_{k=1}^N g_{j_k-i_k}(c,d) \nn\\
&=&
\frac{1}{N!} 
\varepsilon_{i_1 i_2\ldots i_N}\varepsilon_{j_1 j_2\ldots j_N}
 \left(\frac{d}{c}\right)^{\sum_{l=0}^N (i_l - j_l)/2}
\prod_{k=1}^N I_{j_k-i_k}\left(2\sqrt{c d}\right) \nn\\
&=& \det\left[ I_{j-i}\left(2\sqrt{c d}\right)\right]_{1\le i,j\le N}.
\eea 
Here the quantities inside the determinant are to be interpreted as
the $(i,j)$-th entry of an $N\times N$ matrix. 
Now to calculate the corresponding SU($N$) result the restriction $\det U = 1$,
which is equivalent to $\sum_{k=1}^N \phi_k = 0\,{\rm
mod}\, 2\pi$ in terms of the $\phi_k$ variables, must
be built into the integration measure. To do this we follow Brower,
Rossi and Tan~\cite{Brower:1981vt} and incorporate the following delta
function in the integrand of \eqn{Weyl}:
\bea
2\pi \delta\left(\sum_{k=1}^N \phi_k - 0\,{\rm mod}\, 2\pi\right) &=&
\sum_{m=-\infty}^{\infty} 2 \pi \delta\left(\sum_{k=1}^N \phi_k - 2\pi
m\right).
\eea 
This is most conveniently introduced into the integral via its Fourier
transform,
\bea
\sum_{m=-\infty}^{\infty} \exp\left(i m \sum_{k=1}^N \phi_k\right).
\label{fourtran}
\eea
To obtain the SU($N$) integral from the corresponding U($N$) result the
modification is therefore trivial. Including \eqn{fourtran} in the
integrand of \eqn{Weyl} leads to the general SU($N$) result,
\bea
G_{{\rm SU}(N)}(c,d) &=& \int_{{\rm SU}(N)} dU e^{c {\rm Tr} U + d {\rm
Tr} U^\dagger}\nn\\
&=& \sum_{m=-\infty}^{\infty} \det
\left[g_{m+j-i}(c,d)\right]_{1\le i,j\le N}.
\label{simplealmost}
\eea
This expression can be manipulated to factor the $d/c$ dependence out of the
determinant as follows,
\bea
\det \left[\left(\frac{d}{c}\right)^{(l+j-i)/2}I_{l+j-i}\left(2\sqrt{c
d}\right)\right]_{1\le i,j\le N} \hspace{-0.5cm}&=\!\!\!&
\frac{1}{N!}\varepsilon_{i_1 i_2\ldots i_N}\varepsilon_{j_1 j_2\ldots
j_N}\left(\frac{d}{c}\right)^{l N/2 +\sum_k(j_k-i_k)/2}\nn\\
&&\times
\prod_{m=1}^{N}I_{l+j_m-i_m}\left(2\sqrt{c d}\right) \nn\\
&\!\!\!=\!\!\!& \left(\frac{d}{c}\right)^{l N/2} 
\det \left[I_{l+j-i}\left(2\sqrt{c d}\right)\right]_{1\le i,j\le N}.
\eea
Making use of this result in \eqn{simplealmost} leads to the SU($N$)
generating function,
\bea
G_{{\rm SU}(N)}(c,d) 
&=&
\sum_{l=-\infty}^{\infty} \left(\frac{d}{c}\right)^{l N/2}
\det \left[ I_{l+j-i}\left(2\sqrt{cd}\right)\right]_{i\le i,j \le N}.
\label{coolsum}
\eea
For the case of SU(2) we can show that this reduces to the standard
result of Arisue given by \eqn{su2gen}. To do this we need the
recurrence relation for modified Bessel functions of the first kind,
\bea
I_{n-1}(x)-I_{n+1}(x) = \frac{2n}{x} I_n(x).
\label{irecurr}
\eea
Recall that for SU(2) the Mandelstam constraint is $\Tr U = \Tr
U^\dagger$, so the case $G_{{\rm SU}(2)}(c,c)$ can be considered without
loss of generality;
\bea
G_{{\rm SU}(2)}(c,c) &=& \sum_{l=-\infty}^{\infty}
\left[I_l(2c)^2-I_{l-1}(2c)I_{l+1}(2c) \right] \nn\\
&=& I_0(4c) - I_2(4c). 
\eea
Here we have used the standard addition
formula~\cite{Gradshteyn:1994} for modified Bessel functions. Employing the recurrence relation of \eqn{irecurr} gives
\bea
G_{{\rm SU}(2)}(c,c) &=& \frac{1}{2 c} I_1(4 c),
\label{su2simple}
\eea
which is the standard result of Arisue given by \eqn{su2gen}.\\

With an analytic form for SU($N$) in hand we can attempt to find
simpler expressions for $G_{\rm{SU}(N)}(c,d)$ analogous to \eqn{su2simple}.
To our knowledge no general formulas are available for the simplification
of the determinants appearing in \eqn{coolsum}. Without such formulas
we can resort to the crude method of analysing series expansions and
comparing them with known expansions of closed form expressions. Since
the determinants appearing in the generating function are nothing more
than products of modified Bessel functions we expect that if a closed
form expression for the general SU($N$) generating function exists,
it will involve the generalised hypergeometric function. With this
approach we have limited success. The SU(3) result of \eqn{Y} is
recovered numerically but the SU(4) result does simplify analytically.\\

When analysing the series expansion of $G_{\rm{SU}(4)}(c,c)$ we notice that
it takes the form of a generalised hypergeometric function. In
particular we find the following result:  
\bea
G_{{\rm SU}(4)}(c,c) &=& {}_2F_3\left[\begin{array}{c} \frac{3}{2}, \frac{5}{2}
\\ 3 , 4 , 5\end{array}; 16 c^2\right].\label{su4hyper}
\eea
Here the generalised hypergeometric function is defined by
\bea
{}_pF_q\left[ \begin{array}{c} 
a_1, a_2,\ldots ,a_p \\
b_1,b_2,\ldots ,b_q
\end{array}; x\right] = \sum_{k=0}^\infty
\frac{(a_1)_k (a_2)_k\ldots
(a_p)_k}{(b_1)_k (b_2)_k\ldots (b_q)_k} \frac{x^k}{k!},
\eea 
where $(x)_k=x(x+1)\ldots(x+k-1)$ is the rising factorial or 
Pochhammer symbol.
In addition to \eqn{su4hyper} we find the following results for matrix
elements derived from the SU(4) generating function:
\bea
\langle Z_1 \rangle &\!\!\!=\!\!\!& \frac{c\, {}_2F_3\left[\!\begin{array}{c} \frac{5}{2}, \frac{7}{2}
\\ 4 , 5,  6\end{array}; 16 c^2\right]}{ {}_2F_3\left[\!\begin{array}{c} \frac{3}{2}, \frac{5}{2}
\\ 3 , 4 , 5\end{array}; 16 c^2\right]},
\eea
\bea
\hspace{-1cm}\langle Z_1^2 \rangle &\!\!\!=\!\!\!& \frac{\frac{3}{2}c^2 \, {}_2F_3\left[\!\begin{array}{c} \frac{5}{2}, \frac{7}{2}
\\ 5, 6, 7\end{array}; 16 c^2\right] + \frac{2}{3} c^4\, {}_2F_3\left[\!\begin{array}{c} \frac{7}{2}, \frac{9}{2}
\\ 6, 7, 8\end{array}; 16 c^2\right] +\frac{1}{15} c^6\, {}_2F_3\left[\!\begin{array}{c} \frac{9}{2}, \frac{11}{2}
\\ 7, 8, 9\end{array}; 16 c^2\right] }{ {}_2F_3\left[\!\begin{array}{c} \frac{3}{2}, \frac{5}{2}
\\ 3, 4, 5\end{array}; 16 c^2\right]}
\eea
and 
\bea
\hspace{-1cm}\langle Z_1\bar{Z}_1 \rangle &\!\!\!=\!\!\!& \frac{ {}_2F_3\left[\!\begin{array}{c} \frac{3}{2}, \frac{5}{2}
\\ 3, 5, 6\end{array}; 16 c^2\right] + \frac{4}{3} c^2\, {}_2F_3\left[\!\begin{array}{c} \frac{5}{2}, \frac{7}{2}
\\ 4, 6, 7\end{array}; 16 c^2\right] +\frac{4}{9} c^4\, {}_2F_3\left[\!\begin{array}{c} \frac{7}{2}, \frac{9}{2}
\\ 5, 7, 8\end{array}; 16 c^2\right]
}{ {}_2F_3\left[\!\begin{array}{c} \frac{3}{2}, \frac{5}{2}
\\ 3, 4, 5\end{array}; 16 c^2\right]}.
\eea
We stress that these results are nothing more than observations based
on series expansions. Despite some effort analogous expressions for $N>4$ have not been found.\\

The generating functions, $G_{{\rm SU}(N)}(c,d)$  and $G_{{\rm
U}(N)}(c,d)$, are not only of interest in Hamiltonian LGT.  
By differentiating \eqn{coolsum} appropriately with respect to $c$
and $d$ and afterward setting $c$ and $d$ to zero, we obtain the
number of singlets in a given product representation of SU($N$). This
was discussed for the special case of SU(3) in \sect{su3integrals}.  
We now consider the general case in the calculation of
$T_k(n)$; the number of singlets in the SU($k$) product representation,
\bea
\underbrace{(\mathbf{k} \otimes \bar{\mathbf{k}})\otimes \cdots\otimes (\mathbf{k} \otimes \bar{\mathbf{k}})}_{n}.
\eea 
As a group integral $T_k(n)$ is given by
\bea
T_k(n) = \int_{{\rm SU}(k)} d U(|\Tr U |^2  )^n.
\eea
Integrals of this kind are studied in combinatorics, in particular the study of increasing
subsequences of permutations. An increasing subsequence is a sequences
$i_1<i_2 <\cdots < i_m $ such that $\pi(i_1) < \pi(i_2)< \cdots
<\pi(i_m)$, where $\pi$ is a permutation of $\{1,2,\ldots,k\}$. It has been shown that the number of permutations $\pi$ of
$\{1,2,\ldots,k\}$ such that $\pi$ has no increasing subsequence of
length greater than $n$ is $T_k(n)$ \cite{Rains:1998}. 
In addition it is possible to prove that $T_k(n)$ is the
number of pairs of Young tableaux of size $k$ and maximum height
$n$ via the Schensted correspondence~\cite{vanMoerbeke:2001,Rains:1998}. \\

Making use of \eqn{coolsum} we see that only the $l=0$ term
contributes to $T_k(n)$. Letting $x=cd$ we have:
\bea
T_k(n) &=&\left(\frac{\partial^2}{\partial c\partial d}\right)^n \det \left[I_{j-i}(2\sqrt{c
d})\right]_{1\le i,j \le k}\Bigg|_{c=d=0} \nn\\
&=& n! \frac{d^n}{d x^n} \det
\left[I_{j-i}(2\sqrt{x})\right]_{1\le i,j \le k} \Bigg|_{x=0}.
\eea
Hence the generating function for $T_k(n)$ is given by
\bea
\sum_{n=0}^\infty \frac{T_k(n) x^n}{n!^2} &=& \det \left[I_{j-i}(2\sqrt{x})
\right]_{1\le i,j \le k}, 
\eea 
a result first deduced by Gessel~\cite{Gessel:1990}.
The first few $T_k(n)$ sequences  are available as A072131, A072132,
A072133 and A072167 in Sloane's on-line encyclopedia of integer
sequences~\cite{Sloane:OE}.

\subsection{A More Complicated Integral}
\label{morecomplicated}

We now move on to the more complicated integral
\bea
H_m(c,d) &=& \int_{{\rm SU}(N)} dU \exp\left[c ({\rm Tr} U + {\rm Tr}
U^\dagger)+ d{\rm Tr}(U^m)\right]\quad \forall m\in {\Bbb Z}^+ . 
\eea
This integral is of interest as a generating function for the
calculation of integrals such as
\bea
\int_{{\rm SU}(N)} dU {\rm Tr}(U^m) e^{c ({\rm Tr} U + {\rm Tr}U^\dagger)}.
\label{egint}
\eea
For the simple case of SU(3) we can use the Mandelstam constraint, ${\rm Tr}(U^2) =  ({\rm Tr}U)^2-2 {\rm Tr} U^\dagger$, to reduce such
integrals to those obtainable from the generating function of the
previous section. However for higher dimensional gauge
groups not all trace variables can be written in
terms of ${\rm Tr} U$ and  ${\rm Tr} U^\dagger$. For these gauge
groups one must introduce the generating function, $H_m(c,d)$, to
calculate integrals similar to  \eqn{egint}.  \\

To calculate $H_m(c,d)$ we start with the corresponding U($N$)
generating function and follow the procedure of \sect{simpleintegral}
to obtain
\bea
h_m(c,d) &=& \int_{{\rm U}(N)} dU \exp\left[c ({\rm Tr} U + {\rm Tr}
U^\dagger)+ d{\rm Tr}(U^m)\right]\nn\\
&&\hspace{-1.5cm}= \frac{1}{N!} 
\varepsilon_{i_1\ldots i_N}\varepsilon_{j_1\ldots j_N}
\prod_{k=1}^N \int_0^{2
\pi}\frac{d\phi_k}{2\pi}\exp\left[i(j_k-i_k)\phi_k+ 2 c 
\cos\phi_k+ d e^{m i \phi_k}\right].
\label{ungen}
\eea
To proceed we need the following integral,
\bea
\int_{0}^{2\pi} \frac{dx}{2\pi} \exp(i n x + a \cos x + b e^{i m x})
&=&  \sum_{k=0}^{\infty} \frac{b^k}{k!} \int_{0}^{2\pi}
\frac{dx}{2\pi} e^{i (n+m k) x + a \cos x}\nn\\
&=&  \sum_{k=0}^{\infty} \frac{b^k}{k!} I_{n+m k}(a). 
\label{needthis}
\eea
Making use of \eqn{needthis} in \eqn{ungen} leads to the following
expression for the U($N$) generating function, 
\bea
h_m(c,d) &=& \det \left[ \lambda_{m;j-i}(c,d)\right]_{1\le i,j\le N},
\eea
with 
\bea
\lambda_{m;n}(c,d) = \sum_{k=0}^{\infty} \frac{d^k}{k!} I_{n+m k}(2 c).
\label{lam}
\eea
Extending to SU($N$) following the prescription of 
\sect{simpleintegral} 
we arrive at the corresponding SU($N$) generating function,
\bea
H_m(c,d) &=& \sum_{l=-\infty}^{\infty}\det
\left[\lambda_{m;l+j-i}(c,d) \right]_{1\le i,j\le N} .
\label{coolersum}
\eea
An example of an SU($N$) integral derived from this
generating function is the following:
\bea
\int_{{\rm SU}(N)} dU {\rm Tr}(U^m) e^{c ({\rm Tr} U + {\rm Tr}
U^\dagger)} &\!\!\!=\!\!\!& 
\frac{\partial H_m(c,d)}{\partial d}\Bigg|_{d=0} \nn\\
&\!\!\!=\!\!\!& \frac{\partial}{\partial
d}\sum_{l=-\infty}^{\infty}\det\left[I_{l+j-i}(2c)+d I_{l+j-i+m}(2c)\right]\Bigg|_{d=0}.
\label{coolintegral}
\eea
Only two terms need to be kept in the $k$-sum of \eqn{lam} here
because higher order powers of $d$ vanish when the derivative with
respect to $d$ is taken and $d$ set to zero.\\
%

\subsection{Another Integral}
\label{anotherintegral}

Another integral that is calculable using the procedure of
\sect{simpleintegral} is 
\bea
J_m(c,d) &=& \int_{{\rm SU}(N)} dU e^{c {\rm Tr} U^m + d {\rm Tr}
U^{-m}}.
\label{startingpoint}
\eea
We start with the corresponding $U(N)$ integral and again use Weyl's
parameterisation for U($N$) to give 
\bea
j_m(c,d) &\!\!\!=\!\!\!&  \int_{{\rm U}(N)} dU e^{c {\rm Tr} U^m + d {\rm Tr}
U^{-m}} \label{startingpoint-thisone}\\
&&\hspace{-1.5cm}=\frac{1}{N!} 
\varepsilon_{i_1\ldots i_N}\varepsilon_{j_1\ldots j_N}
\prod_{k=1}^N \int_0^{2
\pi}\frac{d\phi_k}{2\pi}\exp\left[i(j_k-i_k)\phi_k+ c e^{m i \phi_k}+
d e^{- m i \phi_k}\right]. 
\label{moreustuff}
\eea
To simplify this further we need the following integral: 
\bea
g_{m;n}(c,d) &=& \int_{0}^{2 \pi} \frac{dx}{2\pi} e^{i n x + c e^{i m x} + d  e^{-i m
x}}  = \sum_{k=0}^\infty  \sum_{l=0}^\infty \frac{c^k d^l}{k!
l!}\int_0^{2\pi} \frac{d x}{2 \pi} e^{i x\left[m(k-l) +n\right]} \nn\\
&=& 
\left\{\begin{array}{lcl}  \displaystyle  \sum_{k=0}^\infty 
\frac{c^k d^{k+ n/m}}{k! \left(k+ n/m\right)!} & & {\rm if}\, m | n
, \\
\displaystyle 0 & & {\rm otherwise}. \end{array}\right. \nn\\
&=& 
\left\{\begin{array}{lcl}  \displaystyle  g_{\frac{n}{m}}(c,d) & & {\rm if}\,  m | n
, \\
\displaystyle 0 & & {\rm otherwise}. \end{array}\right.
\label{ifsandbuts}
\eea
Making use of \eqn{ifsandbuts} in \eqn{moreustuff} leads to 
\bea
j_m(c,d) &=& \det\left[g_{m;j-i}(c,d)\right]_{1\le i,j\le N}.
\label{ungen-1}
\eea
The corresponding SU($N$) result is then 
\bea
J_m(c,d) &=& \sum_{l=\infty}^\infty 
\det\left[g_{m;l+j-i}(c,d)\right]_{1\le i,j\le N}.
\eea
Due to the form of $g_{m;n}(c,d)$ the U($N$) generating function
simplifies significantly for $m \ge N$.
For this case we have, 
\bea
j_m(c,d) &=& \frac{1}{N!} \varepsilon_{i_1\ldots i_N}\varepsilon_{j_1\ldots
j_N} g_{m;j_1-i_1}(c,d)\cdots g_{m;j_N-i_N}(c,d) \nn\\
&=& g_0(c,d)^N,
\label{simplifiedgen}
\eea
since the only non-zero terms in the determinant occur when
$j_l-i_l=0$ for $l=1,\ldots,N$. We can make use of this result in the
calculation of 
\bea
f^{(m)}_{nN} = \int_{{\rm U}(N)} dU \left| \Tr( U^m)^n\right|^2,
\eea 
a quantity which has been studied in the context of combinatorics~\cite{Rains:1998} and random matrices~\cite{Diaconis:1994}.
Making use of our generating function we see that 
\bea
f^{(m)}_{nN} = \left(\frac{\partial^2}{\partial c\partial d}\right)^n \det\left[g_{m;j-i}(c,d)\right]_{1\le i,j\le N}\Bigg|_{c=d=0}.
\eea
For the special case of $ m\ge N$ we can make use
of~\eqn{simplifiedgen} to express $f^{(m)}_{nN} $ as the
sum of squares of multinomial coefficients as follows:
\bea
f^{(m)}_{nN} &=& \left(\frac{\partial^2}{\partial c\partial
d}\right)^n g_0(c,d)^N \Bigg|_{c=d=0}  =
\left(\frac{\partial^2}{\partial c\partial d}\right)^n \left(\sum_{k=0}^\infty \frac{(cd)^k}{k!^2} \right)^N \Bigg|_{c=d=0}\nn\\
&=& 
\left(\frac{\partial^2}{\partial c\partial d}\right)^n \sum_{k=0}^\infty
\sum_{{l_1,\ldots ,l_N \ge 0 \atop l_1+\cdots +
l_N =k}} \frac{(c d)^k}{(l_1!)^2\cdots (l_N!)^2}\Bigg|_{c=d=0} \nn\\
&=& \sum_{{l_1,\ldots ,l_N \ge 0 \atop l_1+\cdots +
l_N =n}} \frac{n!^2 }{(l_1!)^2\cdots (l_N!)^2}.
\eea
With the help of Mathematica we can express
$f^{(m)}_{nN}$ with $ m\ge N$ in terms of sums of binomial
coefficients for  $N = 2,3,4$:
\bea
f^{(m)}_{n,2} &=& \sum_{k=0}^n {n \choose k}^2 = {2n \choose n}
\nn\\
f^{(m)}_{n,3} &=& \sum_{k=0}^n {n \choose k}^2 {2k \choose k} 
\nn\\
f^{(m)}_{n,4} &=& \sum_{k=0}^n {n \choose k}^2 {2n-2k \choose n-k} {2k
\choose k}.
\eea
We have not found similar expressions for $N\ge 5$.\\

A further exercise is to explore the large $N$ limit of the generating
function, $j_m(c,d)$.  Let us expand $j_m(c,d)$, in the form of \eqn{startingpoint-thisone}, in $c$ and $d$ about
$c=d=0$:
\bea
j_m(c,d) &=& \sum_{k,l\ge 0}\frac{c^k d^l}{k! l!} \int_{{\rm U}(N)} dU \Big[{\rm Tr}(U^m)\Big]^k
\left[\overline{{\rm Tr}(U^m)}\right]^l .
\label{expansion}
\eea
A result due to Diaconis and Shahshahani~\cite{Diaconis:1994},
\bea
\int_{{\rm U}(N)} dU
\Big[\Tr( U^m)\Big]^k \Big[\overline{\Tr( U^m)}\Big]^l &=& \delta_{kl} k! m^k \quad \forall
N\ge m k ,
\label{diashah}
\eea
then allows us to access the large $N$ limit.
In the $N\rightarrow\infty$ limit, for each term of the sum in
\eqn{expansion} we have $N \ge m k$ and hence, using \eqn{diashah}, we have,
\bea
j_m(c,d) \xrightarrow[N\rightarrow\infty]{} \sum_{k=0}^\infty
\frac{(cd)^k}{(k!)^2} k! m^k = e^{m c d}. 
\eea
This result is consistent with the $c=d$ and $m=1$ case proved
in \rcite{Baik:2001},
\bea
\lim_{N\rightarrow\infty}\det\left[ I_{i-j}(2 c)\right]_{1\le i,j \le
N} = e^{c^2},
\eea
which is valid for all real $c\ge 0$.\\

For the case of $N = m_0 m$, where $m_0\in {\Bbb N}$, we also find
an interesting simplification for $j_m(c,d)$. In this case the
determinant appearing in \eqn{ungen-1} becomes a block
determinant, which leads to the simplification, 
\bea
\det \left[ g_{m;j-i}(c,d) \right]_{1\le i,j,\le m_0 m} &=& 
\det \left[g_{j-i}(c,d) \right]_{1\le i,j\le m_0}.
\eea
This result is equivalent to $j_m(c,d)\Big|_{N=m_0 m} =
j_1(c,d)\Big|_{N=m_0}$.

\section{Conclusion} 
\label{analyticsummary}

In this chapter we have calculated a number of SU($N$) group
integrals. As generating
functions they allow the analytic calculation of
all expectation values that appear in variational calculations of
SU($N$) glueball masses in 2+1 dimensional Hamiltonian LGT. Our
derivations have made use of techniques for the calculation of U($N$)
integrals which have become standard
tools in the fields of random matrices and combinatorics. The
resulting generating functions generalise the analytic results
that have been available for SU(2) for many years to the general case
of SU($N$).\\

The extension of the techniques presented here to 3+1 dimensions is 
highly non-trivial. The
starting point for all derivations presented in this chapter was the
fact that in 2+1 dimensions the transformation from link to plaquette
variables has unit Jacobian. This allows the matrix elements of
functions of plaquette variables to be calculated analytically by
differentiating a generating function appropriately. 
The viability of applying analytic techniques in 3+1 dimensions is
discussed in \chap{3+1dimensions}.\\

Having derived analytic results for a selection of SU($N$) generating 
functions, in the next chapter we apply them in a variational
calculation of SU($N$)
glueball masses, with $N=2$, 3, 4 and 5, in 2+1 dimensions. The
calculations make use of the simplest classically improved and
tadpole improved Hamiltonians derived in \chap{constructingandimproving} as well as the standard Kogut-Susskind
Hamiltonian. 

\chapter[SU($\bm{N}$) Massgaps in 2+1 Dimensions]{SU($\bm{N}$) Massgaps in 2+1 Dimensions}
\label{sunmassgaps}
\section{Introduction}

In this chapter we calculate the lowest lying glueball masses, or
massgaps, for SU($N$) LGT in 2+1 dimensions, with $N=2$, 3, 4 and 5. 
We use the  
Kogut-Susskind, classically improved and tadpole improved Hamiltonians 
in their calculation and rely heavily upon the
results of the previous chapter for the calculation of the required 
expectation values.\\

The outline of this chapter is as follows. After defining our notation
in \sect{prelims} we fix the variational vacuum wave function
in \sect{fixingthevariational}. Following that we calculate
lattice specific heats in \sect{specheats-1} before studying 
SU($N$) glueball masses in \sect{massgaps}. In \sect{chap5-results} we
present our results and compare them to similar calculations in both
the Lagrangian and Hamiltonian formulations. \sect{concl} contains a 
discussion of further work and our conclusions.
    
\section{Preliminaries}
\label{prelims}

Before fixing the variational ground state we introduce the following
convenient notation.
We define the general order $a^2$ improved lattice Hamiltonian 
for pure SU($N$) gauge theory, 
\bea
\tilde{\LH}(\kappa,u_0) &=& \frac{g^2}{a}\sum_{\boldx,i} \Tr\left[(1-\kappa)\LE_i(\boldx)^2 + \frac{\kappa}{u_0^2} \LE_i(\boldx) U_i(\boldx) \LE_i(\boldx+a \boldsymbol{i}) U^\dagger_i(\boldx)\right]\nn\\
&&  + \frac{2N}{a g^2} \sum_{\boldx, i<j}\left\{(1+4\kappa) P_{ij}(\boldx) -  
\frac{\kappa}{2} \left[R_{ij}(\boldx)+R_{ji}(\boldx)\right]\right\},
\label{genham}
\eea
where the plaquette and rectangle operators are given in
\eqn{tadplaqrect}. The simplest Hamiltonians derived in
\chap{constructingandimproving} can be expressed 
in terms of $\tilde{\LH}$ as
follows. The Kogut-Susskind (\eqn{KSdef}) and ${\cal O}(a^2)$ classically 
improved (\eqn{classimpham}) Hamiltonians are given by $\tilde{\LH}(0,1)$ 
and $\tilde{\LH}(1/6,1)$ respectively.   
The tadpole improved Hamiltonian (\eqn{tadimp}) 
is given by $\tilde{\LH}(1/6,u_0)$,
where $u_0$ is defined self-consistently as a function of $\beta = N/g^2$ as
described in \sect{tadpoleimprovement-1}. \\

With this new notation the vacuum energy density is given by
\bea
\epsilon_0 &=& \frac{a}{N_p} \langle \tilde{\LH} \rangle \nn\\
&=& \left[(1-\kappa)\left(\frac{N^2-1}{2\beta}\right) c - \frac{2\beta(1+4\kappa)}
{Nu_0^4}
\right]\left\langle \plaquette \right\rangle \nn\\
&&\hspace{2cm}+\frac{2\kappa \beta}{N u_0^6}
\left\langle \rectangleA\right\rangle + 2(1+3 \kappa)\beta,
\label{epsilon}
\eea
where the expectation values, as usual, are taken with respect to the 
one-plaquette
trial state defined in \eqn{oneplaquette}. The variational parameter,
$c$, is fixed as a function of
$\beta$ by minimising the vacuum energy density. For the calculation
of the expectation values we use the generating functions of
\chap{analytictechniques} with all infinite sums truncated. 
Once the variational parameter is fixed the trial state is completely
defined as a function of $\beta$.

\section{Fixing the Variational Trial State}
\label{fixingthevariational}
\subsection{Introduction}
In this section we fix the SU($N$) trial state for $2\le N\le 5$, 
making use of the generating functions derived in
\chap{analytictechniques}. These generating functions allow the
analytic calculation of the plaquette and rectangle expectation values
appearing in \eqn{epsilon}. The approach
we take follows \sect{prelims-3}. For the Kogut-Susskind and
classically improved cases, we simply minimise $\epsilon_0$ for a given
value of $\beta$. The value of $c$ at which $\epsilon_0$ takes its minimum
defines $c$ as a function of $\beta$. The tadpole improved calculation
is more complicated because the mean plaquette, from which
$u_0$ is calculated, depends on the variational state, $|\phi_0\rangle$. 
To complicate
matters, the variational state in turn depends on the energy density
and hence $u_0$. For this case we make use of the iterative technique
procedure introduced in \sect{tadpoleimprovement-1}.\\

\subsection{Results}
\begin{figure}
\centering
\subfigure[SU(2)] 
                     {
                         \label{esu2}
                         \includegraphics[width=7cm]{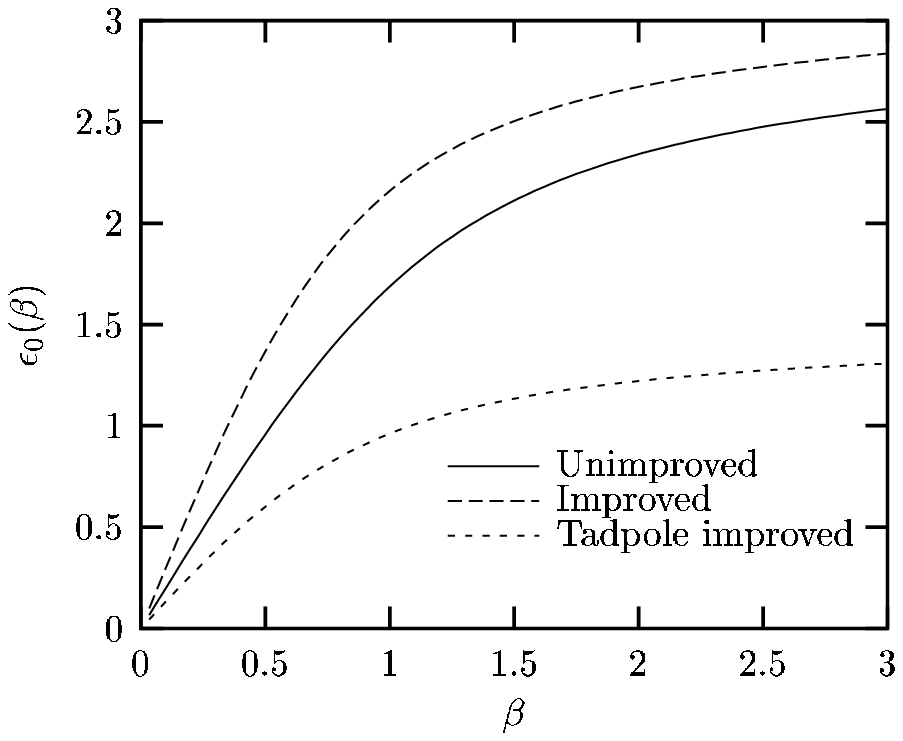}
                     }                   
 \subfigure[SU(3)] 
                     {
                         \label{esu3}
                         \includegraphics[width=7cm]{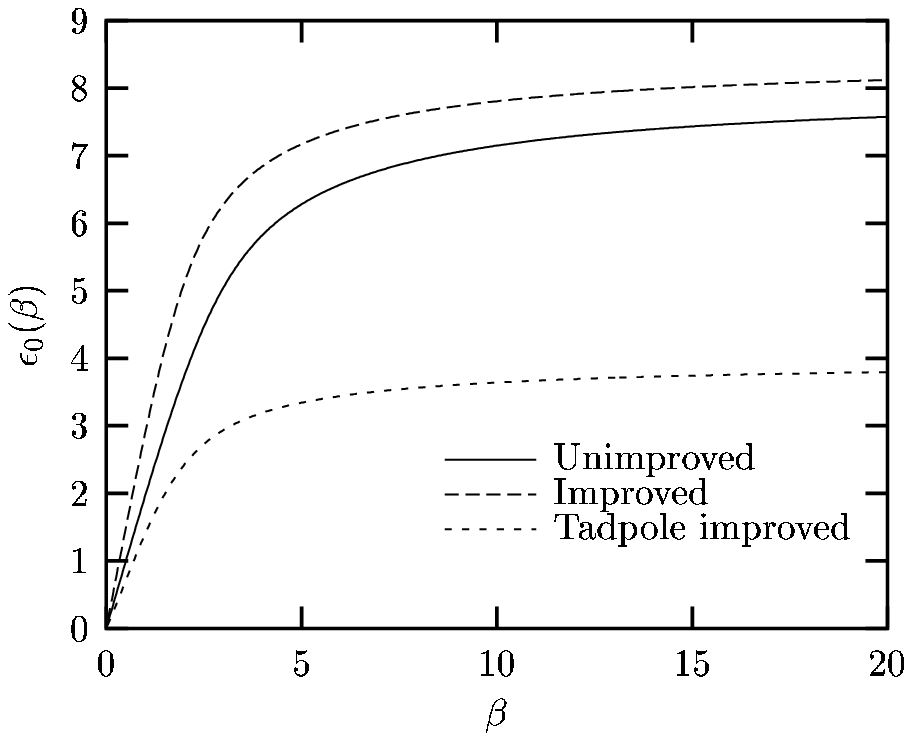}
                     }\\
        \subfigure[SU(4)] 
                     {
                         \label{esu4}
                         \includegraphics[width=7cm]{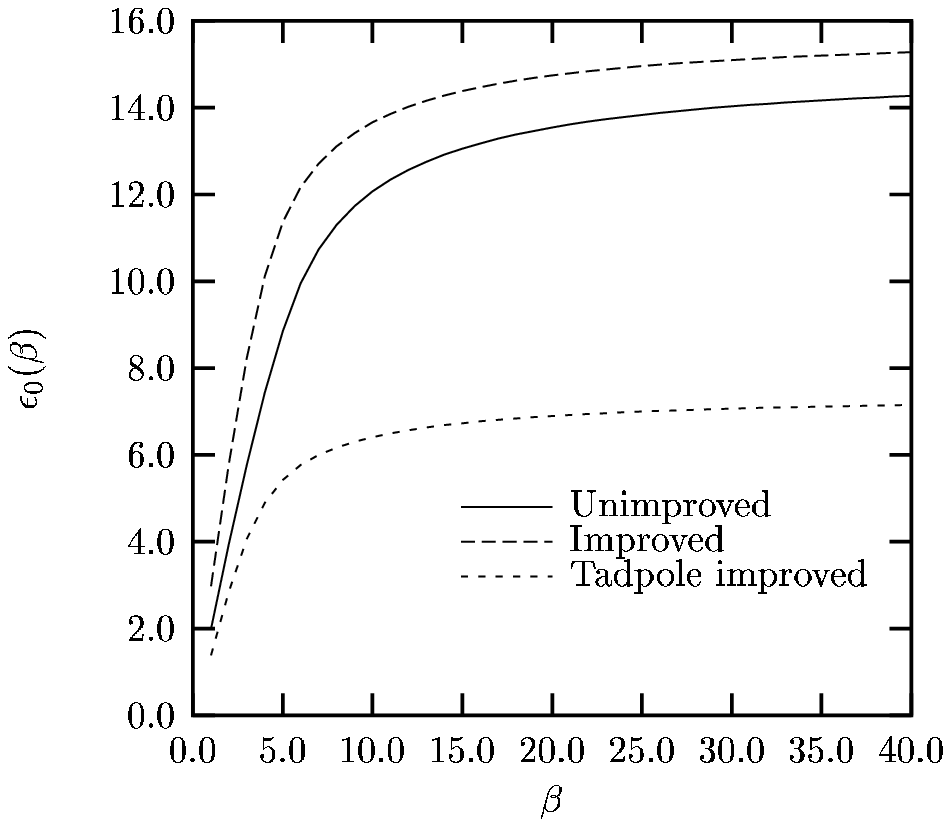}
                     }
\subfigure[SU(5)] 
                     {
                         \label{esu5}
                         \includegraphics[width=7cm]{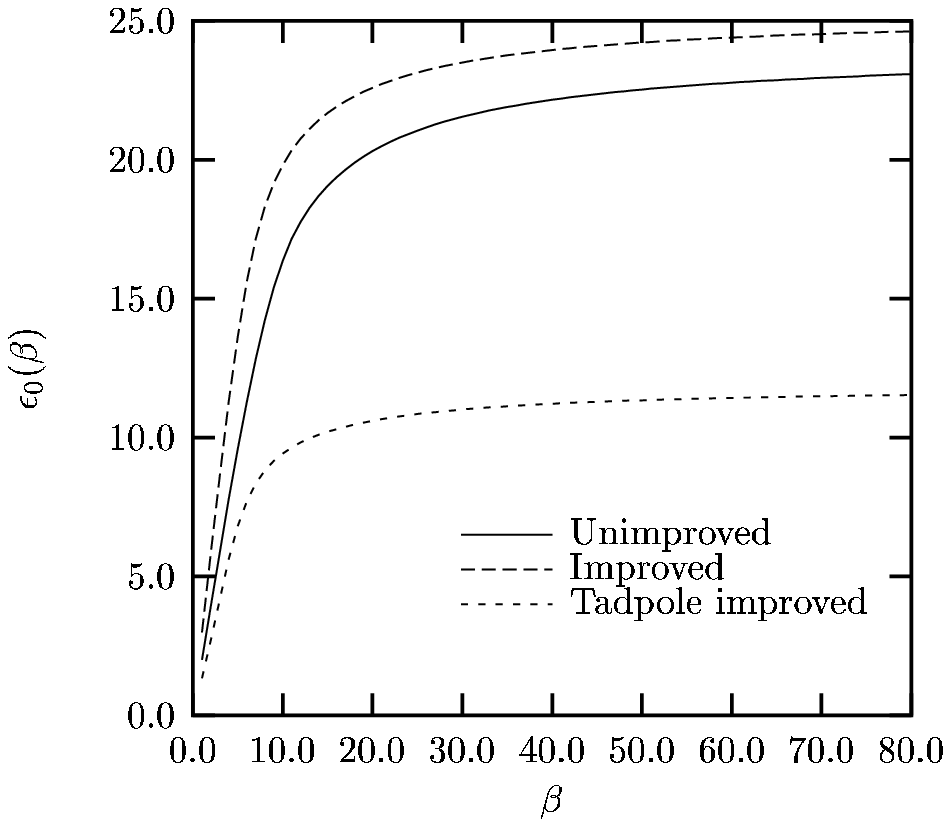}
                     }
\caption{Analytic calculation of the 2+1 dimensional 
unimproved, improved and tadpole
                         improved vacuum energy density in units of
                         $1/(a N_p)$ for SU(2), SU(3), SU(4) and SU(5).}
\label{edens}
\end{figure}

\begin{figure}
\centering
\subfigure[SU(2)] 
                     {
                         \label{csu2}
                       \includegraphics[width=7cm]{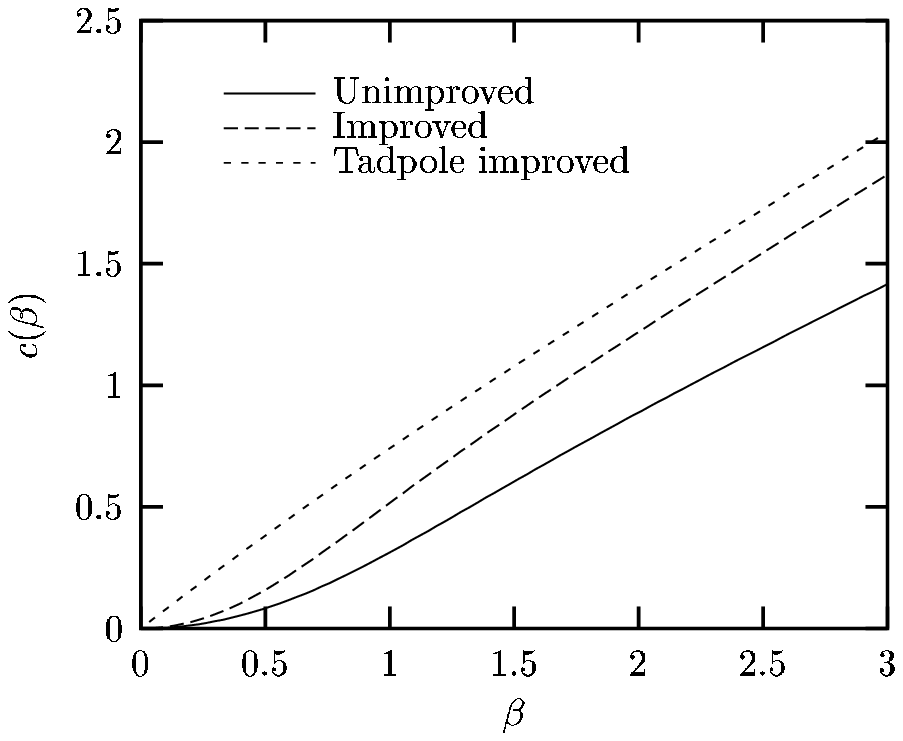}
                     } \hspace{0.25cm}                   
 \subfigure[SU(3)] 
                     {
                         \label{csu3}
                         \includegraphics[width=7cm]{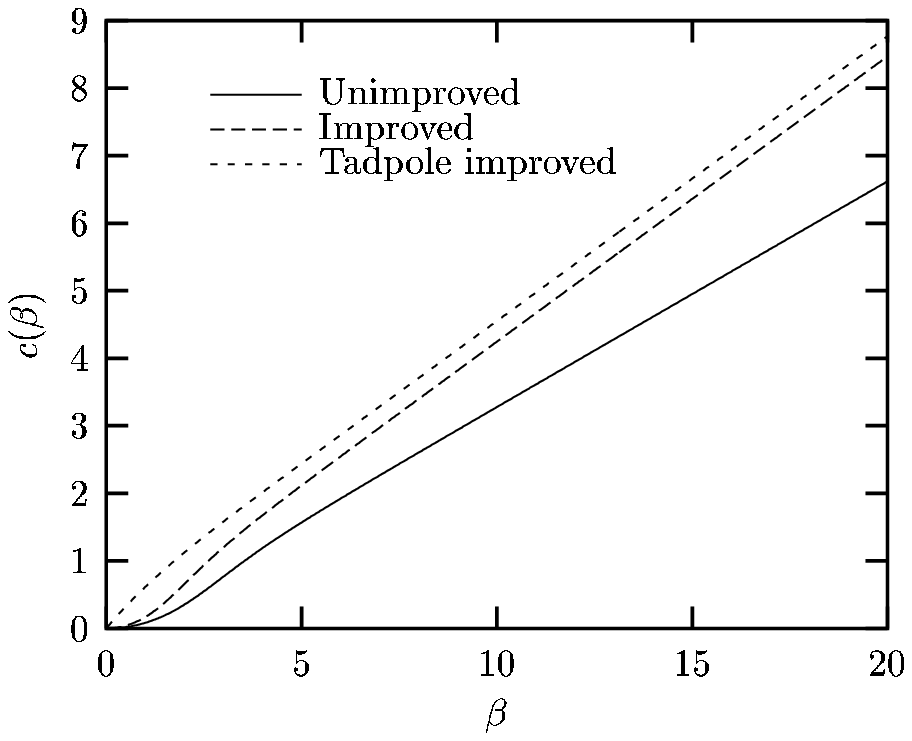}
                     }
\subfigure[SU(4)] 
                     {
                         \label{csu2}
                       \includegraphics[width=7cm]{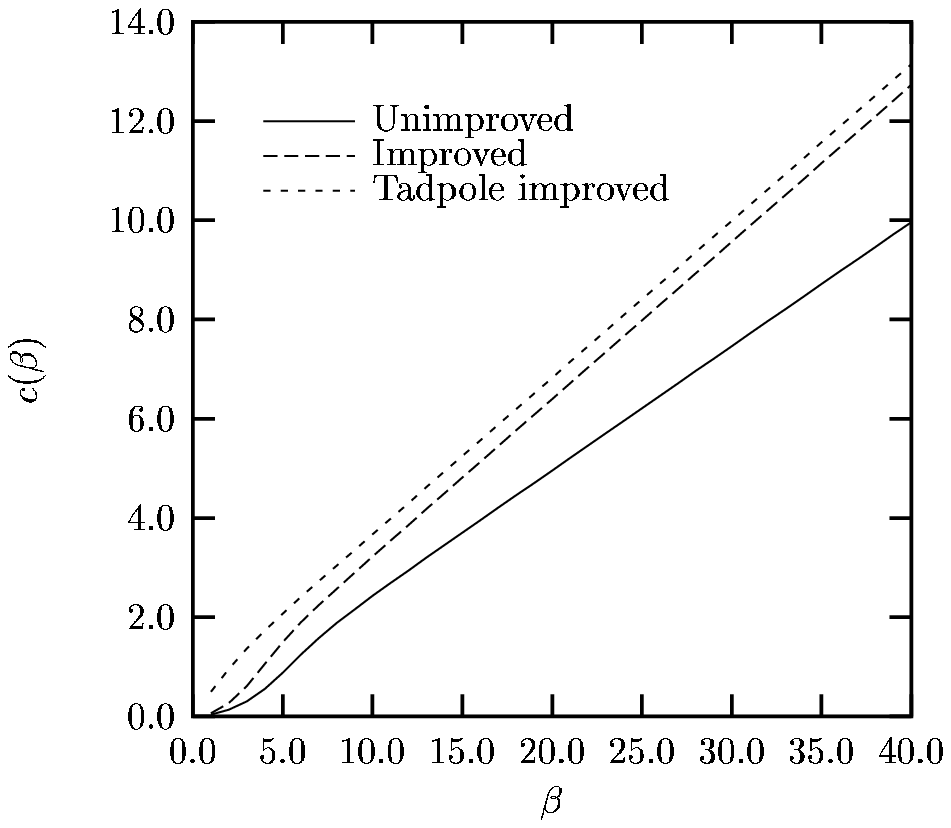}
                     } \hspace{0.25cm}                   
 \subfigure[SU(5)] 
                     {
                         \label{csu3}
                         \includegraphics[width=7cm]{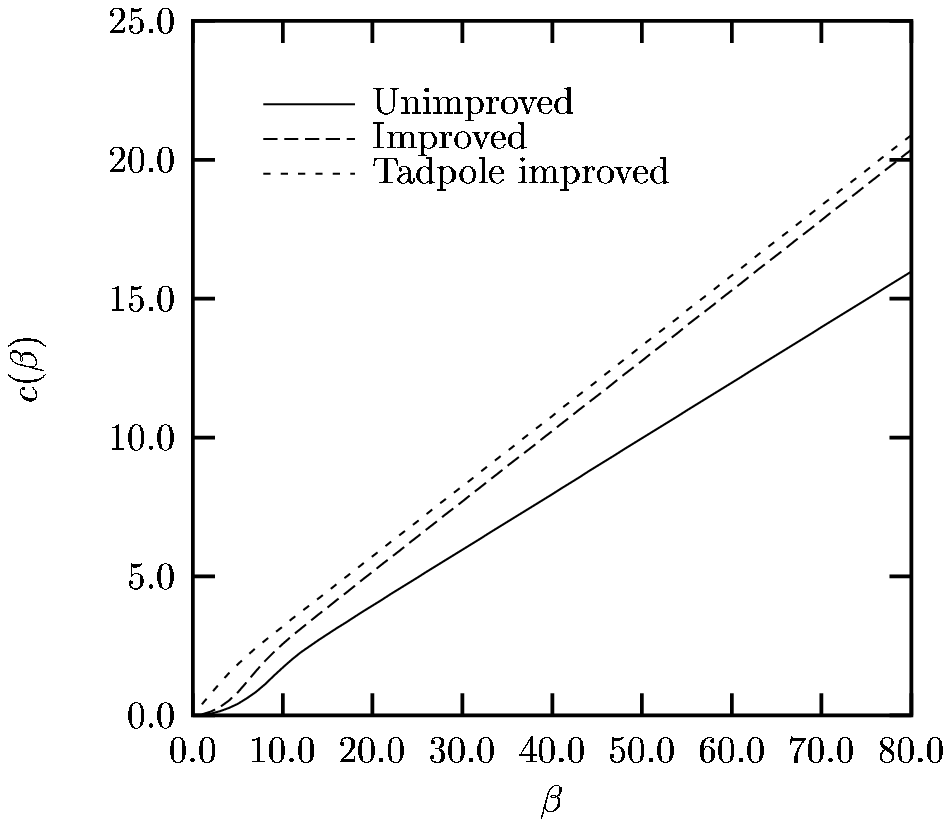}
                     }
\caption{Analytic calculation of the unimproved, improved and tadpole
                         improved variational parameter in 2+1
                         dimensions for SU(2), SU(3), SU(4) and SU(5).}
\label{cofbeta}
\end{figure}

The
results of the Kogut-Susskind, classically improved and tadpole
improved  SU(2), SU(3), SU(4) and SU(5) vacuum energy density calculations are shown
in \fig{edens}. The corresponding variational parameters, $c(\beta)$, 
 are shown in \fig{cofbeta}. For SU(3) the generating function of
\eqn{Y} is used to calculate the required plaquette and rectangle
expectation values.  The generating function
of \eqn{coolsum} is used for SU(4) and SU(5).\\

The familiar strong and weak coupling behavior from variational
calculations is observed in each case.
The differing gradients for the improved and Kogut-Susskind
variational parameters for a given $N$ in the weak coupling limit highlight the fact that when using an improved Hamiltonian one is using a different renormalisation scheme to the unimproved case.\\


\subsection{Dependence on Truncation}
\label{dependenceontruncation}
\begin{figure}
\centering

\includegraphics[width=10cm]{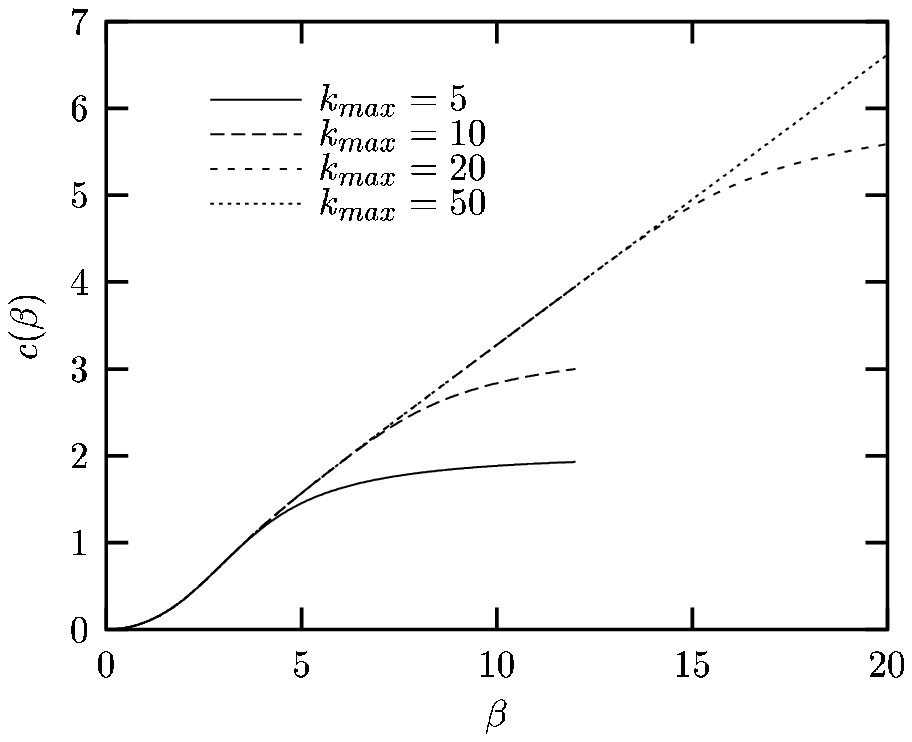}

\caption{Analytic calculation of the unimproved SU(3) variational parameter in 2+1 dimensions, truncating the $k$-sum of ${\cal Y}(c,d)$ at $k=k_{max}$.} 
\label{cconvergence}  
\end{figure}

In practice the $k$-sum appearing in the SU(3) generating function (\eqn{Y}) is
truncated at a maximum value, $k_{max}$.  The dependence of the variational parameter on various truncations of the $k$-sum 
is shown in \fig{cconvergence}. We see that convergence is achieved up
to $\beta \approx 13$ when keeping 20 terms. Further calculations show
that when keeping 50 terms convergence up to $\beta \approx 30$ is
achieved.\\

\begin{figure}
\centering

\subfigure[SU(3)] 
                     {
                         \label{detmaxsu3}
                       \includegraphics[width=7cm]{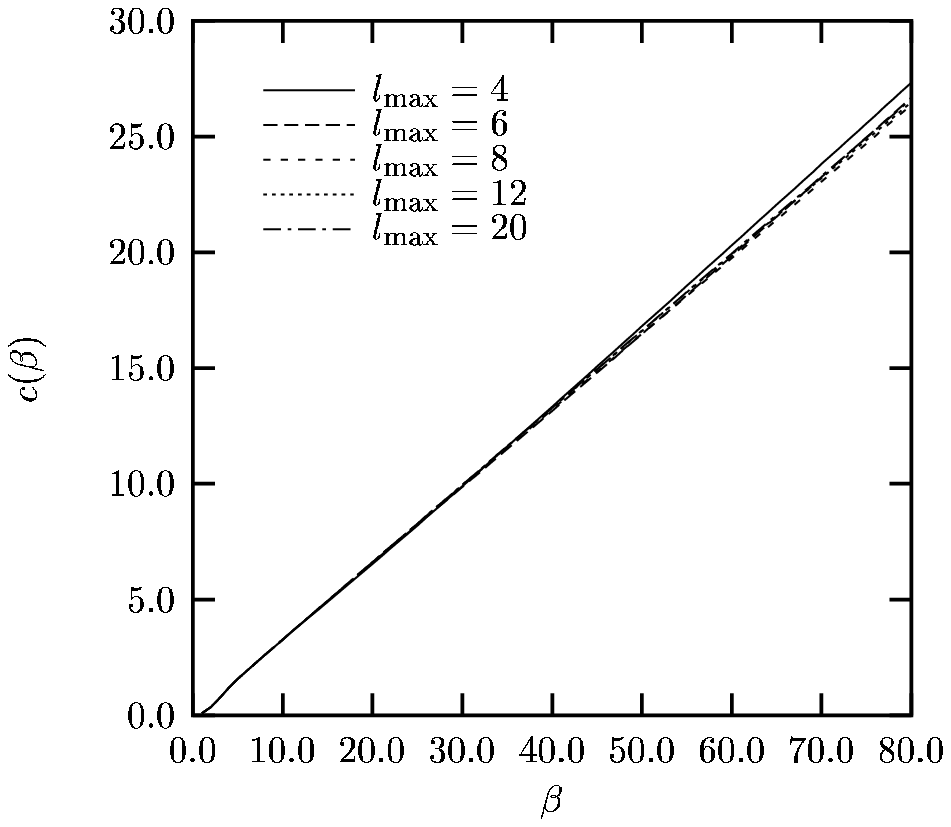}
                     } \hspace{0.25cm}                   
\subfigure[SU(4)] 
                     {
                         \label{detmaxsu4}
                         \includegraphics[width=7cm]{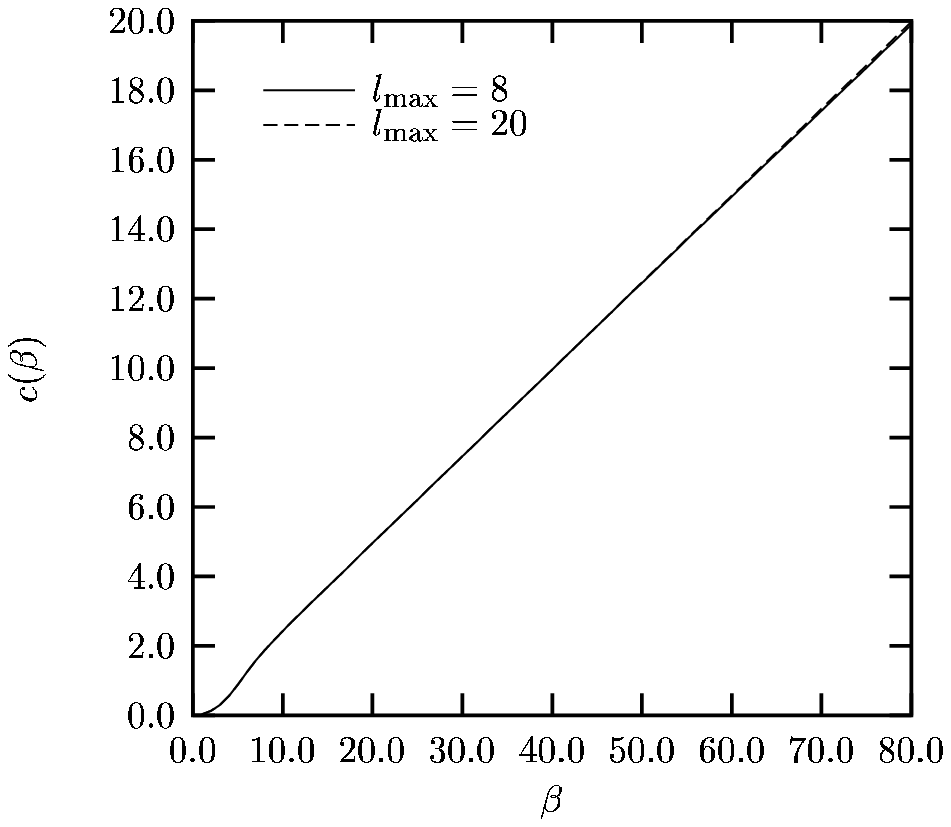}
                     }

\caption{Analytic calculation of the unimproved SU(3) and SU(4)
                         variational parameters in 2+1 dimensions,
                         truncating the $l$-sum of \eqn{coolsum} at
                         $l=\pm l_{\rm max}$.} 
\label{detmaxconvergence}  
\end{figure}

The $l$-sum appearing in the general SU($N$) generating function of
\eqn{coolsum} is also truncated in practice. We replace the infinite
sum over $l$ by a sum from $-l_{\rm max}$ to $l_{\rm max}$.  The dependence of
the SU(3) and SU(4) variational parameters on $l_{\rm max}$ is shown
in \fig{detmaxconvergence}. From the graphs we see that convergence is achieved
quickly as $l_{\rm max}$ increases for both SU(3) and SU(4). The
results for $l_{\rm max}\ge 8$ are barely distinguishable up to $\beta =80$ with the scale used in the plots.  

\section{Lattice Specific Heat}
\label{specheats-1}
\begin{figure}
\centering
\subfigure[SU(2)] 
                     {
                         \label{cvsu2}
                         \includegraphics[width=7cm]{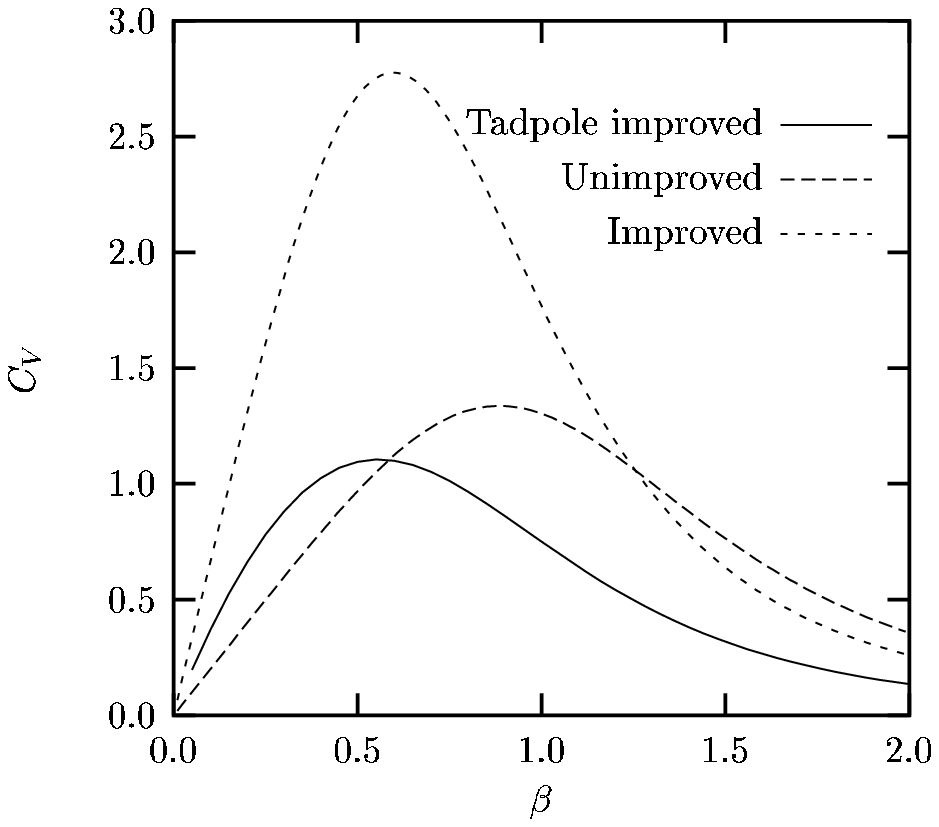}
                     }                   
 \subfigure[SU(3)] 
                     {
                         \label{cvsu3}
                         \includegraphics[width=7cm]{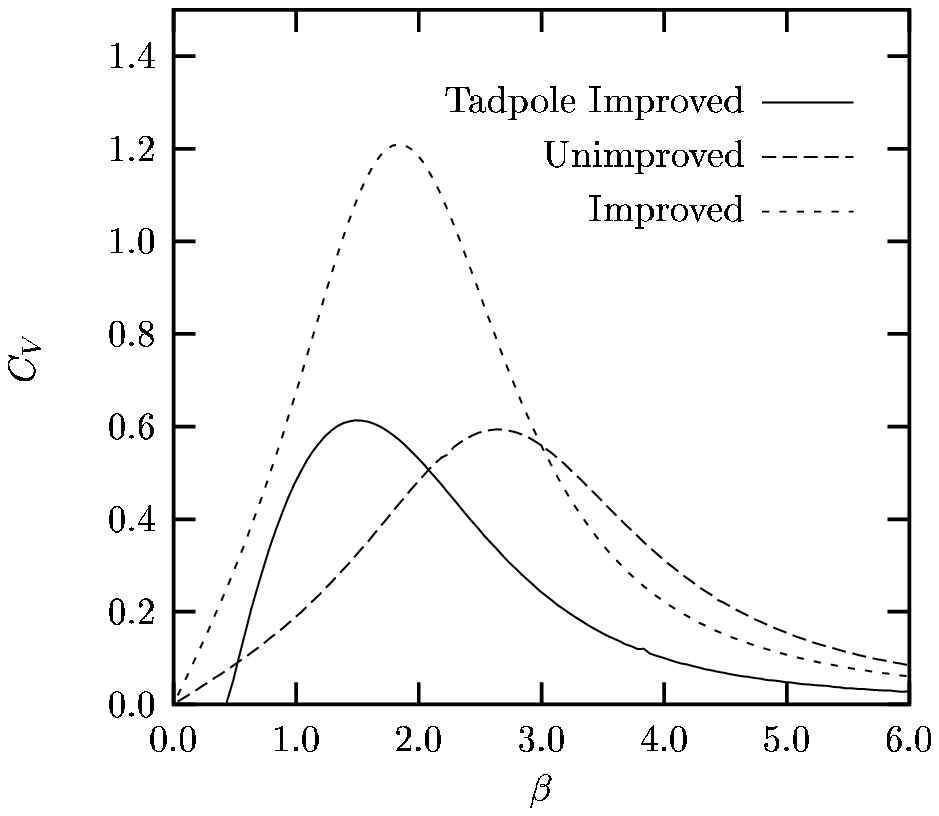}
                     } \\
\subfigure[SU(4)] 
                     {
                         \label{cvsu4}
                         \includegraphics[width=7cm]{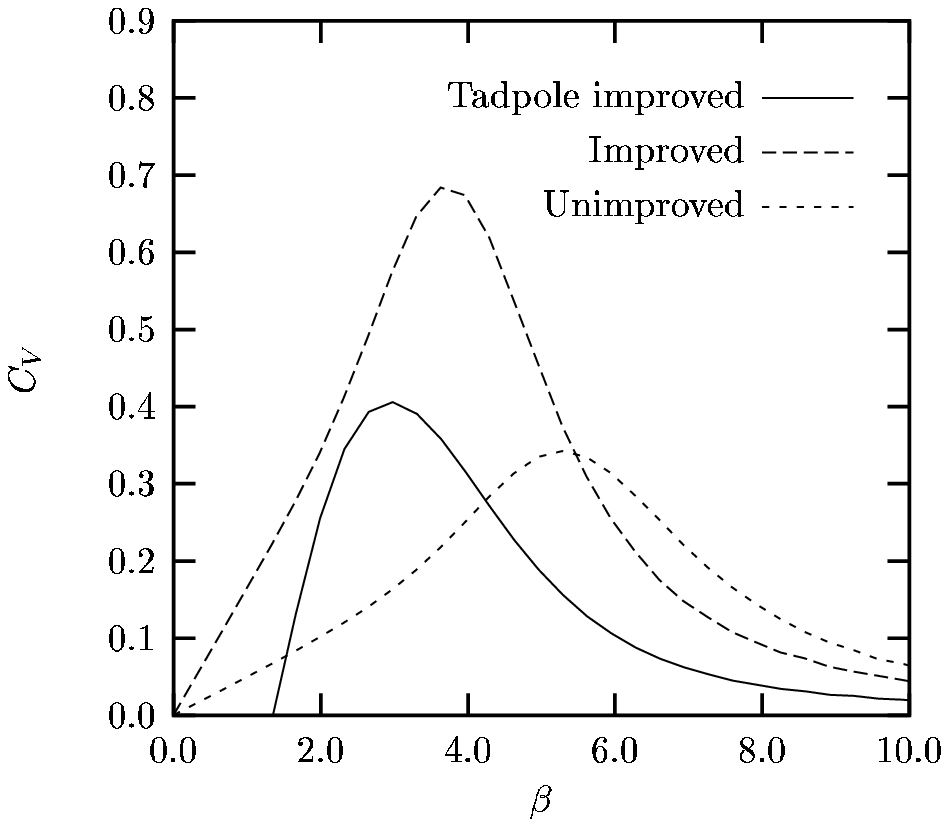}
                     }                   
 \subfigure[SU(5)] 
                     {
                         \label{cvsu5}
                         \includegraphics[width=7cm]{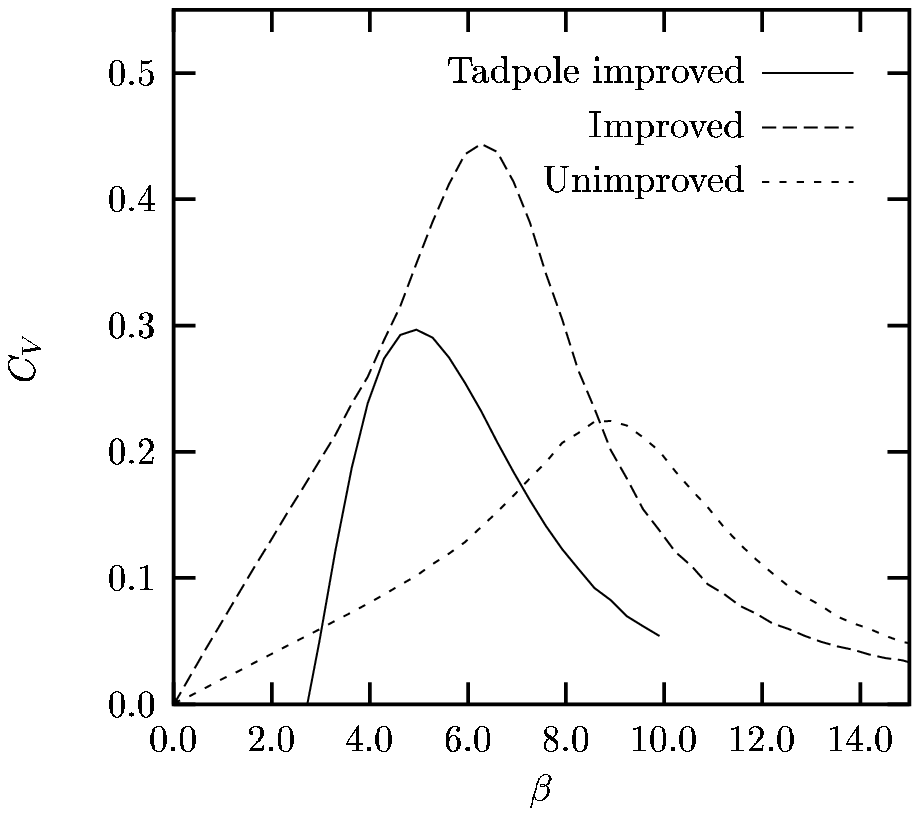}
                     }
\caption{The unimproved, improved and tadpole improved lattice specific heat
                         in 2+1 dimensions for SU(2), SU(3), SU(4) and
                         SU(5).}
\label{cv}
\end{figure}

In addition to the vacuum energy density we can also calculate the 
lattice specific heat defined in \eqn{latticespecificheat}.
The results for SU(2), SU(3), SU(4) and SU(5) are shown in
\fig{cv}. The SU(2) and SU(3) results are calculated with the aid of
\eqns{su2gen}{Y} with the $k$-sum of \eqn{Y} truncated at $k_{max}=50$.  The SU(4) and SU(5) results are obtained using
\eqn{coolsum} to calculate the required matrix elements. For these
cases the infinite
$l$-sum is truncated at $l_{\rm max}=4$, for which the generating function
has converged on the range of couplings used. We recall from 
\chap{testingimprovement} that the location of the peak
indicates the region of transition from strong to weak coupling. For an improved
calculation one would expect the peak to be located at a smaller
$\beta$ (corresponding to a larger coupling) than for the equivalent
unimproved calculation. We see that this is indeed the case for each
example, with the tadpole improved Hamiltonian demonstrating the largest degree of improvement.    

\section{Massgaps}
\label{massgaps}

\subsection{Introduction}
Having fixed the one-plaquette vacuum wave function, in this section we turn to investigating excited states.
Our aim is to calculate the lowest lying energy eigenstates
of the Hamiltonians described by \eqn{genham} for SU($N$) with $2\le N \le 5$. \\

We follow Arisue~\cite{Arisue:1990wv} and expand the excited state
$|\phi_1\rangle$ in the basis consisting of all $n\times m$ rectangular Wilson
loops $\{|n,m\rangle\}_{n,m=1}^{L_{max}}=
\{|l\rangle\}_{l=1}^{L_{max}^2}$ that fit in a given square 
whose side length $L_{max}$
defines the order of the calculation. Enumerating the possible
overlaps between rectangular loops is relatively simple and so a basis
consisting of rectangular loops is an ideal starting point. However,
for an accurate picture of the glueball spectrum we will need to
extend the rectangular basis to include additional smaller area
loops. Without such small area nonrectangular loops, it is possible
that some of the lowest mass states will not appear in the variational
calculation described here.  In order to ensure the orthogonality of $|\phi_0\rangle$ and $|\phi_1\rangle$ we parameterise the excited state as follows
\bea
|\phi_1\rangle &=& \sum_{n,m=1}^{L_{max}} s_{n,m}|n,m\rangle= \sum_{l=1}^{L_{max}^2} s_{l}|l\rangle, 
\eea
with,
\bea
 |l\rangle &=&  \sum_{\boldx} \left[W_l(\boldx) - \langle W_l(\boldx) 
\rangle\right]|\phi_0\rangle.
\eea
Here $\langle W_l(\boldx) \rangle$ is the expectation value of
$W_l(\boldx)$ with respect to the ground state $|\phi_0\rangle$ and
the convenient label $l=(n-1)L_{max}+m$ has been defined to label the
$n\times m$ rectangular state, $|n,m\rangle$.  We define the particular form of
$W_l(\boldx)$ to reflect the symmetry of the sector we wish to
consider.  For SU($N$) we take $W_l(\boldx)= \Tr[w_l(\boldx)\pm
w^\dagger_l(\boldx)]$ for the symmetric ($0^{++}$) and antisymmetric
($0^{--}$) sectors.  To avoid over-decorated equations, the particular 
$W_{l}(\boldx)$ in use is to be deduced from the context.
Here $w_{l}(\boldx)$ is the rectangular Wilson loop joining
the lattice sites $\boldx$, $\boldx+na\boldsymbol{i}$,
$\boldx+na\boldsymbol{i}+ma\boldsymbol{j}$ and
$\boldx+na\boldsymbol{j}$, with
\bea
n = \left[ \frac{l-1}{L_{max}}\right]+1 \quad{\rm and}\quad m = l- L_{max}\left[ \frac{l-1}{L_{max}}\right] .
\eea
Here $[k]$ denotes the integer part of $k$.
In order to calculate excited state energies we minimise the massgap (the difference between the excited state and ground state energies) over the basis defined by a particular order $L_{max}$. To do this we again follow Arisue~\cite{Arisue:1990wv} and define the matrices
\bea
N_{l l'} &=& \frac{1}{N_p}\langle l|\tilde{H} - E_0|l'\rangle,
\label{Nl'l}
\eea
where $E_0$ is the ground state energy, and
\bea
D_{l l'} &=& \frac{1}{N_p}\langle l|l'\rangle 
= \sum_{\boldx}\left[\langle W^\dagger_l(\boldx) W_{l'}(\boldsymbol{0})
\rangle  - \langle W_l(\boldx) \rangle^\ast 
\langle W_{l'}(\boldsymbol{0}) \rangle\right].     
\label{d}
\eea 
Extending the calculation to the general improved Hamiltonian $\tilde{H}(\kappa,u_0)$ and making use of \eqnss{com1}{com4} from Appendix~\ref{commutationrelations} gives
\bea
N_{l l'} &=& -\frac{g^2}{2a}\sum_{i,\boldx}\sum_{\boldx'}\Bigg\{
(1-\kappa) \left\langle \left[E^\alpha_i(\boldx),W^\dagger_{l}(\boldx') \right]\left[E^\alpha_i(\boldx),W_{l'}(\boldsymbol{0})\right]\right\rangle \nn\\
&& \hspace{2cm}+ \frac{\kappa}{u_0^2}\left\langle \left[E^\alpha_i(\boldx),W^\dagger_{l}(\boldx')\right]\left[\tilde{E}^\alpha_i(\boldx+a\boldsymbol{i}),W_{l'}(\boldsymbol{0})\right]\right\rangle\Bigg\}.
\label{N}
\eea 
To minimise the massgap over a basis of a given order we make use of following
 diagonalisation technique~\cite{Tonkin:1987nh}. We first diagonalise $D$, with 
\bea
S^\dagger D S = V^{2}, \label{diag}
\eea
where $V$ is diagonal.
The $n$-th lowest eigenvalue of the modified Hamiltonian,
\bea
H' = V S^\dagger N S V , 
\eea  
then gives an estimate for the massgap corresponding to the 
$n$-th lowest eigenvalue of the Hamiltonian, $\Delta m_n$.\\

\subsection{Classification of States}
\label{classificationofstates}

States constructed from only gluon degrees of freedom can be
classified in the continuum by their $J^{PC}$ quantum numbers. In
\sect{choosingawavefunction} we discussed the assignment of $P$
and $C$ quantum  numbers. In this section we discuss the important
topic of building states with particular continuum spins on the
lattice. Difficulties arise when the continuous rotation group of the
continuum is broken down to the group of lattice rotations. The most
serious difficulty to arise is an ambiguity in the assignment of
continuum spins to states built from lattice operators. 
It is this ambiguity that we
discuss in what follows.\\ 

Let us start with an eigenstate, $|\phi\rangle $, of the Hamiltonian. In
the continuum, following standard quantum mechanics, one can construct
a state with spin $J$ by taking the superposition
\bea
| \phi_J \rangle = \int_0^{2 \pi} e^{i J \theta} {\cal R}_{\theta} |\phi \rangle ,
\label{spinstate}
\eea 
where  ${\cal R}_{\theta}$ is the operator which rotates by
an angle $\theta $. This follows from the fact that $\hat{J}$ is the generator of
rotations. On a square lattice continuous rotations are not
available. All rotations must be modulo $\pi/2$ and so the lattice analogue of \eqn{spinstate} is
\bea
| \phi_J \rangle = \sum_{n=0}^3 e^{\frac{i J n \pi}{2}} {\cal R}_{\frac{n \pi}{2}} |\phi \rangle.
\label{spinlattice}
\eea
Here the state labelled $| \phi_J \rangle $ is no longer a pure spin $J$
state. It also includes states with spins $J\pm 4, J\pm 8, \ldots
$ since these spins produce the same phases as $J$ on the square
lattice as shown in \eqn{spinlattice}. To clarify this point continuum
states with spin $J\pm 4, J\pm 8, \ldots
$ will couple to the lattice state labelled $J=0$. Thus we need to be careful when
using continuum terminology to label lattice wave functions.\\

To give a specific example, suppose we construct a wave function, $|\phi \rangle$, on the lattice  
 with lattice spin $J=0$; a
state built from Wilson loops which are unchanged by rotations of 
$n \pi/2$ for all integers $n$. As explained above, this state  is not
a pure $J=0$ state; it also contains $J=4,8,\ldots $ states. Using a
variational approach we can obtain estimates of the lowest energy
 eigenvalues of states with lattice spin $J=0$. When the continuum
limit is taken we obtain estimates of the lowest continuum energy
eigenvalues for the states with spin $0,4,8,\ldots$. \\

In the Lagrangian approach it is possible to suppress the unwanted
spin $J\pm 4,J\pm 8,\ldots $ states in a given spin $J$ calculation. By
``smearing'' links, one can confidently construct lattice states which
do not couple with the unwanted higher spin continuum states, at least for the
lowest energy eigenvalues~\cite{Teper:1998te}. The technique of
``smearing'' links has not to our knowledge been applied in the
Hamiltonian approach. \\

 Another way to clarify ambiguities in spin assignment 
is to attempt to construct a $J=4$ state on the lattice, devoid
of $J=0$ contributions, and similarly a $J=0$ state, devoid of unwanted
$J=4$ contributions. The construction of an exact $J=4$ state is
impossible on a square lattice due to the unavailability of $\pi/4$
rotations. One can however attempt to construct states that are
approximately symmetric under
rotations by $\pi/4$.  In the Lagrangian approach in 2+1 dimensions, it has been demonstrated, for the simple case of SU(2),
 that such states can be chosen on a square
lattice and that the approximate $\pi/4$ symmetry becomes
exact in the continuum limit~\cite{Johnson:1998ev}. 
This technique is readily applicable in
the Hamiltonian approach but has not yet been attempted. \\

Thus in our Hamiltonian calculation a lattice spin $J$ state will
correspond to a continuum state with spin $J, J\pm 4, J\pm 8,\ldots
$. Using a variational approach we can obtain estimates of the lowest
mass states in the continuum with these spin values. To improve the
spin identification in the continuum beyond modulo $4$ requires more
work. Perhaps the multiple scaling regions visible in our massgap
calculations highlight this ambiguity.
Either way, it will prove interesting to
compare the masses calculated here to that of Teper who has been
careful to identify continuum spins correctly, at least for the lowest
mass excitations.

\subsection{Calculating Matrix Elements}

Having described the minimisation process
we now detail the calculation of
 the matrix elements $N_{l l'}$ and $D_{l l'}$. Our aim is to reduce
$N_{ll'}$ and $D_{ll'}$ to polynomials of one plaquette matrix elements.
This, again has been done for the case of SU(2) by
Arisue~\cite{Arisue:1990wv}. Here we retrace his calculations for the
general case of SU($N$) and extend them to incorporate improved Hamiltonians. We start with $D_{l l'}$.\\

Taking elementary plaquettes as our independent variables, it is easy to show that the only non-zero contributions to $D_{l l'}$ occur when the rectangles $l$ and $l'$ overlap. As an example of a contribution to $D_{l l'}$, consider $\Delta D_{l l'}$; the case where $N_{l\cap l'}$ plaquettes are contained by both rectangles (these are the overlap plaquettes) and $N_{l}$ plaquettes are contained by the rectangle $l$.
In order to calculate such matrix elements we rely heavily on the orthogonality properties of SU$(N)$ characters. We are interested in calculating SU$(N)$ 
integrals of the form
\bea
\int dU_p e^{S(U_p)}\chi_r(U_p V), 
\label{interest}
\eea 
where $U_p$ is a SU$(N)$ plaquette variable and $V$ is a product of any number of plaquettes not including $U_p$. Here $\chi_r(U)$ denotes the character corresponding to the representation $r$. For SU(2), $r = 0,1/2,3/2,\ldots$ and for SU(3), $r= (\lambda,\mu)$ where $\lambda$ denotes the number of boxes in the first row of the Young tableau describing the representation and $\mu$ is the number of boxes in the second row. Similarly, for SU$(N)$, $r=(r_1,r_2,\ldots,r_{N-1})$.\\

Performing a character expansion of the exponential in \eqn{interest} gives:
\bea
\int dU_p e^{S(U_p)} \chi_{r}(U_p V) &=& \sum_{r'} \int dU_p
c_{r'} \chi_{r'}(U_p)\chi_{r}(U_p V).
\eea
This is simply a generalisation of a Fourier expansion.
Here, the coefficient $c_{r'}$ is given by:
\bea
c_{r'} = \int dU_p \chi_{r'}(U_p) e^{S(U_p)}.
\eea
Now, using the orthogonality relation,
\bea
\int dU_p
\chi_{r'}(U_p V) \chi_{r}(U_p) = \frac{1}{d_r} \delta_{r'r} \chi_r(V),
\label{charorthog}
\eea
where $d_r$ is the dimension of the representation $r$, we obtain:
\bea
\int dU_p e^{S(U_p)} \chi_{r}(U_p V) &=& 
\frac{1}{d_r} \chi_r(V)\int dU_p \chi_{r}(U_p) e^{S(U_p)}.
\label{integrate}
\eea
This result allows us to integrate out a single plaquette from an
extended Wilson loop in a given representation $r$. To complete the
calculation we need to relate SU($N$) characters to traces of group
elements. This can be done using Weyl's character
formula~\cite{Bars:1980yy}. For SU($N$), according to Bars~\cite{Bars:1980yy}, the dimensions and characters corresponding to the first few representations are given by:
\bea
\begin{array}{rclrcl}
\displaystyle \chi_{1}(U) &=& \Tr U &
\displaystyle d_{1}(U) &=& N \vspace{0.1cm}\\
 \chi_{2}(U) &=& \displaystyle \frac{1}{2}\left[(\Tr U)^2 + \Tr U^2\right] &
d_{2}(U) &=& \displaystyle \frac{1}{2}N(N+1)\vspace{0.1cm}\\
\displaystyle\chi_{1 1}(U) &=&\displaystyle \frac{1}{2}\left[(\Tr U)^2 - \Tr U^2\right] &
d_{11}(U) &=&\displaystyle \frac{1}{2}N(N-1)\vspace{0.1cm} \\
\chi_{21}(U) &=&\displaystyle \frac{1}{3}\left[(\Tr U)^3 - \Tr U^3\right] &
d_{21}(U) &=&\displaystyle \frac{1}{3}(N-1)N(N+1)\vspace{0.1cm} \\
\displaystyle \chi_{1^{N-1}}(U) &=& \displaystyle \Tr U^\dagger &
\displaystyle d_{1^{N-1}}(U) &=& N
\end{array}
\label{sunchars}
\eea
Here we have adopted the convention of dropping all zeros in the
character labels.  
The Mandelstam constraints for the gauge group in question allows all 
characters to be expressed in terms of a minimal set of trace variables. 
For
example, for SU(3) we make use of the Mandelstam constraint,
\bea
\Tr (A^2 B) = \Tr A \Tr (A B) - \Tr A^\dagger \Tr B + \Tr (A^\dagger B),
\label{su3mand}
\eea  
where $A\in {\rm SU}(3)$ and $B$ is any $3\times 3$ matrix, to express all
characters in terms of $\Tr U$ and $\Tr U^\dagger$. For example,
for the case of SU(3), \eqn{sunchars} simplifies to  
\bea
\begin{array}{rclcrcl}
 \displaystyle \chi_{1}(U) &=& \Tr U &\qquad& d_{1} &=& 3\\ 
\vspace{0.1cm}
\displaystyle \chi_{11}(U) &=& \displaystyle
\displaystyle \frac{1}{2}\left[(\Tr U)^2-\Tr(U^2)\right]= \Tr U^\dagger
&\qquad & d_{11} &=& 3\\ 
\displaystyle \vspace{0.1cm}
\chi_{2}(U) &=& \displaystyle
\displaystyle\frac{1}{2}\left[(\Tr U)^2+\Tr(U^2)\right]= (\Tr U)^2-
\Tr U^\dagger  &\qquad & d_{2} &=& 6\\
\displaystyle  \chi_{21}(U) &=& \displaystyle
\displaystyle \frac{1}{3}\left[(\Tr U)^3-\Tr(U^3)\right]= \Tr U \Tr U^\dagger -1 & \qquad & d_{21} &=& 8
\end{array}
\label{su3chars}
\eea     
However, for general SU($N$) the Mandelstam constraints are
difficult to calculate.
In what follows we will need expressions for $\Tr U$, $\Tr U^\dagger$,
$\Tr (U^2)$, $(\Tr U)^2$, and $\Tr U \Tr U^\dagger$ as linear
combinations of characters for SU($N$). 
It is possible to do this without the use
of the Mandelstam constraint. Such expressions are necessary in order 
to make
use of \eqn{integrate} in
the calculation of expectation values of trace variables. For $\Tr U$,
$\Tr U^\dagger$, $\Tr (U^2)$ and $(\Tr U)^2$ the necessary expressions 
are easily obtained by rearranging
\eqn{sunchars},
\bea
\Tr U &=& \chi_{1}(U) \nn\\
(\Tr U)^2 &=&\chi_{2}(U)+ \chi_{11}(U) \nn\\
\Tr U^2 &=&\chi_{2}(U)-\chi_{11}(U) \nn\\
\Tr U^\dagger &=& \chi_{1^{N-1}}(U).
\label{tracevars-1}
\eea 
To express the remaining expression,
$\Tr U \Tr U^\dagger$, as a linear combination of characters is not as
easily done. For SU(3) one can simply rearrange
the expression for $\chi_{21}(U)$ in \eqn{su3chars}. 
For the general $N$ case 
it is simplest to consider Young tableaux. In terms of Young tableaux we 
have
\bea
\Tr U \Tr U^\dagger  \equiv \square \,\otimes \left.\begin{array}{c} \square\vspace{-0.18cm}\\
\square\vspace{-0.18cm} \\ \vdots \\
\square \end{array}\right\}N-1 .
\eea 
Performing the product representation decomposition gives
\bea
\square \,\otimes \left.\begin{array}{c} \square\vspace{-0.2cm}\\
\square\vspace{-0.2cm} \\ \vdots \\
\square \end{array}\right\}N-1 &=& \left.
\begin{array}{c} \square\vspace{-0.2cm}\\\square\vspace{-0.2cm}\\\vdots\\\square \end{array}\right\}N +
\left.\begin{array}{c c} \square\vspace{-0.2cm} &
\hspace{-0.375cm}\square\\\square\vspace{-0.2cm}\\\vdots\\\square \end{array}
\right\}N-1
\eea
Converting back into the notation of characters and traces gives
\bea
\Tr U \Tr U^\dagger &=&  1 + \chi_{2 1^{N-2}}(U).
\label{uudag}
\eea

\eqn{uudag} together with \eqns{tracevars-1}{integrate}
allow the analytic calculation of each contribution to $D_{ll'}$ for
all $N$. For the case of $\Delta D_{l l'}$ described earlier, we have 
\bea
\Delta D_{l l'} = \frac{2}{N} F_{Z_1}(N_l+ N_{l'} - 2 N_{l\cap l'})
\left[F_{Z_1^2}(N_{l\cap l'}) +F_{Z_1\!\bar{Z}_1}(N_{l\cap l'}) \right] 
-4 F_{Z_1}(N_l)F_{Z_1}(N_{l'}),
\eea
where the character integrals are given by:
\bea
\label{charfunctionsstart}
F_{Z_1}(n) &\!\!\!=\!\!\!& 
\left(\frac{1}{N}\right)^{n-1} 
\langle Z_1 \rangle^n ,  \\
F_{Z_1^2}(n) &\!\!\!=\!\!\!& 
 \frac{1}{2}\left[\frac{1}{N(N+1)}\right]^{n-1}\langle Z_1^2
+ Z_2 \rangle^n  +
 \frac{1}{2}\left[\frac{1}{N(N-1)}\right]^{n-1}\langle Z_1^2
- Z_2 \rangle^n , \\
F_{Z_2}(n) &\!\!\!=\!\!\!& 
 \frac{1}{2}\left[\frac{1}{N(N+1)}\right]^{n-1}\langle Z_1^2
+ Z_2 \rangle^n  -
 \frac{1}{2}\left[\frac{1}{N(N-1)}\right]^{n-1}\langle Z_1^2
- Z_2 \rangle^n  , \\
F_{Z_1\bar{Z}_1}(n) &\!\!\!=\!\!\!& 
1 +
\left[\frac{1}{(N-1)(N+1)}\right]^{n-1} \left(\langle Z_1 \bar{Z}_1
\rangle - 1 \right)^n.
\label{charfunctions}
\eea
Here we have made use of the notation, $Z_n := \Tr\left( U^n\right)$, 
to denote the trace
variables occupying a single plaquette, $U$.
The expectation values appearing in
\eqnss{charfunctionsstart}{charfunctions} are easily
expressed in terms of the generating functions of
\eqns{coolsum}{coolersum}. Differentiating the generating functions
appropriately gives
\bea
\langle Z_1 \rangle &=&  \frac{1}{G_{{\rm SU}(N)}} \frac{\partial
G_{{\rm SU}(N)}}{\partial c}\Bigg|_{d=c} \nn\\
\langle Z_1 \bar{Z}_1\rangle &=& \frac{1}{G_{{\rm SU}(N)}} \frac{\partial^2
G_{{\rm SU}(N)}}{\partial c \partial d}\Bigg|_{d=c} \nn\\
\langle Z_1^2 \pm  Z_2\rangle &=& \frac{1}{G_{{\rm SU}(N)}} \frac{\partial^2
G_{{\rm SU}(N)}}{\partial c^2}\Bigg|_{d=c} \pm \frac{1}{H_{2}}  \frac{\partial
H_{2}}{\partial d}\Bigg|_{d=0}. 
\eea
In practice, we do not need to calculate all of
these matrix elements. 
We see from \eqn{simplealmost} that $\langle Z_1^2\rangle$ and
$\langle Z_1 \bar{Z}_1 \rangle$ are related by 
\bea
\langle Z_1^2 \rangle &=& \frac{1}{2 G_{{\rm SU}(N)}(c,c)}\frac{d^2 G_{{\rm SU}(N)}(c,c) }{dc^2}-\langle Z_1 \bar{Z}_1 \rangle . 
\eea 
This follows from the fact that a group integral does not depend on
the choice of direction for the links. To be more precise, the result
\bea
\int_{{\rm SU}(N)} d U f(U) =  \int_{{\rm SU}(N)} d U^\dagger
f(U^\dagger) = \int_{{\rm SU}(N)} d U f(U^\dagger) ,
\eea
follows from the fact that $dU$ and $d U^\dagger$ each define invariant Haar
measures on SU($N$) which, by uniqueness, must be equal.\\
 
We now move on to the calculation of $N_{ll'}$. It is easy to show
that the only non-zero contributions occur when there is at least one
common link and an overlap between the rectangles. The improvement
term (the second term in \eqn{N}) only contributes when the two
rectangles share at least two neighbouring links in a given
direction. Consider the contribution $\Delta N_{l l'}$ to $N_{ll'}$ in
which there are $L_1$ common links and $L_2$ common strings of two
links in a given direction. Again we suppose $N_l$ plaquettes are
enclosed by rectangle $l$ and that there are $N_{l\cap l'}$ common plaquettes. 
Making use of \eqn{integrate} and \eqnss{com1}{com4} from Appendix~\ref{commutationrelations} we obtain
\bea
\Delta N_{ll'} = \frac{L}{N} F_{Z_1}(N_l\!+\!N_{l'}\!-\!2N_{l\cap l'})\left[ 
F_{Z_2}(N_{l\cap l'})
\!-\!\frac{1}{N} F_{Z_1^2}(N_{l\cap l'})
\!-\! N \!+\! \frac{1}{N} F_{Z_1\!\bar{Z}_1}(N_{l\cap l'})\right],
\eea
with
\bea
L = (1-\kappa)L_1 + \frac{\kappa}{u_0^2}L_2.
\eea
For the case of SU(3) we can simplify this using \eqn{su3mand} to 
\bea
\Delta N_{ll'} = \frac{L}{3} F_{Z_1}(N_l\!+\!N_{l'}\!-\!2N_{l\cap l'})\left[ 
\frac{2}{3}F_{Z_1^2}(N_{l\cap l'})
\!-\!2 F_{Z_1}(N_{l\cap l'})
\!-\! 3 \!+\! \frac{1}{3} F_{Z_1\!\bar{Z}_1}(N_{l\cap l'})\right].
\eea

Having determined individual contributions to $D_{l l'}$ and $N_{l
l'}$, to complete their calculation the possible
overlaps between states $l$ and $l'$ of a given type must be counted.

\subsection{Choosing an Appropriate Vacuum State}
\label{choosinganappropriate}
In \sect{fixingthevariational} we calculated variational vacuum
wave functions for pure SU($N$) gauge theory for $N=2$, 3, 4 and 5. Our
motivation was to use these wave functions as inputs to calculations
of SU($N$) massgaps. We obtained wave functions with a variational
parameter that was proportional to $\beta$ in the large $\beta$ limit
and $\beta^2$ in the small $\beta$ limit. However, this behaviour is
incompatible with the exact continuum vacuum wave function. For the 
one-plaquette trial state, given by \eqn{oneplaquette}, to be compatible
with the exact continuum vacuum wave function in 2+1 dimensions, one must have
$c\propto \beta^2$ in the large $\beta$ limit~\cite{Greensite:1987rg}.
This result is
independent of the dimension of the gauge group in
question. To understand this result we must return to continuum
non-Abelian gauge theory in 2+1 dimensions 
for which Greensite~\cite{Greensite:1979yn} argued that the infrared
behaviour of the vacuum can be described by the wavefunction,
\bea
\exp\left( \mu \int d^2 x \Tr F_{ij}^2 \right).\label{gscont}
\eea
The simplest lattice analogue of \eqn{gscont} is the one-plaquette trial state
given by \eqn{oneplaquette}. Expanding the exponent of
\eqn{oneplaquette} about $a$ using \eqn{plaqexp}, we see that for the
\eqn{oneplaquette} to reduce to \eqn{gscont} in the continuum limit
and for $\mu$ to remain finite, the variational parameter $c$ must be
proportional to $\beta^2$. \\

For the
case of SU(2) in 2+1 dimensions, in the scaling
region it has been shown~\cite{Greensite:1987rg} that for compatibility with the exact SU(2) 
continuum vacuum wave function we must use the Greensite vacuum wave function, 
\bea
 | \phi_0 \rangle
   &=&\exp\left(\frac{0.81\pm0.02}{g^4}\sum_{\bms{x},i<j}\Real\plaquette \right) | 0\rangle.
\label{greensite}
\eea
It is this vacuum wave function that was used by Arisue~\cite{Arisue:1990wv}
in the calculation 
that we generalise in this chapter. 
It would thus seem that
using a variational wave function is not appropriate in the calculation
of massgaps. However, this is not entirely the case as we will now
show.\\

Define the following quantities,
\bea
E[\phi] &=& \frac{\langle \phi| H |\phi \rangle}{\langle \phi|\phi
\rangle}\quad {\rm and} \nn\\
|\phi_1\rangle &\equiv &
|\phi_1(c(\beta),d_1(\beta),\ldots,d_n(\beta))\rangle \equiv
F(d_1(\beta),\ldots,d_n(\beta))|\phi_0(c(\beta))\rangle .
\eea
Here $E$ is an energy functional, $H$ a Hamiltonian, $|\phi_0\rangle$ a trial vacuum
state, defined in \eqn{oneplaquette}, dependent on the coupling only though the dependence of $c$ on
$\beta$.
The $d_k(\beta)$ are determined variationally by minimising $E[\phi_1]-E[\phi_0]$
for each $\beta$ and the function $F$ is chosen so that
$|\phi_1\rangle$ and $|\phi_0\rangle$ are orthogonal.\\

Suppose that we calculate $E[\phi_1]-E[\phi_0]$. We see from
\eqns{d}{N} that this quantity depends on the coupling only through $c$,
$d_1,\ldots,d_n$ and a multiplicative factor of $g^2$:
\bea
 a \Delta m(c(\beta),d_1(\beta),\ldots,d_n(\beta))=E[\phi_1]-E[\phi_0] 
\eea
Suppose that this calculation exhibits the appropriate scaling behavior on some
$\beta$ interval $S$:
\bea
\frac{a}{g^2} \Delta m \equiv \mu(c(\beta),d_1(\beta),\ldots,d_n(\beta)) = \Delta m_0  \quad
\forall \beta \in S.  
\eea
Here $\Delta m_0$ is a constant.
The variational parameters may be rescaled with respect to $\beta $
and the same constant result is obtained but for a different
range of couplings $T$: 
\bea
\mu(c(f(\beta)),d_1(f(\beta)),\ldots,d_n(f(\beta))) = \Delta m_0  \quad
\forall \beta \in T.
\eea 
Here $T$ is the interval for which $b\in T \Rightarrow f(b) \in S$
(i.e. $T$ is the image of $S$ under $f^{-1}$). If we choose
$f(\beta)$ such that $c(f(\beta))$ gives the correct scaling
form (for example $f(\beta) \approx 1.62 \beta^2/8 $ for SU(2) in 2+1
dimensions) we then have
\bea
a \Delta m(c(f(\beta)),d_1(f(\beta)),\ldots,d_n(f(\beta)))\approx 
E[\phi_1]-E_0 \quad \forall \beta \in T, 
\eea
where $E_0$ is the exact ground state energy. 
This follows from the fact that $|\phi_0(c(f(\beta)))\rangle$ has the
correct scaling form and so in the continuum limit we have $H |\phi_0(c(f(\beta)))\rangle =
(E_0+\epsilon)|\phi_0(c(f(\beta)))\rangle $. Here $\epsilon$ is a
small correction term that would become smaller if we improved our trial
vacuum by adding more complicated Wilson loops in the exponent.\\

Since $|\phi_1\rangle $
is constructed orthogonal to the ground state we have
\bea
a \Delta m(c(f(\beta)),d_1(f(\beta)),\ldots,d_n(f(\beta)))\ge E_1-E_0+\epsilon \quad \forall \beta \in T, 
\eea
where $E_1$ is the exact energy of the first excited state, and so 
\bea
g^2 \Delta m_0 \ge E_1-E_0+\epsilon \quad \forall \beta \in T.
\eea  
We cannot therefore say that we have an upper bound on the massgap,
only an approximation. \\

This argument demonstrates that if one obtains scaling with a given
$c(\beta)=c_1(\beta)$ in the vacuum wave function, the same result will be obtained using a
different function, $c(\beta)=c_2(\beta)$, although in a different region of
coupling, provided that there exists a function $f$ such that
$c_1(\beta) = c_2(f(\beta))$.   For this reason we expect that the
calculation of massgaps using the variational trial state will agree
with a calculation using the Greensite vacuum wave function.\\

\begin{figure}
\centering
                       
\subfigure[$c(\beta) = 1/\beta$ compared with $c(\beta) = \beta^2$] 
                     {
                         \label{cequalsb^2andb^-1}
                       \includegraphics[width=7cm]{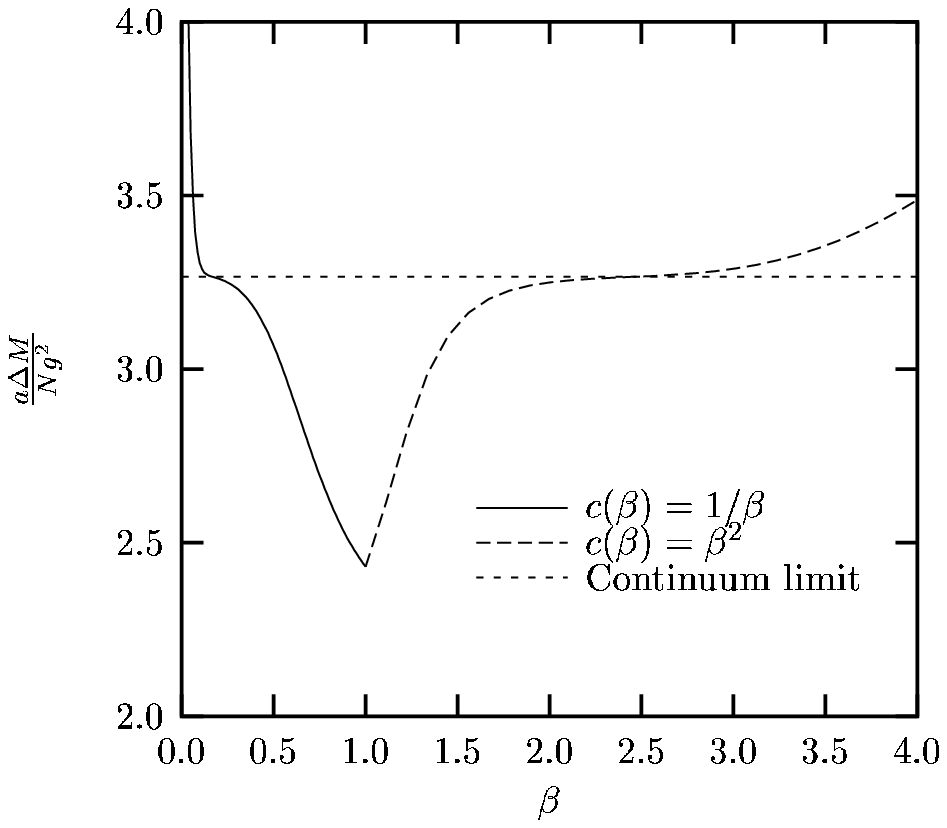}
                     } \hspace{0.25cm}                   
\subfigure[$c(\beta) = \beta$ compared with $c(\beta) = \beta^2$]
                     {
                         \label{cequalsb^2andb}
                         \includegraphics[width=7cm]{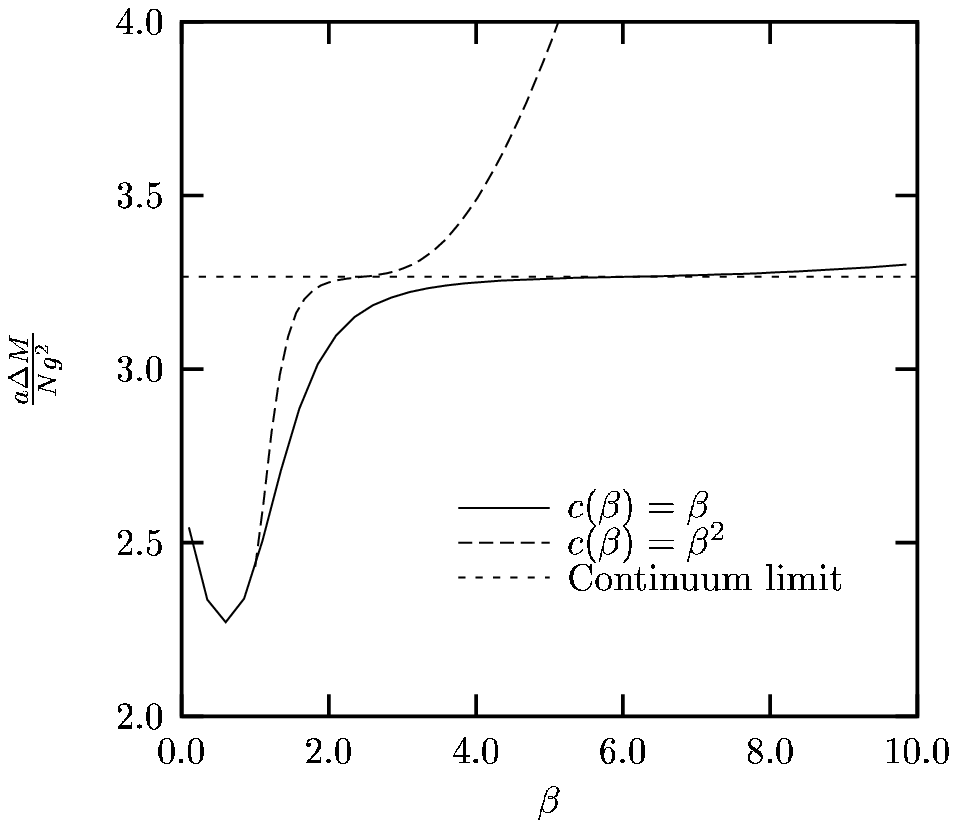}
                     }
\caption{An order 8 calculation of the lowest SU(3) glueball mass in
                         2+1 dimensions using various functional forms
                         for $c(\beta)$. The horizontal line is the continuum limit result obtained in \sect{chap5-results}.}
\label{cofbetavarious}
\end{figure}

The argument above also demonstrates that the precise form of $c(\beta)$ is
unimportant in the calculation of massgaps in this context. A
numerical demonstration of this for the case of 2+1 dimensional SU(3)
gauge theory is shown in
\fig{cofbetavarious}. We see that each glueball mass calculation
approaches the same constant scaling value (labelled by ``Continuum
limit'') but at different values of $\beta$. As it turns out, to calculate variational wave functions
for large dimension gauge groups is cumbersome. Numerical precision
becomes a factor in the minimisation of the energy density. This
problem is magnified in the calculation of tadpole improved results. We thus abandon the use
of variational wave functions in the calculation of massgaps beyond SU(5). Instead
we make use of the one-plaquette wave function of \eqn{oneplaquette} and
define a simple dependence $c(\beta)$, which most often will be
$c(\beta) =\beta$. Calculations for $N>5$ are presented in \chap{largenphysics}.

\section{SU(2), SU(3), SU(4) and SU(5) Massgap Results}
\label{chap5-results}
 
In this section we present glueball mass results for SU($N$) pure
gauge theory in 2+1 dimensions with $N=2$, 3, 4 and 5.
For each SU(3) calculation we keep 80 terms in the $k$-sum of
\eqn{Y} giving convergence up to $\beta = 50$.  For $N>3$ the
truncation $l_{{\rm max}} = 20$ is used. The generation of 
$N_{ll'}$ and $D_{ll'}$  and 
implementation of the minimisation process is accomplished 
with a Mathematica code. \\

\begin{figure}
\centering
\subfigure[SU(2)] 
                     {
                         \label{su2conv}
                         \includegraphics[width=7cm]{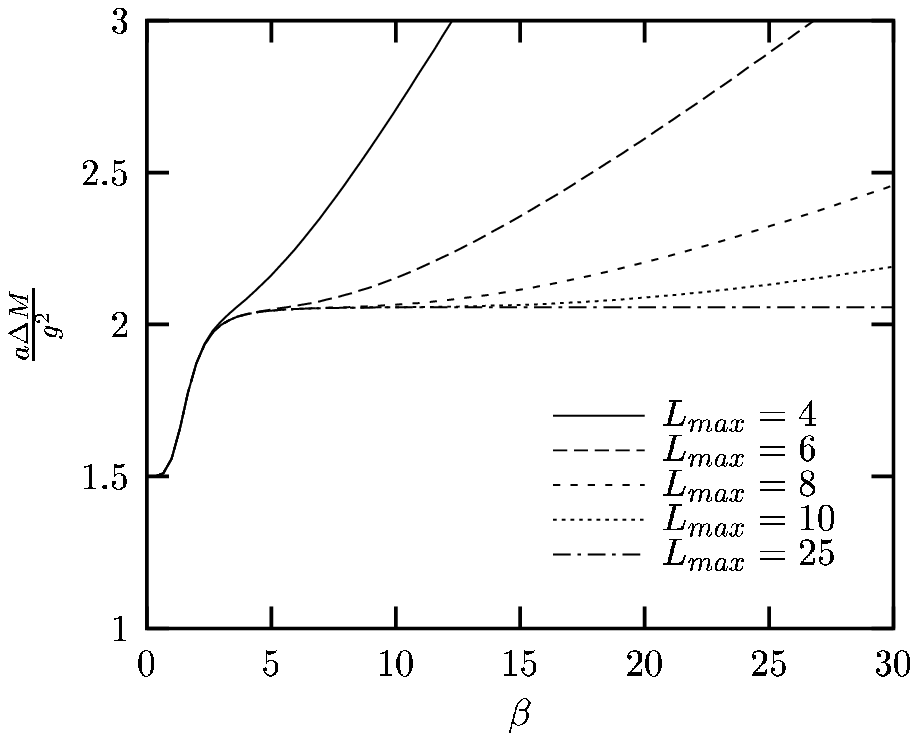}
                     }                   
 \subfigure[SU(3)] 
                     {
                         \label{su3conv}
                         \includegraphics[width=7cm]{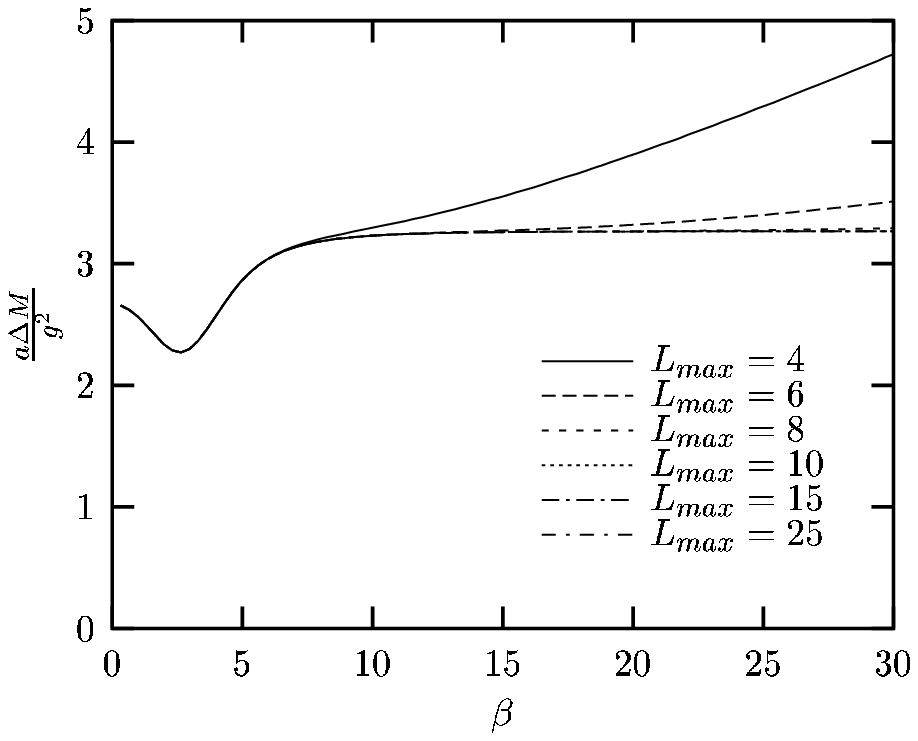}
                     }\\
 \subfigure[SU(4)] 
                     {
                         \label{su4conv}
                         \includegraphics[width=7cm]{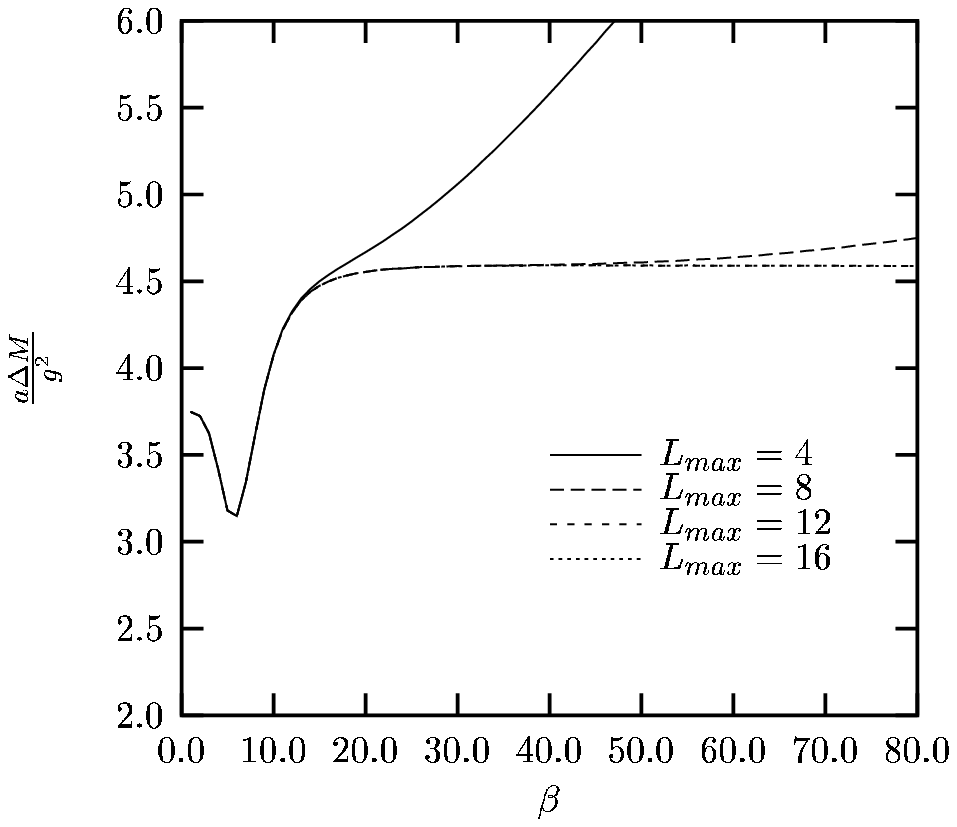}
                     }
\subfigure[SU(5)] 
                     {
                         \label{su5conv}
                         \includegraphics[width=7cm]{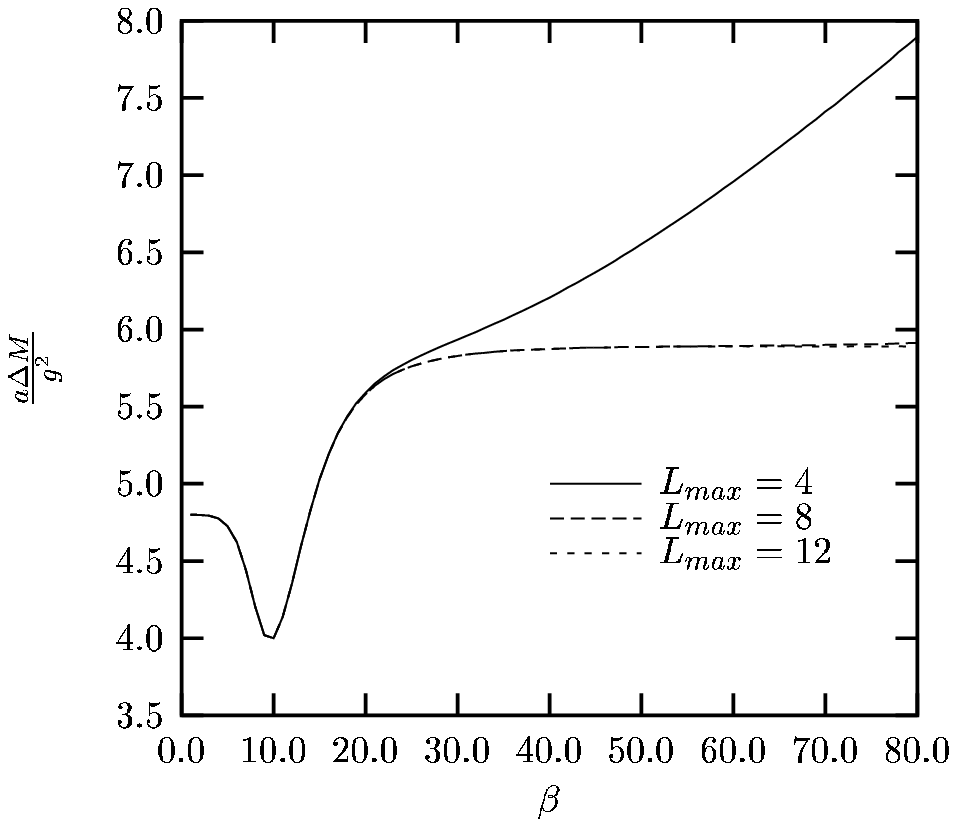}
                     }\caption{The unimproved 2+1 dimensional 
                     symmetric massgaps for SU(2), SU(3), SU(4) and SU(5).}

\label{mgconv}
\end{figure}

For the case of 2+1 dimensions we expect $a \Delta m/g^2$ to become constant in the scaling region. 
The convergence of the massgaps with $L_{max}$ is illustrated in
\fig{mgconv}. We notice that for $N>2$ only small improvements to the
scaling behaviour are gained by extending the calculation beyond order
8 on the range of couplings shown. This suggests that a more complicated basis (including, for
example, nonrectangular loops) is required to simulate SU($N$)
excited states with $N>2$ than for the case of SU(2).\\

\begin{figure}
\centering
\subfigure[SU(2)] 
                     {
                         \label{su2sym}
                        \includegraphics[width=7cm]{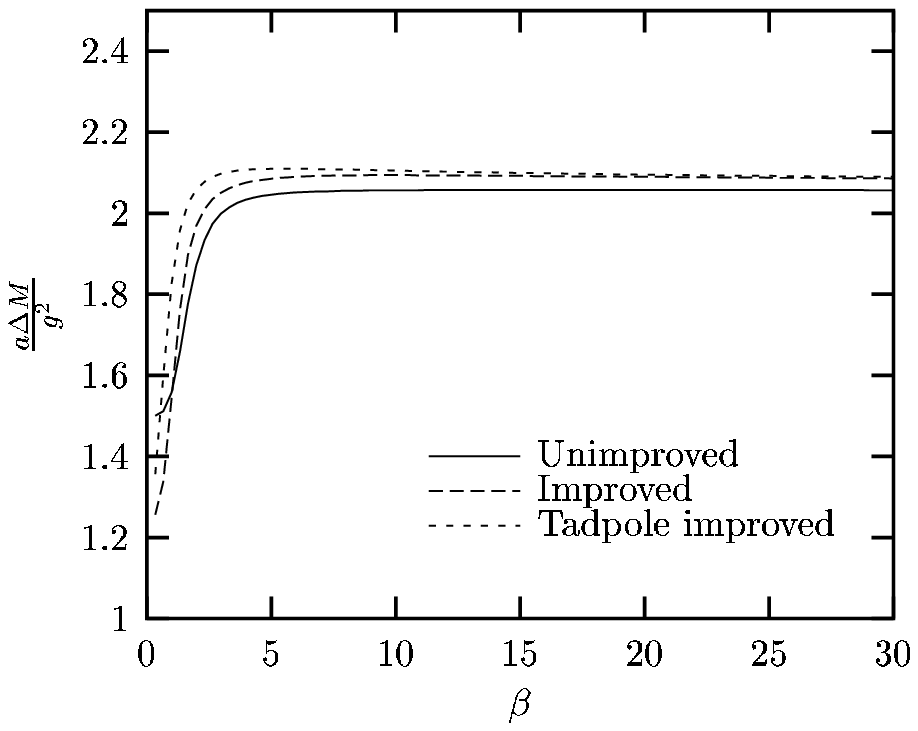}
                     }                   
 \subfigure[SU(3)] 
                     {
                         \label{su3sym}
        \includegraphics[width=7cm]{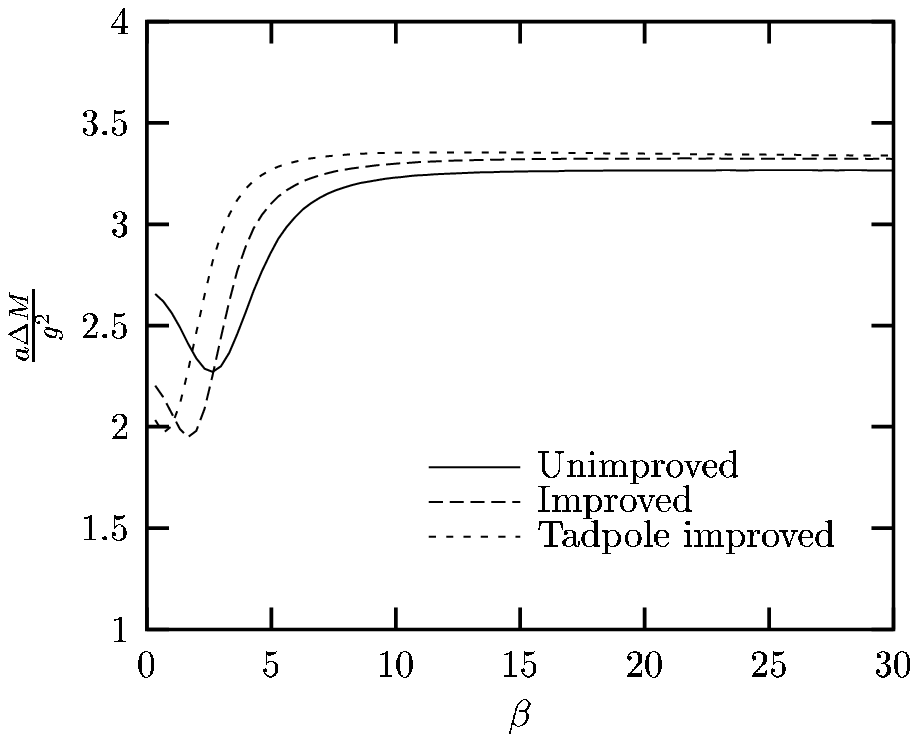}
                     }\\
\subfigure[SU(4)] 
                     {
                         \label{su4sym}
                        \includegraphics[width=7cm]{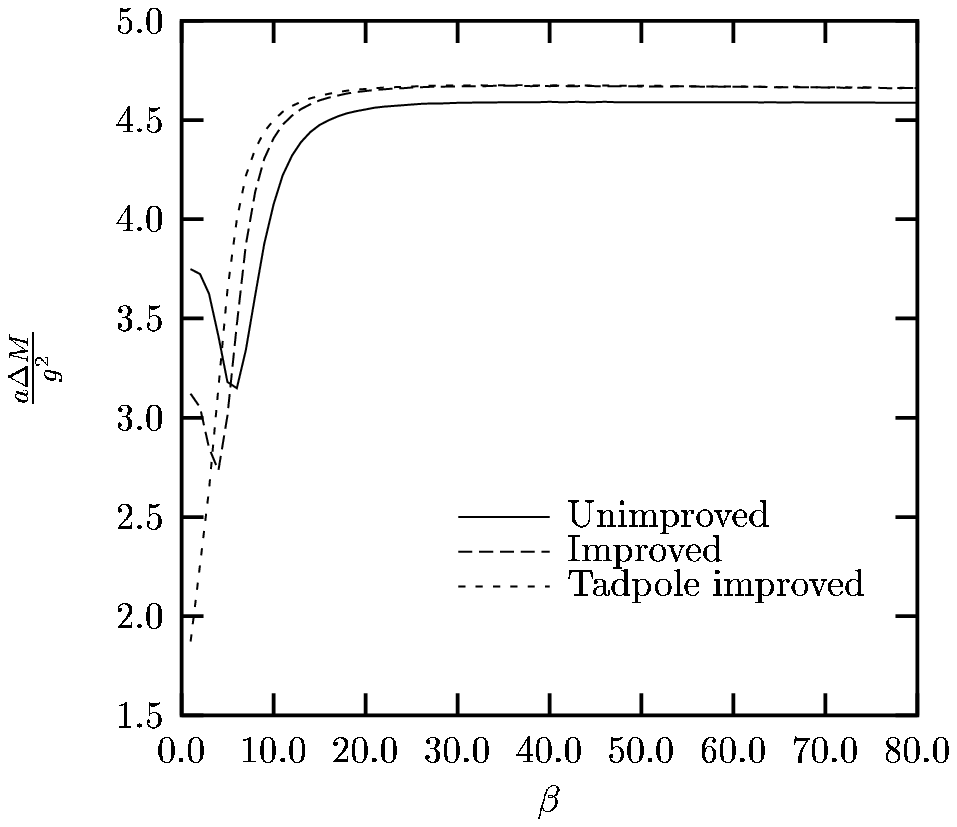}
                     }                   
 \subfigure[SU(5)] 
                     {
                         \label{su5sym}
                         \includegraphics[width=7cm]{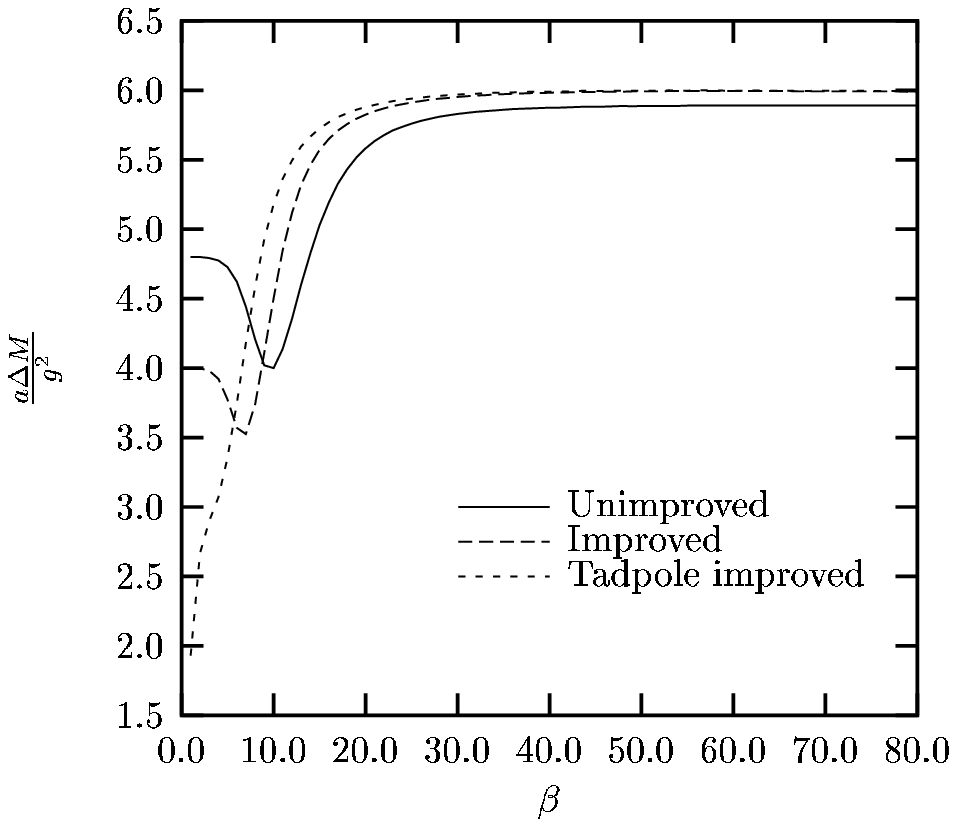}
                     }\\
\caption{The lowest lying 2+1 dimensional symmetric massgaps for 
SU(2), SU(3) (both with $L_{max}=25$), SU(4) (with $L_{max}=16$) and 
SU(5) (with $L_{max} = 12$).}
\label{mgcompare}
\end{figure}

\begin{figure}
\centering
\subfigure[SU(3)] 
                     {
                         \label{su3asym}
                        \includegraphics[width=10cm]{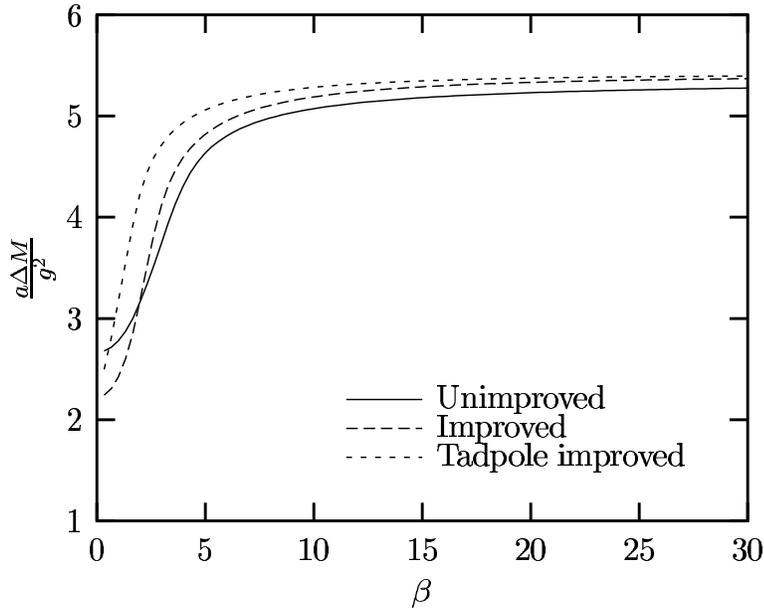}
                     } \\
\subfigure[SU(4)] 
                     {
                         \label{su4asym}
                        \includegraphics[width=7cm]{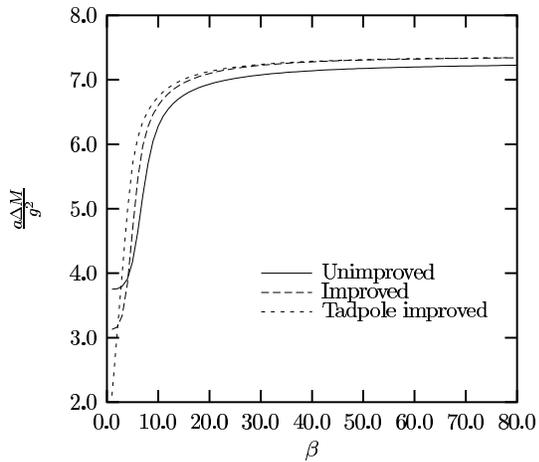}
                     } 
 \subfigure[SU(5)] 
                     {
                         \label{su5antisym}
                         \includegraphics[width=7cm]{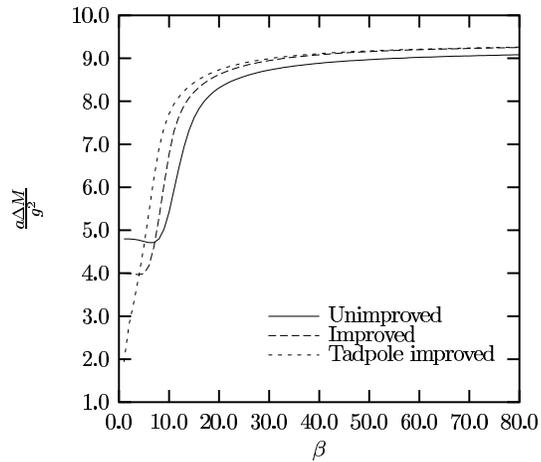}
                     }
\caption{The lowest lying 2+1 dimensional antisymmetric massgaps for 
SU(3) (with $L_{max}=25$), SU(4) and SU(5) (both with $L_{max}=12$).}
\label{asymmgcompare}
\end{figure}

In \fig{mgcompare} results for the lowest lying glueball mass, 
calculated with Kogut-Susskind, improved and
tadpole Hamiltonians, are shown.  
We see that $a \Delta m^S_1/g^2$ is approximated
well by a constant, in very large scaling regions, for the lowest lying
eigenstates for all $N$ considered. The scaling behavior becomes
significantly worse for the antisymmetric sector which is shown in
\fig{asymmgcompare} and for higher energy eigenvalues. This is because
the simplistic form of our excited state wave function is not
sufficient to reproduce the plaquette correlations required to
simulate these higher order states. One would expect the simulation of
higher order eigenstates to improve by including more complicated
loops in our expansion basis or by using a more complicated ground
state. The continuum limit excited states results for SU($N$) are given in \tabss{mgg}{mgmgmg}.\\

For the unimproved SU(2) case, the masses of the lowest two
eigenstates agree closely with 
the calculations of Arisue~\cite{Arisue:1990wv} (respectively $2.056\pm
0.001$ and $3.64\pm 0.03$ in units of $e^2$, which is related to 
$g^2$ by \eqn{couplings-relation}) in which the Greensite
vacuum wave function of \eqn{greensite} was used. This serves as a
check on our counting in calculating the possible overlaps of excited
states. Our calculation is in disagreement with that of Arisue at the
third eigenstate, for which Arisue calculates a mass $(5.15\pm
0.1)e^2$. Our fourth eigenstate is close in mass to Arisue's third and our
third eigenstate does not appear in his results. The reasons for this
are not clear. \\

The lowest mass unimproved SU(2) results (in units of
$e^2$) lies between the previous
Lagrangian Monte Carlo calculation of Teper, $1.582 \pm
0.017$~\cite{Teper:1998te}, and the Hamiltonian strong coupling expansion
result, $2.22\pm 0.05$~\cite{Hamer:1992ic}. It should be made clear
that the strong coupling expansion result quoted here assumes that
asymptotic weak coupling behaviour is already established in the
range, $1.4 < \beta < 2.2$. Data obtained from an extrapolation of the
strong coupling expansion are fit by a straight line in that region to
give the quoted result. For completeness, we must also mention that
the coupled cluster estimates agree on an increasing glueball mass up
to $\beta \approx 2$ but become variable after that \cite{LlewellynSmith:1993ig,Chen:1994ri}. Convergence with
increasing order appears to be a serious problem in such calculations.\\      

The results for the SU(3) symmetric massgap (in units of
$e^2$)
are to be compared to calculations by Luo and Chen $2.15 \pm 0.06$~\cite{Luo:1996ha}, Samuel
$1.84 \pm 0.46$~\cite{Samuel:1997bt} and Teper $2.40 \pm
0.02$~\cite{Teper:1998te}. Our result of $3.26520 \pm 0.00009$ is
considerably higher than all existing comparable results. By including
more complicated loops in the expansion basis one would expect to
reduce this estimate. This is emphasised by the fact that when using
only square basis states our result is considerably higher. To explain
the discrepancy between our results and others it is important to
note that since we use a basis of rectangles, we exclude the
contribution of many nonrectangular small area diagrams that are
included in the calculations of Teper and that of Luo and Chen. For
this reason it may be the case that what we have interpreted as the
lowest glueball mass in this chapter may, in fact, be a higher order
excited state. Teper has calculated the masses of the three lowest
mass glueballs for SU(3) in
the $0^{++}$ sector~\cite{Teper:1998te}: $0^{++}$, $0^{++*}$ and $0^{++**}$,  with the respective results, in units of
$e^2$: $2.40\pm 0.02$, $3.606 \pm  0.063$ and $4.55 \pm
0.11$. It is interesting to note that our result is closer to Teper's
first excited state. In the same study Teper also calculated glueball masses in the $0^{++}$ sector for $N=4$, 5 and 6. The mass, in units of $e^2$, 
of his $0^{++*}$ state for SU(4) is $4.84 \pm 0.12$ and $5.99 \pm
0.16$ for SU(5). We notice that as $N$ is increased the results
presented here move closer to the mass of Teper's $0^{++*}$ state, with the improved
results being closer than the unimproved. In fact for SU(5), the
results presented  here, improved and unimproved, are consistent with Teper's $0^{++*}$ mass.
This forces us to question the interpretation of the large $\beta$
plateaux in \fig{mgcompare} as scaling regions for the 
lowest mass glueballs. It is
possible that the minima present in \fig{mgcompare} in the small
$\beta$ region are possible scaling regions. It is possible that 
our vacuum wave function and minimisation basis are insufficient to 
extend this scaling region over a wide range of couplings and that as
our approximation breaks down we observe a level crossing effect. 
We examine the possibility of the small $\beta$ minima being scaling
regions in \chap{largenphysics}.\\

The antisymmetric results presented here can also be compared with
those of Teper~\cite{Teper:1998te}. While each of our results is
considerably higher than the masses of Teper's $0^{--}$ and $0^{--*}$
states, Teper's $0^{--**}$ state is close in mass to the lowest
mass state calculated here. Teper obtains the following masses, in
units of $e^2$, for the $0^{--**}$ state: $5.42\pm 0.16$ for
SU(3), $6.98 \pm 0.26$ for SU(4) and $9.18\pm 0.45$ for
SU(5). It is interesting to note that our corresponding 
lowest unimproved glueball
masses are consistent with these results. Our improved results
show better agreement with Teper's $0^{--**}$ state for SU(3) and
SU(5) than the corresponding unimproved results. The improved SU(4)
results are also consistent with Teper's SU(4) $0^{--**}$ mass
although the agreement is closer for the unimproved result. \\

\begin{table}[t]
\begin{center}
\begin{tabular}{clll}
\hline
 & Unimproved & Improved & Tadpole Improved   \\
\hline
$\Delta m^S_1$ & $2.05691 \pm 0.00002$ & $ 2.0897\pm 0.0003$ 
& $2.0965\pm 0.0006$ \\
$\Delta m^S_2$ & $3.645 \pm 0.001$ & $ 3.685\pm 0.001$ 
& $3.6953\pm 0.0009$ \\
$\Delta m^S_3$ & $4.5202 \pm 0.0004$ & $4.574\pm 0.004$ 
& $4.583\pm 0.004$ \\
$\Delta m^S_4$ & $5.133\pm 0.003 $ & $5.177\pm 0.004$ 
& $5.189\pm 0.004 $ \\
$\Delta m^S_5$ & $5.867\pm 0.006$ & $5.932\pm 0.008$ 
& $5.943\pm 0.008$ \\
\hline
\end{tabular}
\caption{Estimates of the lowest lying SU(2) glueball masses (in units of $e^2$) computed with various Hamiltonians in 2+1 dimensions. The unimproved, improved and tadpole results are calculated in the respective scaling regions $13.5 \le \beta \le 30.0$, $9.9 \le \beta \le 30.0$ and $9.25 \le \beta \le 30.0$.} \label{mgg}
\end{center}
\end{table} 

\begin{table}[t]
\begin{center}
\begin{tabular}{clll}
\hline
 & Unimproved & Improved & Tadpole Improved   \\
\hline
$\Delta m^S_1$ & 
$3.265868  \pm 0.000042$ &
$3.32365  \pm  0.00012$ &
$3.32580  \pm  0.00015$ \\
$\Delta m^S_2$ &
$6.23903  \pm  0.00065$ &
$6.30391  \pm  0.00083$ &
$6.31192  \pm  0.00084$ \\
$\Delta m^S_3$ & 
$7.5767  \pm  0.0025$ &
$7.6466  \pm  0.0030$ &
$7.6498 \pm  0.0030 $ \\
$\Delta m^S_4$ & 
$8.9462  \pm  0.0029 $ &
$9.0118  \pm  0.0044 $ &
$9.0206  \pm  0.0045 $\\
$\Delta m^S_5$ &
$10.0778 \pm  0.0071 $ &
$10.1546  \pm  0.0094 $ &
$10.1628  \pm  0.0094 $ \\
\hline
\end{tabular}
\caption{Estimates of the lowest lying symmetric SU(3) glueball masses (in
 units of $e^2$) computed with various Hamiltonians in 2+1
 dimensions. The results are calculated in the scaling region which
 minimises the standard error in each case.}\label{mgmg} 
\end{center}
\end{table} 

\begin{table}[t]
\begin{center}
\begin{tabular}{clll}
\hline
 & Unimproved & Improved & Tadpole Improved   \\
\hline
$\Delta m^S_1$ &
$4.59121\pm 0.00007  $     & $4.6720\pm  0.0001 $    & $4.6754\pm  0.0001 $ \\
$\Delta m^S_2$ &
$ 8.8122\pm 0.0012 $       & $8.9276\pm  0.0016 $     & $8.9284\pm  0.0017 $\\
$\Delta m^S_3$ &
$10.5889\pm 0.0051 $        & $10.7807\pm  0.0051 $     & $10.7794\pm  0.0051 $\\
$\Delta m^S_4$ &
$12.5527\pm 0.0048 $        & $12.6266\pm  0.0081 $     & $12.6138\pm  0.0080 $\\
$\Delta m^S_5$ &
$ 14.052\pm 0.012 $  & $14.165\pm  0.016  $     & $14.157\pm  0.016 $\\
\hline
\end{tabular}
\caption{Estimates of the lowest lying symmetric SU(4) massgaps (in
units of $e^2$) computed with various Hamiltonians in 2+1
dimensions. The results are calculated in the scaling region which
minimises the standard error in each case.}\label{mgmg} 
\end{center}
\end{table} 

\begin{table}[t]
\begin{center}
\begin{tabular}{clll}
\hline
 & Unimproved & Improved & Tadpole Improved   \\
\hline
$\Delta m^S_1$ &
$ 5.8903\pm  0.0001$ &  $5.99434\pm  0.00009$&
$5.9983\pm 0.0002
 $ \\
$\Delta m^S_2$ &
$11.2335\pm  0.0036 $&  $  11.3696\pm  0.0050 $&  $11.3731 \pm0.0049 $ \\
$\Delta m^S_3$ &
$13.340\pm  0.011$&  $ 13.658\pm  0.012 $&
$13.663  \pm 0.011 $ \\
$\Delta m^S_4$ &
$15.881\pm  0.012$ &  $ 15.890\pm  0.019$&  $
15.890\pm 0.019 $ \\
$\Delta m^S_5$ &
$17.564\pm  0.025$ &  $17.676\pm  0.035 $&  $
17.682\pm 0.035 $ \\
\hline
\end{tabular}
\caption{Estimates of the lowest lying symmetric SU(5) massgaps (in
units of $e^2$) computed with various Hamiltonians in 2+1
dimensions. The results are calculated in the scaling region which
minimises the standard error in each case.}\label{mgmg} 
\end{center}
\end{table}

\begin{table}[t]
\begin{center}
\begin{tabular}{clll}
\hline
 & Unimproved & Improved & Tadpole Improved   \\
\hline
$\Delta m^A_1$ &$ 5.32750 \pm 0.00047 $&$5.39661 \pm  0.00034
$&$5.39864 \pm  0.00032 $ \\
$\Delta m^A_2$ &$ 7.9389 \pm  0.0021 $&$8.0142 \pm  0.0028 $&$8.0145
\pm 0.0028 $ \\
$\Delta m^A_3$ &$ 8.9319 \pm  0.0045 $&$9.0092 \pm  0.0056 $&$9.0087 \pm  0.0055 $\\
$\Delta m^A_4$ &$ 10.4711 \pm 0.0058 $&$10.5514 \pm 0.0085 $&$10.5502 \pm  0.0085 $\\
$\Delta m^A_5$ &$ 11.304 \pm 0.011 $&$11.384 \pm 0.015 $&$11.381 \pm  0.015 $\\
\hline
\end{tabular}
\caption{Estimates of the lowest lying antisymmetric SU(3) massgaps
(in units of $e^2$) computed with various Hamiltonians in 2+1
dimensions. The results are calculated in the scaling region which
minimises the standard error in each case.} 
\label{mgmgmg}
\end{center}
\end{table}

\begin{table}[t]
\begin{center}
\begin{tabular}{clll}
\hline
 & Unimproved & Improved & Tadpole Improved   \\
\hline
$\Delta m^A_1$ &
$7.21479\pm  0.0012 $ &$7.3310\pm  0.0011$ &$7.33586\pm
0.00077$ \\
$\Delta m^A_2$ &
$10.9117\pm  0.0033$ & $11.0099\pm  0.004909$&$11.0617\pm
0.0046$  \\
$\Delta m^A_3$ &
$12.121\pm  0.007$ &$12.2779\pm  0.0088$&$12.292\pm
0.009$ \\
$\Delta m^A_4$ &
$14.5012\pm  0.0089$ &$14.592\pm  0.014$&$14.574\pm
0.015$ \\
$\Delta m^A_5$ &
$15.4521\pm  0.0173$ & $15.545\pm  0.023$&$15.555\pm
0.023$ \\
\hline
\end{tabular}
\caption{Estimates of the lowest lying antisymmetric SU(4) massgaps
(in units of $e^2$) computed with various Hamiltonians in 2+1
dimensions. The results are calculated in the scaling regions which
minimise the standard error in each case.} \label{mgmgmg}
\end{center}
\end{table}

\begin{table}[t]
\begin{center}
\begin{tabular}{clll}
\hline
 & Unimproved & Improved & Tadpole Improved   \\
\hline
$\Delta m^A_1$ &
$ 9.067\pm  0.003$& $ 9.239\pm  0.002$ &$ 9.248 \pm 0.002$ \\
$\Delta m^A_2$ &
$13.717\pm  0.008$& $ 13.89\pm  0.01$& $ 13.89 \pm 0.01$ \\
$\Delta m^A_3$ &
$15.054\pm  0.015 $&$ 15.32\pm  0.02$& $ 15.327 \pm 0.019$ \\ 
$\Delta m^A_4$ &
$18.08\pm  0.02$&$ 18.12\pm  0.03$& $ 18.12 \pm 0.03$ \\
$\Delta m^A_5$ &
$19.084\pm  0.036 $&$ 19.19\pm  0.05$ &$ 19.20 \pm 0.05$ \\ 
\hline
\end{tabular}
\caption{Estimates of the lowest lying antisymmetric SU(5) massgaps
(in units of $e^2$) computed with various Hamiltonians in 2+1
dimensions. The results are calculated in the scaling regions which
minimise the standard error in each case.} \label{mgmgmg}
\end{center}
\end{table}

When compared to equivalent unimproved calculations, the improved and
tadpole improved massgaps approach scaling faster as $\beta$ is
increased. This is evident in \figs{mgconv}{mgcompare} and is expected since,
for an improved calculation one is closer to the continuum limit when
working at a given coupling. However, for most improved calculations
the scaling behaviour is marginally less precise than the equivalent unimproved
calculation. A possible reason for this is that the one-plaquette trial state
used here does not allow for direct contributions from the improvement term in
the kinetic Hamiltonian. For this term to contribute directly one would need a
trial state which includes Wilson loops extending at least two links
in at least one direction. \\

The improved SU(2) massgap can be compared to the coupled cluster 
calculation of Li et al~\cite{Li:2000bg}. Their result (in units of $e^2$),
$\Delta m^S_1 = 1.59$, is again significantly lower than our result
$2.0897\pm0.0003$. The difference is again attributable to the different
choices of Wilson loops used in the simulation of states. While our
calculation makes use of the simple one-plaquette ground state and a
minimisation basis with only rectangular loops, the coupled cluster 
calculation of Li et al uses a more accurate ground state wave
function consisting of an exponential of a sum of extended loops which
are not necessarily rectangular. Without including additional small
area Wilson loops we cannot be confident that the lowest mass state
accessible with our minimisation basis is in  fact the lowest mass
state of the theory. Clearly there is scope for more work here.\\

\section{Conclusion}
\label{concl}
In this chapter we have extended the analytic techniques of 2+1
dimensional Hamiltonian LGT, traditionally used for SU(2), to general SU($N$).  Impressive scaling is achieved over an extremely wide range
of couplings for the lowest energy eigenstates in the symmetric and
antisymmetric sectors.  Our calculations use a one-plaquette trial state
and a basis of rectangular states over which excited state energies
are minimised.  Such choices allow the use of analytic techniques in
SU($N$) calculations. \\

The results of this chapter give estimates of
the lowest unimproved, improved and tadpole improved SU($N$) 
glueball masses all of 
which are above current estimates. We suspect that the reason for the
discrepancy is a lack of small area nonrectangular states in our 
minimisation basis. A basis of rectangular states was used for
simplicity. The inclusion of nonrectangular states is
straightforward and only complicates the counting of overlaps between
diagrams of a particular type.  When not including sufficient small area
diagrams it is possibly that the lowest mass states of the theory are not
accessible over a large range of couplings. A further improvement to
our calculation would involve the use of an improved trial vacuum
state. Such a state would possibly include several 
extended Wilson loops in its exponent. Without the development of new
techniques for performing the required integrals the use of such a
vacuum wave function would require the use of Monte Carlo techniques
for the calculation of expectation values. In this scenario many
advantages of the Hamiltonian approach presented here would be lost.\\

In the next chapter we extend the calculations presented here to
SU(25) in an attempt to explore the mass spectrum in the large $N$
limit of  pure SU($N$) gauge theory.

\chapter{Large $\bm{N}$ Physics}
\label{largenphysics}
\section{Introduction}

In this chapter we study the glueball mass spectrum for large $N$ pure
gauge theory in 2+1 dimensions. Our primary motivation lies in recent developments in string
theory where a direct connection between pure gauge theory in the large
$N$ limit and certain string theories has been conjectured. An independent
test of this conjecture is provided by large $N$ calculations of the
glueball mass spectrum in LGT. Impressive Monte Carlo calculations by
Teper~\cite{Teper:1998te} and Lucini and Teper~\cite{Lucini:2002wg,Lucini:2001ej} have progressed to the stage where
accurate values of glueball masses have been calculated up to $N=6$. This
has allowed an extrapolation to the $N\rightarrow \infty$ limit. With the
analytic techniques of the Hamiltonian approach it is possible to extend
these calculations to much larger $N$, allowing a more reliable
$N\rightarrow \infty$ extrapolation.\\

The outline of this chapter is as follows. We start in \sect{largenbackground} with
a historical background, where the original motivations for large $N$
physics are discussed and the recent developments in string theory
introduced. In \sect{directcalculation} we consider the
exploration 
of the large $N$
limit directly by considering the $N\rightarrow \infty$ limit of the
expectation values that appear in the variational calculation of massgaps.
We then move on to the extrapolation technique in \sect{extrapolation}, where
various glueball masses are calculated at finite values of $N$ up to
$N = 25$ in some cases. These results are then extrapolated to explore
the large $N$ limit of the 2+1 dimensional glueball spectrum. The 
$N\rightarrow \infty$ limits obtained here are compared to the
results of the analogous Monte Carlo studies in the Lagrangian approach.

\section{Background}
\label{largenbackground}
The nonperturbative region of QCD has been examined predominantly with
numerical techniques. While the numerical Monte Carlo simulations of 
Lagrangian LGT have made significant progress, there have been
few developments in analytic techniques. One analytic technique which has 
received considerable attention in recent years is the large $N$ limit. \\

In 1974 't Hooft proposed a study of SU($N$) gauge theories in the
large $N$ limit rather than the physically interesting $N=3$ case~\cite{'tHooft:1974jz}. It
was hoped that the SU($N$) theory could be solved analytically and
would be, in some sense, close to SU(3). 
Indeed it was shown that the large $N$ theory simplifies
drastically~\cite{'tHooft:1974jz} but still captures at least some of the complexity of
the SU(3) theory~\cite{Witten:1979kh, Eguchi:1982nm}. 
Motivated by the fact that the QCD coupling constant, $g^2$, is a
poor expansion parameter because its value depends on the energy scale
of the process under consideration, t' Hooft considered an expansion
in another dimensionless parameter of SU($N$) QCD, $1/N$. Based on an ingenious
organisation of Feynman diagrams, 't Hooft was able to show that in
the large $N$ limit, keeping the 't Hooft coupling, $g^2 N$, fixed, only planar diagrams
remain. While the remaining planar theory can be solved exactly in
two dimensions, the three and four dimensional theories have not yet
been solved. \\

Recent developments in string theories have attempted to address this
problem. In 1997 Maldacena conjectured a correspondence, in the
duality sense, between
superconformal field theories and string theory propagating in a
non-trivial geometry~\cite{Maldacena:1998re}.  The idea
behind duality is
that a single theory may have two (or more) descriptions with the
property that when one is strongly coupled the other is weakly
coupled. A technique for breaking the conformal
invariance and supersymmetry constraints of Maldacena's original
proposal was later developed by Witten~\cite{Witten:1998qj}. 
This led to the hope that
nonperturbative pure SU($N$) gauge theories could be described
analytically in 3+1 dimensions by their string theory dual. To develop
this hope into a concrete solution would require perturbative
expansions within the dual string theory. Such developments would seem to be some
way off.\\

In order to test the viability of using string theories 
to probe nonperturbative QCD independent
tests are needed. The glueball mass spectrum of QCD provides a perfect
laboratory. Recent Monte Carlo calculations have provided stable
estimates of glueball masses up to $N=6$~\cite{Teper:1998te,Lucini:2001ej} in 2+1 dimensions and up to $N=5$~\cite{Lucini:2001ej} in
3+1 dimensions. From these estimates an $N\rightarrow
\infty$ limit can be extrapolated in each case. Convincing evidence of 
$\ord(1/N^2)$ finite $N$ corrections have been demonstrated in 2+1 dimensions,
verifying a specific prediction of 't Hooft's $1/N$ expansion in the
quarkless case.  From the string theory side, present
calculations require a strong coupling limit ($N, g^2 N\rightarrow
\infty $) to be taken. In this limit the string theory reduces to classical supergravity and 
the calculation of glueball masses, $m(J^{PC})$, is straightforward. In this limit the 
glueball is represented by a dilaton
field whose mass can be extracted by solving the dilaton wave
equation~\cite{Csaki:1998qr,deMelloKoch:1998qs,Zyskin:1998tg}. 
Despite
the extreme approximations required surprising quantitative agreement
with the weak coupling results from LGT were obtained. Later it was
realised that, in 2+1 dimensions, lower mass states can be constructed from the graviton
field~\cite{Brower:1999nj}. 
With this approach the quantitative agreement with LGT is
lost but qualitative agreement in the form of the prediction 
\bea
m(0^{++}) < m(2^{++}) < m(1^{-+})
\eea   
still holds. The qualitative agreement between
string theories and LGT persists
in 3+1 dimensions~\cite{Brower:2000rp} with
 \bea
m(0^{++}) < m(2^{++}) < m(0^{-+}),
\eea  
for both theories. It appears that the identification of glueball states
remains a problem. A serious difficulty is the removal of unwanted states which have masses of the same magnitude as the glueballs. Recipes for removing these unwanted states have been
proposed~\cite{Russo:1998mm,Csaki:1999vb} but new parameters need to be
introduced.\\

The exploration of large $N$ gauge theories outside of string theory 
has not been restricted to
traditional Monte Carlo methods. Recent studies using the light-front 
Hamiltonian of transverse LGT have shown agreement with the Monte
Carlo calculations of Teper~\cite{Dalley:2000ye}. 
In this approach explicit calculations at
$N\rightarrow \infty$ are possible without the need for
extrapolation. Continuum calculations, without the use of string
theory, have also commenced. An impressive series of papers~\cite{Karabali:1996je,Karabali:1998wk,Karabali:1998yq}
has led to specific predictions for the string tension for all $N$ in
2+1 dimensions~\cite{Karabali:1998yq}. These predictions lie within $3\%$ of lattice
calculations up to $N=6$ beyond which LGT results are not available. A
recent review is available in \rcite{Nair:2002uj}.

\section{Direct Calculation of Asymptotics}
\label{directcalculation}
In this section we consider the direct calculation of the asymptotics
of the generating functions defined in
\chap{analytictechniques}. In this way it is hoped that
analytic forms for the asymptotics of the matrix elements on which the
glueball masses depend will be obtained. We start with the generating
function of \eqn{coolsum}.\\

To calculate the asymptotic form of \eqn{coolsum} in the large $N$
limit we make use of a differential equation technique and 
some elementary linear algebra~\cite{Krattenhaler}. Consider an
$N\times N$ matrix, $M_{N l}(x)$, depending on 
parameters $x$ and $l$, for which we wish
to calculate the determinant. Provided an $N\times N$ matrix $T_{N l}(x)$ exists such that
\bea
\frac{d M_{N l}}{dx}(x) = T_{N l}(x) M_{N l}(x), 
\label{mateqn}
\eea  
the determinant of $M_{N l}(x)$ satisfies the simple first order
differential equation
\bea
\frac{d \det M_{N l}}{dx}(x) = \Tr\left[ T_{N l}(x)\right] \det M_{N l}(x).
\label{detdiff}
\eea
The problem of finding the determinant reduces to one of finding a
suitable $T_{N l}(x)$ and solving the first order differential
equation in \eqn{detdiff}. \\

Consider the $N\times N$ matrix $M_{N l}(x)$ whose $(i,j)$-th entry is
defined as follows
\bea
\left[ M_{N l}(x) \right]_{ij} = I_{\pm l+j-i}(2 \sqrt{x}).
\eea
Making use of Mathematica we calculate a suitable $T_{N l}(x)$ such that
$M_{N l}(x)$ satisfies \eqn{mateqn} for various $N$ and $l$. In this way a
general form for $T_{N l}(x)$ may be guessed. Once a form is guessed the proof that it satisfies \eqn{mateqn} for all $N$ and $l$ is simply a matter of substitution. 
Calculations for $N= 2,\ldots, 10$ and $l= 0,\ldots ,6$ suggest a
general form for $T_{N l}(x)$ which has a trace given by, 
\bea
\Tr T_{N 0}(x) &=& 1 - \frac{x^N}{(N+1)!(N+2)!} + 
\frac{2 x^{N+1}}{(N+1)!(N+3)!}+ \cdots \nn\\
\Tr   T_{N l}(x) &=& \frac{N l}{2 x} + \frac{N}{N+l} + \cdots \quad
l\ne 0,
\label{tracet}
\eea
with the higher order terms vanishing in the $N\rightarrow \infty$
limit. To find an asymptotic form for $\det M_{N l}$ we must however do better than this and 
find the asymptotic form for the matrix $T_{Nl}(x)$, not simply its trace. At this
stage we have not found such an asymptotic form. Without
such a result we can only conjecture an asymptotic form for $\det M_{N l}(x)$. 
However all is not lost. The success or failure of any asymptotic form that we conjecture
can be determined numerically.\\

\begin{figure}
\centering

\subfigure[$l=2$] 
                     {
                         \label{l2}
                       \includegraphics[width=7cm]{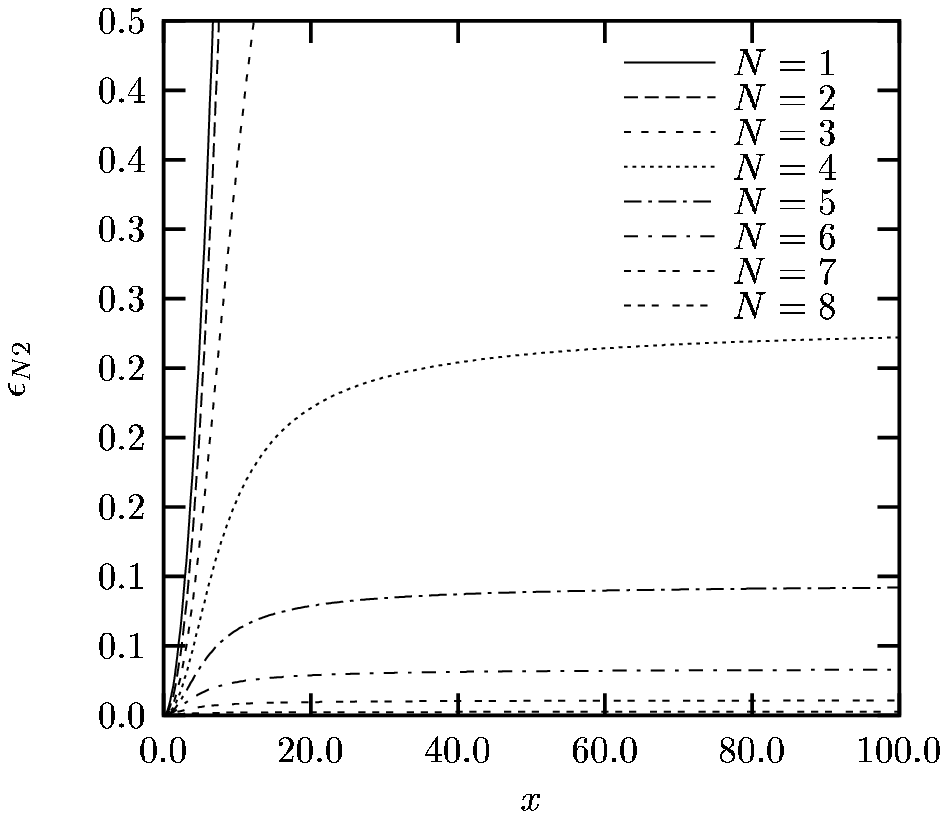}
                     } \hspace{0.25cm}                   
\subfigure[$l=-4$] 
                     {
                         \label{l-4}
                         \includegraphics[width=7cm]{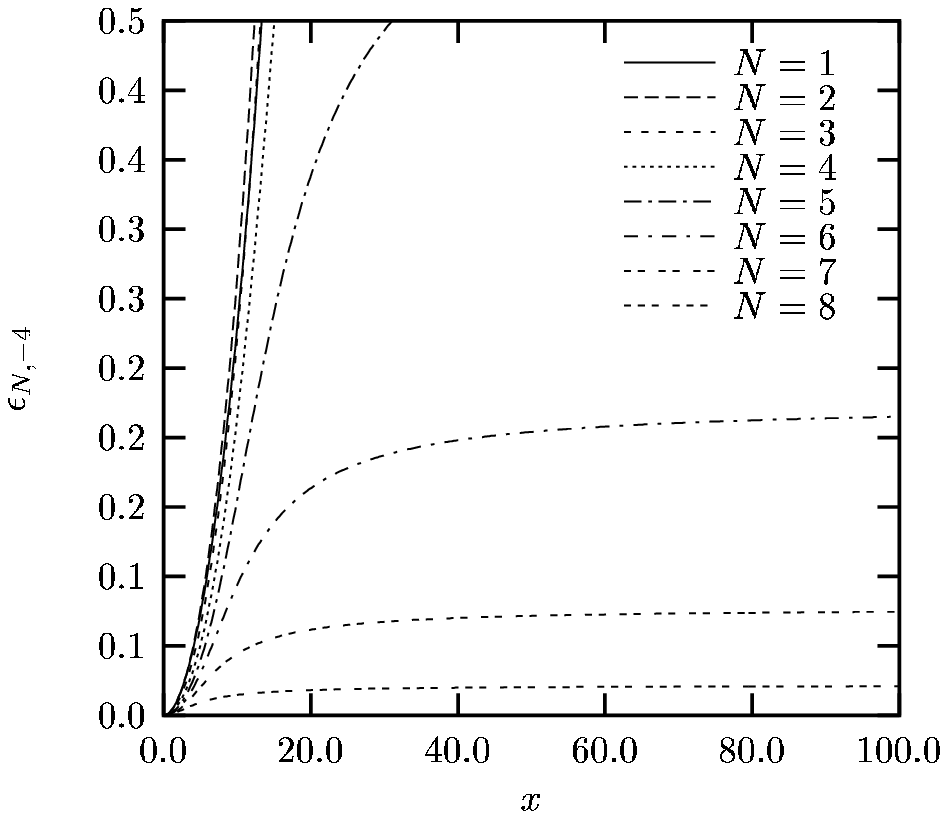}
                     }

\caption{A plot of $\epsilon_{N l}(x)$  versus $x$, for $l=2$ and $l=-4$
demonstrating the convergence of the exact result $\det M_{N 2}(x)$ to
the asymptotic form in the large $N$ limit, \eqn{asymptoticform}.} 
\label{asymptotic-convergence}  
\end{figure}

Substituting \eqn{tracet} in \eqn{detdiff} and solving the resulting
differential equation leads to the asymptotic form,
\bea
\det M_{N l}(x) \sim C_{N l} x^{N |l|/2} \exp\left(\frac{Nx}{N +|l|}\right).
\eea
Here $C_{N l}$ does not depend on $x$. We determine $C_{N l}$ by noticing that
\bea
 C_{N l} = \lim_{x\rightarrow 0}x^{-N |l|/2} \exp\left(-\frac{Nx}{N+|l|}\right) \det M_{N l}(x).
\eea
Again we use Mathematica and calculate $C_{N l}$ for various $N$ and
$l$. Some numerical experimentation 
leads to the conjecture that, in the large $N$ limit,
\bea
\det M_{N l}(x) \sim \frac{1!2!\ldots (|l|-1)!}{N!(N+1)!\ldots (N+|l|-1)!}
  x^{N |l|/2} \exp\left(\frac{N x}{N + |l|}\right).
\label{asymptoticform}
\eea
We denote the fractional error of this expansion by $\epsilon_{N
l}(x)$. To verify this result numerically we plot $\epsilon_{N l}$ for
various $N$ and $l$ as a function of $x$. 
The typical examples of $l=2$ and $l=-4$ with $N=1,\ldots ,
8$ are shown in \fig{asymptotic-convergence}. Fast convergence
over a large range of $x$ ans $N$ is increased is clear. Similar results are obtained for other values of $l$. In each case it is clear that convergence hastens for $N > |l|$.
Making use of \eqns{coolsum}{asymptoticform} we obtain the large $N$
asymptotics of the SU($N$) generating function,
\bea
G_{{\rm SU}(N)} &\sim & e^{cd} \sum_{l=-\infty}^{\infty}
\left(\frac{d}{c}\right)^{N l/2} \frac{1!2!\ldots (|l|-1)!}{N!(N+1)!\ldots (N+|l|-1)!}
  (c d)^{N |l|/2} \nn\\
&\sim &  e^{cd}\left( 1 + \frac{c^{N}+d ^{N}}{N!}\right) \sim  e^{cd}.
\eea 
We now move on to the second generating function of
\chap{analytictechniques}, that of \eqn{coolersum}. For this
case we define
\bea
P_{N l} (x,y) = \det \left[I_{l+j-i}(2 x)+ yI_{l+j-i+2}(2 x)
\right]_{1\le i,j\le N}.
\eea
For the purpose of calculating the matrix element $\langle Z_2
\rangle$ in the large $N$ limit, we are interested in the large $N$ 
asymptotic form of
\bea
\frac{\partial}{\partial y} \det P_{N l}(x,y)\Bigg|_{y=0}.
\eea
 Again we perform numerical experiments
using Mathematica and conjecture that there exists a $S_{N l}(x,y)$ such
that 
\bea
\frac{\partial}{\partial y} P_{N l}(x,y) &=& S_{N l}(x,y) P_{N l}(x,y),
\eea
with 
\bea
\Tr S_{N 0}(x,y)\Bigg|_{y=0} &\sim & \frac{x^{2 N}}{(N-1)(N+1)}\nn\\
\Tr S_{N l}(x,y)\Bigg|_{y=0} &\sim & \frac{N |l|
x^2}{(N+|l|-1)(N+|l|)(N+|l|+1)} \quad {l\ne 0}.
\eea
We then have,
\bea
\frac{\partial}{\partial y} \det P_{N l}(x,y)\Bigg|_{y=0} &\sim & \Tr\left[ S_{N
l}(x,y)\right] \det P_{N l}(x,y) \Bigg|_{y=0}  \nn\\
&& \hspace{-4cm}\sim \frac{N |l|
x^2}{(N+|l|-1)(N+|l|)(N+|l|+1)} \prod_{m=0}^{|l|-1} \frac{m!}{(N+m)!}
x^{N |l|} \exp\left(\frac{N x^2}{N+|l|}\right).
\label{asymstuff}
\eea 
Here we have recognised that $P_{N l}(x,0)=M_{N l}(x)$ and used 
\eqn{asymptoticform}. Making use of \eqn{coolintegral} we can sum
\eqn{asymstuff} over $l$ to obtain
the asymptotic form of the following SU($N$) integral,
\bea
\int_{{\rm SU}(N)} dU {\rm Tr}(U^2) e^{c ({\rm Tr} U + {\rm Tr}
U^\dagger)} &\sim& \frac{c^{2 N}}{(N-1)!(N+1)! } e^{c^2}  
\nn\\
&&\hspace{-4cm} +\sum_{l=1}^{\infty} \frac{2 N l
c^2}{(N+l-1)(N+l)(N+l+1)} \prod_{m=0}^{l-1} \frac{m!}{(N+m)!}
c^{N |l|} \exp\left(\frac{N c^2}{N+|l|}\right) \nn\\
&& \hspace{-4cm}\sim \frac{2 c^{N+2} e^{c^2}}{(N+1)(N+2)N!}\sim 0 . 
\eea

Compiling the above results we expect the following simple asymptotic forms 
for the matrix elements on which the glueball masses depend:
\bea
\langle Z_2\rangle &\sim & 0 \nn\\
\langle Z_1^2\rangle &\sim & c^2 \nn\\
\langle Z_1 \bar{Z}_1 \rangle &\sim & 1+c^2 \nn\\
\langle Z_1 \rangle &\sim& c .
\label{asymme}
\eea
These limits are verified numerically in direct calculations of
the matrix elements as $N$ is increased but only for small $c$. 
For finite $N$, glueball mass scaling occurs at
large $c$, outside the range of $c$ for which the asymptotic forms
listed in \eqn{asymme} hold.  
The problem can be explained as follows.  
In the above we have taken the naive $N\rightarrow
\infty$ limit. The limit in which we are interested is the one in which
$N\rightarrow \infty$ with $g^2 N$ held constant. In practice $c$ is
given a dependence on $g^2$ and hence the requirement that $g^2 N$
remain fixed forces a  functional dependence of $c$ on $N$. For
instance if we have $c\propto N/g^{2}$ then $c \propto N^2$ in the large $N$ limit if $g^2 N$ is held constant. The large $N$ limit of the matrix elements therefore cannot be analysed 
 until the functional dependence of $c$ on $\beta$ is fixed. 
It would therefore seem that an accurate description of the expectation values arising the glueball mass calculation in the large $N$ limit, requires many more terms than calculated here.
  
\section{Extrapolation}
\label{extrapolation}

\subsection{Introduction}

It this section we calculate 2+1 dimensional SU($N$) massgaps for various $N$ with the method of
\sect{massgaps}, with the intention of exploring the large $N$
limit of the results obtained. With this method we can
extend the calculation of the lowest energy massgaps to SU(25) on a
desktop computer. This should be compared to current results in the
Lagrangian approach which are limited, at present, to SU(6). With large values of $N$
available a more reliable extrapolation to the $N\rightarrow \infty$
limit is possible.\\

For the minimisation process, we use the same basis of rectangular states as in
\sect{massgaps}. With this basis the states $J^{++}$ and
$J^{--}$, with $J=0$ or 2, are accessible. The $J=2$ states require the use of a minimisation basis that contains no square states so that the excited states are invariant under rotations by $\pi$ but not $\pi/2$. The dependence of the
massgaps on the coupling does not change significantly as $N$ is
increased. We do however observe the appearance of what could possibly
be additional low $\beta$ scaling regions as $N$ is increased. \\


\subsection[Convergence with $l_{{\rm max}}$]{Convergence with $\bm{l_{{\rm max}}}$}
\label{convwithl}

\begin{figure}
\centering
\subfigure[SU(4) $\Delta M^{++}$. ] 
                     {
                         \label{symconv}
                         \includegraphics[width=7cm]{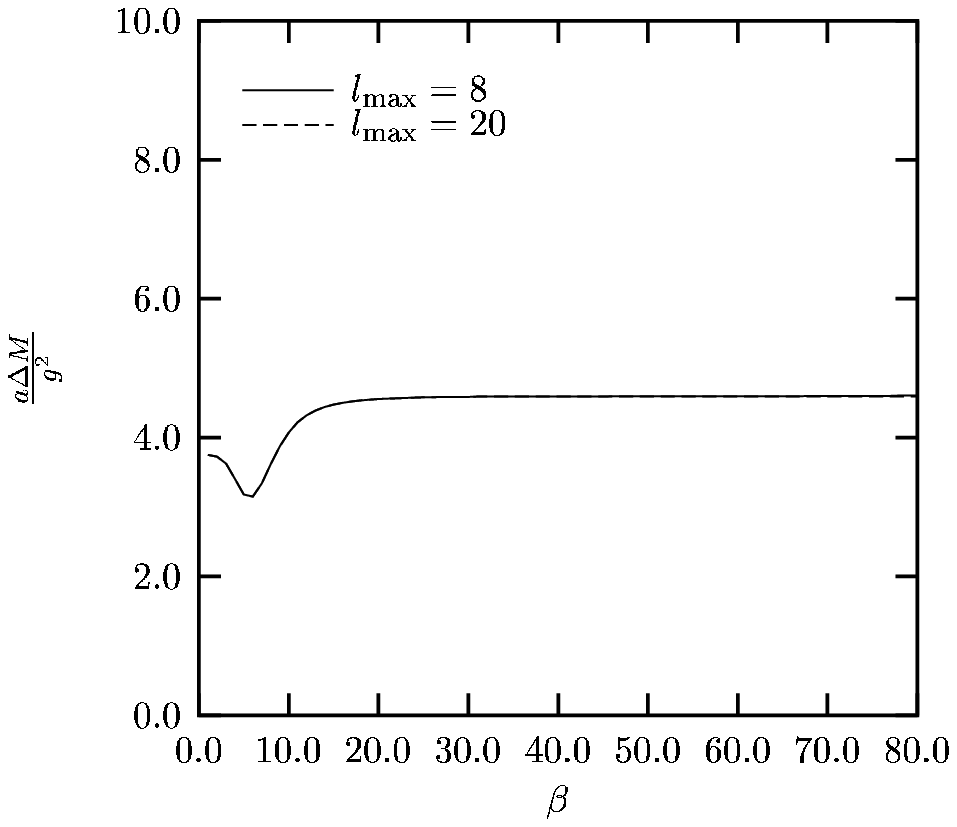}
                     }    \hspace{0.25cm}                     
\subfigure[SU(3) $\Delta M^{--}$.] 
                     {
                         \label{asymconv}
                       \includegraphics[width=7cm]{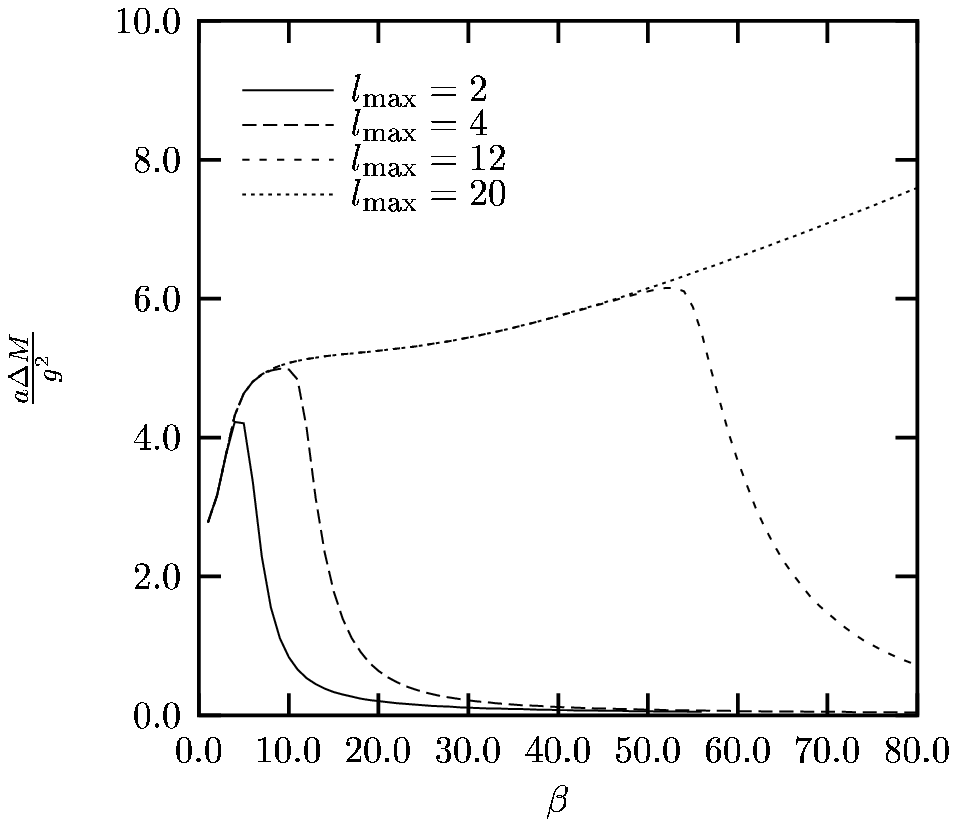}
                     }                  

\caption{Example spin 0 massgaps in units of $g^2/a$  demonstrating 
         different convergence properties when truncating the
         $l$-sum of \eqn{coolsum} at $l=\pm l_{\rm max}$.} 
\label{massgapconvergence}  
\end{figure}

In \sect{dependenceontruncation} the dependence of the
variational parameter on the truncation of the $l$-sum in \eqn{coolsum}
was considered. The calculation of the variational parameter depends
only on the plaquette expectation value. The dependence of the
massgaps on the truncation should also be explored as their
calculation incorporates on additional matrix elements. We find that fast
convergence is achieved as $l_{\rm max}$ is increased for the $0^{++}$ massgaps. A typical example
is shown in \fig{symconv}, where the massgaps are almost
indistinguishable up to $\beta =80$ on the scale of the plot. For
$0^{--}$ the convergence is not as fast. A typical example is shown in
\fig{asymconv}. We find that in order to obtain convergence of
the antisymmetric massgaps up to $1/g^2 \approx 350$ for SU(25) we
need a truncation of $l_{\rm max}=30$.

\subsection[The Small $\beta$ Minima]{The Small $\bm{\beta}$ Minima}

The massgaps for $N\ge 2$, in units of $N g^2/a$ do not differ
significantly as functions of $1/g^2$. The lowest lying $0^{++}$
glueball masses have
the characteristic form shown in \fig{symconv} for each $N$. We observe a
minimum at small $\beta$ and a scaling plateau at
large $\beta$. \\

In this section we examine the minima occurring in the lowest $0^{++}$
and $0^{--}$ glueball masses and consider the possibility that they
may correspond to a scaling region. It appears that for SU(3) this
region has been interpreted as a scaling region by some authors~\cite{Fang:2002ps,Chen:1995ca,ConradPhD}. From the scaling
arguments of \sect{choosinganappropriate}, by changing the functional
dependence of the parameter appearing in the vacuum state the scaling
region can be modified. In this way it is possible to choose the
variational parameter's dependence on $\beta$ such that the region in
which the massgap takes its minimum is stretched over a large
$\beta $ interval. In the same way, the scaling plateau occurring
for large $\beta $ may be compressed. With this in mind, it is
unclear which scaling region we should take to be the correct one.
For this reason we analyse both potential scaling regions. We start, 
in this section, with the small $\beta$ minima and consider the large
$\beta$ plateaux in the next section.\\

\begin{figure}
\centering

\includegraphics[width=10cm]{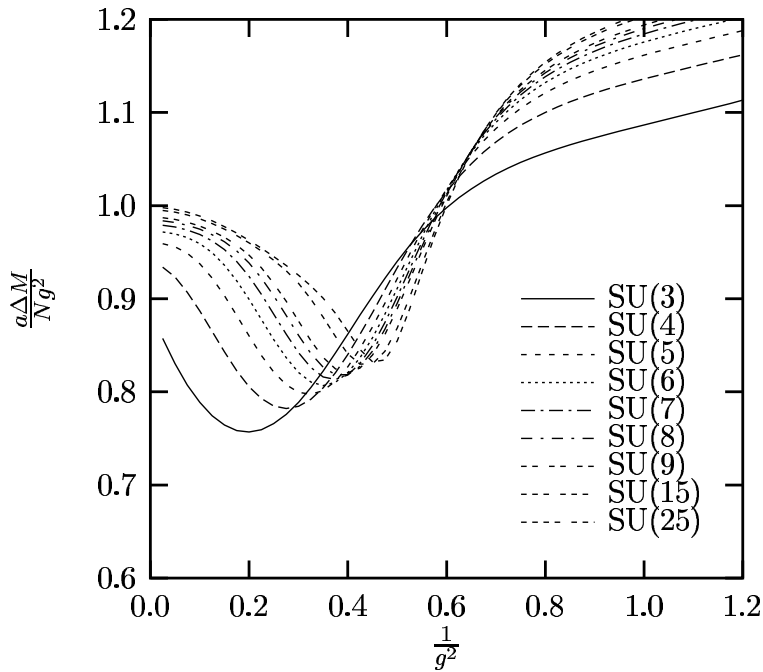}

\caption{$L_{max} = 4$  
SU($N$) lowest mass symmetric massgaps in 2+1 dimensions in units of
$g^2 N/a$ as functions of $1/g^2$. The $l$-sum of
\eqns{coolsum}{coolersum} truncated at $l_{{\rm max}}=5$. } 
\label{smallbetamin}  
\end{figure}

\fig{smallbetamin} shows the lowest lying SU($N$) $0^{++}$
glueball masses in 2+1 dimensions in units of $g^2 N/a$ as functions of
$1/g^2$ for various $N\in[3,25]$. It is apparent that the minima
depend only weakly on $N$. We see that the minima appear to approach a
finite limit from below as $N\rightarrow \infty$. On the scale of
 \fig{smallbetamin} the minima corresponding to $N=15$ and
$N=25$ are barely distinguishable.\\

\begin{figure}
\centering
                       
\includegraphics[width=10cm]{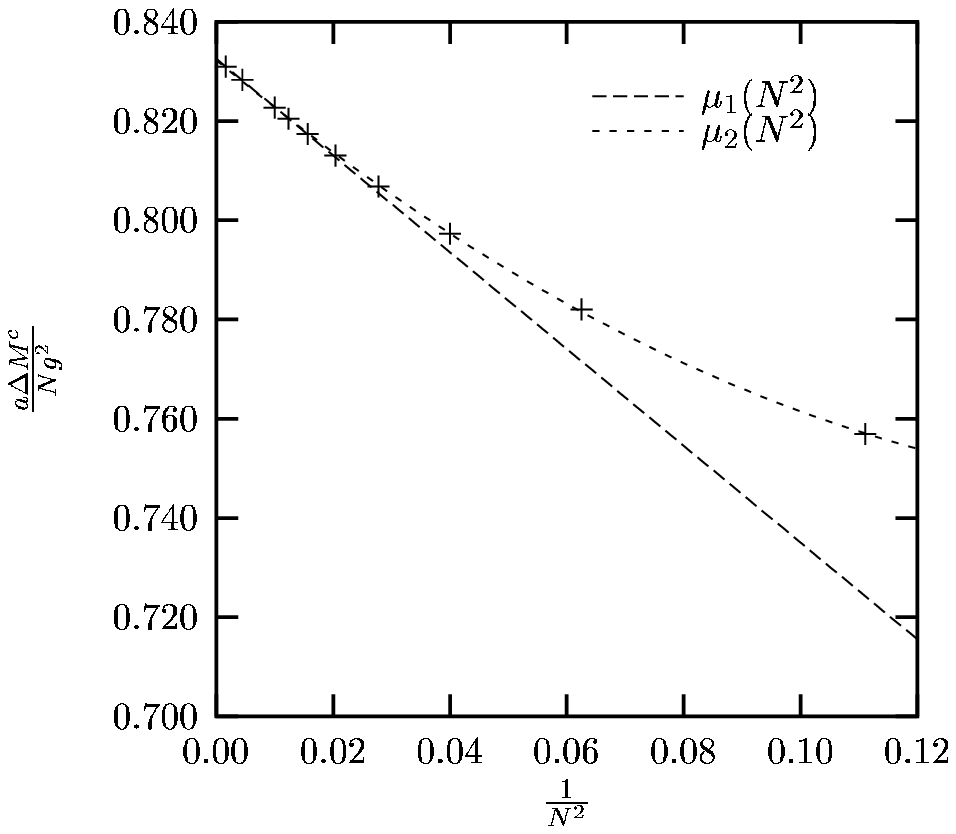}

\caption{ The continuum limit lowest mass symmetric 2+1 dimensional massgaps in units
of $N g^2/a$ as a
function of $1/N^2$ taken from the small $\beta$ minima. The dashed lines are fits to \eqns{smallbetafit1}{smallbetafit2}.}
\label{smallbetalargen}  
\end{figure}

Let us denote the minima in \fig{smallbetamin} by $a\Delta
M^c/(N g^2)$ and consider them as a functions of
$N$. Fitting $a\Delta
M^c/(N g^2)$ to the model 
\bea
\frac{a \Delta M^c}{N g^2} = \gamma_1 + \frac{\gamma_2}{N^{2 \gamma_3}} 
\eea
for $N \ge 8$ gives best fit parameters
\bea
\gamma_1  &=& 0.83262 \pm 0.00022  \nn\\
\gamma_2  &=& -0.97 \pm 0.13      \nn\\
\gamma_3  &=& 0.990 \pm 0.035 .
\label{fitone-1}
\eea
We fit on the data $N \ge 8$ to minimise pollution from next to
leading order corrections. 
If we assert that the power of the leading order correction for
finite $N$ must be an integer power of $1/N$, then from \eqn{fitone-1}
that order must be 2 in agreement with the predictions of large $N$ QCD. 
Fitting to a model with $1/N^2$ corrections for $N\ge 8$ gives
\bea
\mu_1(N^2) = 0.83256 \pm 0.00007 - \frac{0.9753 \pm 0.0072}{N^2}.
\label{smallbetafit1}
\eea 
We can go further and attempt to find the next to leading order
corrections by fitting to the model
\bea
\frac{a \Delta M^c}{N g^2} = \gamma_1 + \frac{\gamma_2}{N^{2}} +
\frac{\gamma_3}{N^{2 \gamma_4}}. 
\label{smallbetafit2}
\eea
The best fit parameters when fit on the whole data set are 
\bea
 \gamma_1 &=&      0.83287     \pm    0.00011   \\
\gamma_2 &=& -1.116      \pm    0.028     \\
\gamma_3 &=&  1.715     \pm     0.085     \\
\gamma_4 &=& 1.627      \pm    0.050   
\eea
which is in disagreement with integer power next to leading order
corrections. A good fit however is achieved to the model with $1/N^4$
next to leading order corrections:
\bea
\mu_2(N^2) &=& 0.83233 \pm 0.00018
 - \frac{0.986 \pm 0.010}{N^2} + \frac{2.772 \pm 0.090}{N^4}.
\label{smallbetafit2}
\eea
The precise locations of the minima, $a\Delta
M^c/(N g^2)$, are plotted as a function of $1/N^2$ in \fig{smallbetalargen}
along with the fitted models of \eqns{smallbetafit1}{smallbetafit2}.\\

From \eqn{smallbetafit1}, our best fit result for the
$N\rightarrow \infty$ limit of  $a\Delta M^c/(N g^2)$ is  $0.83256 \pm
0.00007$. The error here is purely statistical. We should, as
always, expect a significant systematic error attributable to our
choice of ground state and minimisation basis. This result may
be compared to the Monte Carlo calculation of Lucini and Teper who obtain the 
$N\rightarrow \infty$ limit of the lowest $0^{++}$ glueball mass as $0.8116\pm
0.0036$~\cite{Lucini:2002wg}. This is comparable to the result
presented here. The result of Lucini and Teper is based on a linear extrapolation to
the $N\rightarrow \infty$ limit of $2 \le N \le 6$ data. 
Their result differs significantly from the one presented here at the 
leading order finite $N$ corrections. The correction term  obtained by Lucini and Teper, 
$-(0.090\pm 0.028)/N^2$, has the same sign but is significantly smaller than the
one presented here. When our
data is fit on the range $3\le N \le 6$ (we don't obtain a small
$\beta$ minimum for SU(2)) our slope is halved but is still
significantly larger than that of Lucini and Teper. It should be
pointed out that Lucini and Teper's calculation was 
performed in the Lagrangian approach in which the coupling is the
so called Euclidean coupling $g_E$. The Euclidean coupling and the
Hamiltonian coupling, $g^2$, are equal up to order $g_E^2$
corrections. The precise relation between the couplings is 
only known for small $g_E^2$~\cite{Hamer:1996ub}, a case which does not apply for the
small $\beta $ minima. It appears that our simple calculation induces
a level crossing whereby the lowest mass glueball state switches 
to a higher mass state beyond the small $\beta$ minimum. Although we do
not have an explanation for this, presumably, by including
additional states in the minimisation basis or implementing a more
complicated vacuum state, the level crossing would no longer appear
and the small $\beta$ minima would extend 
into large $\beta$ scaling regions. This should be checked
in further studies. A glueball mass
extracted from a large $\beta$
scaling region can be confidently compared to a corresponding
Lagrangian calculation, for in that region of couplings the ratio of
$g_E^2$ to $g^2$ is unity. \\ 

\begin{figure}
\centering

\includegraphics[width=10cm]{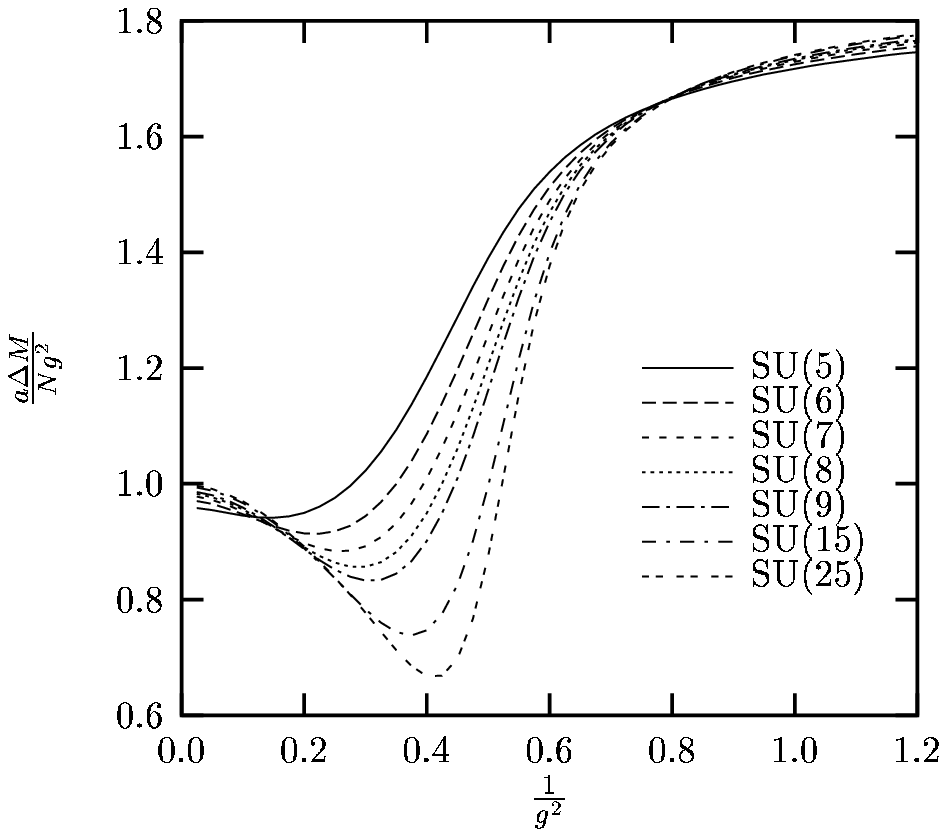}

\caption{$L_{max} = 6$ 
SU($N$) lowest mass antisymmetric massgaps in 2+1 dimensions in units of
$g^2 N/a$ as functions of $1/g^2$. The $l$-sum of
\eqns{coolsum}{coolersum} truncated at $l_{{\rm max}}=5$. } 
\label{smallbetamin-asym}  
\end{figure}

\begin{figure}
\centering
                       
\includegraphics[width=10cm]{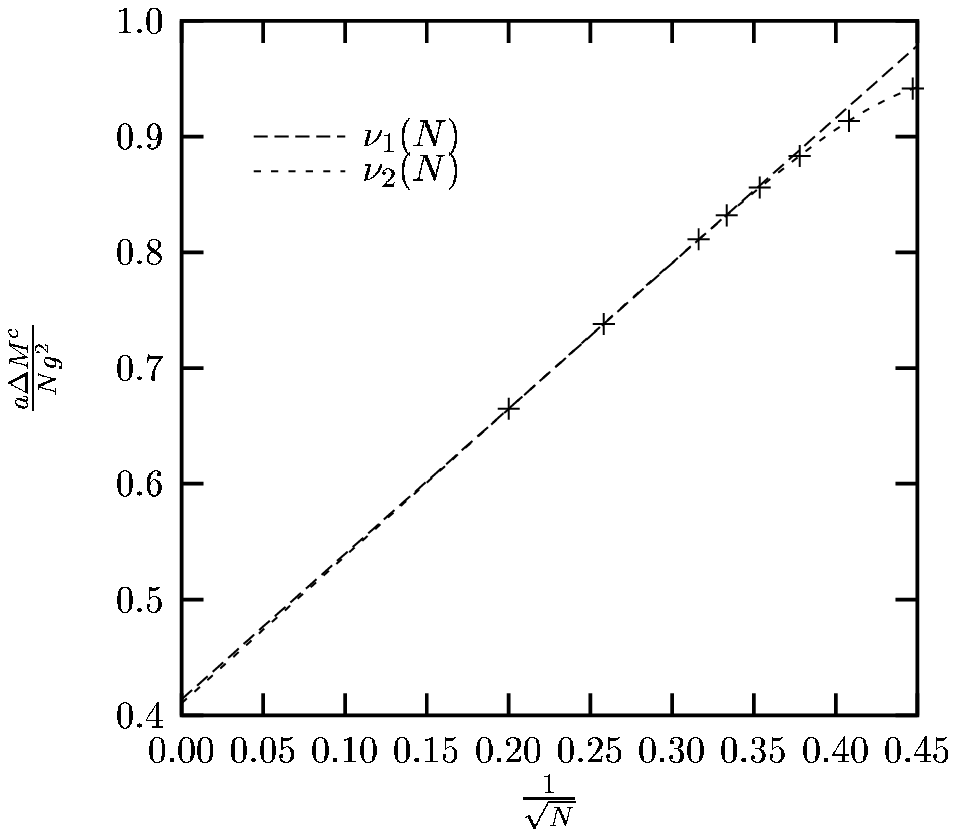}

\caption{ The continuum limit lowest mass antisymmetric 2+1
dimensional massgaps in units of $N g^2/a$ as a
function of $1/\sqrt{N}$ taken from the small $\beta$ minima. The dashed lines are fits to \eqns{smallbetafit1-asym}{smallbetafit2-asym}.}
\label{smallbetalargen-asymmetric}  
\end{figure}

We obtain small $\beta$ minima for the $0^{--}$ massgaps but only
for $N\ge 5$. Plots of these massgaps with $N\ge 5$ are
shown in \fig{smallbetamin-asym}. The behaviour is
significantly different to that observed for the symmetric
massgap. By $N=25$ the minima do not appear close to
convergence. Indeed if convergence is occurring at all, it is
significantly slower than was observed for the $0^{++}$ state. 
To explore this further, in 
\fig{smallbetalargen-asymmetric} 
we plot the minima of
\fig{smallbetamin-asym}, which we again denote by $a\Delta
M^c/(N g^2)$, as a  function of $1/\sqrt{N}$. We observe linear
behaviour in the large $N$ limit. Fitting the $N\ge 9$ data to the model
\bea
\frac{a \Delta M^c}{N g^2} = \gamma_1 + \frac{\gamma_2}{N^{\gamma_3 /2}}, 
\eea
gives best fit parameters
\bea
\gamma_1  &=& 0.41 \pm 0.01  \nn\\
\gamma_2  &=& 1.25 \pm 0.02      \nn\\
\gamma_3  &=& 0.98 \pm 0.04 ,
\label{fitone}
\eea
which is consistent with $\gamma_3 = 1$. Fitting the $N \ge 9$ data to a model with
leading order finite $N$ corrections starting at $1/\sqrt{N}$ gives
\bea
\nu_1(N) = 0.41390 \pm 0.00007 + \frac{1.255 \pm 0.003}{\sqrt{N}}.
\label{smallbetafit1-asym}
\eea 
As was done for the symmetric case we can attempt to go further and
find the form of the next to leading order finite $N$ corrections.
Fitting the $N\ge 5$ data to the model
\bea
\nu_2(N) = \gamma_1 + \frac{\gamma_2}{\sqrt{N}} +
\frac{\gamma_3}{N^{\gamma_4 /2}}. 
\label{smallbetafit2-asym}
\eea
gives best fit parameters 
\bea
 \gamma_1 &=&      0.410     \pm    0.001  \\
\gamma_2 &=&  1.2718      \pm    0.0045     \\
\gamma_3 &=&  -76     \pm     23     \\
\gamma_4 &=& 9.5      \pm    0.4  
\eea
indicating vanishingly small next to leading order corrections for
$N\ge 5$. The models $\nu_1$ and $\nu_2$ are plotted against the $N\ge
5$ data in \fig{smallbetalargen-asymmetric}. \\

The small $\beta$ minima appear to converge to a non-zero
$N\rightarrow \infty$ limit. The convergence is significantly slower
than for the small $\beta$ minima in the symmetric case. It is also
clear that the leading order finite $N$ corrections are not the
expected $\ord(1/N^2)$ but rather $\ord(1/\sqrt{N})$. The
$N\rightarrow \infty$ limit of $0.41 \pm 0.01$ is approximately half
the corresponding $0^{++}$ result. Such a state does not appear in the
calculation of Teper. \\

It is possible that the small $\beta$ minima are spurious. Dimensional
analysis gives the expected scaling form but 
does not allow us to decide which scaling region is
preferable, or if one is a lattice artifact. 
However, based on lattice calculations to date, 
 we expect the $0^{++}$ glueball to be lighter than the $0^{--}$ in the continuum limit. 
Spurious scaling regions have been observed in other lattice calculations~\cite{WichmannPhD}.
It is clear that additional analysis is needed
here. Again, an important step will be to include additional nonrectangular
states in the minimisation basis.

\subsection[The Large $\beta$ Plateaux]{The Large $\bm{\beta}$ Plateaux}

Having considered the small $\beta$ minima as possible scaling regions
we now move on to the large $\beta$ plateaux. This scaling  region
appears for all $N$ and for all states considered and so its
interpretation as a genuine scaling region is less
dubious. Furthermore, the glueball mass results extracted from these scaling
regions may be confidently compared to the corresponding Lagrangian
calculations since the ratio of the Hamiltonian to Euclidean coupling
is unity up to small $\ord(g_E^2)$ corrections.\\

\begin{figure}
\centering
                       
\includegraphics[width=10cm]{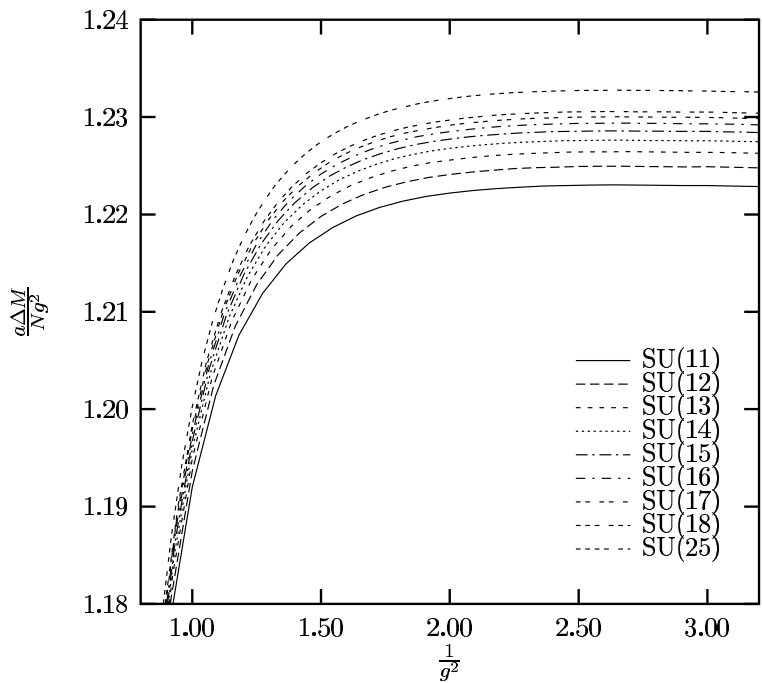}

\caption{ Lowest mass symmetric 2+1 dimensional massgaps in units
of $N g^2/a$ as a function of $1/g^2$.}
\label{S-ev1-hib-CONV}  
\end{figure}

\begin{figure}
\centering
                       
\includegraphics[width=10cm]{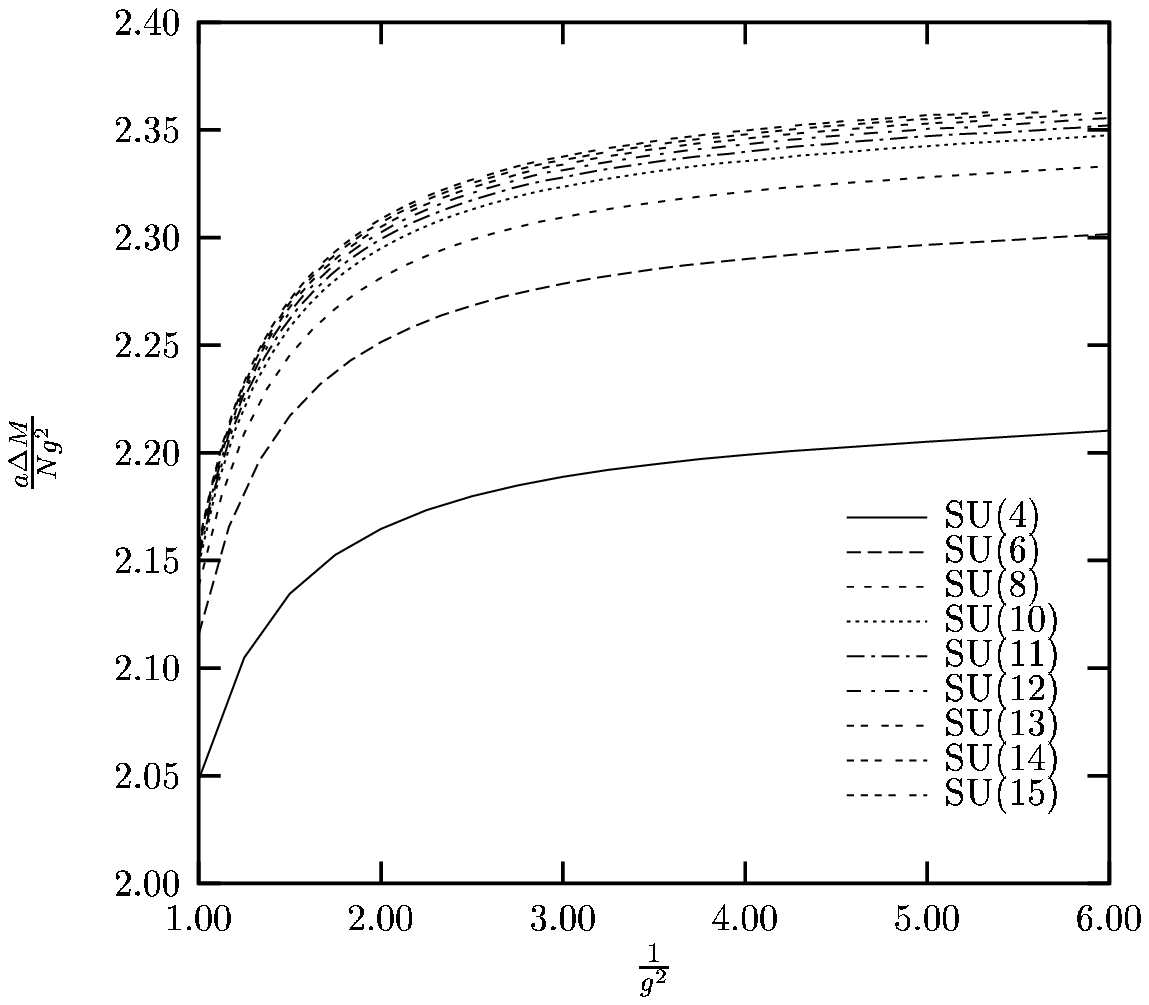}

\caption{ Second lowest mass symmetric 2+1 dimensional massgaps in units
of $N g^2/a$ as a function of $1/g^2$.}
\label{S-ev2-hib-CONV}  
\end{figure}

We start with the $0^{++}$ state for which the best scaling behaviour
is obtained. We calculate the five lowest lying massgaps corresponding
to the five lowest glueball masses accessible with our choice of 
ground state and minimisation
basis. We consider values of $N$ in the range $3\le N \le 25$. For each $N$ considered 
we find a large $\beta$ scaling plateau for each of the five lowest mass states. 
The lowest lying massgap is shown in \fig{S-ev1-hib-CONV} for a range
of $N$. We observe that in units of $N g^2/a$ the massgaps do not
depend strongly on $N$ and that in the scaling region they appear to
approach a finite limit. Similar observations can be made for the
four higher mass states obtained, although the scaling behaviour is
less precise. We show the second lowest glueball mass in \fig{S-ev2-hib-CONV} as an example. A continuum limit for each massgap
is obtained in the scaling region by fitting to a constant. For each
fit we use a region of at least 10 data points which minimises the
standard error. The continuum limit results for the five lowest lying
massgaps, denoted by  $a
\Delta M^c /(N g^2)$, are shown as functions of $1/N^2$ in 
\figs{S-ev1-hib-CONV}{s-largen-conv}. Also shown in
the plots are fits to models with leading order large $N$ corrections
starting $1/N^2$,
\bea
\kappa_i^{++} = p_i + \frac{q_i}{N^2} +  \frac{r_i}{N^4}.
\label{kappa++}
\eea 
Here the subscript $i$ labels the $i$-th lowest glueball mass. 
The values of the best fit parameters are given in \tab{kappa++fit}.\\

\begin{table}[t]
\begin{minipage}{\linewidth}
\renewcommand{\thefootnote}{\thempfootnote}
\begin{center}
\begin{tabular}{c||lll}
\hline
$i$  &  \multicolumn{1}{c}{$p_i$}   &  \multicolumn{1}{c}{$q_i$}  & \multicolumn{1}{c}{$r_i$}   \\
\hline
1  &  $1.23526\pm 0.00023$    & $-1.540\pm 0.018$      &  $1.97 \pm  0.16$    \\
2 & $2.36924 \pm 0.00072$      & $-2.478\pm 0.047$      & $-1.62 \pm 0.42 $      \\
3 & $ 2.88446 \pm 0.00071 $     &$ -3.435 \pm 0.017$      &0\footnote{Set to zero to obtain a stable fit.}      \\
4 & $  3.35422 \pm 0.00047 $    & $ -3.476 \pm 0.012$     & 0\footnotemark[\value{mpfootnote}]      \\
5 & $  3.7667 \pm 0.0013  $   & $ -4.114 \pm 0.092$     & $1.88 \pm
0.83$ \\
\hline
\end{tabular}
\end{center}
\end{minipage}
\caption{The best fit parameters for the five lowest energy $0^{++}$ massgaps
when fitting \eqn{kappa++} to the available data.}
\label{kappa++fit}
\end{table} 

\begin{figure}
\centering
                       
\includegraphics[width=10cm]{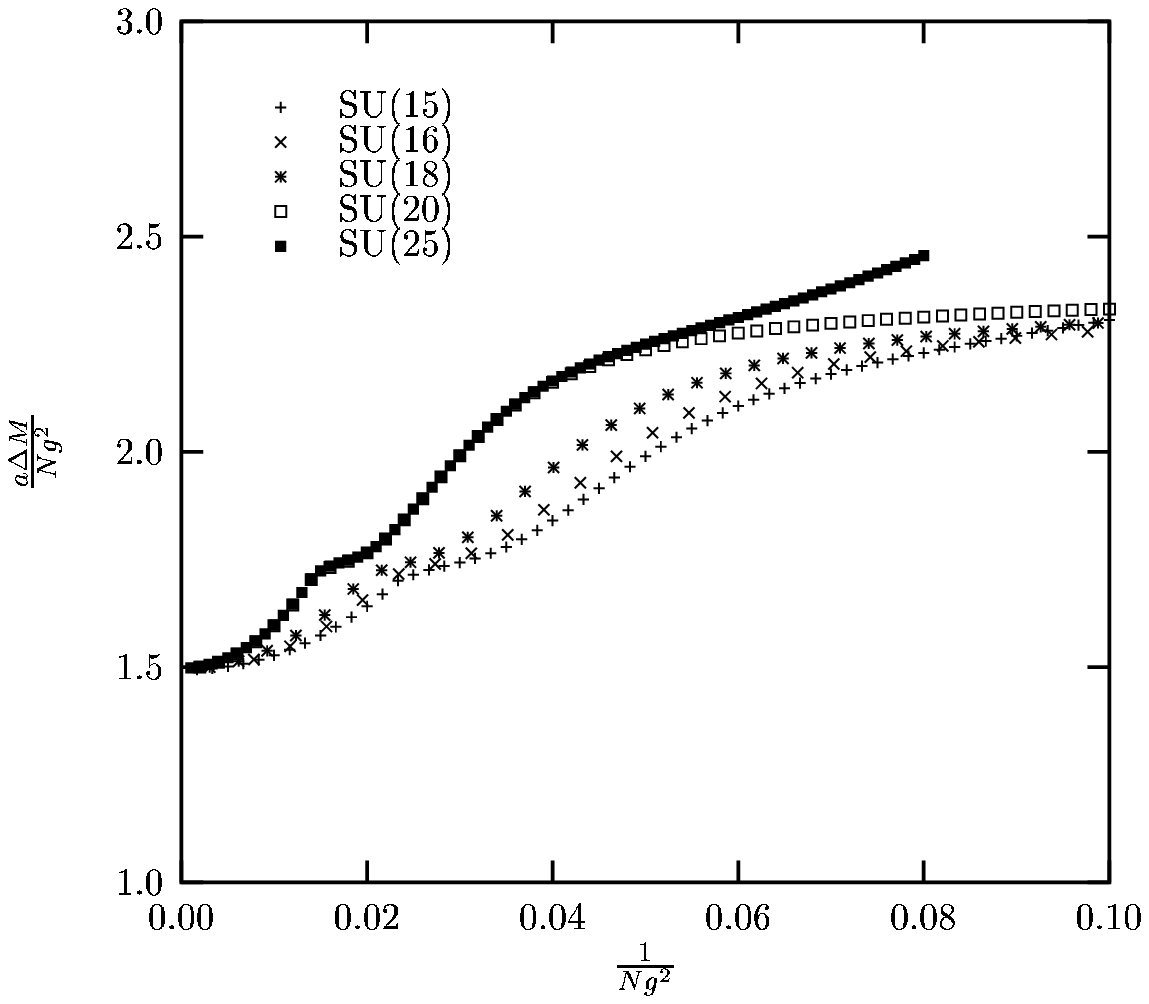}

\caption{Second lowest mass symmetric 2+1 dimensional massgaps in units
of $N g^2/a$ as a function of $1/(N g^2)$.}
\label{S-ev2-lob-CONV}  
\end{figure}

These results should be compared with those of Lucini and Teper who
obtain large $N$ glueball masses in units of $N g^2/a$ in the $0^{++}$ sector with linear fits given by
\bea
 0^{++}:&& 0.8116(36) - \frac{0.090(28)}{N^2} \nn\\
 0^{++*}:&& 1.227(9) - \frac{0.343(82)}{N^2} \nn\\
 0^{++**}:&& 1.65(4) -\frac{2.2(7)}{N^2}.
\eea
The fit for the $0^{++**}$ state is obtained using $4\le N \le 6$
data. The remaining fits are obtained using $2\le N\le 6$ data.
We find that the lowest glueball mass extracted from our large $\beta$
plateaux is consistent, in the $N\rightarrow \infty$ limit, with the
state which Lucini and Teper label $0^{++**}$. The slopes of the fits are of the same
sign but differ significantly in magnitude. The $0^{++**}$ state of
Lucini and Teper does not appear in our data in the form of a large
$\beta$ scaling plateau. There is however a hint of an approach to
scaling in the vicinity of their result in our second lowest massgap data. This
effect is only visible in our data for $N \ge 13$ and occurs for small $\beta$
as shown in \fig{S-ev2-lob-CONV}. The values of $\beta$ for
which this effect appears are quite close to those for which the small
beta minima are observed in the lowest mass eigenstate. 
Similar effects are observed
in the $0^{--}$ data but the effect is much less convincing with the
data currently available.\\

\begin{figure}
\centering
                       
\includegraphics[width=10cm]{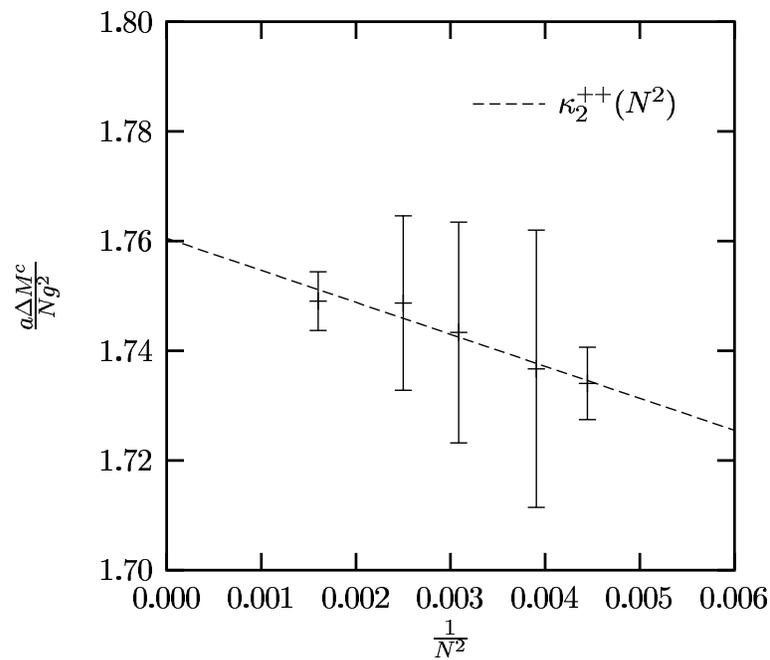}

\caption{Continuum limit extrapolations (in units
of $N g^2/a$ as a function of $1/N^2$) derived from the low $\beta$ scaling
region that appears for $N\ge 13$ in the second lowest mass $0^{++}$
SU($N$) massgap. The dashed line is the fit to the
quadratic model of \eqn{kappa++lob}.}
\label{S-ev2-lob}  
\end{figure}

Let us consider this small $\beta$ region as a possible scaling
region. We fit a constant to the available $N\ge 15$ data on a range
of at least 4 data points chosen so that the standard error is
minimised. These scaling values are plotted as a function of $1/N^2$
in \fig{S-ev2-lob}. Also shown is the best fit linear model
\bea
\kappa_2^{++} = 1.7605\pm 0.0032 - \frac{5.83 \pm 0.97}{N^2}. 
\label{kappa++lob}
\eea
This produces an $N\rightarrow \infty$ limit which is close to, but
inconsistent with, the result obtained by Lucini and Teper for their 
$0^{++**}$ state.\\

\begin{figure}
\centering
                       
\includegraphics[width=10cm]{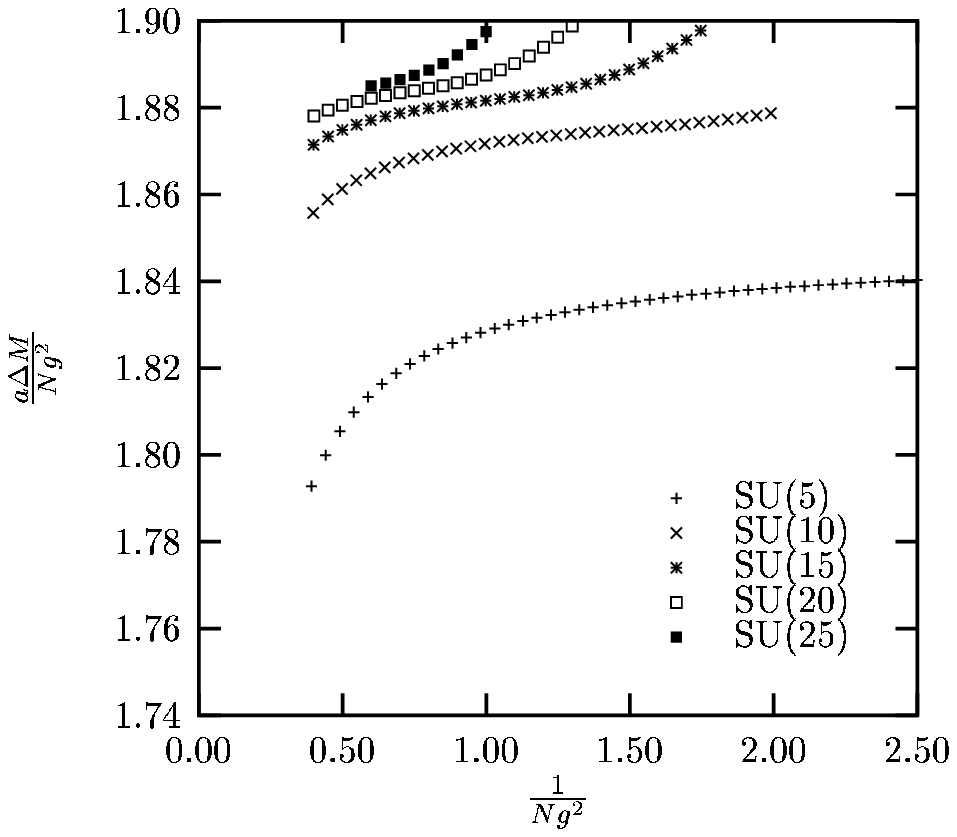}

\caption{ Lowest mass $0^{--}$ 2+1 dimensional massgaps in units
of $N g^2/a$ as a function of $1/(N g^2)$.}
\label{AS-ev1-hib-CONV}  
\end{figure}

\begin{figure}
\centering
                       
\includegraphics[width=10cm]{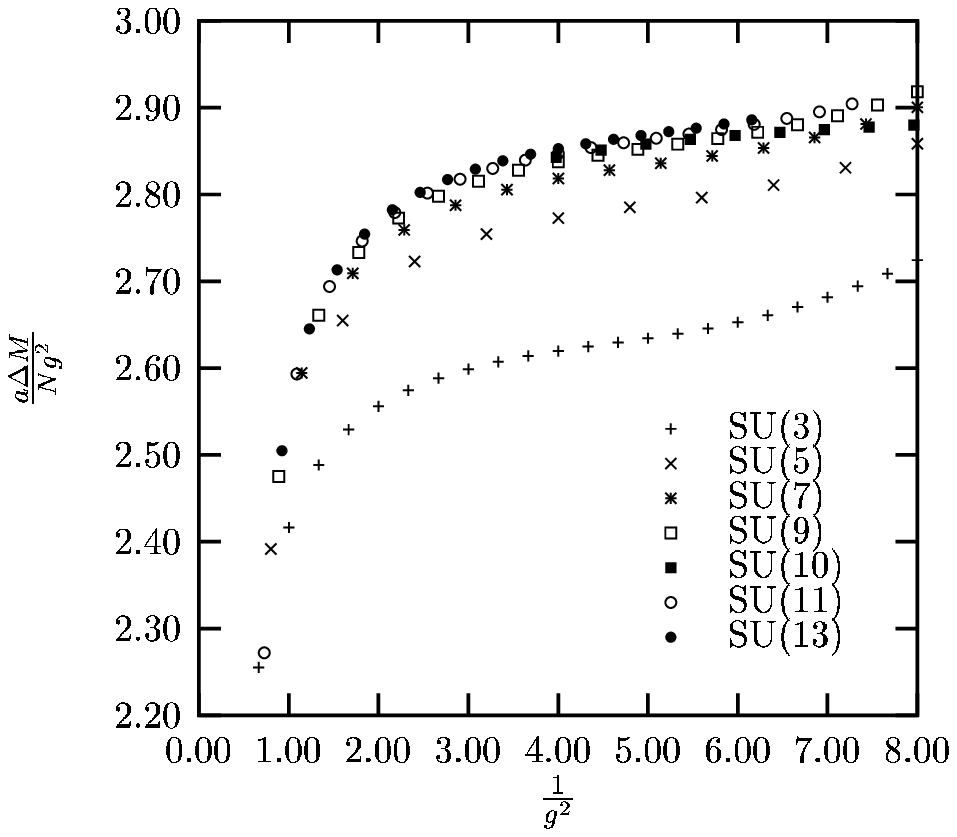}

\caption{Second lowest mass $0^{--}$ 2+1 dimensional massgaps in units
of $N g^2/a$ as a function of $1/g^2$.}
\label{AS-ev2-hib-CONV}  
\end{figure}

\begin{figure}
\centering
                       
\includegraphics[width=10cm]{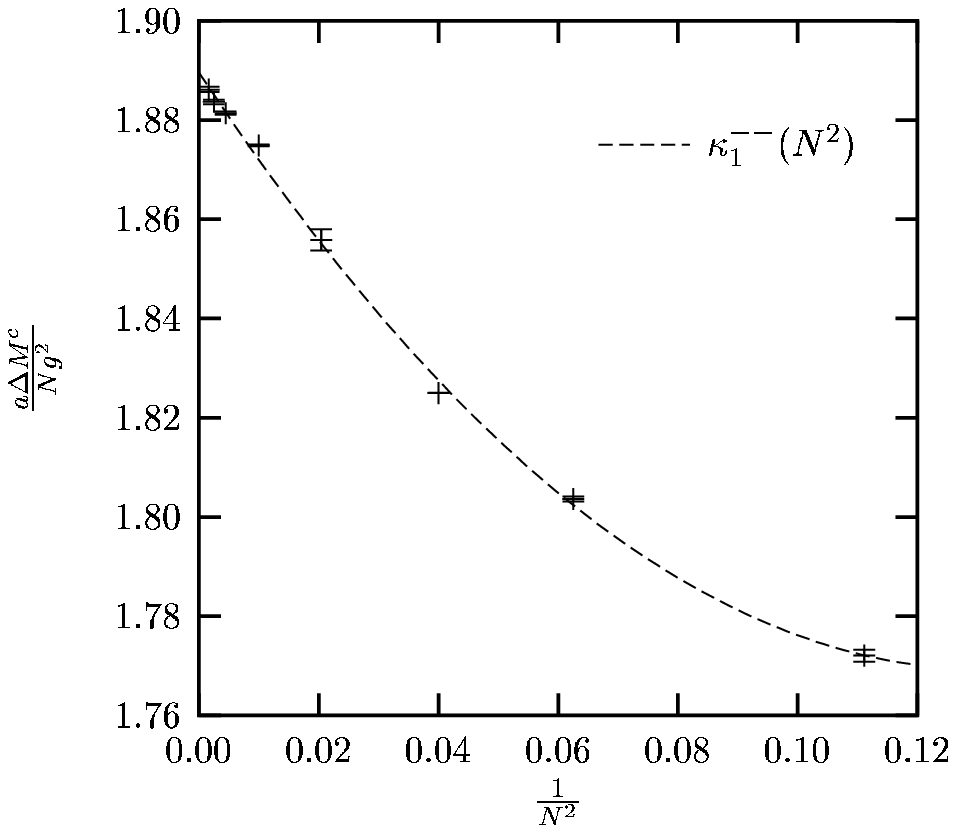}

\caption{The 2+1 dimensional continuum limit lowest $0^{--}$ SU($N$)
glueball mass
 in units
of $N g^2/a$ as a function of $1/N^2$. The dashed line is the fit to the
quadratic model of \eqn{kappa--}.}
\label{AS-ev1-hib}  
\end{figure}

We now move on to the $0^{--}$ states. For these states the scaling
behaviour is not as precise as that obtained for the $0^{++}$ states.  
Again we calculate the five lowest lying massgaps corresponding
to the five lowest mass glueballs accessible with our choice of minimisation
basis and ground state. For each $N$ considered up to 25 we find a
large $\beta$ scaling plateau for each of the five states. The lowest
lying massgap is shown in \fig{AS-ev1-hib-CONV} for a range
of $N$. The second lowest energy massgap is shown in
\fig{AS-ev2-hib-CONV}. Considerably less data has been obtained
for the $0^{--}$ states due to the large $l_{\rm max}$ required for
convergence as discussed in \sect{convwithl}. Despite this, the results qualitatively replicate those
of the $0^{++}$ states. We observe that in units of $N g^2/a$ the massgaps do not
depend strongly on $N$ and that in the scaling region they appear to
approach a finite limit. Similar observations can be made for the
four higher mass states, although the scaling behaviour is less
precise. The continuum limit values extracted
are plotted as functions of $1/N^2$ in
\figs{AS-ev1-hib}{as-largen-conv}. The dashed lines show fits of the
continuum limit values to the model 
\bea
\kappa_i^{--} = p_i + \frac{q_i}{N^2} +  \frac{r_i}{N^4},
\label{kappa--}
\eea
with the parameters for each excited state given in
\tab{kappa--fit}.\\

\begin{table}[t]
\begin{minipage}{\linewidth}
\renewcommand{\thefootnote}{\thempfootnote}
\begin{center}
\begin{tabular}{c||lll}
\hline
$i$  &  \multicolumn{1}{c}{$p_i$}   &  \multicolumn{1}{c}{$q_i$}  & \multicolumn{1}{c}{$r_i$}   \\
\hline
1  &  $1.8896 \pm 0.0011$    & $-1.829 \pm 0.068$      &
$6.95 \pm 0.62$    \\
2  &  $2.8930 \pm 0.0044 $    & $-2.96 \pm 0.23$      &
$6.6\pm 2.0$    \\
3  &  $3.20871 \pm 0.0047$    & $-3.35 \pm 0.22$      &
$8.8 \pm 1.9$    \\
4  &  $3.83647 \pm 0.0024 $    & $-3.41\pm 0.05$      &  0\footnote{Set to zero to obtain a stable fit.}    \\
5  &  $4.0759 \pm 0.0019 $    & $-2.45 \pm 0.12$      &
$-7.4\pm 1.1$    \\
\hline
\end{tabular}
\end{center}
\end{minipage}
\caption{The best fit parameters for the five lowest energy $0^{--}$ massgaps
when fitting \eqn{kappa--} to the available data.}
\label{kappa--fit}
\end{table} 

These results should be compared again with those of Lucini and Teper who
obtain large $N$ glueball masses in units of $N g^2/a$ in the $0^{--}$ sector with linear fits given by
\bea
 0^{--}:&& 1.176(14) + \frac{0.14(20)}{N^2} \nn\\
 0^{--*}:&& 1.535(28) - \frac{0.35(35)}{N^2} \nn\\
 0^{--**}:&& 1.77(13) + \frac{0.24(161)}{N^2}.
\eea
The result for $0^{--**}$ is from \rcite{Teper:1998te}. All
fits were obtained using $3\le N \le 6$ data. The $N\rightarrow
\infty$ limit of our lowest lying massgap
extracted from the large $\beta$ plateaux is consistent with Teper's
$0^{--**}$ result.\\

Having considered spin 0 states we now move on to spin 2, the only
other spin accessible when using a basis of rectangular states. For
the case of spin 2 the scaling behaviour of the massgaps is
significantly worse than for spin 0. More troublesome is the fact that 
the convergence of the $2^{++}$
massgaps with increasing $l_{\rm max}$ is slower than the case of $0^{--}$. 
The situation for the $2^{--}$ state is markedly better with the
convergence of massgaps with increasing $l_{\rm max}$ being no different to that of
$0^{++}$. For this reason we concentrate on the $2^{--}$ sector here.\\

\begin{figure}
\centering
                       
\includegraphics[width=10cm]{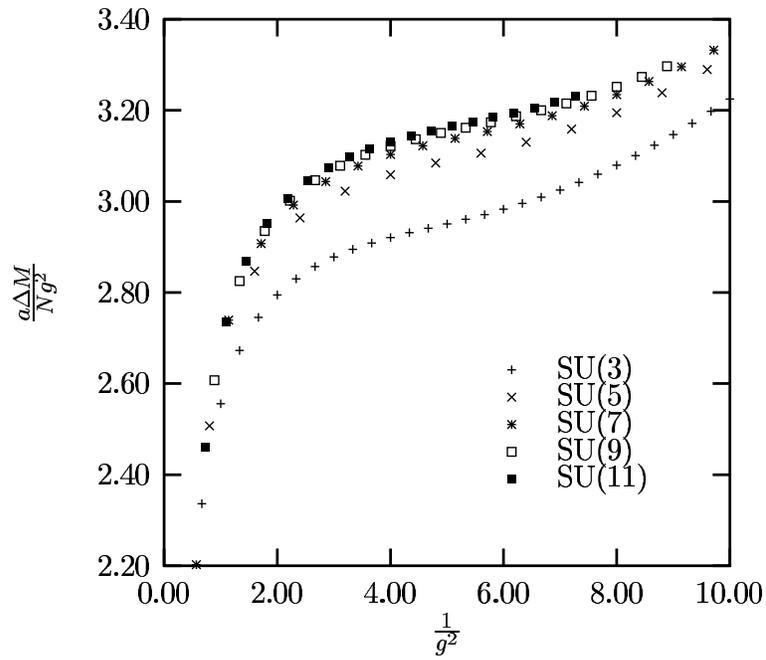}

\caption{The lowest lying 2+1 dimensional $2^{--}$ SU($N$) massgaps in units
of $N g^2/a$ as a function of $1/g^2$.}
\label{AS-ev1-hib-spin2-CONV}  
\end{figure}

\begin{figure}
\centering
                       
\includegraphics[width=10cm]{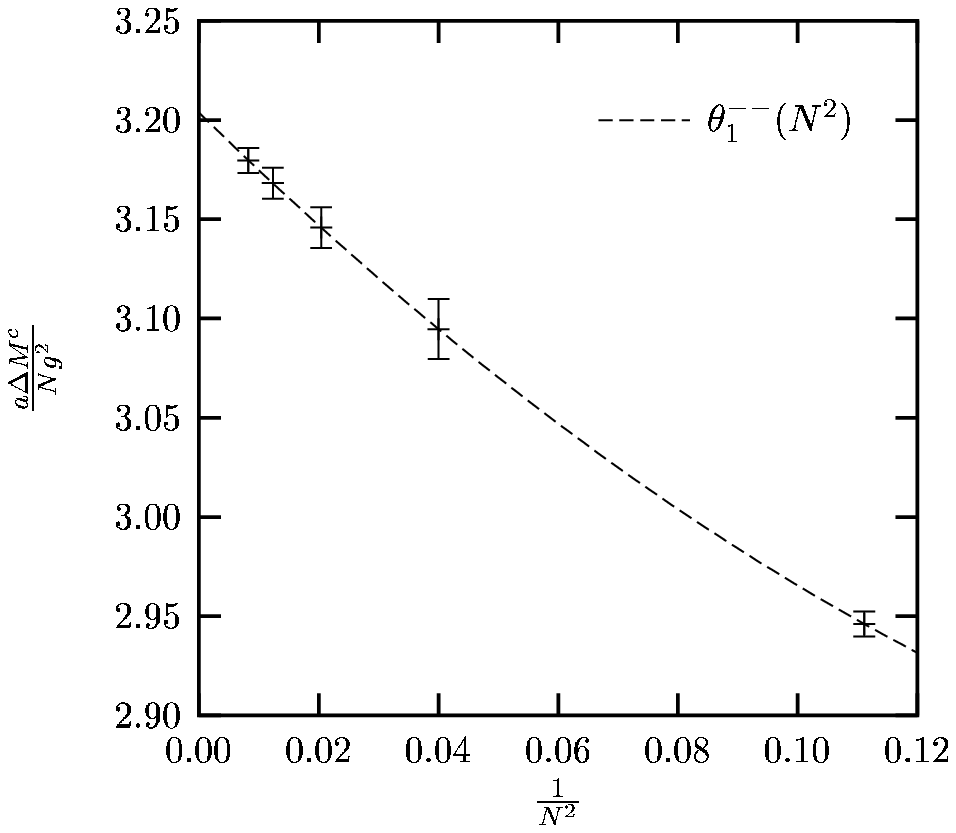}

\caption{The 2+1 dimensional 
continuum limit lowest $2^{--}$ SU($N$) glueball mass
 in units
of $N g^2/a$ as a function of $1/N^2$. The dashed line is the fit to the
quadratic model of \eqn{theta--}.}
\label{AS-ev1-hib-spin2}  
\end{figure}

The $2^{--}$ states produce less precise scaling
behaviour than the spin 0 states. However large $\beta$ plateaux
appear for each of the five lowest lying massgaps. We use these
regions to estimate their respective continuum limits. For these
states small $\beta$ minima do not appear.  
The lowest lying $2^{--}$ massgap is shown in
\fig{AS-ev1-hib-spin2-CONV} for a range of $N$.  
Once again we observe that in units of $N g^2/a$ the massgaps do not
depend strongly on $N$ and that in the scaling region they appear to
approach a finite limit. Similar observations can be made for the
four higher energy states obtained, although the scaling behaviour
worsens as the mass of the state increases. 
The extracted continuum limit values are plotted as functions of $1/N^2$ in \figs{AS-ev1-hib-spin2}{as-largen-conv-spin2}. The dashed lines show fits to the model 
\bea
\theta_i^{--} = p_i + \frac{q_i}{N^2} +  \frac{r_i}{N^4}.
\label{theta--}
\eea
The best fit parameters for each excited state are given in \tab{theta--fit}. \\

\begin{table}[t]
\begin{minipage}{\linewidth}
\renewcommand{\thefootnote}{\thempfootnote}
\begin{center}
\begin{tabular}{c||lll}
\hline
$i$  &  \multicolumn{1}{c}{$p_i$}   &  \multicolumn{1}{c}{$q_i$}  & \multicolumn{1}{c}{$r_i$}   \\
\hline
1  &  $3.20379\pm 0.00006$    & $-2.9571\pm 0.0032$      &
$5.73912\pm 0.02585$    \\
2  &  $4.07376\pm 0.00066$    & $-2.189\pm 0.050$      &
$-9.46 \pm 0.43$    \\
3  &  $4.96259 \pm 0.00090$    & $-4.947 \pm 0.045$      &
$1.99\pm 0.35$    \\
4  &  $5.26717\pm 0.00040$    & $-6.221\pm 0.031$      & $15.92
\pm 0.26$    \\
5  &  $5.744 \pm 0.024$    & $-5.13\pm 0.56 $      & 0\footnote{Set to zero to obtain a stable fit.}    \\
\hline
\end{tabular}
\end{center}
\end{minipage}
\caption{The best fit parameters for the five lowest energy $2^{--}$ massgaps
when fitting \eqn{theta--} to the available data.}
\label{theta--fit}
\end{table} 

Once again these results should be compared with those of Lucini and Teper who
obtain large $N$ glueball masses in units of $N g^2/a$ in the $2^{--}$ sector with linear fits derived from
$3\le N \le 6$ data given by
\bea
 2^{--}:&& 1.615(33) - \frac{0.10(42)}{N^2} \nn\\
 2^{--*}:&& 1.87(12) - \frac{0.37(200)}{N^2}. 
\eea
The $2^{--*}$ result is from \rcite{Teper:1998te}. The 
$N\rightarrow \infty$ results presented here are significantly higher. This time 
no correspondence between our results and those of Lucini and Teper can
be obtained.   \\

\begin{figure}
\centering
                       
\includegraphics[width=10cm]{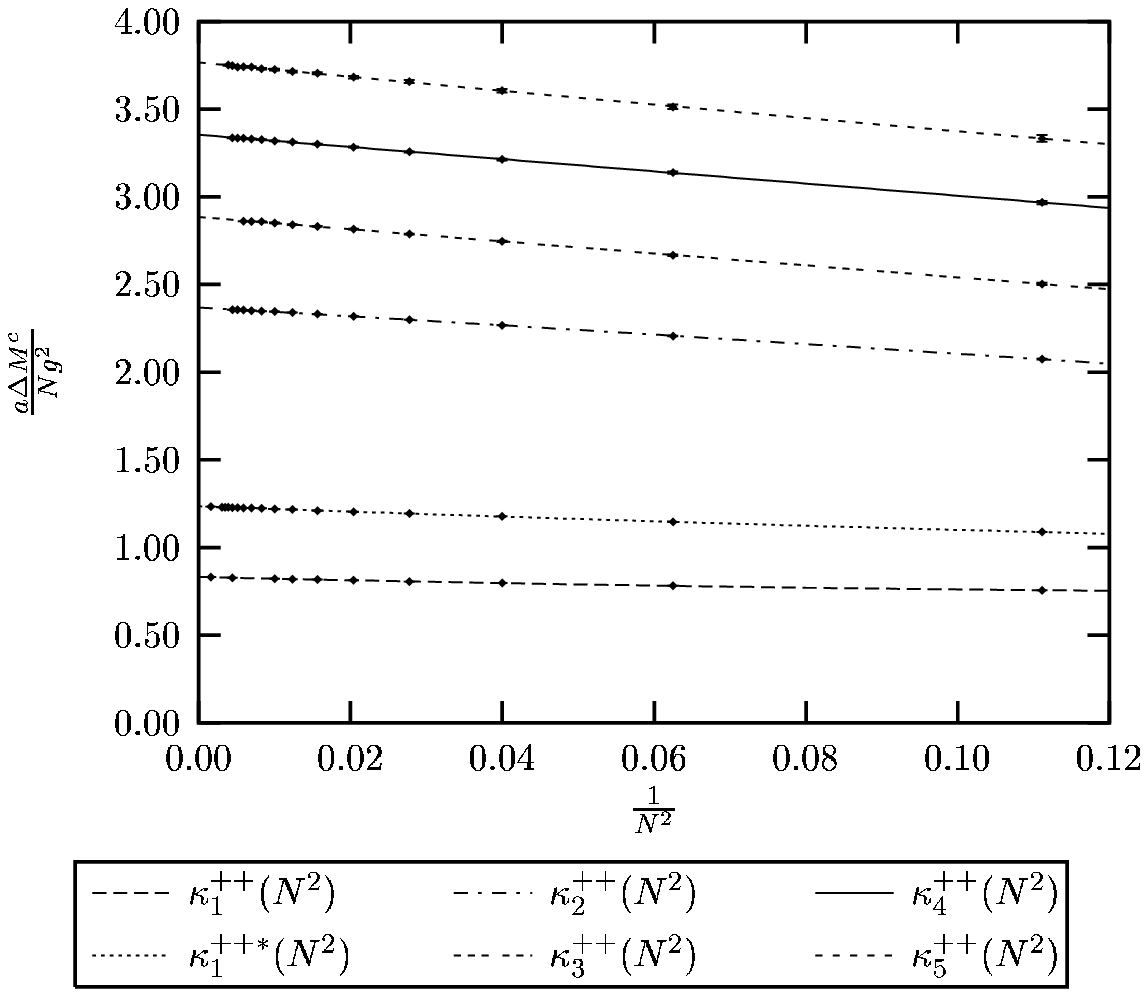}

\caption{An estimate of the continuum limit mass spectrum for 
$0^{++}$ SU($N$) glueballs in units
of $N g^2/a$ as a function of $1/N^2$. The lines are fits to the 
models of \eqn{kappa++}.}
\label{S-spectrum}  
\end{figure}

We finish this section by presenting the mass spectra obtained for the
$0^{++}$, $0^{--}$ and $2^{--}$ sectors. The results are summarised by
the plots in \figsss{S-spectrum}{AS-spectrum}{AS-spin2-spectrum}.

\begin{figure}
\centering
                       
\includegraphics[width=10cm]{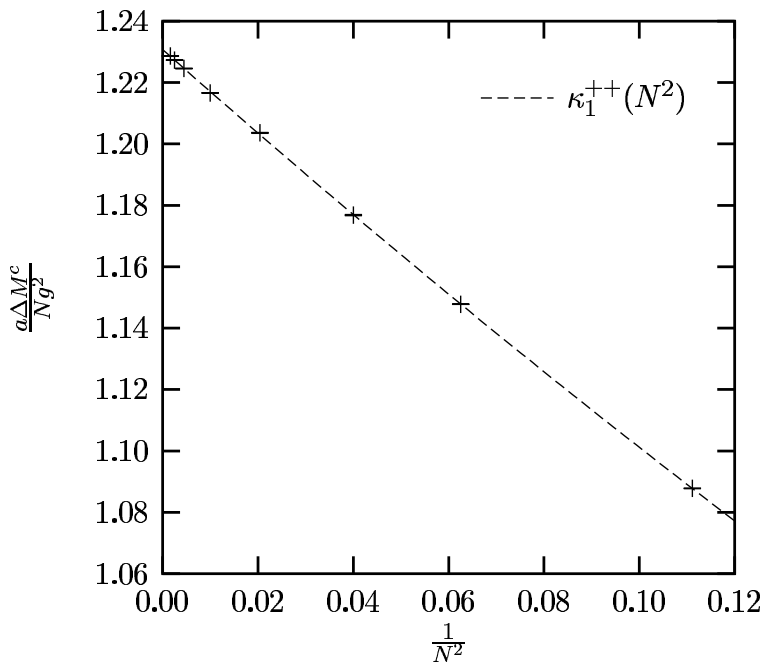}

\caption{Continuum limit lowest mass $0^{++}$ SU($N$) massgap
 in units
of $N g^2/a$ as a function of $1/N^2$. The dashed line is the fit to the
quadratic model of \eqn{kappa++}.}
\label{S-ev1-hib}  
\end{figure}

\begin{figure}
\centering
                             
\subfigure[2nd eigenvalue] 
                     {
                         \label{S-ev2-hib}
                         \includegraphics[width=7cm]{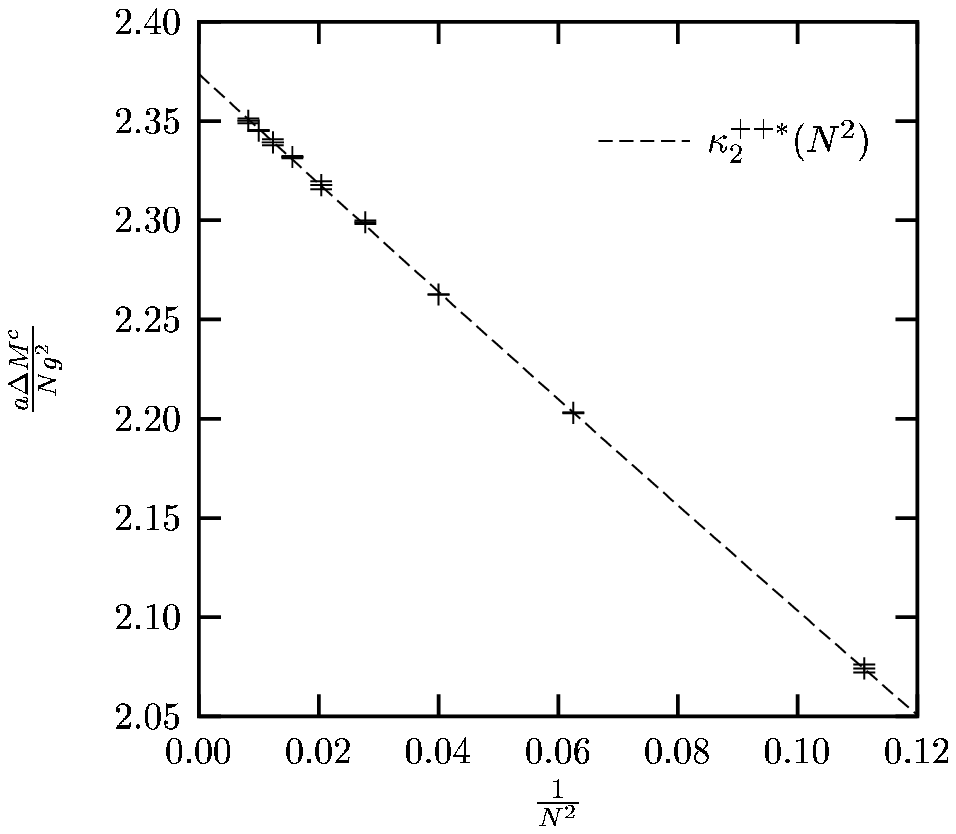}
                     }\hspace{0.25cm}
\subfigure[3rd eigenvalue] 
                     {
                         \label{S-ev3-hib}
                         \includegraphics[width=7cm]{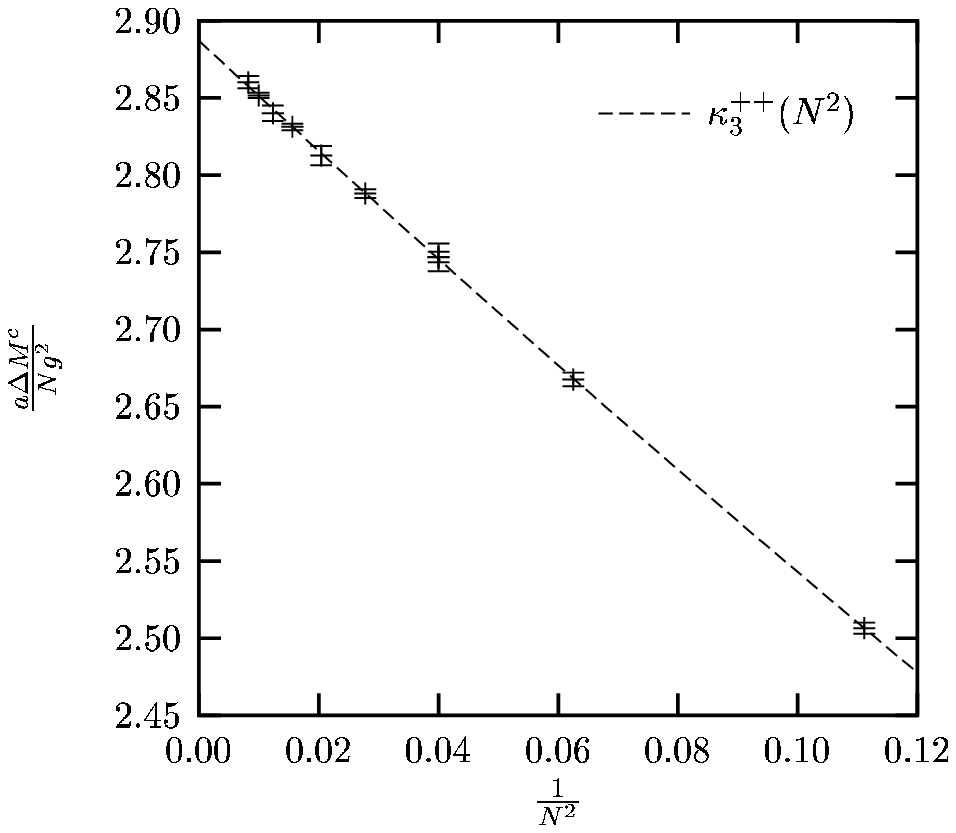}
                     }\\ 
\subfigure[4th eigenvalue] 
                     {
                         \label{S-ev4-hib}
                         \includegraphics[width=7cm]{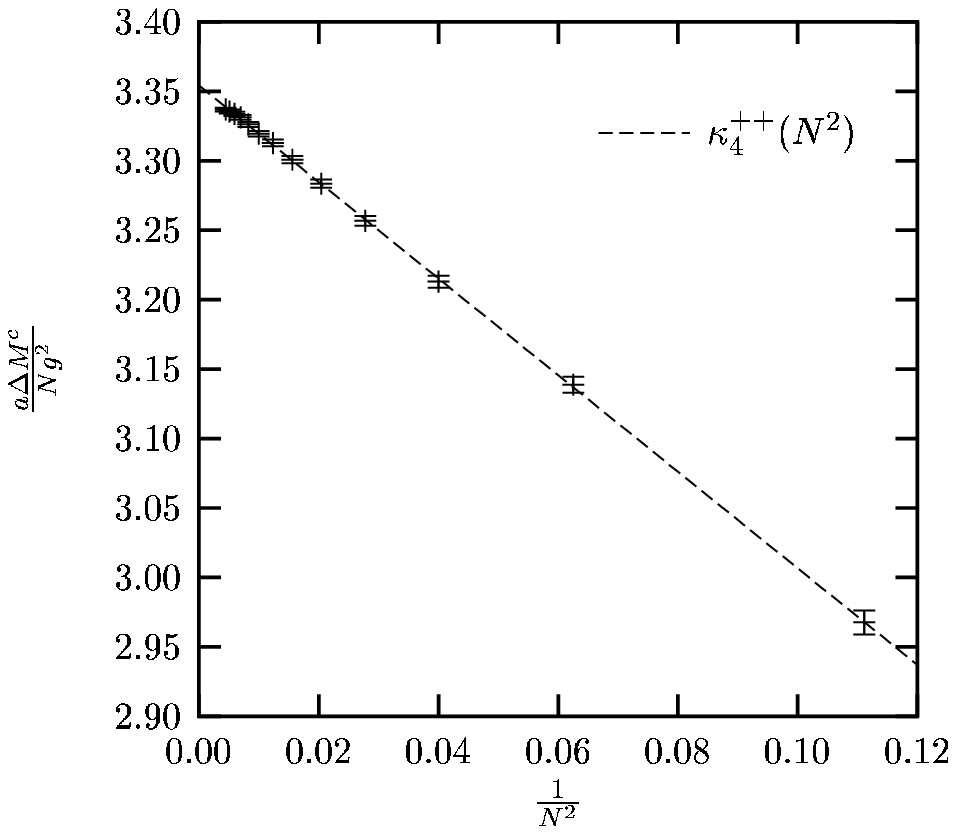}
                     }\hspace{0.25cm}
\subfigure[5th eigenvalue] 
                     {
                         \label{S-ev5-hib}
                         \includegraphics[width=7cm]{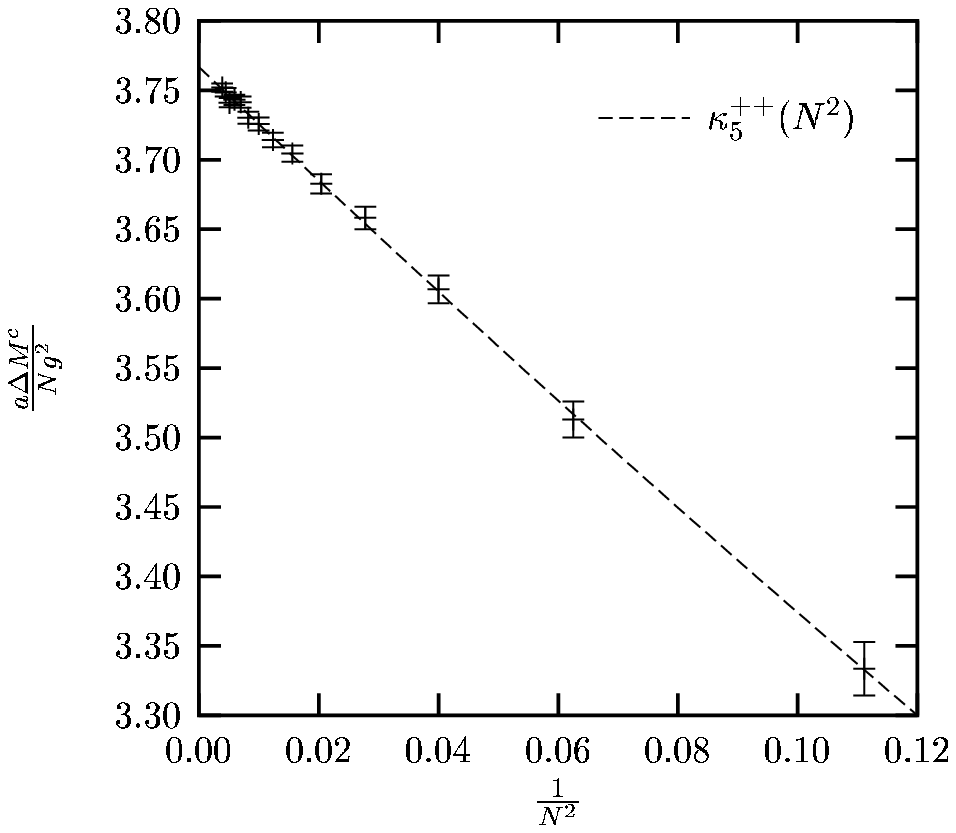}
                     }
\caption{Continuum limit $0^{++}$ SU($N$) massgaps in units of $N
 g^2/a$ as functions of $1/N^2$. The dashed lines are fits given in
\eqn{kappa++}.} 
\label{s-largen-conv}  
\end{figure}

\begin{figure}
\centering
               
\subfigure[2nd eigenvalue] 
                     {
                         \label{asymconv-2}
                         \includegraphics[width=7cm]{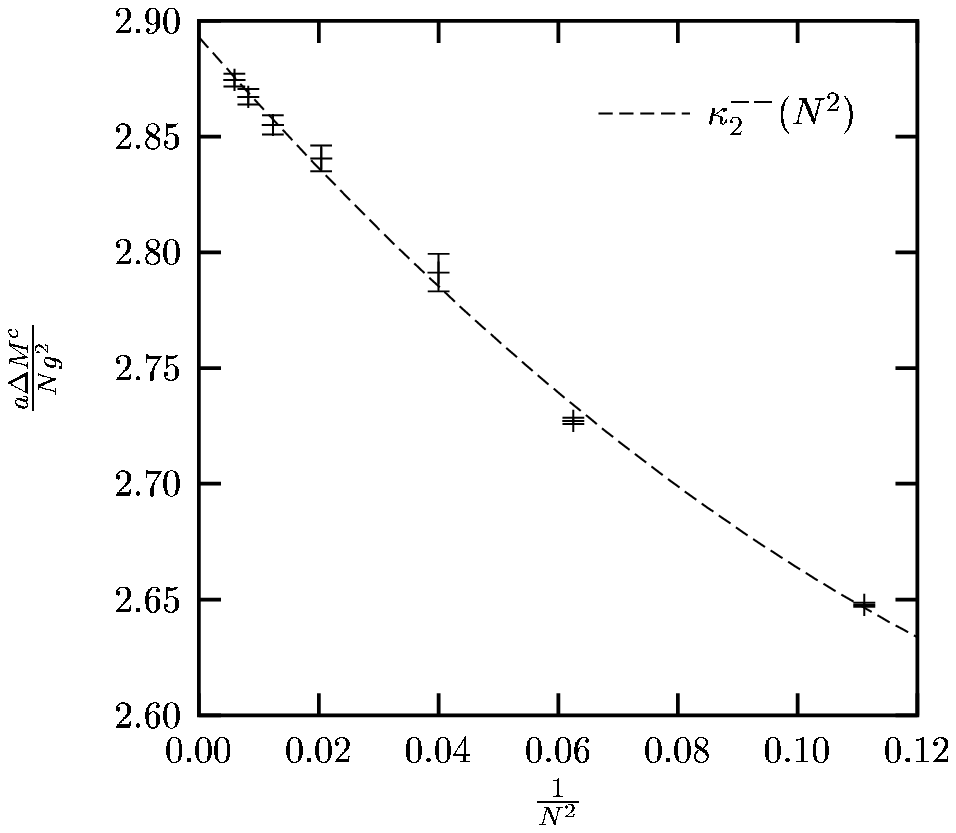}
                     }\hspace{0.25cm}
\subfigure[3rd eigenvalue] 
                     {
                         \label{asymconv-3}
                         \includegraphics[width=7cm]{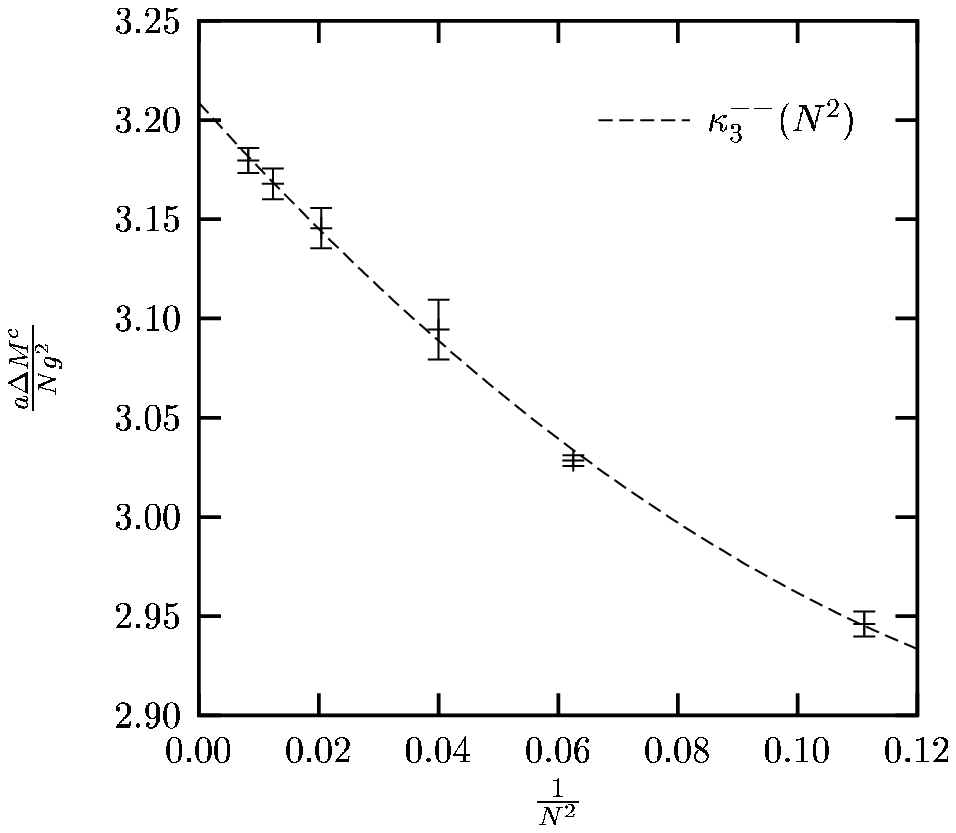}
                     }\\
\subfigure[4th eigenvalue] 
                     {
                         \label{asymconv-4}
                         \includegraphics[width=7cm]{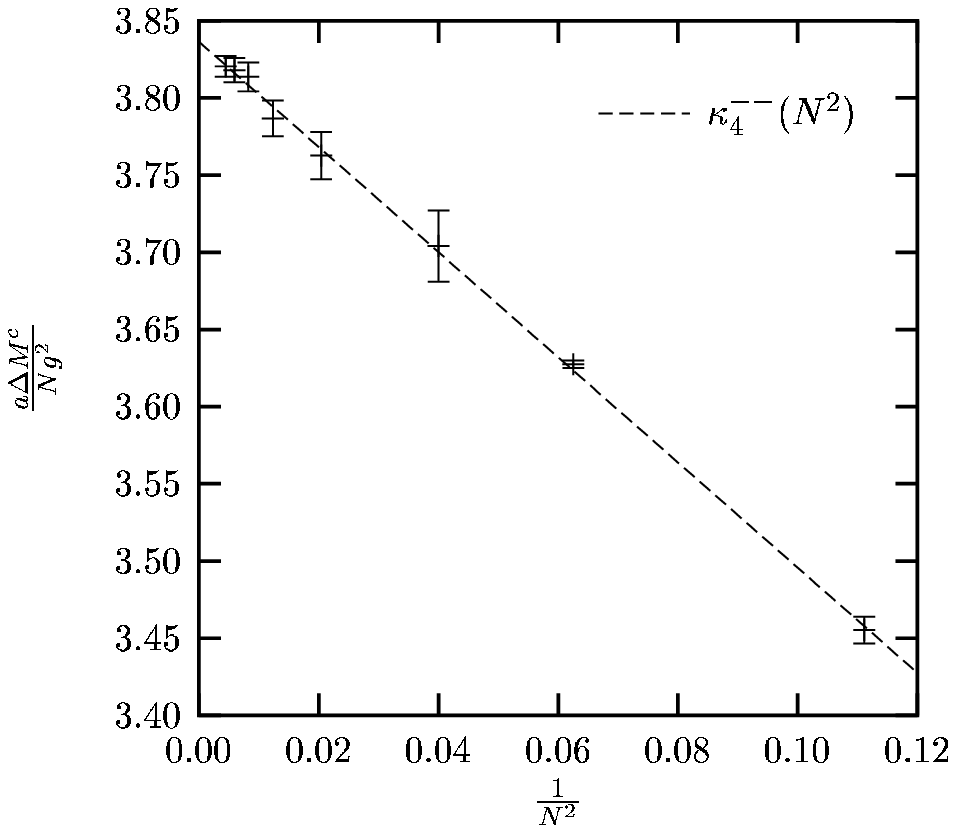}
                     }\hspace{0.25cm}      
\subfigure[5th eigenvalue] 
                     {
                         \label{asymconv-5}
                         \includegraphics[width=7cm]{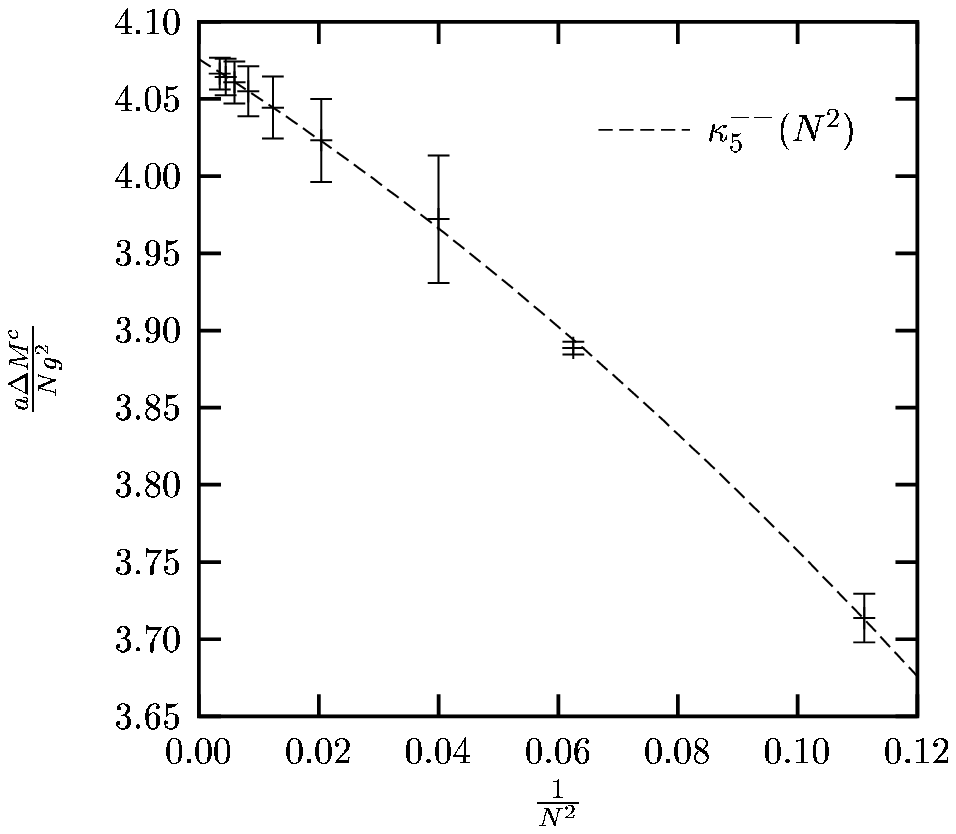}
                     }
\caption{Continuum limit $0^{--}$ SU($N$) massgaps in units of $N
 g^2/a$ as functions of $1/N^2$. The dashed lines are fits given in
 \eqn{kappa--}.} 
\label{as-largen-conv}  
\end{figure}

\begin{figure}
\centering
               
\subfigure[2nd eigenvalue] 
                     {
                         \label{asymconv-2}
                         \includegraphics[width=7cm]{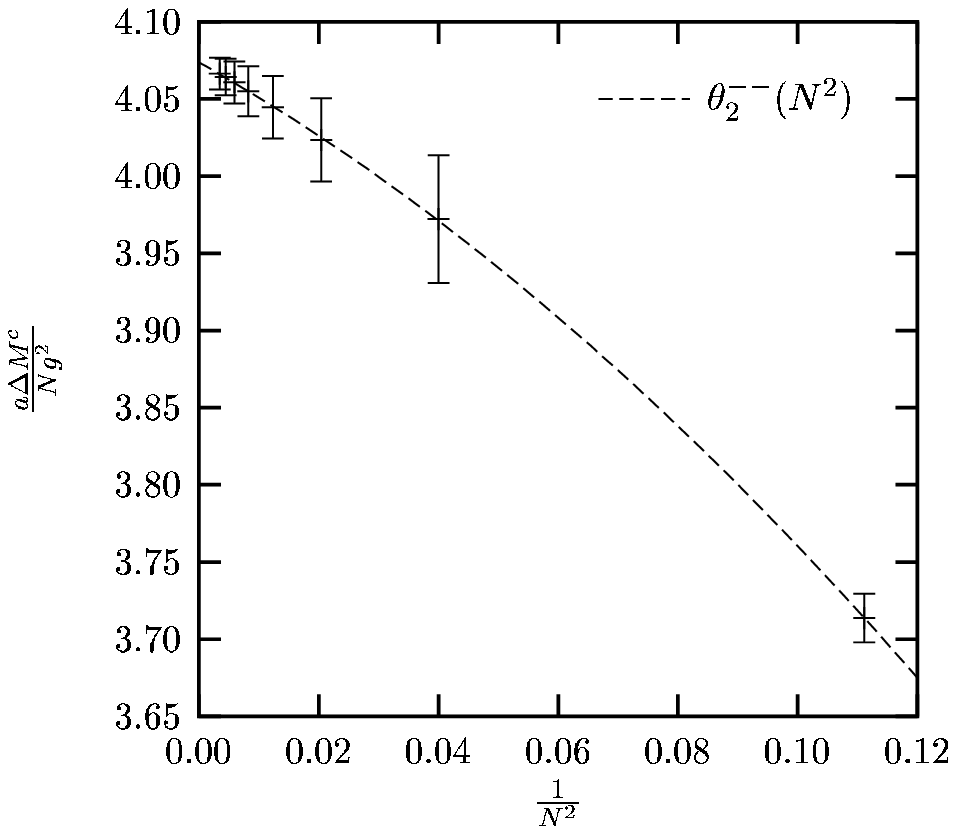}
                     }\hspace{0.25cm}
\subfigure[3rd eigenvalue] 
                     {
                         \label{asymconv-3}
                         \includegraphics[width=7cm]{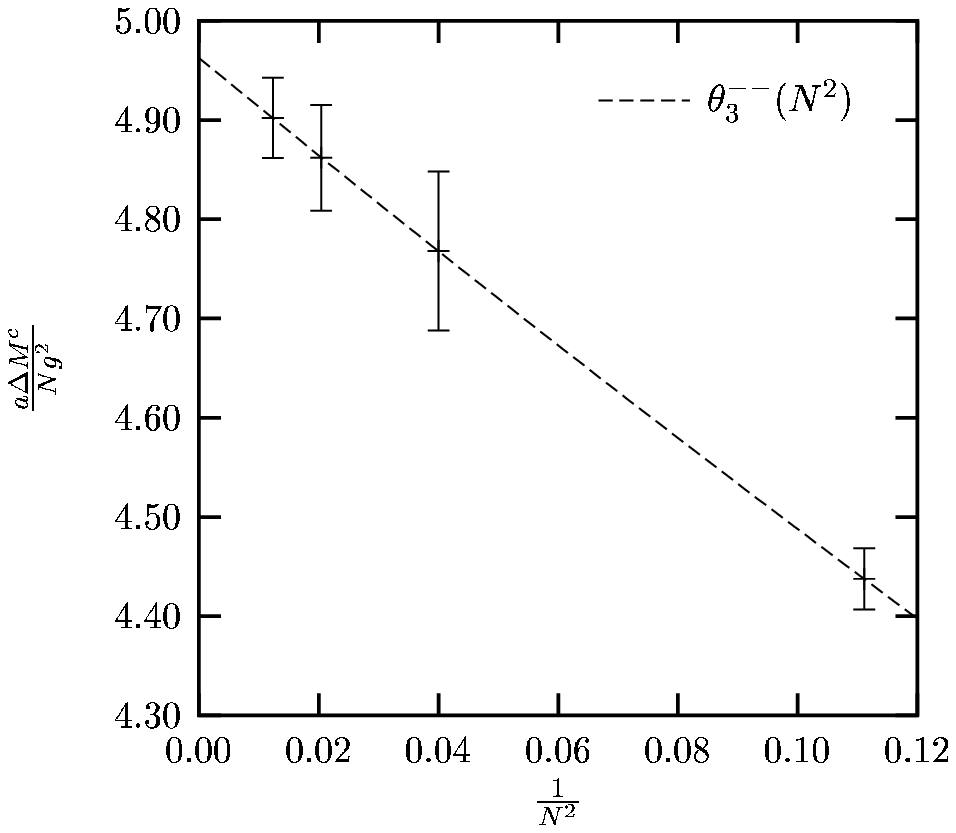}
                     }\\
\subfigure[4th eigenvalue] 
                     {
                         \label{asymconv-4}
                         \includegraphics[width=7cm]{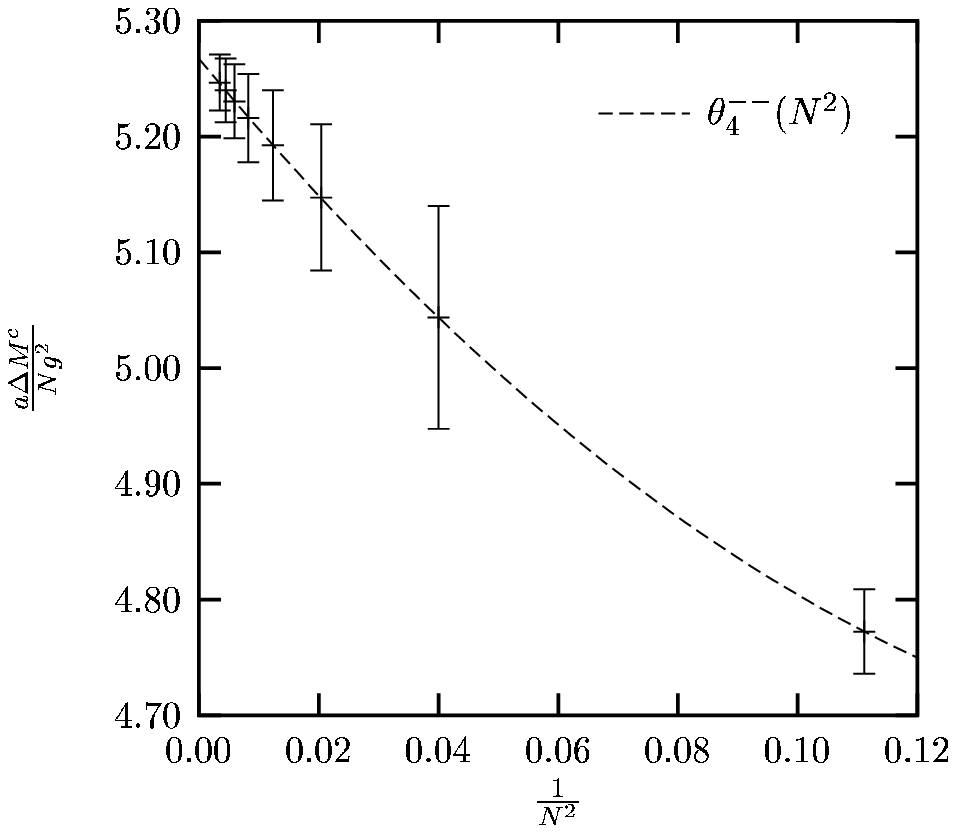}
                     }\hspace{0.25cm}      
\subfigure[5th eigenvalue] 
                     {
                         \label{asymconv-5}
                         \includegraphics[width=7cm]{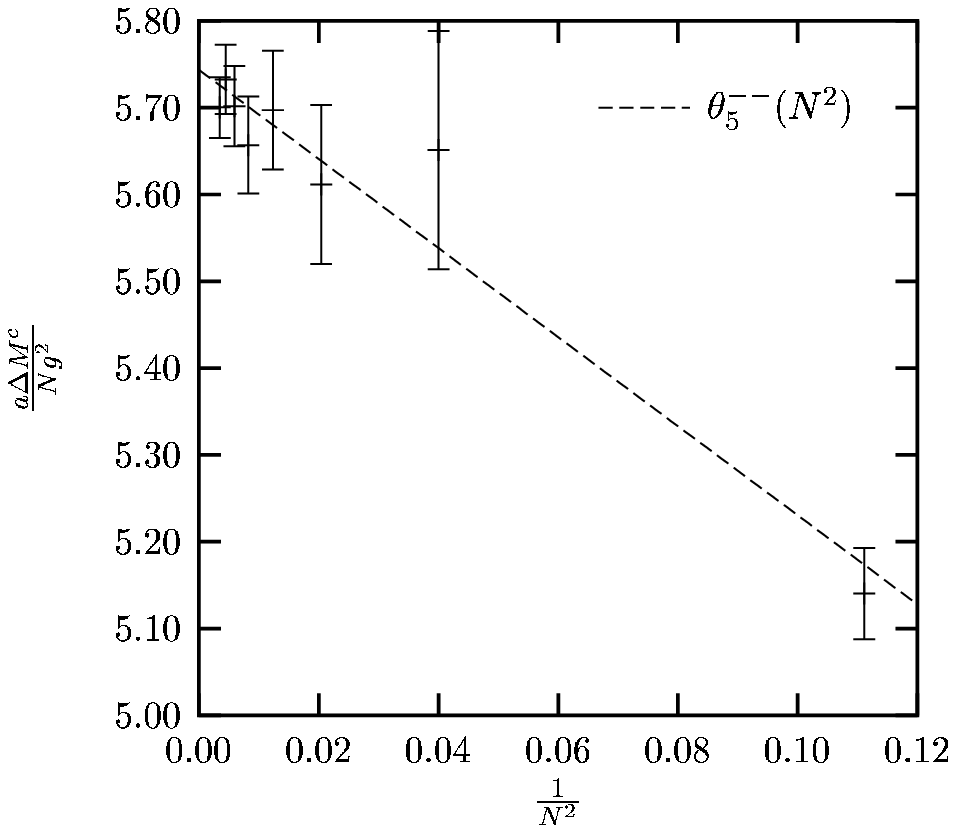}
                     }
\caption{The 2+1 dimensional continuum limit $2^{--}$ SU($N$) massgaps in units of $N
 g^2/a$ as functions of $1/N^2$. The dashed lines are fits given in
 \eqn{theta--}.} 
\label{as-largen-conv-spin2}  
\end{figure}

\begin{figure}
\centering
                       
\includegraphics[width=10cm]{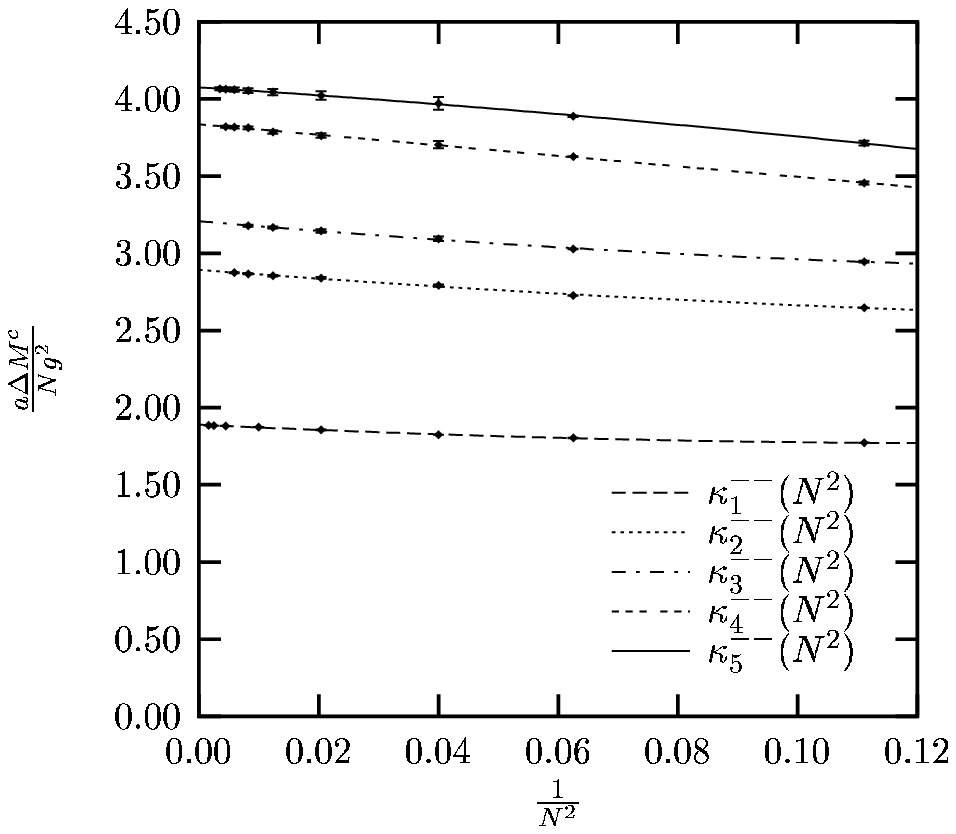}

\caption{An estimate of the continuum limit mass spectrum for 
$0^{--}$ SU($N$) glueballs in units
of $N g^2/a$ as a function of $1/N^2$. The lines are fits to the 
models of \eqn{kappa--}.}
\label{AS-spectrum}  
\end{figure}

\begin{figure}
\centering
                       
\includegraphics[width=10cm]{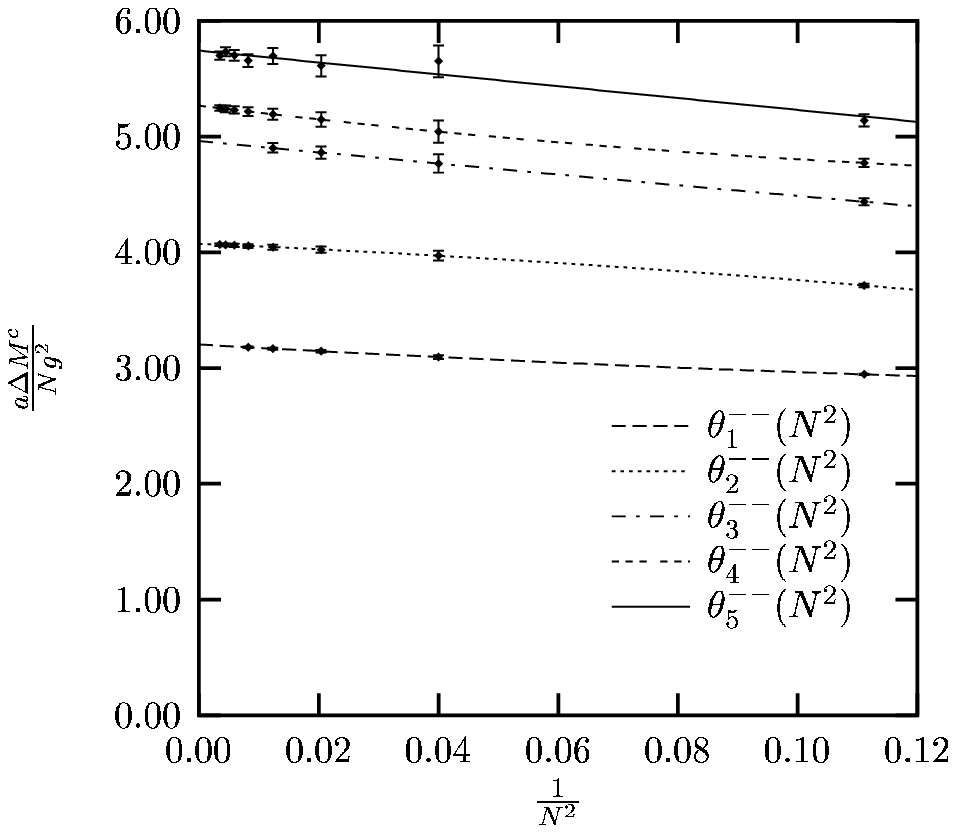}

\caption{An estimate of the continuum limit mass spectrum for $2^{--}$ SU($N$) glueballs in units
of $N g^2/a$ as a function of $1/N^2$. The lines are fits to the 
models of \eqn{theta--}.}
\label{AS-spin2-spectrum}  
\end{figure}

\subsection{An Empirical Observation}

\begin{figure}
\centering
                       
\includegraphics[width=10cm]{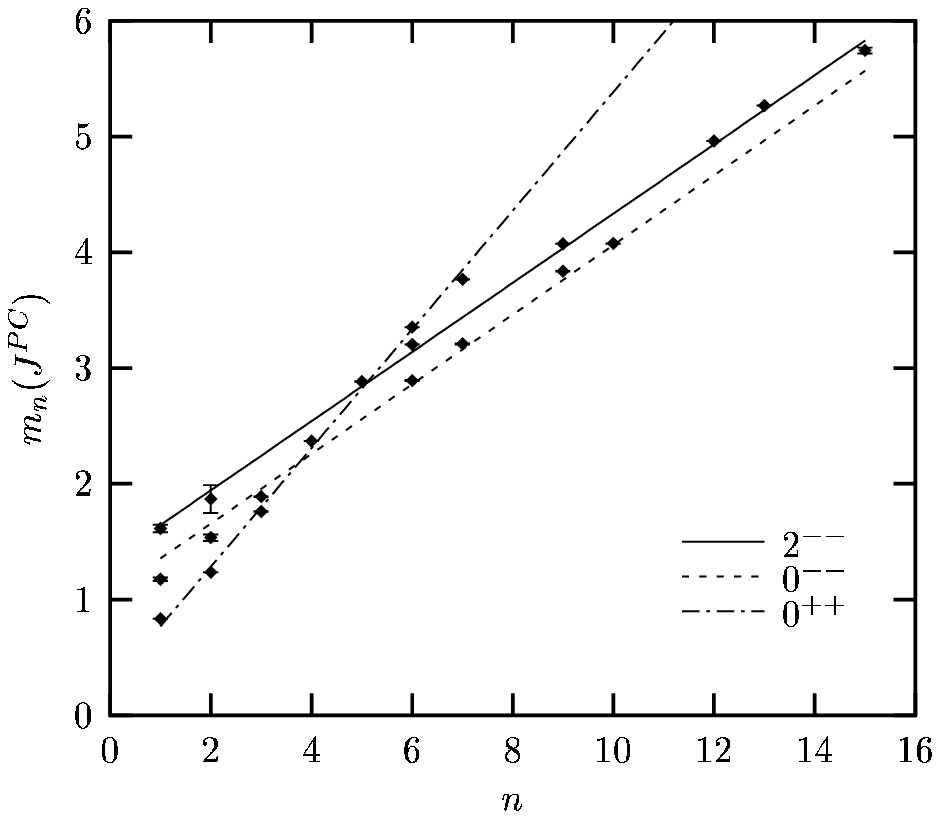}

\caption{A choice of enumeration of glueball masses and the fits of \eqn{spec-fits}.}
\label{spectra+model}  
\end{figure}

In this subsection we present an empirical observation that could
possibly be useful in the construction of simple glueball
models. Since Lucini and Teper include many small area Wilson loops in
their construction of excited states, we assume that they calculate
the lowest few glueball states without omission. We assume the
calculations presented here give higher mass glueball
states. Presumably, the correct enumeration of these states is not
possible without the inclusion of nonrectangular states in the
minimisation basis. For example, the third lowest mass state in a given sector calculated here could be the true seventh lowest 
mass state. However, when our results are combined with those of Lucini and Teper an interesting empirical observation can be made. We can choose a labelling of the $0^{++}$, $0^{--}$ and $2^{--}$ excited states such that the large $N$ limit of their masses  lie on a straight line as shown in \fig{spectra+model}. We have included the results of Lucini and Teper for $0^{--}$, $0^{--*}$, $2^{--}$ and  $2^{--*}$ which do not correspond to any of the states 
calculated here. The straight lines are fits to the model,
\bea
m_n(J^{PC}) = \gamma_1 (2 n + \gamma_2)  
\eea
where $\gamma_1$ and $\gamma_2$ are parameters and $\gamma_2$ is restricted to integer values. For the $J^{PC}$ states considered we obtain the following best fit models
\bea
m_n(0^{++}) &=& (0.256\pm 0.002)(2 n + 1) \nn\\
m_n(0^{--}) &=& (0.151\pm 0.002)(2 n + 5) \nn\\
m_n(2^{--}) &=& (0.1495\pm 0.0008)(2 n + 7).
\label{fitmodspec}
\eea
Here $n\ge 1$ labels the $n$-th lowest mass state. We note that the fit improves in accuracy as $n$ is increased. It is interesting to note that the slopes of the $0^{--}$ and $2^{--}$ models are consistent suggesting that the constant of proportionality does not depend on $J$. Another interesting observation can be made by recasting the models in the form 
\bea
m_n(0^{++}) &=& (0.256\pm 0.002)(2 n + 0 + 1) \nn\\
m_n(0^{--}) &=& (0.151\pm 0.002)(2 n + 4 + 1) \nn\\
m_n(2^{--}) &=& (0.1495\pm 0.0008)(2 n + 6+ 1).
\label{spec-fits}
\eea  
We notice a similarity with the two dimensional harmonic oscillator spectrum,
\bea
E_n \propto 2 n + J + 1,
\eea
when we take into consideration the ambiguity modulo 4 of spin identification on the lattice.
With this in mind we propose a simple model for the $J^{PC}$ spectrum
\bea
m_n(J^{PC}) = \zeta_{PC}\left[2n + \gamma(J^{PC}) + 1\right], 
\label{conjmod}
\eea 
where $\zeta_{PC}$ is a spin independent parameter and
$\gamma(J^{PC})$ is an integer for which $\gamma(J^{PC}) = J\, {\rm
mod}\, 4$. From \eqn{fitmodspec} we have $\zeta_{--} \approx 0.15$ and
$\zeta_{++} = 0.256\pm 0.002$. To check this simple model we can
attempt to predict the lowest lying states obtained by Lucini and
Teper in the sectors that have not been considered in this chapter. We
start with the $2^{++}$ sector in which Teper and Lucini obtain
$1.359(12)$ and $1.822(62)$ for the $N\rightarrow \infty$ limit of the
$2^{++}$ and $2^{++*}$ glueball masses in units of $N g^2/a$. The
predictions of \eqn{conjmod}, with $\gamma(2^{++}) = 2$ and
$\zeta_{++} = 0.256$, are 1.28 and 1.79 for the 1st and 2nd lowest
glueball masses in units of $N g^2/a$. As we would expect the
prediction is better for the higher mass state. We can also consider
the $1^{++}$ sector which has an equivalent spectrum to $1^{-+}$ due
to the phenomenon of parity doubling~\cite{Teper:1998te}. Lucini and
Teper obtain $1.98(8)$ for the mass of the $1^{++}$ glueball in units
of $N g^2/a$. The model of \eqn{conjmod}, with $\gamma(1^{++})=5$ and
$\zeta_{++} = 0.256$, gives 2.048. In the $1^{--}$ sector the
agreement is not as good, with Lucini and Teper obtaining $1.85(13)$
for the mass of the $1^{--}$ glueball and the model of \eqn{conjmod},
with $\gamma(1^{--})= 9$ and $\zeta_{--} = 0.15$, giving 1.812. \\

It will be interesting to see if the model suggested here stands up to
further calculations with an extended minimisation basis and in other
$J^{PC}$ sectors. The analytic techniques presented here are at a
great advantage to the standard Monte Carlo techniques of Lagrangian
LGT for the purpose of testing glueball models. High order excited
states are readily accessible; in a variational calculation with $s$
states in the minimisation basis, $s$ glueball states are
accessible. This is in contrast to competing Lagrangian calculations
in which only $3$ states are currently accessible in some $J^{PC}$ sectors.

\subsection{Discussion}

In this section we have calculated variational mass spectra for pure SU($N$)
gauge theory with a simple basis of rectangular states in 2+1
dimensions. 
Such a basis
is easy to work with computationally and is therefore a good
starting point. However to accurately explore the pure gauge 
mass spectra additional states need to be included. Perhaps a more
suitable basis would be the set of all states with an area less than some 
maximum value which would define the order of the calculation. 
The next stage beyond this would be to include
higher representation states in the form of Wilson loops in which some
links are covered more than once. In this way 
multiple glueball states could presumably be explored. 
The inclusion of additional small area states in the basis, in
particular states which are not symmetric about reflections in
coordinate axes, would also allow the calculation of spin $1$
glueball masses and also massgaps in the $P=-C$ sector. \\


It is important to point out that the inclusion of additional states in
our minimisation basis does not present a significant challenge. The only
complication would be in counting the possible overlaps of particular
states. While more difficult than the counting required
here, the process could presumably be automated using symbolic programming and
techniques from graph theory. The analytic techniques used here would
still be applicable until higher representation states were included
in the minimisation basis. At
that stage further character integrals would be required to handle the 
additional integrals arising in the calculation.\\

A final extension of the method presented here would be to make use of
an improved ground state; a ground state which includes
additional Wilson loops in the exponent. However, the
advantage of using analytic techniques would then be lost, unless new technology
was developed for tackling the resulting integrals. One would need
to calculate the required matrix elements via Monte Carlo simulation
on small lattices, losing the advantage of working on an infinite
lattice with analytic expressions that we have here.

\section{Conclusion}

In this chapter we have applied the analytic techniques developed in
\chap{analytictechniques}, in a study of the large $N$ glueball
mass spectrum in 2+1 dimensions.\\

A direct attempt at calculating asymptotic expansions of the
generating functions was carried out in
\sect{directcalculation}. While some low order large $N$
expansions were obtained, many terms appear to be needed for a
direct $N\rightarrow \infty$ calculation of the matrix elements
appearing in glueball mass calculations at small couplings.\\

Having tried a direct approach to calculating matrix
elements in the $N\rightarrow \infty$ limit without success, an extrapolation technique was used.
In \sect{extrapolation} we calculated glueball masses at finite
$N$, with $N$ as large as 25 in some cases. This allowed accurate $N\rightarrow
\infty$ extrapolations to be made. Evidence of leading
order $1/N^2$ finite $N$ corrections to the glueball masses was
obtained, confirming a specific prediction of large $N$ gauge theory.
The interpretation of the possible scaling regions was discussed and
agreement with the Lagrangian study of Lucini and
Teper~\cite{Lucini:2002wg} was obtained 
in some cases in the $N\rightarrow \infty$ limit. The discrepancies
are attributable to our use of only rectangular states in the
minimisation basis. Without the inclusion of small
area, nonrectangular states, it is likely that some low energy states are inaccessible. 
Further work will fill in the incomplete
spectra presented here. Interesting empirical observations were made
when the results presented here were combined with those of Lucini and
Teper; the enumeration of excited states can be chosen so that the glueball 
mass spectrum has the structure
of a two dimensional harmonic oscillator. To develop this observation
into a model would require a more complete calculation of the mass
spectrum with nonrectangular states in the minimisation basis and
possibly a more complicated vacuum trial state.\\

Clearly much work remains to be done before an accurate picture of the
2+1 dimensional pure SU($N$) gauge theory spectrum is achieved within a
Hamiltonian variational approach. We have demonstrated however that
the analytic techniques of Hamiltonian LGT can be used for massgap calculations with $N$ as large as 25 on a desktop computer. 
This is significantly closer to
the $N\rightarrow \infty$ limit than is currently possible with the
Lagrangian approach using supercomputers. The inclusion of additional
states in the minimisation basis, while not presenting a major
challenge, will allow a thorough study of the 2+1 dimensional 
pure SU($N$) mass
spectrum at least up to $N=25$. \\

In the next chapter we consider the extension of the analytic
techniques used with success in 2+1 dimensions here, to the physically
interesting case of 3+1 dimensions.

\chapter{3+1 Dimensions}
\label{3+1dimensions}

\section{Outline}
In this chapter we explore the viability of extending the analytic
techniques used with success in \chaps{sunmassgaps}{largenphysics} 
to the calculation of glueball masses in the pure gauge sector in 
3+1 dimensions. 
The primary difficulty lies in the calculation of
expectation values in 3+1 dimensions. In \sect{3+1introduction} we briefly review what has been achieved in Hamiltonian LGT in 3+1 dimensions. We discuss the difficulties faced in 3+1 dimensions and possible
solutions in \sect{bianchiidentities}. In
\sect{gausslaw} we consider the problem of Gauss' law
constraints. This is a topic that has been discussed in the context of
Hamiltonian LGT most recently by Ligterink, Walet and
Bishop~\cite{Ligterink:2000ug} and concerns the constraint equations
that appear when non-abelian gauge theories are canonically quantised. In
\sect{onecube} we move on to the calculation of variational
glueball masses on a single cube using the analytic variational
technique discussed in 
\chaps{sunmassgaps}{largenphysics}. We finish in
\sect{chap7-futurework} with a
discussion of the viability of pursuing analytic techniques for pure
SU($N$) LGT in 3+1 dimensions based on the results of \sect{onecube}.

\section{Introduction}
\label{3+1introduction}

From a renormalisation point of view the key difference between 2+1
and 3+1 dimensional gauge theory lies in the units of the coupling 
constant. As was seen in \chap{constructingandimproving}, 2+1 dimensional
gauge theory has a coupling constant, $e^2$, with the dimensions of
mass and so the coupling constant explicitly sets a mass scale for
calculations on the lattice. In contrast
the 3+1 dimensional coupling constant is dimensionless. This makes the
extraction of continuum physics from lattice calculations more subtle
in 3+1 dimensions than in 2+1, as discussed in \sect{extractingcontinuumphysics}.\\

On a practical level, in
Hamiltonian calculations there is a more serious problem faced in moving
from 2+1 to 3+1 dimensions. The analytic techniques that we have used
with success in \chaps{sunmassgaps}{largenphysics} are no longer applicable.
These techniques rely heavily upon the fact that in 2+1 dimensions 
a change of variables
from links to plaquettes has unit Jacobian. The form for the
equivalent Jacobian in 3+1 dimensions is considerably more
complicated. The most
comprehensive study of the change of variables from links to
plaquettes is due to Batrouni~\cite{Batrouni:1982bg,Batrouni:1983ch}. 
We discuss this change of variables in more detail in \sect{bianchiidentities}.\\

The extension of the techniques used in \chaps{sunmassgaps}{largenphysics} 
to 3+1 dimensions is not
straightforward. There are a number of immediate problems. Firstly,
since plaquettes are not independent variables in 3+1 dimensions one
can not automatically work in the infinite volume limit. In a precise
study one would need to calculate identical quantities on different
sized lattices and extrapolate to the infinite volume limit.
Secondly, in the context of analytic calculations, even on small
lattices the integrals
involved in the calculation of basic matrix elements are considerably
more complicated than those encountered in 2+1 dimensions. Such matrix
elements could in
principle be carried out analytically on small lattices but since the number
of integration variables increases quickly with the volume of the
lattice a calculation on even a $5^3$ lattice would seem
exceedingly difficult. How quickly the infinite volume limit is
reached will therefore determine the worth of pursuing analytic Hamiltonian 
methods in 3+1 dimensions. Finally, there is the complication of Gauss' law which we discuss in \sect{gausslaw}.\\

The only Hamiltonian techniques to have been applied with any
success to the case of SU(3) gauge theory in 3+1 dimensions have been
strong coupling expansions, the $t$-expansion and exponential wave
function methods. Each of which we now summarise.\\

Strong coupling perturbative techniques were used in the early days of
LGT in an attempt to bridge the gap between the
strong and weak coupling limits. Strong coupling expansions of the
Callan-Symanzik $\beta $ function were calculated  to $\ord(g^{-24})$~\cite{Kogut:1979vg,Kogut:1980sg}
and showed signs of interpolating smoothly between the strong and weak
coupling limits. This suggested that the $\beta$ function was a
smooth function of the coupling with its only zero at $g=0$, providing
a strong argument at the time for the continuum limit of LGT
confining quarks. Corresponding strong coupling expressions for
glueball masses did not share the same success. Despite strong
coupling calculations to $\ord(g^{-28})$~\cite{Hamer:1989qm} scaling
was not observed in $0^{++}$, $1^{+-}$ or $2^{++}$ glueball masses. \\

The $t$-expansion was introduced by Horn and
Weinstein~\cite{Horn:1984bq} as an analytic method suitable for the
study of LGT in the Hamiltonian formulation. It has been applied in
the calculation of glueball masses in 3+1 dimensions for 
SU(2)~\cite{Horn:1985ax} and
SU(3)~\cite{VanDenDoel:1986bw,vandenDoel:1987xk,Horn:1991fz} LGT in
the pure gauge sector. More recently it has been used in an attempt to
calculate the lowest hadron masses~\cite{Horn:1993wb}. For each case, 
however, asymptotic scaling of masses was not directly observed. 
Extrapolation techniques
such as Pad{\'e} approximants were required to probe the weak coupling
region. The extrapolated values mass ratio results agreed with Monte
Carlo estimates of the time.  \\

The coupled cluster method and related exponential wave function
techniques have received by far the most attention in
Hamiltonian LGT in recent years. Essentially these techniques aim to
solve the Kogut-Susskind eigenvalue equation by making a suitable
ansatz for the wave function.  The coupled cluster method was
originally constructed with applications in nuclear physics in
mind~\cite{Coester:1958,Coester:1960} but has since found the majority of its
applications in molecular physics~\cite{Bishop:1987}. Its application
in the context of Hamiltonian LGT is described in 
\rcites{Schutte:1997du}{McKellar:2000zk}. The truncated
eigenvalue method, developed by Guo, Chen and Li~\cite{Guo:1994vq},
is another exponential wave function technique to have 
found application in Hamiltonian LGT.  \\

A number of groups have made considerable progress in the application 
of exponential wave function techniques to Hamiltonian LGT in 3+1 dimensions.
While most studies have explored gauge groups other than SU(3) in less
than three dimensions, studies of SU(3) glueballs in 3+1 dimensions 
have commenced. 
Results tangent to the expected scaling form, indicating an approach
to scaling, have been
obtained for SU(3) pure gauge theory in a coupled cluster calculation by
Leonard~\cite{ConradPhD}. However convergence with increasing orders
appears to be slow. Finite order truncation errors appear to be under
more 
control in the truncated eigenvalue method. The first calculations of
3+1 dimensional SU(3) glueball masses with this method~\cite{Hu:1997ys}
 gave a $1^{+-}$ to $0^{++}$ mass ratio that was consistent with the
Monte Carlo results of the time. The agreement was not as good for the
$0^{--}$ to  $0^{++}$ mass ratio. A convincing
demonstration of asymptotic scaling has not yet been produced in a
calculation of glueball masses for SU(3) in 3+1 dimensions.  \\

More promising 3+1 dimensional results 
have been obtained for higher dimensional gauge groups. Ironically
these results have been obtained with much simpler methods than either
of the exponential wave function methods described above. 
Asymptotic scaling of the lowest $0^{++}$ glueball mass has been
demonstrated by Chin, Long and Robson in a variational calculation on a small volume ($6^3$
sites) lattice for SU(5)
and SU(6)~\cite{Chin:1986fe}. This calculation follows the
variational technique described in \chap{sunmassgaps} but uses
Monte Carlo rather than analytic techniques to calculate the required
expectation values. It uses only plaquette states in the minimisation
basis rather than the large basis of rectangular states used in \chaps{sunmassgaps}{largenphysics}. Naively one would hope
that the same method could be explored on similar sized
lattices for larger $N$, with additional states in the minimisation basis, using the analytic techniques of \chaps{sunmassgaps}{largenphysics}. We take this as our motivation for the studies
presented in this chapter. It will become clear however that attempting a
similar calculation to Chin, Long and Robson on a single cube using 
analytic techniques presents a significant challenge.

\section{The Move to 3+1 Dimensions}
\label{bianchiidentities}

In LGT one usually works in 2+1 dimensions to test
a technique with the intention of later extending it to the physically
relevant 3+1 dimensions. This may be
justified in the Lagrangian approach where the time coordinate 
is treated on the same 
footing as the spatial coordinates and adding another dimension equates to nothing more
than an increased load on computer memory. However, in Hamiltonian LGT
two serious technical differences exists between 2+1 and 3+1
dimensions. The first is Gauss' law and the second is related to
constructing a Jacobian for transforming from link to plaquette variables.
\\

To understand the first difference one needs to recall that Hamiltonian LGT
is formulated in the temporal gauge which sets $A_0 =0$. If one starts
with the Yang-Mills Lagrangian and performs the standard equal-time 
quantisation,
one runs into problems because the time derivative of $A_0$ does not
appear in the Lagrangian. The variational principle gives equations of
motion for the space-like components resulting in the standard
Yang-Mills Hamiltonian. 
For the time-like component we obtain a set of algebraic 
constraint equations which are the analogue of Gauss' law. There
is one constraint equation for each colour component of $A_0$, $N^2-1$ for
SU($N$), at each lattice site. As has been pointed out in the context
of Hamiltonian LGT by Ligterink, Walet and 
Bishop~\cite{Ligterink:2000ug}, it is only when one
works with a set of variables whose
number of degrees of freedom matches the number of unconstrained
degrees of freedom in the theory, that one can avoid the
technicalities of constraint equations. For this reason, in
2+1 dimensions we really are quite lucky. The number of plaquette variables
on a square two dimensional lattice is  precisely equal to 
the number of unconstrained variables~\cite{Ligterink:2000ug}. 
For a three dimensional cubic lattice this is not the case. One could envisage
constructing a polyhedral lattice in $d$ dimensions such that the number of
faces (plaquettes) would equal the number of unconstrained variables. 
However such lattices appear to be prohibited
by Euler's equation which relates the number of vertices, edges and
faces of polyhedra.\\

The other serious technical difference between 2+1 and 3+1 dimensions, that of
constructing a Jacobian for transforming from link to plaquette
variables, is
relevant to both the Hamiltonian and Lagrangian formulations of
LGT. It only becomes important when the method of choice relies on
plaquette variables for its implementation. 
The transformation from
link variables to plaquette variables is required in the approach taken
in \chaps{sunmassgaps}{largenphysics} to make use of the 
analytic results available for certain group integrals. 
Such a transformation does not
need to be made if one is happy to use Monte Carlo techniques to
handle the integrals as is the case in \rcite{Chin:1986fe}. \\

The most complete treatment of the
transformation from link to plaquette variables is due to
Batrouni~\cite{Batrouni:1983ch,Batrouni:1982bg} who worked in the Lagrangian formulation. His approach was based
on the continuum
work of Halpern~\cite{Halpern:1979ik} who constructed a field-strength 
formulation of
gauge theory. In Halpern's construction the definition of the field strength $F_{\mu \nu}$ is
inverted to give an expression for $A_\mu$ as a function of the field
strength. To do this requires the choice of a suitable gauge. The
Jacobian of the transformation is precisely the Bianchi identity; a
constraint equation on $F_{\mu \nu}$. For the Abelian case the
Bianchi identity is equivalent to the requirement that the total
magnetic flux leaving a volume is zero. Batrouni developed an equivalent
field strength formulation on the lattice. On the lattice, the link operators, $U_l$, correspond to the vector potentials, $A_\mu$, and the 
plaquette variables correspond to the field strengths, $F_{\mu \nu}$. Batrouni
demonstrated that the Jacobian of the transformation from link
variables to plaquette variables was the lattice analogue of the
Bianchi identity. For Abelian gauge theories the lattice Bianchi
identity can be separated into factors which depend only on the plaquette variables of elementary cubes, with one factor for each cube of the lattice. 
For non-abelian
theories the Bianchi identity has only been found to separate in this
way for special types of lattices, the largest volume example being an infinite
tower of cubes. For the infinite lattice the Bianchi identity is a  
 complicated nonlinear inseparable function of distant plaquette
variables. Interestingly it is the only source of correlations between
plaquette variables in LGT. Mean plaquette methods have been developed to 
deal with the added complications of the non-abelian Bianchi identity but 
have not progressed far~\cite{Batrouni:1982dx}.\\

To summarise, Hamiltonian LGT in 3+1 dimensions
faces some serious problems. Care must be taken in choosing
appropriate variables to work with if constraints on the lattice electric
fields are to be avoided. Additionally, if one wishes to use analytic 
techniques to calculate matrix elements the complications of the
Bianchi identity restricts the calculations to small lattices.

\section{Constraint Equations}
\label{gausslaw}

In this section we focus on one of the problems faced in moving 
from 2+1 to 3+1 dimensions, that of Gauss' law. From the
discussion of \sect{bianchiidentities} it would seem that one
needs either to construct an appropriate set of variables with
precisely the correct number of unconstrained degrees of freedom or be
faced with the problem of building constraint equations on the lattice 
electric fields. As has been pointed out by Ligterink, Walet and
Bishop~\cite{Ligterink:2000ug} there is an alternative. The problem of
satisfying Gauss' law can be solved by working with wave functions that
are annihilated by the generator of Gauss' law. We discuss this matter
in that which follows.\\ 

In the Hamiltonian formulation of LGT we work in the
temporal gauge which sets $A_0=0$. In terms of link variables
this is equivalent to setting all time-like links to the
identity. This however is only a partial gauge fixing; setting
$A_0=0$ does not completely eliminate the arbitrariness in the
definition of the vector potential, it merely reduces the set of
possible gauge transformations to purely space-like ones. To see this
we revert to continuum gauge theory.\\


In the continuum one encounters difficulties when attempting to
canonically 
quantise  SU($N$) pure gauge theory~\cite{Fadeev&Slavnov,Jackiw1993}. 
The difficulties stem from the fact that it is
impossible to define a momentum conjugate to the time-like component
of the vector potential $A_0$. This follows from the non-appearance of
$\partial_0 A^a_0$ in the Yang-Mills Lagrangian. The simplest way
around the problem is to fix to the temporal gauge $A_0 = 0$ before
quantising. The result is a Hamiltonian whose equations of motion do
not mirror those of the Lagrangian; the non-abelian generalisation of
Gauss' Law,
\bea
D_j F^{j 0} =\nabla \cdot \bm{E} = 0,
\eea
is absent. One does not need to impose Gauss' law as a constraint on
the electric fields, $E^i=F^{i 0}$. Defining $G = \nabla \cdot \bm{E}$,
with the usual canonical commutation relations, it can be shown that
$G^a$ commutes with the Hamiltonian. Consequently if one chooses to
work with states $|\psi\rangle$ such that
\bea
G^{a} |\psi\rangle = 0,
\eea
then one recovers Gauss' law ensuring that the physics of the
Lagrangian formulation is recovered. \\
  
On the lattice a similar problem arises~\cite{Ligterink:2000ug}. 
One must therefore ensure
that any state used is annihilated by the generator of Gauss' law, in order 
to recover physical results with certainty. 
The generator of Gauss' law on the lattice can be written as 
~\cite{Kogut:1975ag,Kogut:1980sg},
\bea
{\cal G}^a (\bm{x}) = \sum_{i} \left[\LE^a_i(\bm{x}) + \LE^a_{-i}(\bm{x})\right].
\eea   
In this notation the lattice electric fields satisfy the commutation relations~\cite{Kogut:1980sg}
\bea
\left[ \LE_i^a(\bm{x}) , U_j(\bm{y})\right] &=& \lambda^a U_j(\bm{y}) \delta_{ij}
\delta_{\bms{x}\bms{y}} \label{commutations1} \\
\left[\LE_{-i}^a(\bm{x}+\hbm{i} a) , U_j(\bm{y})\right] &=& -
U_j(\bm{y}) \lambda^a
\delta_{ij} \delta_{\bms{x}\bms{y}} . \label{commutations2}
\eea 
It should be pointed out that in this notation we have
$U^\dagger_i(\bm{x}) = U_{-i}(\bm{x}+\hbm{i} a)$. 
For physical states we must therefore have ${\cal G}^a (\bm{x}) |\psi\rangle =
0$ for each lattice site, $\bm{x}$, and all $a=1,\ldots,N^2-1$. It should be checked that this is
the case for the one-plaquette exponential trial state. Consider first
a state consisting of a single plaquette acting on the strong coupling vacuum,
\bea
|p_{ij}(\bm{x})\rangle  = \Tr \left[ U_i(\bm{x}) U_j(\bm{x}+\hbm{i} a)
U_{-i}(\bm{x}+\hbm{j} a) U_{-j}(\bm{x})\right] |0\rangle.
\eea
Since $|0\rangle $ is annihilated by the electric field, using
\eqns{commutations1}{commutations2} we immediately have
\bea
{\cal G}^a(\bm{y})  |p_{ij}(\bm{x})\rangle = 0,
\eea
for all lattice sites, $\bm{y}$, not lying on the corners of the
plaquette, $p_{ij}(\bm{x})$. 
Consider now sites lying on the corners of the plaquette in
question. In particular consider $\bm{y} = \bm{x}$. Making use of the
commutation relations of \eqns{commutations1}{commutations2},
\bea
{\cal G}^a(\bm{x})  |p_{ij}(\bm{x})\rangle &=& \Tr \left[ \lambda^a U_i(\bm{x}) U_j(\bm{x}+\hbm{i} a)
U_{-i}(\bm{x}+\hbm{j} a) U_{-j}(\bm{x})\right] |0\rangle \nn\\
&&  - \Tr \left[ U_i(\bm{x}) U_j(\bm{x}+\hbm{i} a)
U_{-i}(\bm{x}+\hbm{j} a) U_{-j}(\bm{x}) \lambda^a  \right] |0\rangle \nn\\
&=& 0.
\eea
The same applies for other sites on the plaquette. This argument
is not specific to plaquettes. Gauss' law is found to be satisfied
locally by any closed Wilson loop on the lattice, traced over colour
indices, acting on the strong coupling vacuum. It is easy to extend
this result  to products of such loops acting on $|0\rangle$. To see
this we consider how Gauss' law applies to the product of two plaquettes
\bea
|p_{ij}(\bm{x})p_{ij}(\bm{x}+a\hbm{i})\rangle &=&
 p_{ij}(\bm{x})p_{ij}(\bm{x}+a\hbm{i}) |0\rangle.
\eea
Only at the sites $\bm{x}+a\hbm{i}$
and $\bm{x}+a(\hbm{i}+\hbm{j})$ does this case differ from the single plaquette example . Let us consider Gauss' law at
$\bm{x}_{+}=\bm{x}+a\hbm{i}$. Once again, using the fact that the
lattice electric field annihilates the strong coupling vacuum, and \eqns{commutations1}{commutations2} we have 
\bea
{\cal G}^a(\bm{x}_+) |p_{ij}(\bm{x})p_{ij}(\bm{x}_{+})\rangle
&=& \sum_k \left[ \LE^a_k(\bm{x}_+) +  \LE^a_{-k}(\bm{x}_+),
p_{ij}(\bm{x})p_{ij}(\bm{x}_+)\right]|0\rangle \nn\\
&=& \sum_k\left\{ 
\left[
\LE^a_k(\bm{x}_+)+\LE^a_{-k}(\bm{x}_+),p_{ij}(\bm{x})\right]p_{ij}(\bm{x}_+)|0\rangle
\right. \nn\\
&&\left. + 
p_{ij}(\bm{x})\left[
\LE^a_k(\bm{x}_+)+\LE^a_{-k}(\bm{x}_+),p_{ij}(\bm{x}_+)\right]|0\rangle\right\}
\nn\\
&=& 0.
\eea
This result is easily extended to arbitrary products of closed Wilson
loops, traced over colour indices, acting on $|0\rangle
$. Consequently, any function of such loops acting on $| 0\rangle $
also satisfies
Gauss' law provided it admits a Taylor series expansion. The one
plaquette trial state thus obeys Gauss' law and is therefore suitable
for use in simulating physical states. 

\section{The One Cube Universe}
\label{onecube}

In this section we adapt the analytic techniques used with success in 
\chaps{sunmassgaps}{largenphysics} for use in 3+1 dimensions. 
As a starting point we consider the case of a
single cube. This will serve as a test case to assess the
feasibility of a larger volume study. Our aim is to
use a small basis of states to calculate 3+1 dimensional SU($N$) glueball masses variationally, and check if an approach to the correct scaling form is
observed. We do not expect to achieve scaling with such a small lattice.
However, an approach to scaling would warrant an extended study on larger
lattices. We start with a general description of the approach. The starting point is the choice of trial state. As
in 2+1 dimensions we choose to work with the one-plaquette trial state
of \eqn{oneplaquette}, with the variational parameter again being fixed by
minimising the vacuum energy density of \eqn{epsilon}. In 3+1
dimensions however the expression for the plaquette expectation value in terms of the
variational parameter is significantly more complicated than for the
case of 2+1 dimensions. Having fixed
the variational parameter we then construct a small basis of states,
with each state fitting on a single cube, and minimise the glueball mass over this
basis. The process of minimising the massgap follows precisely
\chap{sunmassgaps}. The key difficulty of working in 3+1 dimensions is the calculation
of the required integrals. In the next section we explain how an analytic approximation to these integrals can be obtained. 

\subsection[SU($N$) Integrals in 3+1 Dimensions]{SU($\bm{N}$) Integrals in 3+1 Dimensions}

The analytic techniques used in \chap{sunmassgaps} rely on
the fact that the transformation from link to
plaquette variables has unit Jacobian in 2+1 dimensions. Batrouni has
calculated the Jacobian for arbitrary numbers of dimensions~\cite{Batrouni:1982bg,Batrouni:1983ch}. For a lattice consisting of a single cube, the result of
\rcite{Batrouni:1982bg} is
\bea
J &=& \delta(P_1 P_2 P_3 P_4 P_5 P_6 -1) \nn\\
&=&  \sum_{r}\frac{1}{d_r^4}
\prod_{i=1}^{6}\chi_{r}(P_i), \label{secondline}
\eea   
where $P_1,\ldots ,P_6$ are the six plaquette variables on the single
cube and the sum is over all characters $\chi_r$ of SU($N$). The
second line is simply a character expansion of the first line. In the 
variational study of glueball
masses, we need to 
calculate the integrals of overlapping trace variables on a single 
cube. It is always possible to reduce these integrals to integrals
involving non-overlapping trace variables using the orthogonality
properties of the characters. For example, consider the expectation
value, on a single cube, of a
twice covered bent rectangle, with respect to the one-plaquette trial
state of \eqn{oneplaquette},
\bea
\langle \phi_0|
\begin{array}{c}\includegraphics[width=1.125cm]{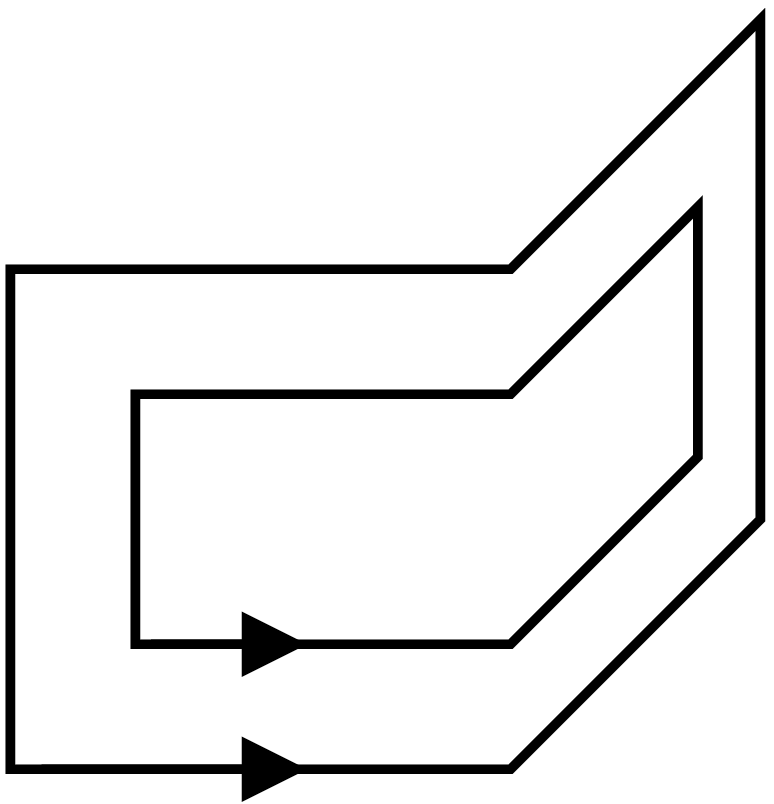}\end{array}|\phi_0\rangle
&=&
\sum_r\frac{1}{d_r^4} \int \prod_{i=1}^4 dP_i \chi_r(P_i) e^{c
\Tr(P_i+P^\dagger_i)}\nn\\
&&\hspace{-3cm}\times \int dP_5 dP_6  \chi_r(P_5) \chi_r(P_6) \left[\chi_2(P_5 P_6)+\chi_{11}(P_5 P_6) \right]e^{c \Tr(P_5+P_6+P^\dagger_5+P^\dagger_6)}.
\label{3dexample}
\eea
To proceed with this integral we need to perform character
expansions. The orthogonality of the characters can then be used to
write characters over two plaquettes in terms of one-plaquette
characters. To demonstrate how this is done we consider the character expansion of one of the
integrals in \eqn{3dexample},
\bea
\int dP \chi_r(P) \chi_2(P P') e^{c \Tr(P +P^\dagger)} &=&
\sum_{r'} c_{r'} \int dP \chi_{r'}(P)  \chi_2(P P').
\eea
Here the $c_{r'}$ are given by
\bea
c_{r'} &=& \int dP \chi_{r'}(P)\chi_r(P) e^{c \Tr(P +P^\dagger)}.
\eea
Making use of the orthogonality of characters, given by
\eqn{charorthog}, we obtain
\bea
\int dP \chi_r(P) \chi_2(P P') e^{c \Tr(P +P^\dagger)} &=&
\frac{c_2}{d_2} \chi_{2}(P)  \chi_2(P P').
\eea
Making use of this and proceeding similarly for the analogous integral
involving $\chi_{11}(P_5 P_6)$, we can reduce \eqn{3dexample} to
\bea
\langle \phi_0|\begin{array}{c}\includegraphics[width=1.125cm]{1.eps}\end{array}|\phi_0\rangle
&=&\sum_r\frac{1}{d_r^4} \left[\int dP \chi_r(P) e^{c
\Tr(P+P^\dagger)}\right]^4\nn\\
&&\hspace{-5cm}\times\Bigg\{
 \frac{1}{d_2} \left[\int dQ \chi_2(Q) \chi_r(Q)
 e^{c\Tr(Q+Q^\dagger)}\right]^2 
+ \frac{1}{d_{11}} \left[\int dQ \chi_{11}(Q) \chi_r(Q)
 e^{c\Tr(Q+Q^\dagger)}\right]^2 \Bigg\}.
\eea
The plaquette matrix element provides a more straightforward example; 
\be
\langle \phi_0|\plaquette | \phi_0\rangle =\sum_r\frac{1}{d_r^4}\left[\int dP \chi_r(P) e^{c
\Tr(P+P^\dagger)}\right]^4 \int dQ \Tr Q \chi_r(Q)
 e^{c\Tr(Q+Q^\dagger)}.
\label{plaqexp-3d}
\ee
To proceed further we need expressions for the integrals over characters that appear in each matrix element of interest. We define these character integrals 
generically by
\bea
{\cal C}_{r_1 r_2\ldots r_n}(c) &=&  \int dP \chi_{r_1}(P)
\chi_{r_2}(P)\cdots \chi_{r_n}(P) e^{c
\Tr(P+P^\dagger)}.
\eea
All SU($N$) integrals encountered in the calculation of glueball masses in 
3+1 dimensions can be expressed in terms of character integrals. 
It is possible to calculate them generally, however three points need to be 
considered. Firstly, if
we are to use the machinery of \chap{analytictechniques}, 
we need to know how
to express a given character in terms of trace variables. Secondly,
as the dimension of the gauge group 
increases the number of independent trace variables increases rapidly. 
In order to contain the number of integrals required for a
given order of approximation, it is necessary to express all high
order trace variables in terms of a basis of lower order
ones. Finally, the collection of SU($N$) integrals
presented in \chap{analytictechniques} must be extended. We now
address each of these points in turn.

The problem of expressing the characters of SU($N$) in terms of trace
variables was solved by Bars~\cite{Bars:1980yy} who showed that for SU($N$),
\bea
\chi_n(U) = \sum_{k_1,\ldots,k_n} \delta\left(\sum_{i=1}^n i k_i -
n\right)\prod_{j=1}^n \frac{1}{k_j! j^{k_j}} (\Tr U^j)^{k_j}. 
\label{singlebox}
\eea   
General characters can then be expressed in terms of $\chi_n(U)$ using
standard techniques from group theory~\cite{Weyl:1946},
\bea
\chi_{r_1 r_2 \cdots r_{N-1}}(U) = \det\left[ \chi_{r_j+i-j}(U)
\right]_{1\le i,j \le N-1}.
\label{chareqn}
\eea
For a given character, \eqn{chareqn} produces a multinomial of
traces of different powers of $U$. However, not all of these 
trace variables are
independent. The so called Mandelstam constraints define the
relationship between dependent trace variables. For SU(2) and SU(3) we have
\bea
\begin{array}{rcll}
\Tr U &=& \Tr U^\dagger & \forall U \in {\rm SU}(2)\quad {\rm and} \\
\Tr (U^2) &=& (\Tr U)^2 - 2 \Tr(U^\dagger) & \forall U \in {\rm SU}(3).
\end{array}
\eea  
Similar relations can be obtained to express all trace variables in
terms of $\Tr U$ for SU(2), and for SU(3) in terms of $\Tr U$ and $\Tr
U^\dagger$. To reduce the calculation of character integrals to a manageable size, we require
expressions for high power trace variables in terms
of a minimal set of lower order  trace variables. To do this we
proceed as follows.
We start with an alternative expression of $\det U = 1$, satisfied for all
SU($N$) matrices $U$,
\bea
\varepsilon_{i_1 i_2 \ldots i_{N}} U_{i_1 j_1} U_{i_2 j_2} \cdots
U_{i_N j_N} = \varepsilon_{j_1 j_2\ldots j_N}.
\label{unimod}
\eea
Here the colour indices of the group elements have been made
explicit and all repeated indices are summed over. Multiplying 
Levi-Civita symbols produces sums of products
of delta functions, the precise form of which depends on how many
pairs of indices are contracted. A standard result from differential
geometry will be useful here,
\bea
\varepsilon_{a_1\ldots a_n}\varepsilon_{b_1\ldots b_n} = \det
\left(\delta_{a_i b_j}\right)_{1\le i,j\le n}.
\label{gendelta}
\eea
Multiplying both sides of \eqn{unimod} by $U^\dagger_{j_N j_{N+1}}$ and
contracting over repeated indices gives
\bea
\varepsilon_{i_1 i_2 \ldots i_{N-1}j_{N+1}} U_{i_1 j_1} U_{i_2 j_2}
\cdots U_{i_{N-1} j_{N-1}}
 &=& \varepsilon_{j_1 j_2\ldots j_N} U^\dagger_{j_N j_{N+1}} \nn\\
\varepsilon_{l_1 l_2 \ldots l_{N-1} i_N}\varepsilon_{i_1 i_2 \ldots i_{N}}
U_{i_1 j_1} U_{i_2 j_2}\cdots U_{i_{N-1} j_{N-1}} &=& \varepsilon_{l_1 l_2 \ldots l_{N-1} i_N} \varepsilon_{j_1 j_2\ldots j_N} U^\dagger_{j_N i_{N}}.
\label{unimod2}
\eea
In the last line we have renamed dummy indices and multiplied both
sides by a Levi-Civita symbol. Making use of \eqn{gendelta} in
\eqn{unimod2} produces a product of delta functions on each side of
the equation. By introducing appropriate combinations of delta 
functions and contracting over repeated indices we can construct trace
identities as we please. We find that identities useful in our context
are produced by the introduction of delta functions and an additional
matrix, $A$, which is not necessarily an element of SU($N$), into \eqn{unimod2} as follows:
\bea
&&\varepsilon_{k_1\cdots
k_{N-1} i_N} \varepsilon_{i_1\cdots i_N} U_{i_1 j_1}A_{j_1 k_1}  \prod_{m=2}^{N-1} \delta_{k_m
j_m} U_{i_m j_m} \nn\\
&&\hspace{5cm} =  \varepsilon_{k_1\cdots
k_{N-1} i_N} \varepsilon_{j_1\cdots j_N}  A_{j_1 k_1}U^\dagger_{j_N i_N} \prod_{m=2}^{N-1} \delta_{k_m
j_m}.
\label{identgen}
\eea  
Here we have again renamed dummy indices. Some examples of 
SU($N$) identities obtained from \eqn{identgen} valid for all $N \times N$
matrices $A$ are as follows:
\bea
-\Tr U  \Tr(U A) + \Tr(U^2 A) = -\Tr A \Tr U^\dagger + \Tr(U^\dagger
 A) &\!\!\!\forall\!\!\!& U\in {\rm SU}(3)  \nn\\
 -(\Tr U)^2 \Tr(U A) + \Tr(U A)\Tr(U^2) + 2 \Tr U \Tr(U^2 A) - 2
 \Tr(U^3 A)\nn\\
 =  -2 \Tr A \Tr U^\dagger + 2 \Tr(U^\dagger A) 
&\!\!\!\forall\!\!\! & U\in {\rm SU}(4).
\eea
We require formulae for $\Tr (U^n)$ in terms of lower order trace
variables for SU($N$).  Such formulae, known commonly as Mandelstam
constraints, are obtained by setting $A =
U^{n-N+1}$ in \eqn{identgen}. In Appendix~\ref{mandelstamconstraints} 
we list the Mandelstam
constraints obtained in this way up to SU(8). We see that for SU($N$)
it is possible to express all characters in terms of $N-1$ trace
variables. In what follows we will choose to express the general
SU($N$) characters in terms of the set of trace variables, $\{\Tr U^\dagger, \Tr U, \Tr U^2,\ldots,
\Tr U^{N-2}\}$. It may well be possible to reduce the size of this
set. This would indeed improve the efficiency of our
technique. However, no
effort has been made to do this at this stage.\\

The procedure for calculating the general character integral, ${\cal
C}_{r_1\ldots r_n}(c)$, is then as follows. We first express the
characters $\chi_{r_1},\ldots,\chi_{r_n}$ in terms of trace variables
using \eqns{singlebox}{chareqn}. We then simplify these
expressions using the Mandelstam constraints of Appendix~\ref{mandelstamconstraints}. The
final step is to perform the integrals over trace variables, which is
an increasingly non-trivial task as one increases the dimension of the
gauge group. For instance, the general character integral
 for SU($N$) involves integrals over powers of the trace variables
$\Tr U^\dagger ,\Tr U, \Tr U^2,\ldots \Tr U^{N-2} $. To proceed we need the following integral,
\bea
{\cal T}_{q_1\ldots q_k}^{p_1\ldots p_k}(c)= \int_{{\rm SU}(N)} dU (\Tr U^{q_1}
)^{p_1}(\Tr U^{q_2} )^{p_2}\cdots  (\Tr U^{q_k} )^{p_{k}}
e^{c\Tr (U+U^\dagger)}.
\eea
For SU($N$), the cases of interest to us here are described by,
$T_{-1,1,2,\ldots, N-2}^{p_1,\ldots,p_{N-1}}(c)$. 
To calculate this integral we need to extend the work of
\chap{analytictechniques} and consider the generating function,
\bea
G_{q_1 \ldots q_k}(c,\gamma_1,\ldots,\gamma_k) = 
\int_{{\rm SU}(N)} dU e^{c\Tr(U+U^\dagger)+ \sum_{i=1}^{k} \gamma_i 
\Tr (U^{q_i})}. 
\eea
Following the procedure of \chap{analytictechniques} we obtain,
\bea
G_{q_1 \ldots q_k}(c,\gamma_1,\ldots,\gamma_k) &=&
\sum_{l=-\infty}^\infty \det\left[ 
\lambda_{l+j-i,q_1,\ldots,q_k}(c,\gamma_1,\ldots,\gamma_k)
\right]_{1\le i,j \le N},
\label{gp1p2}
\eea
with,
\bea
\lambda_{m,q_1,\ldots,q_k}(c, \gamma_1, \ldots,\gamma_k) &=&
\sum_{s_1,s_2,\ldots,s_k = 0}^{\infty} \frac{\gamma_1^{s_1} \cdots
\gamma_k^{s_k}}{s_1! \cdots s_k!} I_{m+s_1 q_1 +\cdots +s_k q_k}( 2 c).
\eea
We then obtain an expression for ${\cal T}_{q_1\ldots q_k}^{p_1\ldots
p_{k}}(c)$ by differentiating $G_{q_1\ldots
q_{k}}$ appropriately;
\bea
{\cal T}^{p_1\ldots p_{k}}_{q_1\ldots q_{k}}
(c) &=& \frac{\partial^{p_1+\cdots+p_{k}}}{\partial
\gamma_{1}^{p_{1}} \cdots  \partial\gamma_{k}^{p_{k}}}
G_{q_1 \ldots q_{k}}(c,\gamma_1,\ldots,\gamma_k)
\Bigg|_{\gamma_{1} =\cdots =\gamma_{k} =0}.
\label{genfunc-3d}
\eea

\subsection{The Variational Ground State}

In this section we fix the variational ground state following the
usual procedure of minimising the unimproved vacuum energy density
given by \eqn{epsilon} with $\kappa = 0$. This equation is independent
of the number of dimensions. The dimensionality of the lattice arises
at the stage of calculating the plaquette expectation value. \\

Making use of \eqn{plaqexp-3d}, the variational parameter can be fixed as
a function of $\beta$ for the one-cube lattice. In practice, the
character sum in \eqn{plaqexp-3d} and the infinite $l$-sum in \eqn{gp1p2} 
need to be truncated. We truncate the infinite $l$-sum at $\pm l_{max}$ and instead of summing over all SU($N$) characters, we sum over 
only those characters, $r = (r_1,r_2,\ldots,r_{N-1})$, with
$r_{max}\ge r_1\ge r_2 \ge \cdots \ge r_{N-1}$. With this truncation
scheme, memory constraints restrict calculations of the variational
SU($N$) ground state to $N\le 7$, when
working with $r_{max}=2$ and $l_{max}=2$ on a desktop computer.\\

\begin{figure}
\centering
                       
\includegraphics[width=10cm]{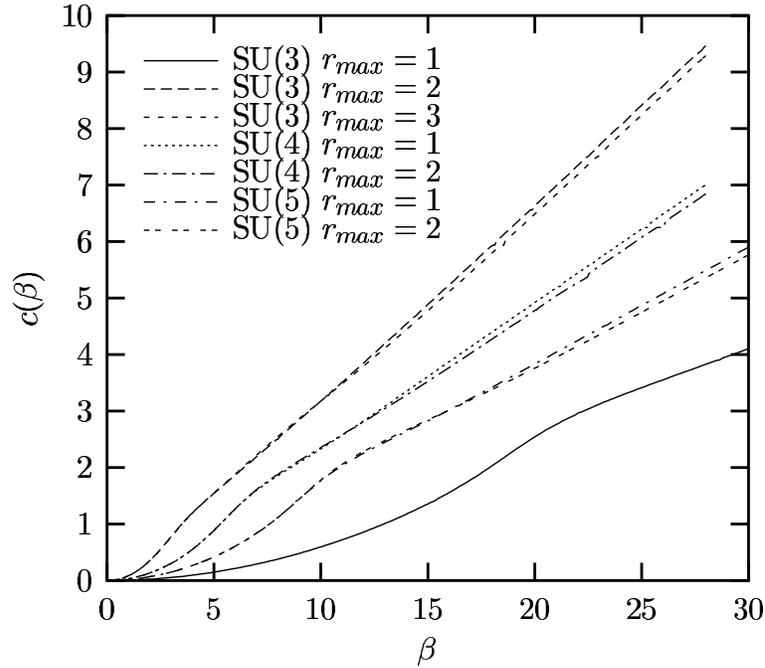}

\caption{The one cube SU($N$) variational parameter as a function of $\beta$ for various $N$ showing the dependence
on the character sum truncation.}
\label{c-convwithchars}  
\end{figure}

The dependence of the variational parameter on $r_{max}$ with
$l_{max}=2$ is shown
for various gauge groups in \fig{c-convwithchars}. As $r_{max}$ increases the 
variational parameter appears to
converge for each $N$ considered. Moreover the convergence appears to improve as the dimension
of the gauge group is increased. With the exception of 
SU(3), the $r_{max}=1$ and 2
results are indistinguishable on the range $0\le \xi\le 0.7$, where
$\xi = 1/(N g^2)$. As this
is the range of interest to us later in the chapter we restrict
further calculations to $r_{max} = 1$. \\

\begin{figure}
\centering
                       
\includegraphics[width=10cm]{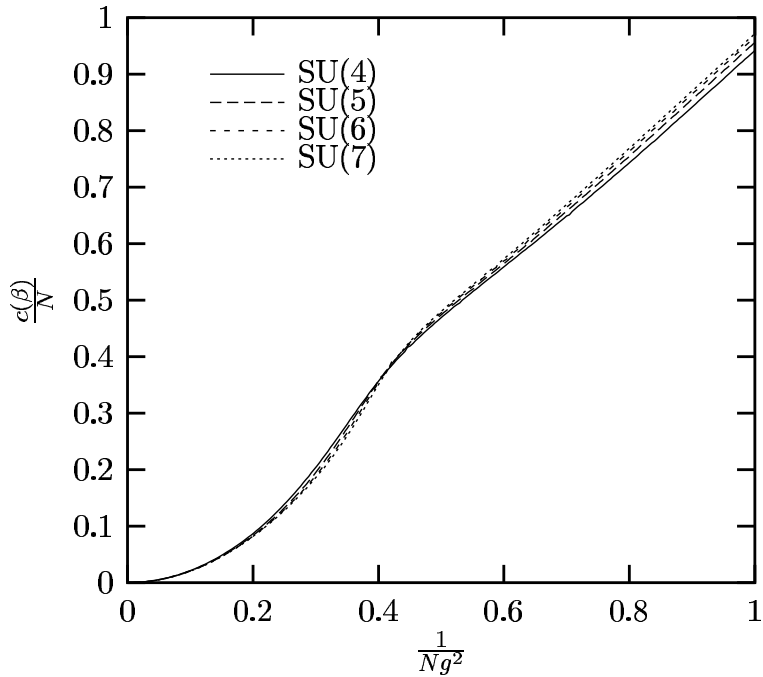}

\caption{The one cube SU($N$) variational parameter in units
of $N$ as a function of $1/(N g^2)$ for various $N$.}
\label{3+1sunvarpar}  
\end{figure}

\begin{figure}
\centering
                       
\includegraphics[width=10cm]{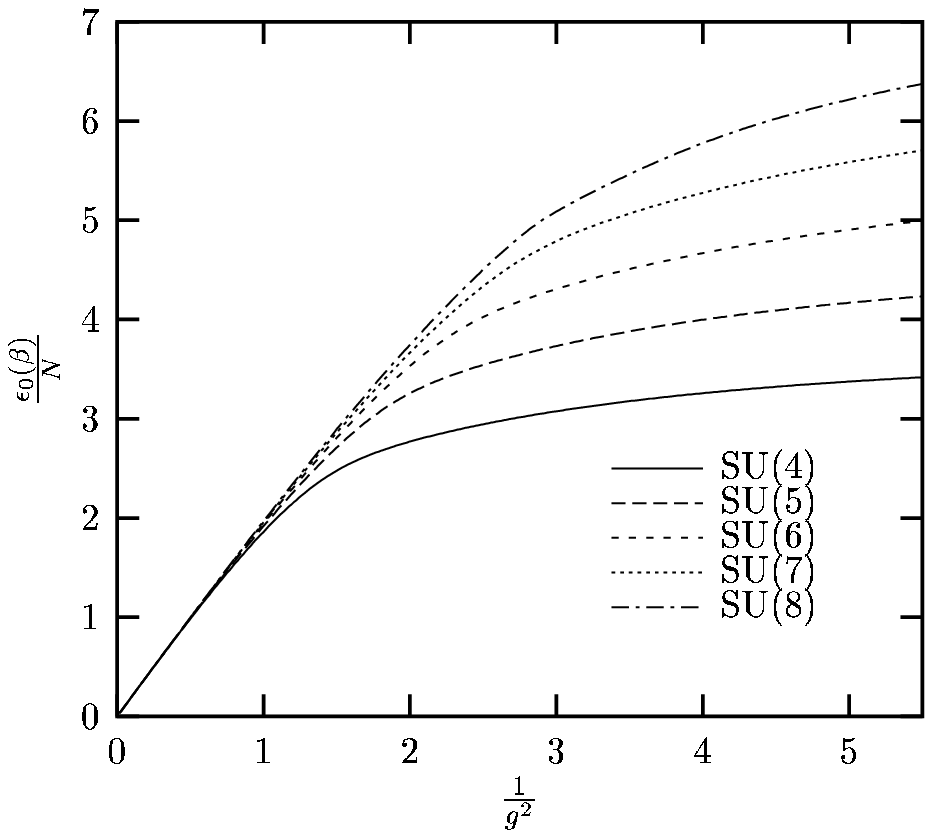}

\caption{The one cube SU($N$) energy density in units
of $N$ as a function of $1/g^2$ for various $N$.}
\label{3+1sunedens}  
\end{figure}

The SU($N$) variational parameters on the one-cube lattice with
$r_{max}=1$ and $l_{max}=2$ are shown for various $N$ in
\fig{3+1sunvarpar}. The results do not differ greatly. 
The corresponding variational energy
densities are shown in \fig{3+1sunedens}.

\subsection{Expressions for the Glueball Mass}

We follow precisely the method described in \sect{massgaps} for
the calculation of 3+1 dimensional glueball masses on a single cube. 
Here, however, we choose a different basis of
states to minimise over. Instead of rectangular loops, we use states
which fit on a single cube. We start with a basis of two states, the
plaquettes and the bent rectangles,
\bea
B &=& \{ |1\rangle ,|2\rangle \},
\label{basis}
\eea
with 
\bea
|i\rangle &=&  \sum_{\bms{x}} \left[F_i(x)-\langle F_i(x)
\rangle\right] |\phi_0 \rangle
\eea 
and 
\bea
F_1(x) &=& \begin{array}{c}\includegraphics[width=0.75cm]{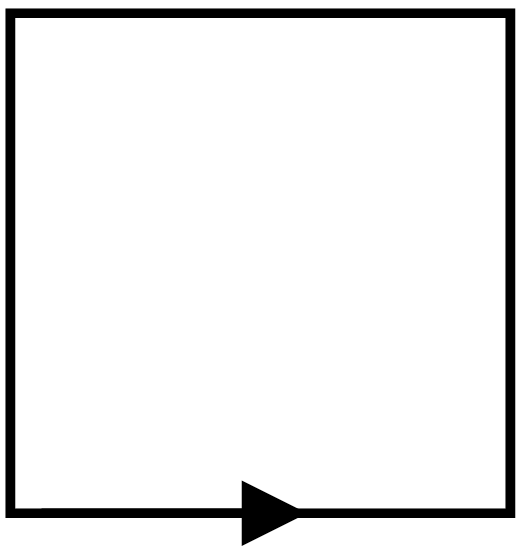}\end{array}
\pm \begin{array}{c}\includegraphics[width=0.75cm]{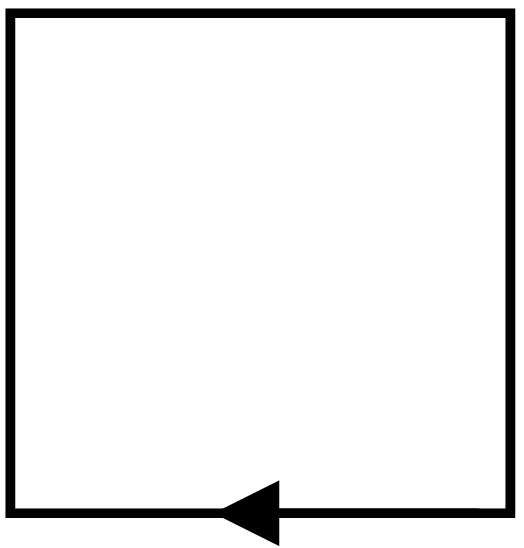}\end{array}\nn\\
F_2(x)
&=&\begin{array}{c}\includegraphics[width=1.125cm]{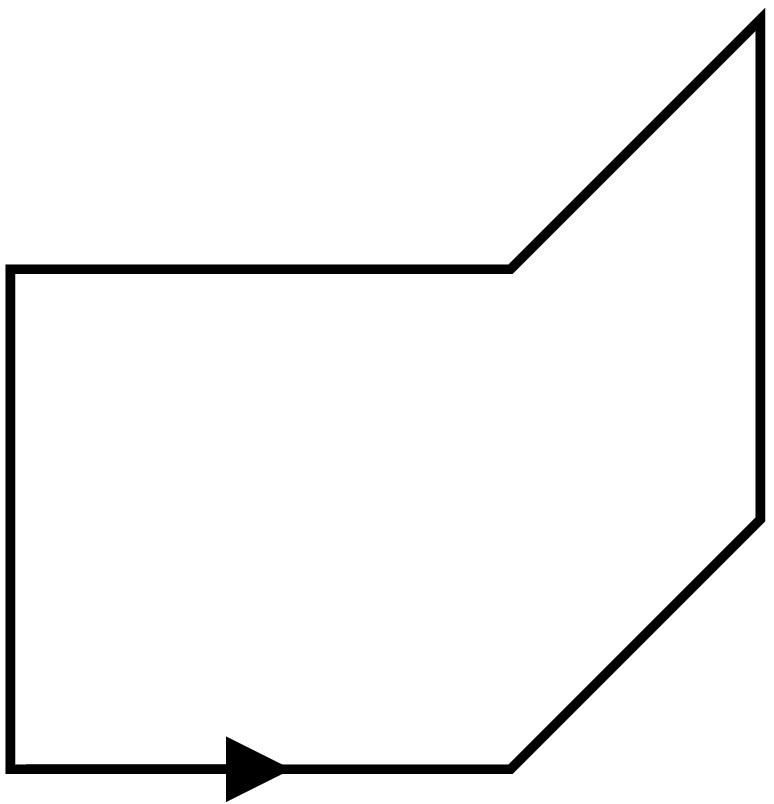}\end{array}
\pm
\begin{array}{c}\includegraphics[width=1.125cm]{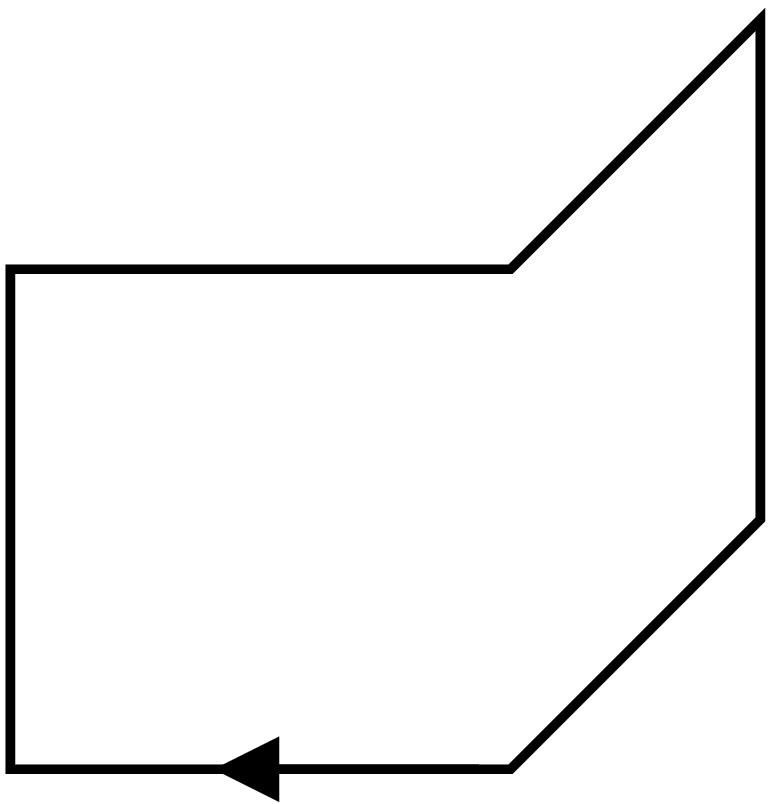}\end{array} .
\eea 
In 3+1 dimensions, the ``+'' sign corresponds to the $0^{++}$ state and 
the ``$-$'' sign corresponds to the $1^{+-}$ state~\cite{Hamer:1989qm}. 
In order to calculate the glueball masses, 
we need expressions for the matrix
elements $N^C_{ii'}$ and $D^C_{ii'}$ of \eqns{Nl'l}{d}
respectively. Here the superscript, $C$, denotes the charge conjugation
eigenvalue, $C=\pm 1$, of the state in question. Making use of \eqn{integrate} and carefully counting the number of possible overlaps between different loops we arrive at
\bea
\frac{2aN^C_{11}}{N_p g^2} &=&
4 \left[ \flangle\grapha\frangle
-\frac{1}{N}\flangle \graphd \frangle+C\left(- N +\frac{1}{N}\flangle \graphdd\frangle\right)
\right]  \nn\\
 \frac{D^C_{11}}{N_p} &=&
2 \flangle\graphd + C \graphdd \frangle
+10 \bflangle \begin{array}{c}\includegraphics[width=1.125cm]{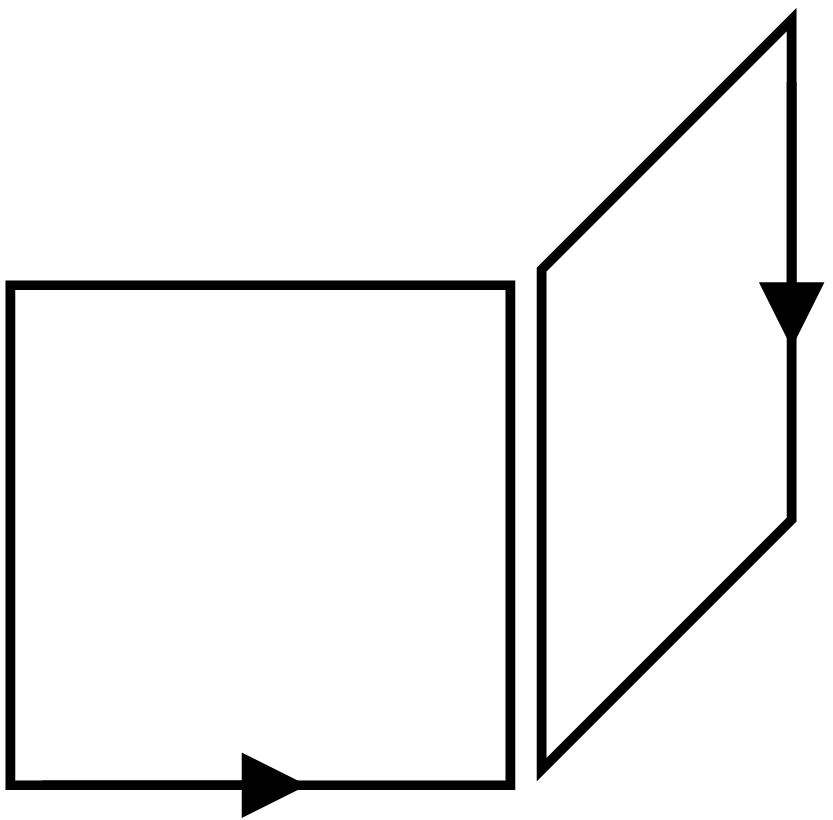}\end{array} 
\!\!+ C \!\!\begin{array}{c}\includegraphics[width=1.125cm]{21.eps}\end{array}\bfrangle
-24\left(\frac{C+1}{2}\right) \flangle\!\plaquette\! \frangle^2 \nn\\
\frac{2aN^C_{21}}{N_p g^2} &=& 12 \bflangle
\frac{1}{N} \begin{array}{c}\includegraphics[width=1.125cm]{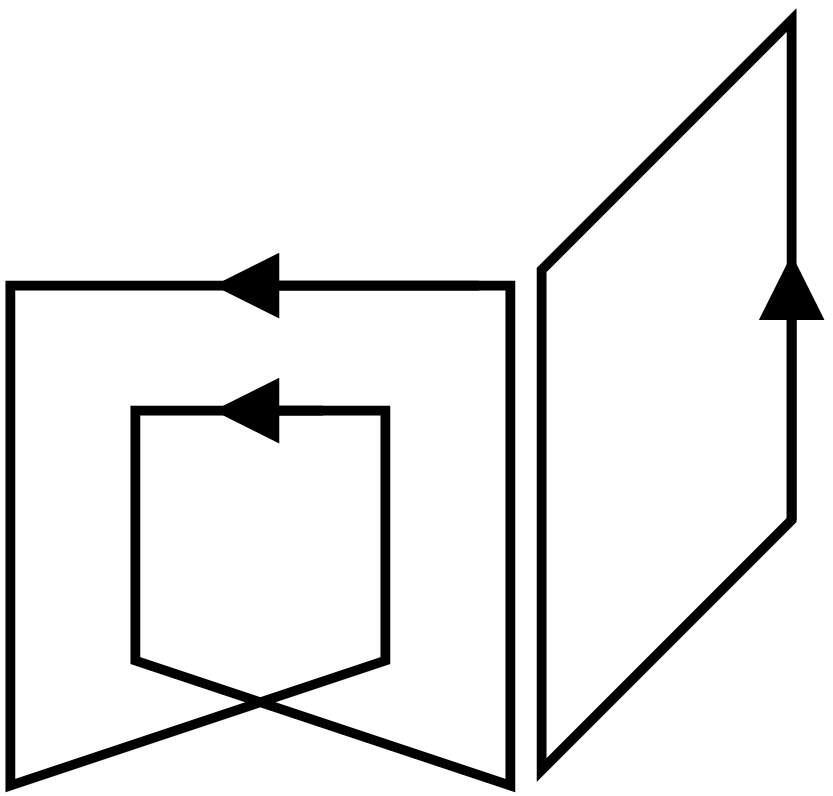}\end{array}
-\frac{1}{N^2} \begin{array}{c}\includegraphics[width=1.125cm]{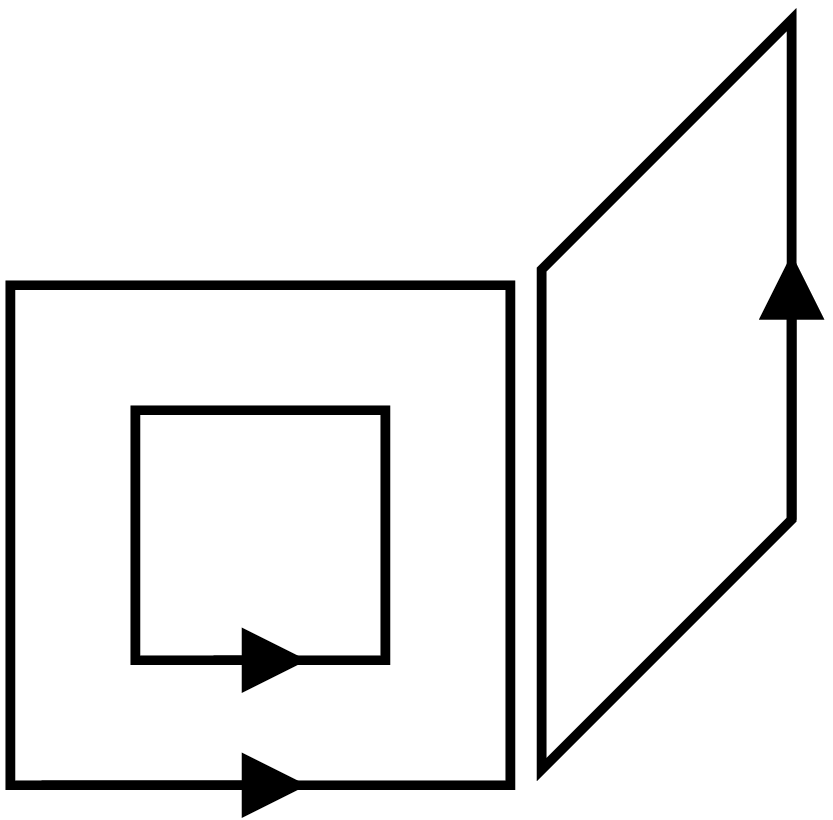}\end{array}
+C\left(-
\plaquette
 +\frac{1}{N^2}
 \begin{array}{c}\includegraphics[width=1.125cm]{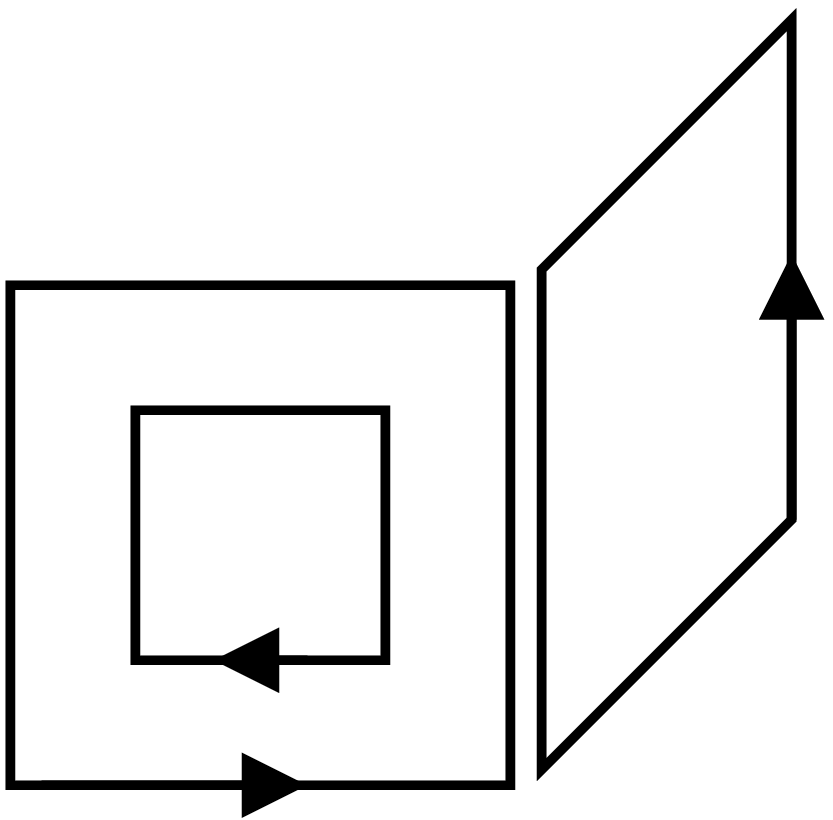}\end{array}\right)\bfrangle\nn\\
 \frac{D^C_{21}}{N_p} &=&
 \frac{8}{N}\bflangle\begin{array}{c}\includegraphics[width=1.125cm]{16.eps}\end{array}+C\begin{array}{c}\includegraphics[width=1.125cm]{17.eps}\end{array}\bfrangle+\frac{16}{N}\bflangle\begin{array}{c}\includegraphics[width=1.125cm]{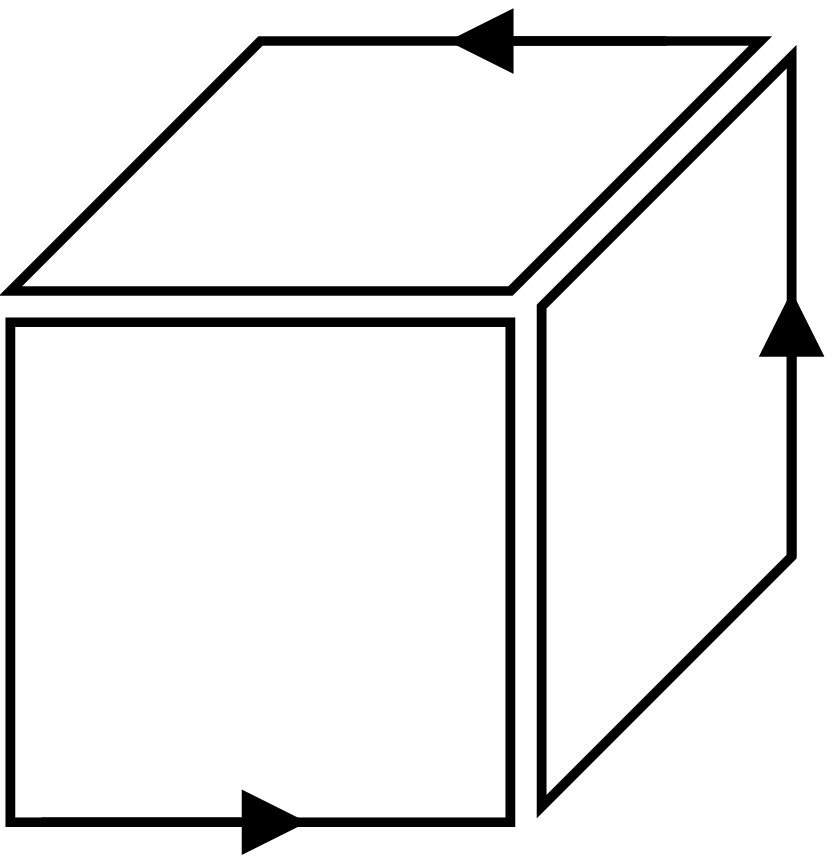}\end{array}+
 C
 \begin{array}{c}\includegraphics[width=1.125cm]{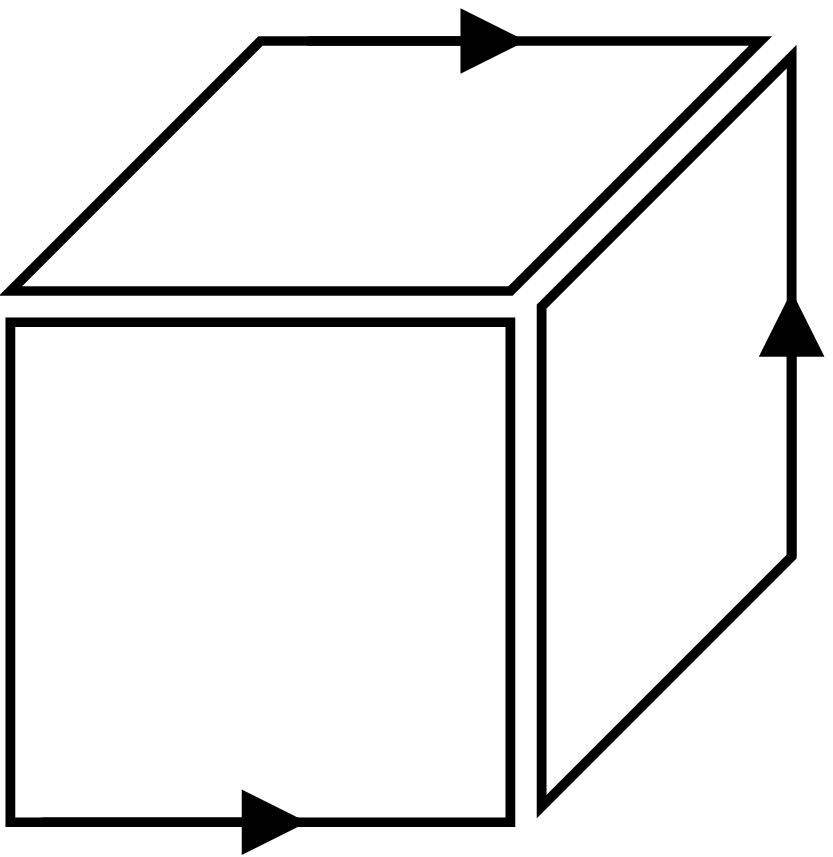}\end{array}\bfrangle
 \nn\\
&& - \frac{48}{N}\left(\frac{C+1}{2}\right) \flangle \plaquette
 \frangle \bflangle
 \begin{array}{c}\includegraphics[width=1.125cm]{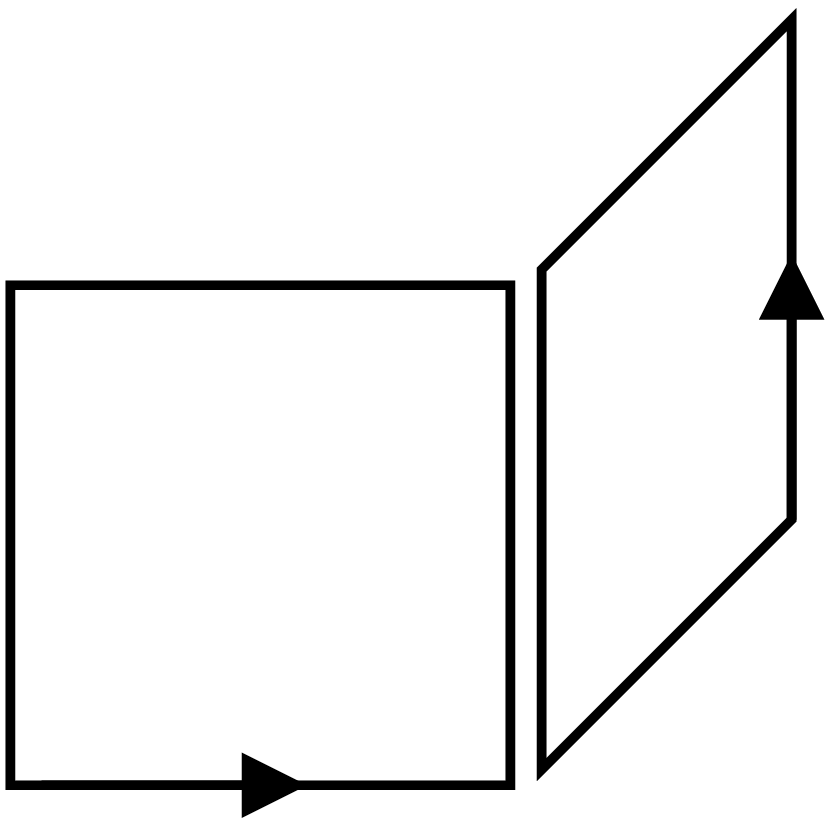}\end{array}\bfrangle .\label{mes-3+1}
\eea
Here we have introduced the notation
\bea
\langle \!\langle O \rangle\!\rangle = \langle \phi_0 | O |\phi_0 \rangle. 
\eea
The combinatorics which lead to the coefficients in the matrix
elements in \eqn{mes-3+1} are a result of counting the possible 
overlaps within a
single cube. The remaining matrix elements $N_{22}$ and $D_{22}$ can
be calculated similarly, resulting in
\bea
\frac{2aN^C_{22}}{N_p g^2} &=& 12\left[\bflangle
 \begin{array}{c}\includegraphics[width=1.125cm]{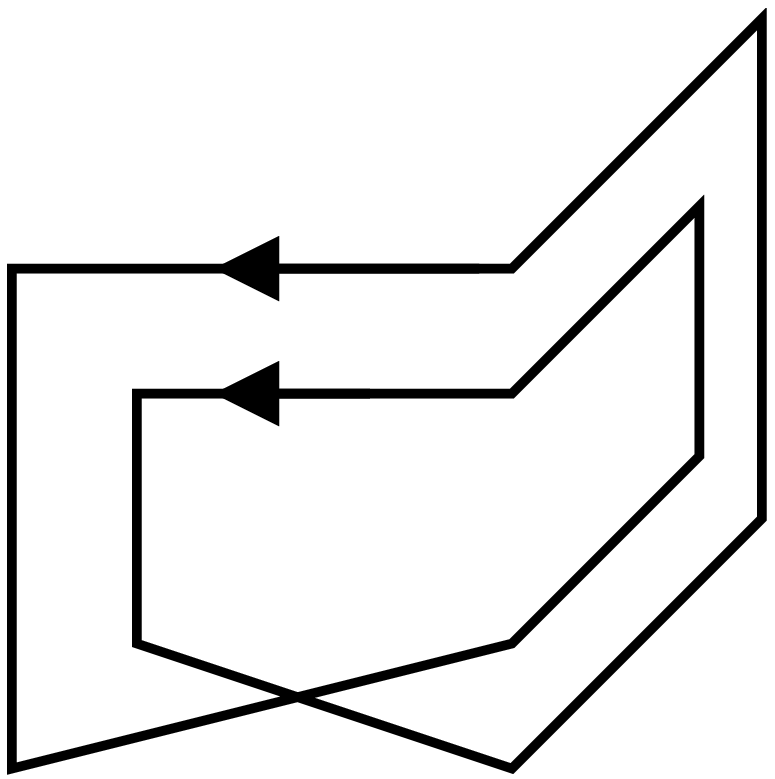}\end{array}\bfrangle
-\frac{1}{N}\bflangle\begin{array}{c}\includegraphics[width=1.125cm]{1.eps}\end{array}\bfrangle 
+C\left(-N+\frac{1}{N}
 \bflangle \begin{array}{c}\includegraphics[width=1.125cm]{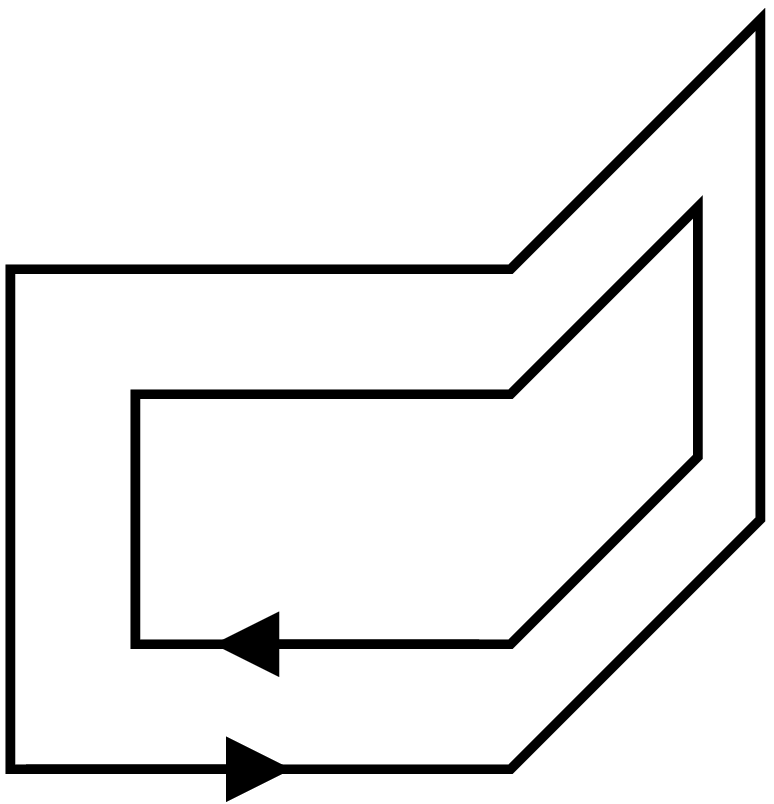}\end{array}\bfrangle\right)
\right] \nn\\
&& + \frac{32}{N} \bflangle
\frac{1}{N}\begin{array}{c}\includegraphics[width=1.125cm]{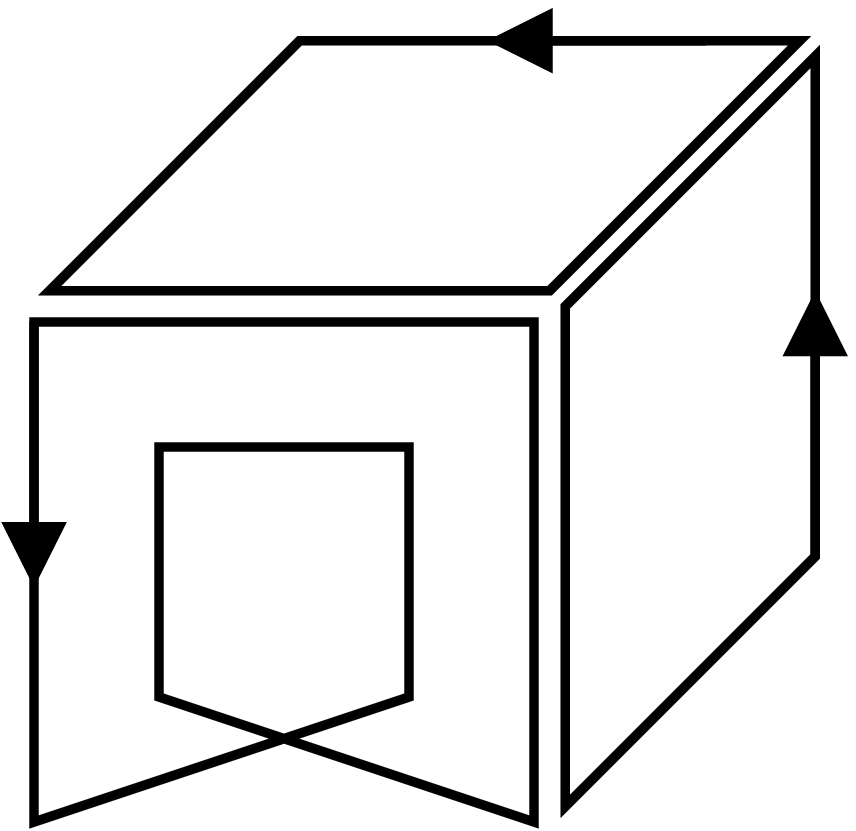}\end{array}
-
\frac{1}{N^2}
\begin{array}{c}\includegraphics[width=1.125cm]{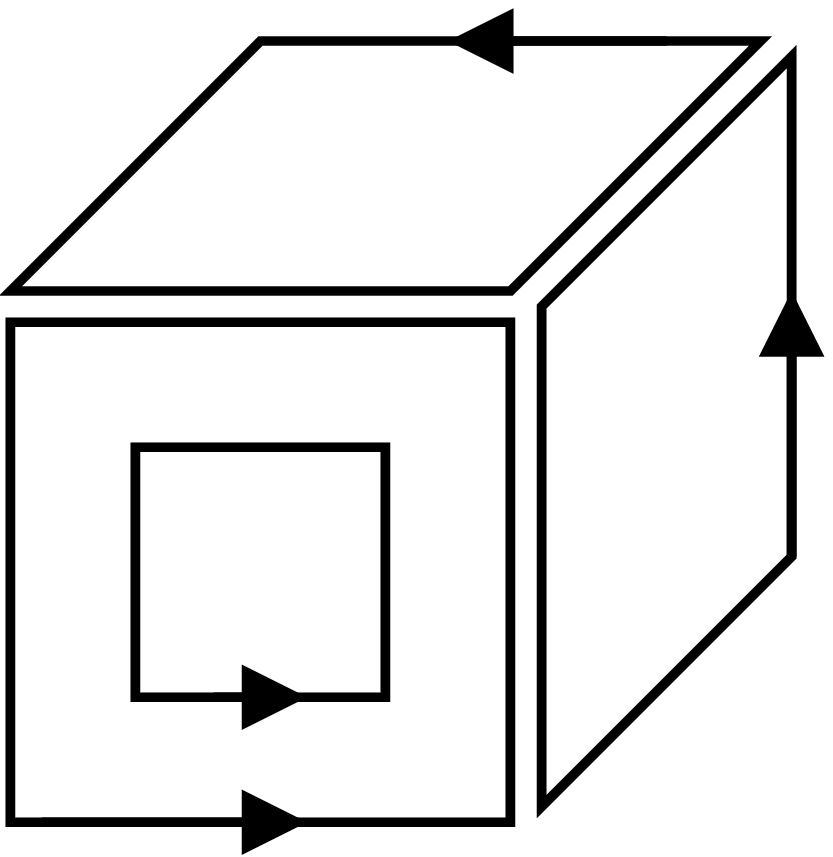}\end{array}
+C\left(-
  \begin{array}{c}\includegraphics[width=1.125cm]{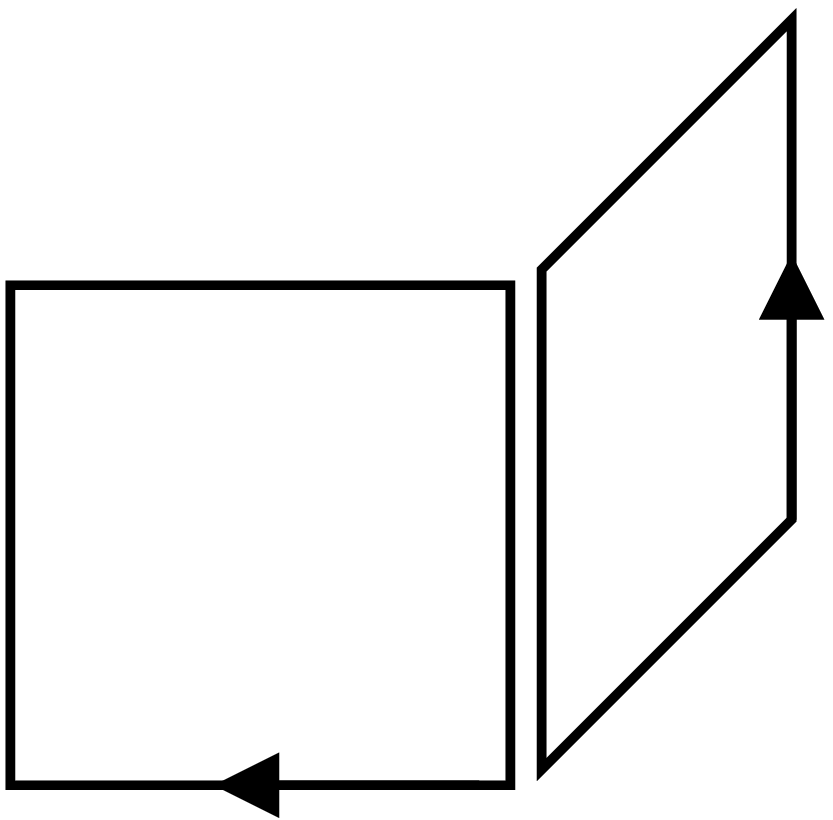}\end{array}
+\frac{1}{N^2}\begin{array}{c}\includegraphics[width=1.125cm]{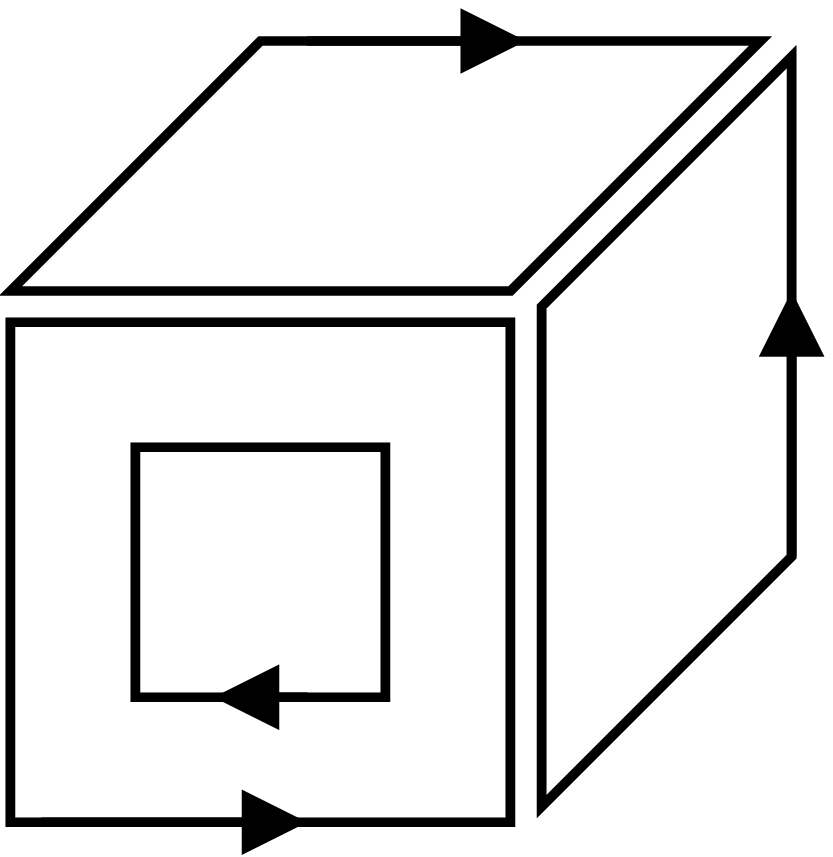}\end{array}
\right)\bfrangle \nn\\
\frac{D^C_{22}}{N_p} &=&
  4\bflangle \begin{array}{c}\includegraphics[width=1.125cm]{1.eps}\end{array}+C
  \begin{array}{c}\includegraphics[width=1.125cm]{2.eps}\end{array}\bfrangle
  + \frac{24}{N^2}\bflangle
\begin{array}{c}\includegraphics[width=1.125cm]{18.eps}\end{array}
  + C \begin{array}{c}\includegraphics[width=1.125cm]{19.eps}\end{array}
\bfrangle \nn\\
&& + \frac{20}{N^2} \bflangle \begin{array}{c}\includegraphics[width=1.125cm]{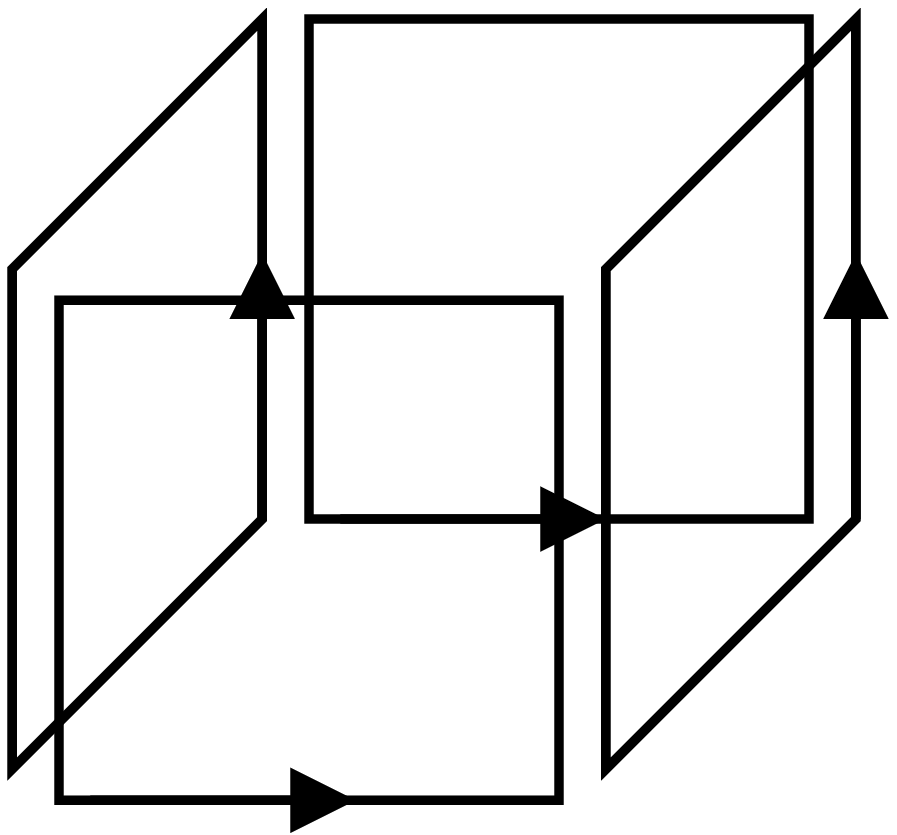}\end{array}
  + C \begin{array}{c}\includegraphics[width=1.125cm]{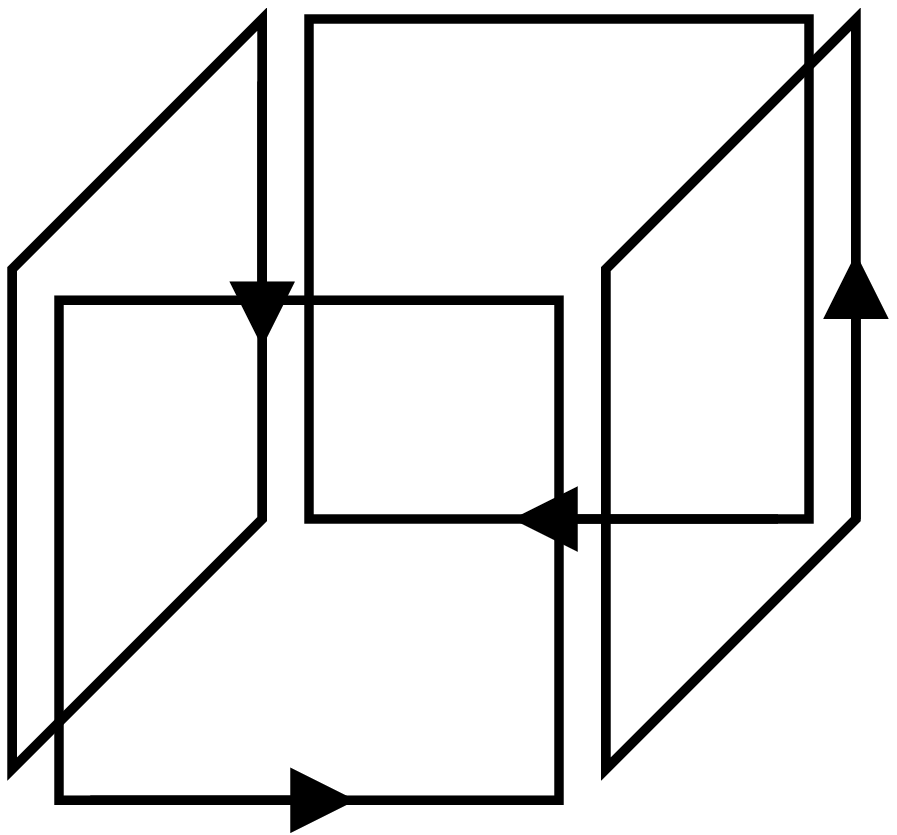}\end{array}
\bfrangle -\frac{96}{N^2}\left(\frac{C+1}{2}\right)\bflangle \begin{array}{c}\includegraphics[width=1.125cm]{20.eps}\end{array} \bfrangle^2.
\eea
Having calculated $N^C_{ii'}$ and $D^C_{ii'}$, we follow the 
minimisation procedure described in \chap{sunmassgaps} to arrive at
the glueball mass, $\Delta M^{PC}$.

\subsection{Results}
\label{results-3+1}
In this section we present calculations of the $0^{++}$ (symmetric) and $1^{+-}$
(antisymmetric) glueball masses on the
one-cube lattice for SU($N$) with $4\le N\le 7$. 
We first define the rescaled glueball mass, $\mu^{PC}$, corresponding
to $\Delta M^{PC}$ as follows:
\bea
 \mu^{PC} = \log( a \Delta M^{PC} \xi^{-51/121}) -\frac{51}{121}\log\left(\frac{48 \pi^2}{11}\right) +\frac{24
 \pi^2}{11} \xi .
\eea
Here, as usual, $P$ and $C$ denote the parity and charge conjugation 
eigenvalues
of the state in question. From the discussion
\sect{extractingcontinuumphysics} asymptotic scaling of a glueball
mass  is observed if the corresponding rescaled glueball mass becomes
constant for some range of couplings.\\ 

\begin{figure}
\centering
               
\subfigure[Calculated with only plaquettes in the minimisation basis] 
                     {
                         \label{S-1-0}
                         \includegraphics[width=7cm]{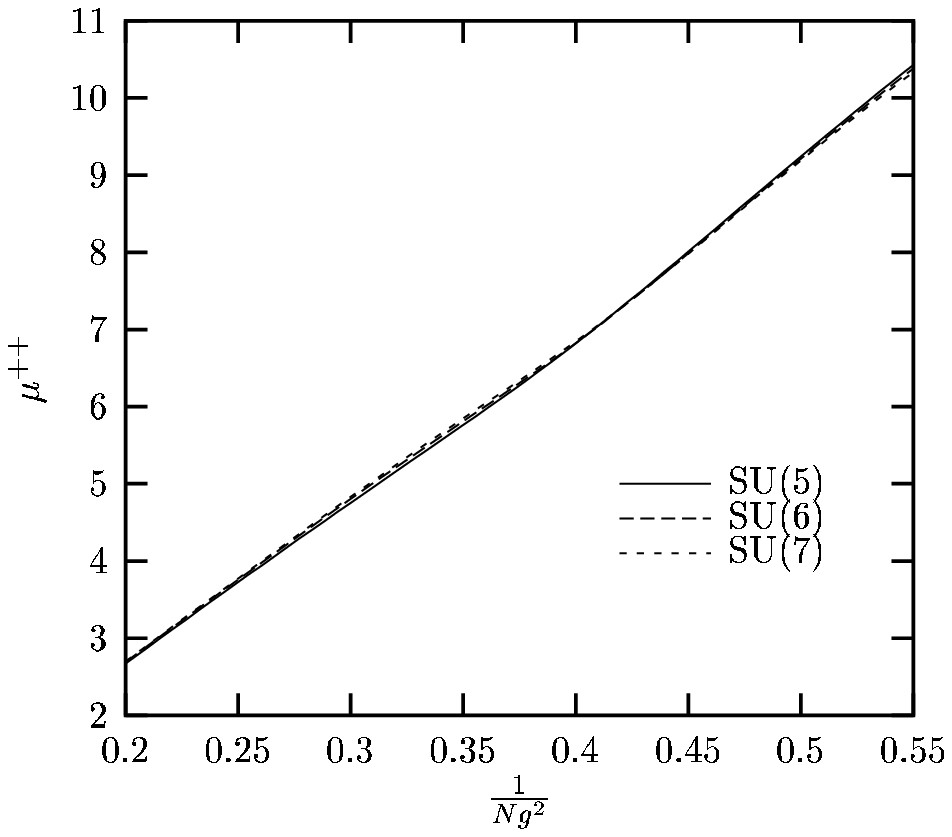}
                     }\hspace{0.25cm}
\subfigure[Calculated with both plaquettes and bent rectangles in the
                         minimisation basis] 
                     {
                         \label{S-1-1}
                         \includegraphics[width=7cm]{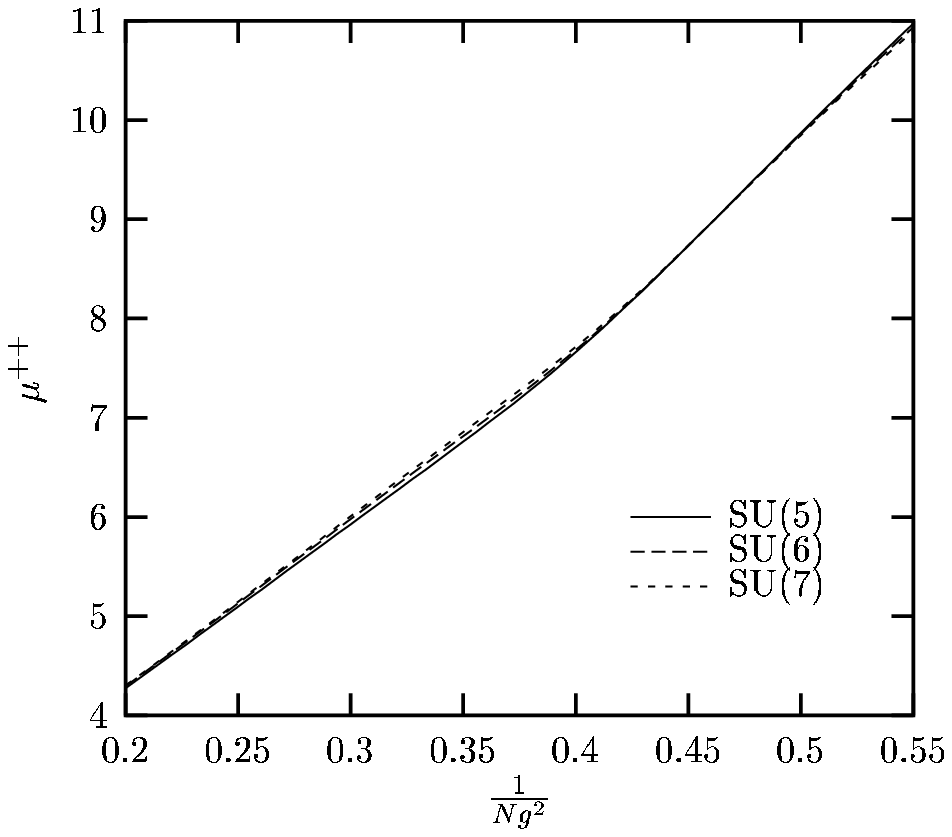}
                     }
\caption{The one cube SU($N$) $0^{++}$ rescaled glueball mass as a
function of $1/(N g^2)$ for various $N$ obtained with different
minimisation bases.}
\label{3+1sunmassgap-S}  
\end{figure}

The results
for the rescaled symmetric glueball mass are shown in
\fig{3+1sunmassgap-S}. \fig{S-1-0} shows the rescaled 
glueball mass calculated with only plaquettes in the minimisation
basis. \fig{S-1-1} shows the same quantity calculated with the
minimisation basis of \eqn{basis}.  The
aim of this exploratory study is to observe whether or not a move
toward scaling is apparent 
as the number of states in the minimisation basis is increased. This
is clearly not the case for the symmetric glueball mass on the range of
couplings explored here. There is no sign of $\mu^{++}$ 
becoming constant on any range of couplings within $0\le \xi \le 0.55$.\\

\begin{figure}
\centering
               
\subfigure[Calculated with only plaquettes in the minimisation basis] 
                     {
                         \label{AS-1-0}
                         \includegraphics[width=7cm]{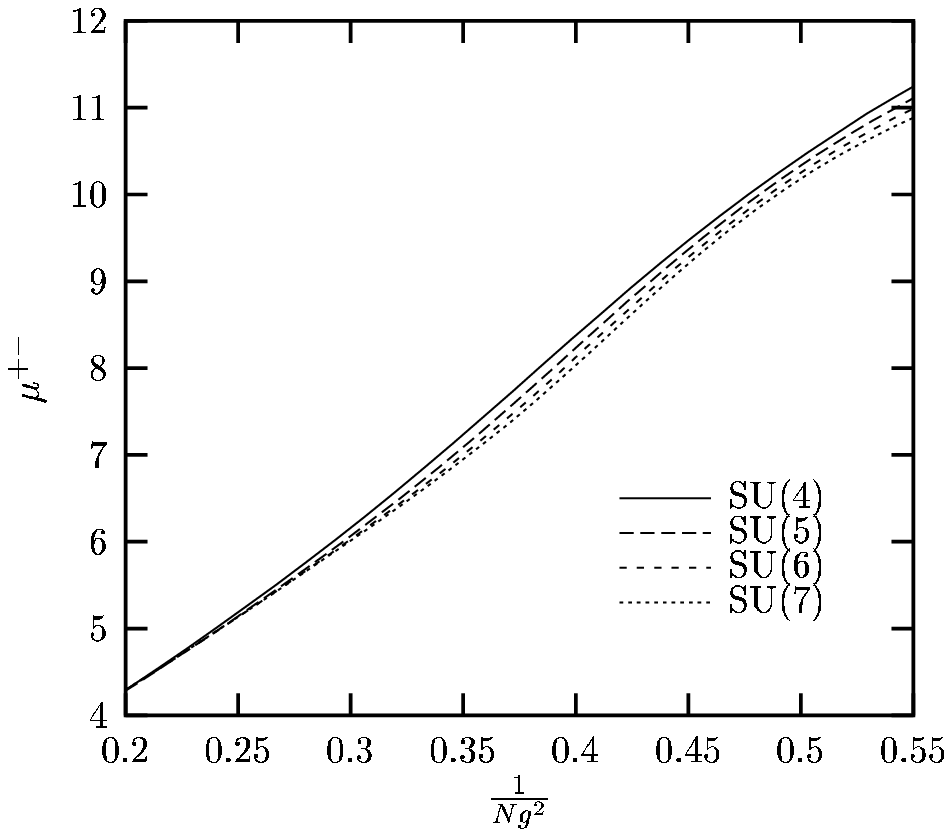}
                     }\hspace{0.25cm}
\subfigure[Calculated with both plaquettes and bent rectangles in the
                         minimisation basis] 
                     {
                         \label{AS-1-1}
                         \includegraphics[width=7cm]{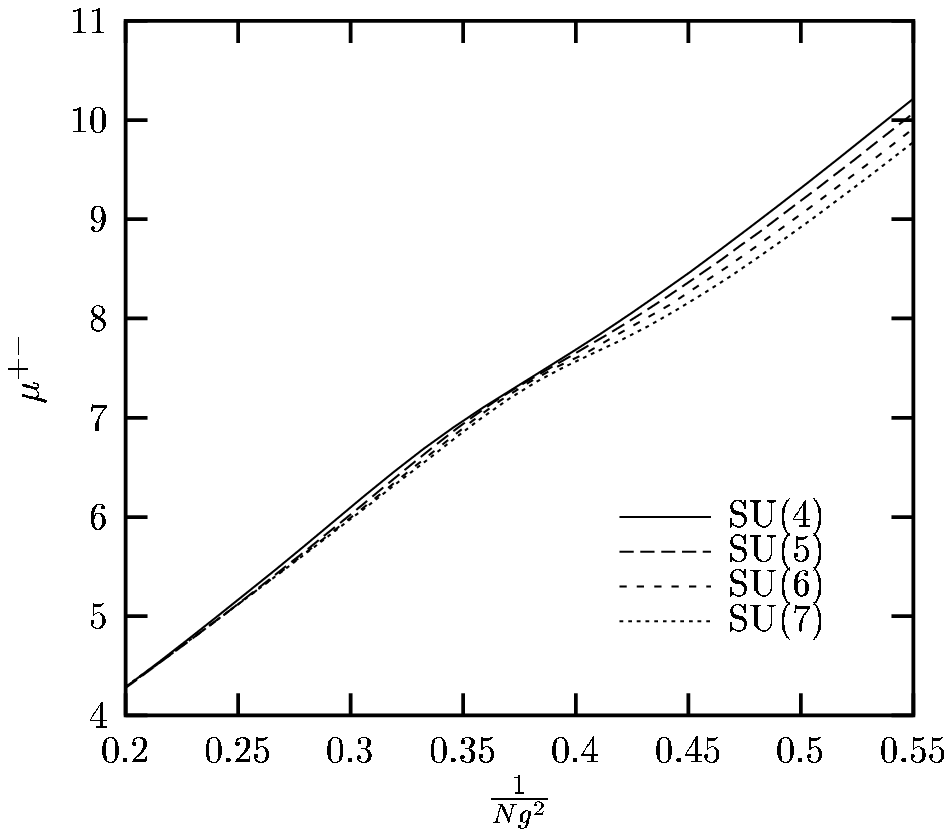}
                     }
\caption{The one cube SU($N$) $1^{+-}$ rescaled glueball mass as a
function of $1/(N g^2)$ for various $N$ obtained with different
minimisation bases.}
\label{3+1sunmassgap-AS}  
\end{figure}

The prospects are marginally better for the antisymmetric case.
The results for the rescaled antisymmetric glueball mass are displayed in
\fig{3+1sunmassgap-AS}. A move toward scaling is observed,
most clearly for the $N=7$ case, near $\xi=0.38$. This matches the region
of couplings for which Chin, Long and Robson observed scaling of the $0^{++}$ glueball mass, for $N=5$ and 6, on a $6^3$ lattice 
using only plaquettes in their
minimisation basis~\cite{Chin:1986fe}.

\subsection{String Tension}

If the exploratory study presented here was to be extended to larger
lattices, at some stage it would be useful to compute the string
tension, $\sigma$. It is common for calculations in Lagrangian LGT to
express results for masses in units of $\sqrt{\sigma}$. This allows
masses calculated on the lattice to be expressed in MeV since the string tension can be 
calculated in MeV
from the decay of heavy quarkonia for example.   In the
Hamiltonian formulation, precise calculations of the string tension have only
been performed in the strong coupling regime. Variational
estimates, at least for SU(2)~\cite{Arisue:1983tt}, have not exhibited asymptotic scaling
when making use of the one-plaquette ground state of \eqn{oneplaquette}. It is possible that the situation may be improved for higher dimensional gauge groups but this has not yet been tested.\\

In this section we calculate the symmetric and antisymmetric
glueball masses in units of $\sqrt{\sigma}$. For the string tension, since
reliable variational results are not available, we use the strong
coupling expansions of Kogut and Shigemitsu~\cite{Kogut:1980pm}. These
are available for SU(5) and SU(6), as well as SU(2) and SU(3)
which we do not consider here.\\

Since the square root of the string tension has units of mass, the ratio of a 
mass to the string tension is constant in a
scaling region. Again, with the crude model presented here, we do not expect
to observe scaling. We seek only an indication of an approach to
scaling. Such behaviour would warrant further study. \\

\begin{figure}
\centering
               
\subfigure[Lowest energy state] 
                     {
                         \label{S-1-both}
                         \includegraphics[width=7cm]{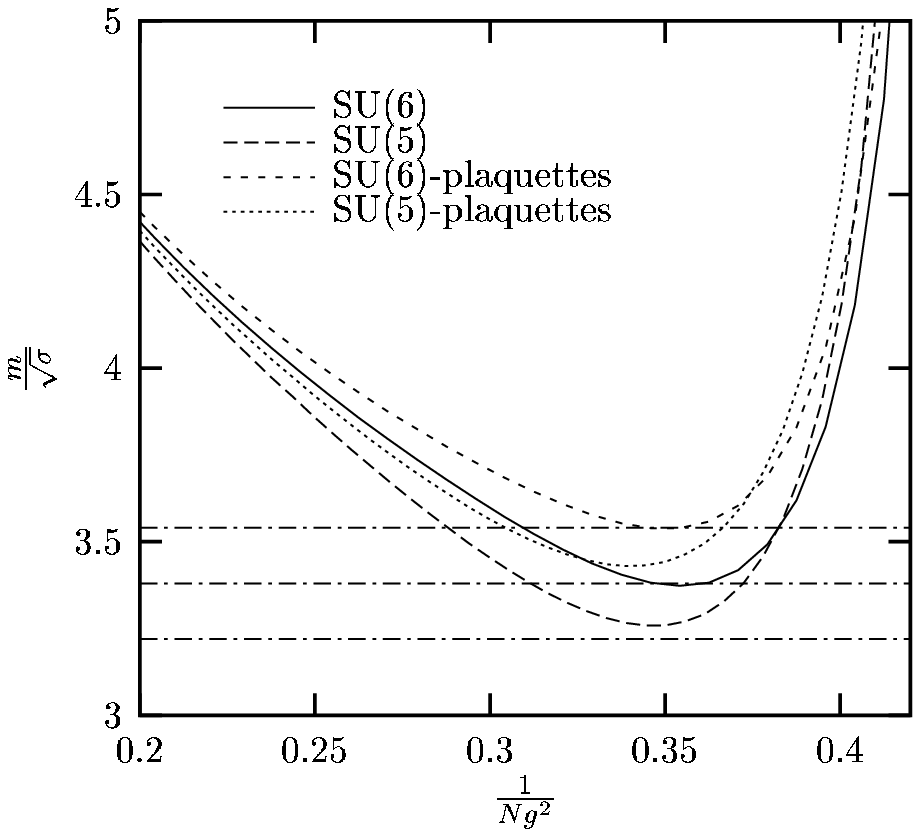}
                     }\hspace{0.25cm}
\subfigure[Second lowest energy state] 
                     {
                         \label{S-2}
                         \includegraphics[width=7cm]{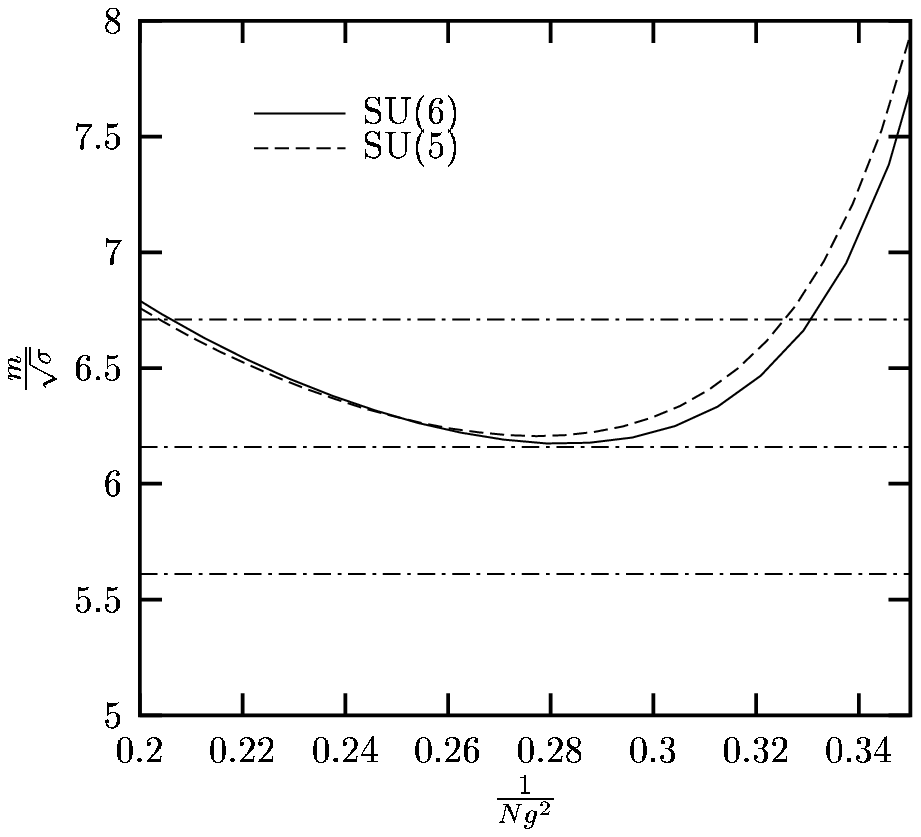}
                     }
\caption{The one cube SU(5) and SU(6) $0^{++}$ 
glueball masses in units of $\sqrt{\sigma}$ as functions of
$1/(N g^2)$. The horizontal lines indicate the result and error bars
of the SU(5)  $0^{++}$ calculation of Lucini and Teper~\cite{Lucini:2001ej}.}
\label{3+1-S-ST}  
\end{figure}

 The results for the symmetric states are
shown in \fig{3+1-S-ST}. Since we perform calculations with two
states in the minimisation basis, the two lowest mass states are
accessible. In \fig{S-1-both} the results for the lowest mass
state are shown. Calculations with different minimisation bases are
shown, with the ``SU$(N)$-plaquettes'' label indicating that only  plaquette states are used. For each case the glueball mass has a
local minimum in the range $0.33\le\xi\le 0.37$. The various results
do not differ greatly. There is no sign of
improved scaling behaviour when either bent rectangles are included or
$N$ is increased. The effect of including bent rectangles is to lower
the local minimum.\\

The second lowest glueball mass is shown in
\fig{S-2}. Again, no improvement in scaling behaviour is seen
as $N$ is increased. In \fig{3+1-S-ST} the 
horizontal lines indicate the 3+1 dimensional SU(5) calculations (and
error bars) of Lucini and Teper~\cite{Lucini:2001ej}. Interestingly,
both masses calculated here with a simplistic model, have minima that lie
within the error bars.\\

\begin{figure}
\centering
               
\subfigure[Lowest energy state] 
                     {
                         \label{AS-1-both}
                         \includegraphics[width=7cm]{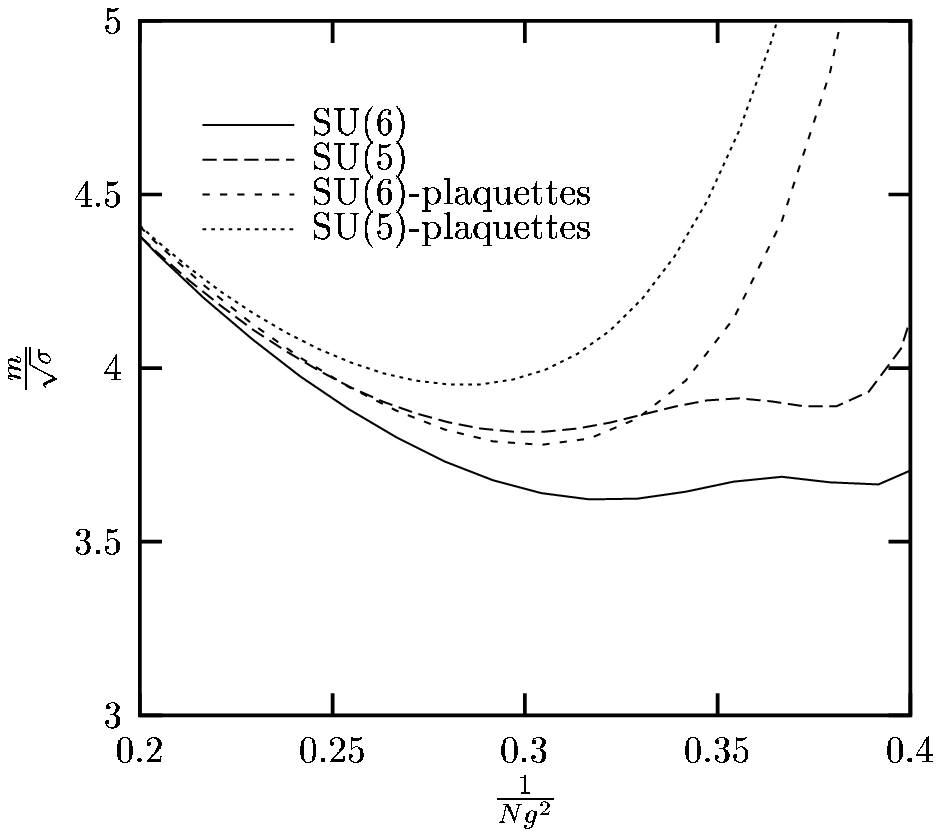}
                     }\hspace{0.25cm}
\subfigure[Second lowest energy state] 
                     {
                         \label{AS-2}
                         \includegraphics[width=7cm]{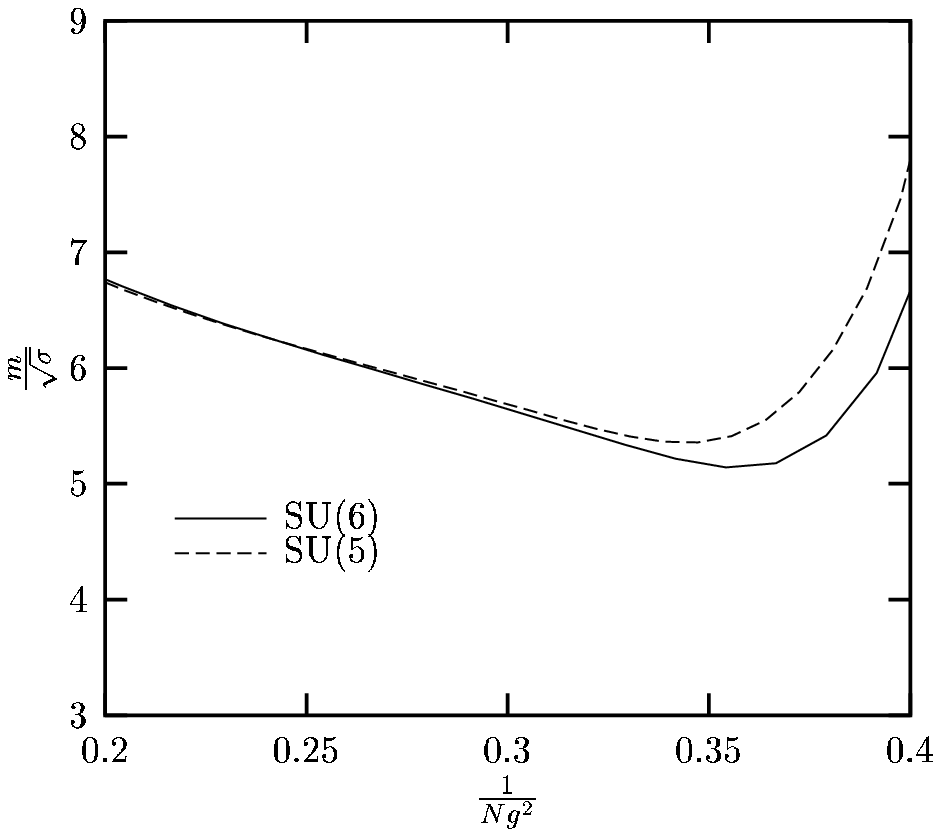}
                     }
\caption{The one cube SU(5) and SU(6) $1^{+-}$ glueball masses in units of $\sqrt{\sigma}$ as functions of
$1/(N g^2)$.}
\label{3+1-AS-ST}  
\end{figure}

We now move on to the antisymmetric states. 
The results for the antisymmetric glueball masses in units of
$\sqrt{\sigma}$ are shown in \fig{3+1-AS-ST}. The results are
more promising than the symmetric case. We see an improved approach to
scaling when bent rectangles are included in the minimisation basis
for the lowest glueball mass. For this case the glueball mass shows
promising signs of becoming constant in the ranges  $0.32\le \xi \le 0.39$ for
SU(6) and $0.3\le \xi \le 0.37$ for SU(5). Only
marginal improvement in scaling behaviour is evident as $N$ is
increased from 5 to 6. For each of these cases
no data is available for comparison to our knowledge. Such promising
signs are not apparent in the second lowest glueball mass in this
sector, as seen in \fig{AS-2}.  \\

It would be interesting to perform analogous calculations for SU(7),
for which the most promising results were displayed in
\sect{results-3+1}. However strong coupling expansions for the
 SU(7) string tension are not available in
\rcite{Kogut:1980pm} and have not been published elsewhere to
our knowledge.

\section{Future Work}
\label{chap7-futurework}
In this chapter we have studied the variational $0^{++}$ and $1^{+-}$ glueball masses
 on a single cube in 3+1 dimensions. 
Our intention was to determine the viability of using analytic
 Hamiltonian methods in 3+1 dimensional glueball mass
 calculations. With such a crude small volume approximation and a
 minimisation basis containing only two states, the observation of
 asymptotic scaling was not expected. Promising signs of an approach
 to asymptotic scaling were displayed by the $1^{+-}$ glueball mass, as $N$ was increased, at couplings close to the 
scaling window observed by a comparable, but larger volume, study by
 Chin, Long and Robson~\cite{Chin:1986fe}. Interesting results were
 also observed for calculations of the glueball masses in units of the
 string tension. With no variational results for the string tension
 available, the strong coupling results of Kogut and
 Shigemitsu~\cite{Kogut:1980pm} were used. Interestingly, the mass of
 both SU(5) $0^{++}$ glueball states calculated had minima which were
 consistent with the Lagrangian calculation of Lucini and Teper~\cite{Lucini:2001ej}. Better scaling behaviour was exhibited by the SU(5) and SU(6) $1^{+-}$ states, although no alternative calculations are 
available for comparison to our knowledge. For the lowest mass $1^{+-}$ states
 calculated, the scaling behaviour was improved by increasing the number of states in the minimisation basis and the dimension of the gauge group. The promising results observed in this chapter warrant further study on larger lattices, with additional states in the minimisation basis.\\

To extend this calculation to larger lattices and minimisation bases, two challenges will be faced. Firstly, the correct implementation of the Bianchi identity, within our plaquette based analytic approach, forces the number of integration variables to grow quickly with the volume. New approaches to handling the general character integrals may need to be developed in order to avoid the inevitable memory restrictions. Secondly, great care will need to be taken in the counting of overlaps between different states. This process could, in principle, be automated using techniques from graph theory and symbolic programming. \\

It would also be of interest to extend the calculation presented here to larger $N$.
To do this a more efficient technique for handling the general character integrals would need to be developed to minimise the memory demands of the calculation. 
The following is just one possibility. It is likely, that by expressing the products of characters appearing in the character integrals as sums of characters, one could reduce the memory requirements of the calculation drastically. The general character integrals could then be expressed in terms of integrals of the form,
\bea
G_r(c) &=& \int_{{\rm SU}(N)} dU \chi_r(U^\dagger) 
e^{c \Tr(U + U^\dagger)},
\eea
where $r = (r_1,r_2,\ldots r_{N-1})$ labels a representation of SU($N$).
Having completed the calculations presented in this chapter, it was discovered that $G_r(c)$ can be handled using the techniques of \chap{analytictechniques}, with the result,
\bea
G_r(c) &=& \sum_{l=-\infty}^\infty \det \left[I_{r_i + l+ j -i}(2 c) \right]_{1\le i,j\le N}.
\eea 
The final stage of this improvement would be to find a convenient way
to express the products of characters appearing in the calculation as a sum of characters. This would be possible with the symbolic manipulation of Young tableaux.

\chapter{Conclusion}
\label{thesisconclusion}

This thesis has been concerned with the study of two topics:
improvement and the use of analytic techniques in Hamiltonian LGT.
In this chapter we summarise our key results and discuss the
directions that further work should take.\\

\chaps{constructingandimproving}{testingimprovement}
were concerned primarily with the issue of improvement. In
\chap{constructingandimproving} a technique allowing the
straightforward construction of lattice Hamiltonians was
developed. Using this direct approach a number of improved
Hamiltonians were derived. The key advantage of our approach is that it
allows the straightforward construction of improved kinetic
Hamiltonians with a finite number of terms, 
a task that has proved nontrivial in the past. In
\chap{testingimprovement} two simple tests of improvement, a strong coupling calculation of the
U(1) Callan-Symanzik $\beta$ function in 3+1 dimensions and a mean
field calculation of the SU(2) lattice specific heat in 2+1 dimensions, were presented. For
each example considered, the tadpole improved Hamiltonians (when they were used) achieved the
greatest level of improvement followed by the classically improved
Hamiltonians and then the Kogut-Susskind Hamiltonians.\\

In \chap{analytictechniques}, techniques which allow the
analytic calculation of SU($N$) matrix elements in 2+1 dimensions were
developed. This chapter generalised the methods that have been employed
in variational SU(2) LGT calculations for almost 20 years, to the
general case of SU($N$). A close connection with the fields of random
matrices and combinatorics was also discussed.\\

In \chap{sunmassgaps}, the analytic techniques of
\chap{analytictechniques} were applied in the  
calculation of $N=2,3,4,$ and 5 glueball masses in 2+1
dimensions. Kogut-Susskind, improved and tadpole improved
Hamiltonians, all of which were derived in
\chap{constructingandimproving}, were used and their results
compared. For each case considered, the lowest mass results obtained were
considerably higher than the lowest glueball masses obtained by comparable
Hamiltonian and Lagrangian calculations. However, the lowest mass
states obtained in the $0^{++}$ and $0^{--}$ sectors were close
in mass to the respective excited states, $0^{++*}$ and  $0^{--**}$, 
of Teper and Lucini~\cite{Teper:1998te,Lucini:2002wg}. In the $0^{++}$ sector this agreement improved as $N$ was increased, and for each $N$ the agreement was closer for the
improved calculations.  This situation was replicated in the $0^{--}$
sector if
the SU(4) results were excluded. For SU(4), the lowest unimproved
glueball mass 
was closer to the mass of Teper and Lucini's $0^{--**}$ state than the improved
results. The problem of enumerating states can be attributed to the
fact that the calculations presented in \chap{sunmassgaps} used
a basis of rectangular states. The competing calculations incorporate additional  
nonrectangular states. Without the inclusion of 
small area, nonrectangular states in the
calculation, it
is possible that some low mass states are inaccessible. Further
calculations, which include nonrectangular states in the minimisation
basis, are needed to clarify this issue. Including such additional
states will also allow spin 1 and $P=-C$ states to be studied, and
permit a more complete study of the spin 0 and 2 glueball mass spectra.\\

In \chap{largenphysics} we presented the first steps of a 
study of the SU($N$)
glueball mass spectrum in the large $N$ limit using the techniques of
\chap{sunmassgaps}. Our motivation was 
the Maldacena conjecture which, in its simplest form,  predicts a precise
correspondence between certain string theories and gauge theories in
the large $N$ limit. Using analytic Hamiltonian techniques,
calculations with $N$ as large as 25 were possible on a desktop
computer, 
in contrast to
comparable Lagrangian calculations which have reached $N=6$. With such
large $N$ glueball masses  accessible, reliable extrapolations to the $N\rightarrow
\infty$ were possible. The results provided evidence for leading
order $1/N^2$ finite $N$ corrections to the glueball masses, verifying
a specific prediction of large $N$ gauge theory. \\

In \chap{3+1dimensions} we moved on to the physically interesting case of 3+1 dimensions. There we studied the viability of adapting the analytic techniques used with success in \chaps{sunmassgaps}{largenphysics} to 3+1 dimensions. A simple variational 
calculation of $0^{++}$ and $1^{+-}$ glueball masses was carried out
on a single cube, with only two states in the minimisation basis. With
such a simplistic approximation asymptotic scaling was not
expected. However, it was hoped that with the use of analytic 
Hamiltonian techniques, large values of $N$ could be reached. This was
not the case, with memory constraints restricting calculations to
$N\le 7$ on a one cube lattice. Our motivation for attempting this
calculation was a variational Hamiltonian study by Chin, Long and 
Robson~\cite{Chin:1986fe} which demonstrated asymptotic scaling of the $0^{++}$ glueball mass on a $6^3$ lattice for $N=5$ and 6, with improved scaling observed as $N$ was increased. 
Promising signs of an approach to asymptotic scaling were observed in \chap{3+1dimensions} for the $1^{+-}$ state, in the same region of couplings as scaling was seen by Chin, Long and Robson. No such signs were observed for the $0^{++}$ state. The extension to larger lattices thus seems warranted. Such extensions will be hampered by memory constraints unless new techniques for handling the general character integrals are developed. Another important improvement, which does not pose a serious problem, will be the inclusion of more states in the minimisation basis. The applicability of 
analytic variational techniques in 3+1 dimensions will thus be determined by how quickly the infinite volume limit is reached.\\

While we have derived various improved Hamiltonians and put them to
use in glueball mass calculations in 2+1 dimensions, their use in 3+1 dimensional non-abelian LGT is untested. 
The next stages of a 3+1 dimensional calculation, within the variational analytic approach presented here, would incorporate improved Hamiltonians and move beyond a single cube. It is possible that in such a calculation the use of an improved Hamiltonian could mean the difference between observing and not observing scaling on the range of couplings available.  \\

The results of this thesis demonstrate that the analytic techniques of Hamiltonian LGT can compete with the traditional Monte Carlo techniques of Lagrangian LGT, at least in the calculation of glueball masses in 2+1 dimensions. The use of the analytic techniques presented here are by no means restricted to variational calculations. It is possible that they will find use in coupled cluster calculations, or in the moment calculations of the plaquette expansion method~\cite{Hollenberg:1993bp}. The move to 3+1 dimensions is formidable but the calculations of \chap{3+1dimensions} suggest that analytic techniques may by feasible, at least for large $N$, provided the infinite volume limit is reached quickly.

\bibliographystyle{apsrev} 
\bibliographystyle{unsrt}

\bibliography{thesisbibliography}

\begin{thebibliography}{143}
\expandafter\ifx\csname natexlab\endcsname\relax\def\natexlab#1{#1}\fi
\expandafter\ifx\csname bibnamefont\endcsname\relax
  \def\bibnamefont#1{#1}\fi
\expandafter\ifx\csname bibfnamefont\endcsname\relax
  \def\bibfnamefont#1{#1}\fi
\expandafter\ifx\csname citenamefont\endcsname\relax
  \def\citenamefont#1{#1}\fi
\expandafter\ifx\csname url\endcsname\relax
  \def\url#1{\texttt{#1}}\fi
\expandafter\ifx\csname urlprefix\endcsname\relax\def\urlprefix{URL }\fi
\providecommand{\bibinfo}[2]{#2}
\providecommand{\eprint}[2][]{\url{#2}}

\bibitem[{\citenamefont{Arisue et~al.}(1983)\citenamefont{Arisue, Kato, and
  Fujiwara}}]{Arisue:1983tt}
\bibinfo{author}{\bibfnamefont{H.}~\bibnamefont{Arisue}},
  \bibinfo{author}{\bibfnamefont{M.}~\bibnamefont{Kato}}, \bibnamefont{and}
  \bibinfo{author}{\bibfnamefont{T.}~\bibnamefont{Fujiwara}},
  \bibinfo{journal}{Prog. Theor. Phys.} \textbf{\bibinfo{volume}{70}},
  \bibinfo{pages}{229} (\bibinfo{year}{1983}).

\bibitem[{\citenamefont{Lucini and Teper}(2001)}]{Lucini:2001ej}
\bibinfo{author}{\bibfnamefont{B.}~\bibnamefont{Lucini}} \bibnamefont{and}
  \bibinfo{author}{\bibfnamefont{M.}~\bibnamefont{Teper}},
  \bibinfo{journal}{JHEP} \textbf{\bibinfo{volume}{06}}, \bibinfo{pages}{050}
  (\bibinfo{year}{2001}), \eprint[http://arXiv.org/abs]{hep-lat/0103027}.

\bibitem[{\citenamefont{Gross and Wilczek}(1973)}]{Gross:1973id}
\bibinfo{author}{\bibfnamefont{D.~J.} \bibnamefont{Gross}} \bibnamefont{and}
  \bibinfo{author}{\bibfnamefont{F.}~\bibnamefont{Wilczek}},
  \bibinfo{journal}{Phys. Rev. Lett.} \textbf{\bibinfo{volume}{30}},
  \bibinfo{pages}{1343} (\bibinfo{year}{1973}).

\bibitem[{\citenamefont{Politzer}(1973)}]{Politzer:1973fx}
\bibinfo{author}{\bibfnamefont{H.~D.} \bibnamefont{Politzer}},
  \bibinfo{journal}{Phys. Rev. Lett.} \textbf{\bibinfo{volume}{{\bf 30}}},
  \bibinfo{pages}{1346} (\bibinfo{year}{1973}).

\bibitem[{\citenamefont{Gribov and Lipatov}(1972)}]{Gribov:1972ri}
\bibinfo{author}{\bibfnamefont{V.~N.} \bibnamefont{Gribov}} \bibnamefont{and}
  \bibinfo{author}{\bibfnamefont{L.~N.} \bibnamefont{Lipatov}},
  \bibinfo{journal}{Yad. Fiz.} \textbf{\bibinfo{volume}{15}},
  \bibinfo{pages}{781} (\bibinfo{year}{1972}).

\bibitem[{\citenamefont{{Hagiwara} et~al.}(2002)\citenamefont{{Hagiwara},
  {Hikasa}, {Nakamura}, {Tanabashi}, {Aguilar-Benitez}, {Amsler}, {Barnett},
  {Burchat}, {Carone}, {Caso} et~al.}}]{PDBook}
\bibinfo{author}{\bibfnamefont{K.}~\bibnamefont{{Hagiwara}}},
  \bibinfo{author}{\bibfnamefont{K.}~\bibnamefont{{Hikasa}}},
  \bibinfo{author}{\bibfnamefont{K.}~\bibnamefont{{Nakamura}}},
  \bibinfo{author}{\bibfnamefont{M.}~\bibnamefont{{Tanabashi}}},
  \bibinfo{author}{\bibfnamefont{M.}~\bibnamefont{{Aguilar-Benitez}}},
  \bibinfo{author}{\bibfnamefont{C.}~\bibnamefont{{Amsler}}},
  \bibinfo{author}{\bibfnamefont{R.}~\bibnamefont{{Barnett}}},
  \bibinfo{author}{\bibfnamefont{P.}~\bibnamefont{{Burchat}}},
  \bibinfo{author}{\bibfnamefont{C.}~\bibnamefont{{Carone}}},
  \bibinfo{author}{\bibfnamefont{C.}~\bibnamefont{{Caso}}},
  \bibnamefont{et~al.}, \bibinfo{journal}{Phys. Rev.}
  \textbf{\bibinfo{volume}{D66}}, \bibinfo{pages}{010001}
  (\bibinfo{year}{2002}), \urlprefix\url{http://pdg.lbl.gov}.

\bibitem[{\citenamefont{Chodos et~al.}(1974)\citenamefont{Chodos, Jaffe,
  Johnson, Thorn, and Weisskopf}}]{Chodos:1974je}
\bibinfo{author}{\bibfnamefont{A.}~\bibnamefont{Chodos}},
  \bibinfo{author}{\bibfnamefont{R.~L.} \bibnamefont{Jaffe}},
  \bibinfo{author}{\bibfnamefont{K.}~\bibnamefont{Johnson}},
  \bibinfo{author}{\bibfnamefont{C.~B.} \bibnamefont{Thorn}}, \bibnamefont{and}
  \bibinfo{author}{\bibfnamefont{V.~F.} \bibnamefont{Weisskopf}},
  \bibinfo{journal}{Phys. Rev.} \textbf{\bibinfo{volume}{D9}},
  \bibinfo{pages}{3471} (\bibinfo{year}{1974}).

\bibitem[{\citenamefont{Cornwall and Soni}(1983)}]{Cornwall:1983zn}
\bibinfo{author}{\bibfnamefont{J.~M.} \bibnamefont{Cornwall}} \bibnamefont{and}
  \bibinfo{author}{\bibfnamefont{A.}~\bibnamefont{Soni}},
  \bibinfo{journal}{Phys. Lett.} \textbf{\bibinfo{volume}{B120}},
  \bibinfo{pages}{431} (\bibinfo{year}{1983}).

\bibitem[{\citenamefont{Isgur and Paton}(1985)}]{Isgur:1985bm}
\bibinfo{author}{\bibfnamefont{N.}~\bibnamefont{Isgur}} \bibnamefont{and}
  \bibinfo{author}{\bibfnamefont{J.}~\bibnamefont{Paton}},
  \bibinfo{journal}{Phys. Rev.} \textbf{\bibinfo{volume}{D31}},
  \bibinfo{pages}{2910} (\bibinfo{year}{1985}).

\bibitem[{\citenamefont{Johnson and Teper}(2002)}]{Johnson:2000qz}
\bibinfo{author}{\bibfnamefont{R.~W.} \bibnamefont{Johnson}} \bibnamefont{and}
  \bibinfo{author}{\bibfnamefont{M.~J.} \bibnamefont{Teper}},
  \bibinfo{journal}{Phys. Rev.} \textbf{\bibinfo{volume}{D66}},
  \bibinfo{pages}{036006} (\bibinfo{year}{2002}),
  \eprint[http://arXiv.org/abs]{hep-ph/0012287}.

\bibitem[{\citenamefont{Wilson}(1974)}]{Wilson:1974sk}
\bibinfo{author}{\bibfnamefont{K.~G.} \bibnamefont{Wilson}},
  \bibinfo{journal}{Phys. Rev.} \textbf{\bibinfo{volume}{D10}},
  \bibinfo{pages}{2445} (\bibinfo{year}{1974}).

\bibitem[{\citenamefont{Kogut and Susskind}(1975)}]{Kogut:1975ag}
\bibinfo{author}{\bibfnamefont{J.~B.} \bibnamefont{Kogut}} \bibnamefont{and}
  \bibinfo{author}{\bibfnamefont{L.}~\bibnamefont{Susskind}},
  \bibinfo{journal}{Phys. Rev.} \textbf{\bibinfo{volume}{D11}},
  \bibinfo{pages}{395} (\bibinfo{year}{1975}).

\bibitem[{\citenamefont{Beneke}(2002)}]{Beneke:2002ks}
\bibinfo{author}{\bibfnamefont{M.}~\bibnamefont{Beneke}}
  (\bibinfo{year}{2002}), \eprint[http://arXiv.org/abs]{hep-lat/0201011}.

\bibitem[{\citenamefont{Ryan}(2002)}]{Ryan:2001ej}
\bibinfo{author}{\bibfnamefont{S.~M.} \bibnamefont{Ryan}},
  \bibinfo{journal}{Nucl. Phys. Proc. Suppl.} \textbf{\bibinfo{volume}{106}},
  \bibinfo{pages}{86} (\bibinfo{year}{2002}),
  \eprint[http://arXiv.org/abs]{hep-lat/0111010}.

\bibitem[{\citenamefont{Gupta}(1997)}]{Gupta:1997nd}
\bibinfo{author}{\bibfnamefont{R.}~\bibnamefont{Gupta}} (\bibinfo{year}{1997}),
  \eprint[http://arXiv.org/abs]{hep-lat/9807028}.

\bibitem[{\citenamefont{Kogut}(1980)}]{Kogut:1980sg}
\bibinfo{author}{\bibfnamefont{J.~B.} \bibnamefont{Kogut}},
  \bibinfo{journal}{Phys. Rept.} \textbf{\bibinfo{volume}{67}},
  \bibinfo{pages}{67} (\bibinfo{year}{1980}).

\bibitem[{\citenamefont{Bali et~al.}(2000)}]{Bali:2000vr}
\bibinfo{author}{\bibfnamefont{G.~S.} \bibnamefont{Bali}} \bibnamefont{et~al.}
  (\bibinfo{collaboration}{TXL}), \bibinfo{journal}{Phys. Rev.}
  \textbf{\bibinfo{volume}{D62}}, \bibinfo{pages}{054503}
  (\bibinfo{year}{2000}), \eprint[http://arXiv.org/abs]{hep-lat/0003012}.

\bibitem[{\citenamefont{Aoki et~al.}(2000)}]{Aoki:1999yr}
\bibinfo{author}{\bibfnamefont{S.}~\bibnamefont{Aoki}} \bibnamefont{et~al.}
  (\bibinfo{collaboration}{CP-PACS}), \bibinfo{journal}{Phys. Rev. Lett.}
  \textbf{\bibinfo{volume}{84}}, \bibinfo{pages}{238} (\bibinfo{year}{2000}),
  \eprint[http://arXiv.org/abs]{hep-lat/9904012}.

\bibitem[{\citenamefont{Yoshie}(2002)}]{Yoshie:2002ru}
\bibinfo{author}{\bibfnamefont{T.}~\bibnamefont{Yoshie}}
  (\bibinfo{collaboration}{CP-PACS}), \bibinfo{journal}{Nucl. Phys. Proc.
  Suppl.} \textbf{\bibinfo{volume}{109A}}, \bibinfo{pages}{25}
  (\bibinfo{year}{2002}).

\bibitem[{\citenamefont{Gregory et~al.}(2000)\citenamefont{Gregory, Guo,
  Kr{\"o}ger, and Luo}}]{Gregory:1999pm}
\bibinfo{author}{\bibfnamefont{E.~B.} \bibnamefont{Gregory}},
  \bibinfo{author}{\bibfnamefont{S.-H.} \bibnamefont{Guo}},
  \bibinfo{author}{\bibfnamefont{H.}~\bibnamefont{Kr{\"o}ger}},
  \bibnamefont{and} \bibinfo{author}{\bibfnamefont{X.-Q.} \bibnamefont{Luo}},
  \bibinfo{journal}{Phys. Rev.} \textbf{\bibinfo{volume}{D62}},
  \bibinfo{pages}{054508} (\bibinfo{year}{2000}),
  \eprint[http://arXiv.org/abs]{hep-lat/9912054}.

\bibitem[{\citenamefont{Bringoltz and Svetitsky}(2002)}]{Bringoltz:2002ug}
\bibinfo{author}{\bibfnamefont{B.}~\bibnamefont{Bringoltz}} \bibnamefont{and}
  \bibinfo{author}{\bibfnamefont{B.}~\bibnamefont{Svetitsky}}
  (\bibinfo{year}{2002}), \eprint[http://arXiv.org/abs]{hep-lat/0209005}.

\bibitem[{\citenamefont{Coester}(1958)}]{Coester:1958}
\bibinfo{author}{\bibfnamefont{F.}~\bibnamefont{Coester}},
  \bibinfo{journal}{Nucl. Phys.} \textbf{\bibinfo{volume}{7}},
  \bibinfo{pages}{421} (\bibinfo{year}{1958}).

\bibitem[{\citenamefont{Coester and K{\"u}mmel}(1960)}]{Coester:1960}
\bibinfo{author}{\bibfnamefont{F.}~\bibnamefont{Coester}} \bibnamefont{and}
  \bibinfo{author}{\bibfnamefont{H.~G.} \bibnamefont{K{\"u}mmel}},
  \bibinfo{journal}{Nucl. Phys.} \textbf{\bibinfo{volume}{17}},
  \bibinfo{pages}{477} (\bibinfo{year}{1960}).

\bibitem[{\citenamefont{Ligterink et~al.}(2000)\citenamefont{Ligterink, Walet,
  and Bishop}}]{Ligterink:2000ug}
\bibinfo{author}{\bibfnamefont{N.~E.} \bibnamefont{Ligterink}},
  \bibinfo{author}{\bibfnamefont{N.~R.} \bibnamefont{Walet}}, \bibnamefont{and}
  \bibinfo{author}{\bibfnamefont{R.~F.} \bibnamefont{Bishop}},
  \bibinfo{journal}{Annals Phys.} \textbf{\bibinfo{volume}{284}},
  \bibinfo{pages}{215} (\bibinfo{year}{2000}),
  \eprint[http://arXiv.org/abs]{hep-lat/0001028}.

\bibitem[{\citenamefont{McKellar et~al.}(2000)\citenamefont{McKellar, Leonard,
  and Hollenberg}}]{McKellar:2000zk}
\bibinfo{author}{\bibfnamefont{B.~H.~J.} \bibnamefont{McKellar}},
  \bibinfo{author}{\bibfnamefont{C.~R.} \bibnamefont{Leonard}},
  \bibnamefont{and} \bibinfo{author}{\bibfnamefont{L.~C.~L.}
  \bibnamefont{Hollenberg}}, \bibinfo{journal}{Int. J. Mod. Phys.}
  \textbf{\bibinfo{volume}{B14}}, \bibinfo{pages}{2023} (\bibinfo{year}{2000}).

\bibitem[{\citenamefont{Hollenberg}(1993)}]{Hollenberg:1993bp}
\bibinfo{author}{\bibfnamefont{L.~C.~L.} \bibnamefont{Hollenberg}},
  \bibinfo{journal}{Phys. Rev.} \textbf{\bibinfo{volume}{D47}},
  \bibinfo{pages}{1640} (\bibinfo{year}{1993}).

\bibitem[{\citenamefont{Hollenberg}(1994)}]{Hollenberg:1994pv}
\bibinfo{author}{\bibfnamefont{L.~C.~L.} \bibnamefont{Hollenberg}},
  \bibinfo{journal}{Phys. Rev.} \textbf{\bibinfo{volume}{D50}},
  \bibinfo{pages}{6917} (\bibinfo{year}{1994}).

\bibitem[{\citenamefont{McIntosh and
  Hollenberg}(2002{\natexlab{a}})}]{McIntosh:2001uk}
\bibinfo{author}{\bibfnamefont{J.~A.~L.} \bibnamefont{McIntosh}}
  \bibnamefont{and} \bibinfo{author}{\bibfnamefont{L.~C.~L.}
  \bibnamefont{Hollenberg}}, \bibinfo{journal}{Nucl. Phys. Proc. Suppl.}
  \textbf{\bibinfo{volume}{106}}, \bibinfo{pages}{257}
  (\bibinfo{year}{2002}{\natexlab{a}}),
  \eprint[http://arXiv.org/abs]{hep-lat/0111045}.

\bibitem[{\citenamefont{McIntosh and
  Hollenberg}(2002{\natexlab{b}})}]{McIntosh:2001fm}
\bibinfo{author}{\bibfnamefont{J.~A.~L.} \bibnamefont{McIntosh}}
  \bibnamefont{and} \bibinfo{author}{\bibfnamefont{L.~C.~L.}
  \bibnamefont{Hollenberg}}, \bibinfo{journal}{Phys. Lett.}
  \textbf{\bibinfo{volume}{B538}}, \bibinfo{pages}{207}
  (\bibinfo{year}{2002}{\natexlab{b}}),
  \eprint[http://arXiv.org/abs]{hep-lat/0111061}.

\bibitem[{\citenamefont{Arisue}(1990)}]{Arisue:1990wv}
\bibinfo{author}{\bibfnamefont{H.}~\bibnamefont{Arisue}},
  \bibinfo{journal}{Prog. Theor. Phys.} \textbf{\bibinfo{volume}{84}},
  \bibinfo{pages}{951} (\bibinfo{year}{1990}).

\bibitem[{\citenamefont{Fritzsch and Gell-Mann}(1972)}]{Fritzsch:1972}
\bibinfo{author}{\bibfnamefont{H.}~\bibnamefont{Fritzsch}} \bibnamefont{and}
  \bibinfo{author}{\bibfnamefont{M.}~\bibnamefont{Gell-Mann}}, in
  \emph{\bibinfo{booktitle}{Proceedings of the XVI International Conference on
  High Energy Physics}}, edited by \bibinfo{editor}{\bibfnamefont{J.~D.}
  \bibnamefont{Jackson}} \bibnamefont{and}
  \bibinfo{editor}{\bibfnamefont{A.}~\bibnamefont{Roberts}}
  (\bibinfo{year}{1972}), vol.~\bibinfo{volume}{2}, p. \bibinfo{pages}{135}.

\bibitem[{\citenamefont{Fritzsch et~al.}(1973)\citenamefont{Fritzsch,
  Gell-Mann, and Leutwyler}}]{Fritzsch:1973pi}
\bibinfo{author}{\bibfnamefont{H.}~\bibnamefont{Fritzsch}},
  \bibinfo{author}{\bibfnamefont{M.}~\bibnamefont{Gell-Mann}},
  \bibnamefont{and}
  \bibinfo{author}{\bibfnamefont{H.}~\bibnamefont{Leutwyler}},
  \bibinfo{journal}{Phys. Lett.} \textbf{\bibinfo{volume}{B47}},
  \bibinfo{pages}{365} (\bibinfo{year}{1973}).

\bibitem[{\citenamefont{Bali}(2001)}]{Bali:2001nc}
\bibinfo{author}{\bibfnamefont{G.~S.} \bibnamefont{Bali}}
  (\bibinfo{year}{2001}), \eprint[http://arXiv.org/abs]{hep-ph/0110254}.

\bibitem[{\citenamefont{Morningstar}(2002)}]{Morningstar:2001nu}
\bibinfo{author}{\bibfnamefont{C.}~\bibnamefont{Morningstar}},
  \bibinfo{journal}{AIP Conf. Proc.} \textbf{\bibinfo{volume}{619}},
  \bibinfo{pages}{231} (\bibinfo{year}{2002}),
  \eprint[http://arXiv.org/abs]{nucl-th/0110074}.

\bibitem[{\citenamefont{Godfrey and Napolitano}(1999)}]{Godfrey:1998pd}
\bibinfo{author}{\bibfnamefont{S.}~\bibnamefont{Godfrey}} \bibnamefont{and}
  \bibinfo{author}{\bibfnamefont{J.}~\bibnamefont{Napolitano}},
  \bibinfo{journal}{Rev. Mod. Phys.} \textbf{\bibinfo{volume}{71}},
  \bibinfo{pages}{1411} (\bibinfo{year}{1999}),
  \eprint[http://arXiv.org/abs]{hep-ph/9811410}.

\bibitem[{\citenamefont{Amsler}(2002)}]{Amsler:2002ey}
\bibinfo{author}{\bibfnamefont{C.}~\bibnamefont{Amsler}},
  \bibinfo{journal}{Phys. Lett.} \textbf{\bibinfo{volume}{B541}},
  \bibinfo{pages}{22} (\bibinfo{year}{2002}),
  \eprint[http://arXiv.org/abs]{hep-ph/0206104}.

\bibitem[{\citenamefont{Amsler and Close}(1996)}]{Amsler:1996td}
\bibinfo{author}{\bibfnamefont{C.}~\bibnamefont{Amsler}} \bibnamefont{and}
  \bibinfo{author}{\bibfnamefont{F.~E.} \bibnamefont{Close}},
  \bibinfo{journal}{Phys. Rev.} \textbf{\bibinfo{volume}{D53}},
  \bibinfo{pages}{295} (\bibinfo{year}{1996}),
  \eprint[http://arXiv.org/abs]{hep-ph/9507326}.

\bibitem[{\citenamefont{Lee and Weingarten}(2000)}]{Lee:1999kv}
\bibinfo{author}{\bibfnamefont{W.-J.} \bibnamefont{Lee}} \bibnamefont{and}
  \bibinfo{author}{\bibfnamefont{D.}~\bibnamefont{Weingarten}},
  \bibinfo{journal}{Phys. Rev.} \textbf{\bibinfo{volume}{D61}},
  \bibinfo{pages}{014015} (\bibinfo{year}{2000}),
  \eprint[http://arXiv.org/abs]{hep-lat/9910008}.

\bibitem[{\citenamefont{Sexton et~al.}(1995)\citenamefont{Sexton, Vaccarino,
  and Weingarten}}]{Sexton:1995kd}
\bibinfo{author}{\bibfnamefont{J.}~\bibnamefont{Sexton}},
  \bibinfo{author}{\bibfnamefont{A.}~\bibnamefont{Vaccarino}},
  \bibnamefont{and}
  \bibinfo{author}{\bibfnamefont{D.}~\bibnamefont{Weingarten}},
  \bibinfo{journal}{Phys. Rev. Lett.} \textbf{\bibinfo{volume}{75}},
  \bibinfo{pages}{4563} (\bibinfo{year}{1995}),
  \eprint[http://arXiv.org/abs]{hep-lat/9510022}.

\bibitem[{\citenamefont{Acciarri et~al.}(2001)}]{Acciarri:2000ex}
\bibinfo{author}{\bibfnamefont{M.}~\bibnamefont{Acciarri}} \bibnamefont{et~al.}
  (\bibinfo{collaboration}{L3}), \bibinfo{journal}{Phys. Lett.}
  \textbf{\bibinfo{volume}{B501}}, \bibinfo{pages}{173} (\bibinfo{year}{2001}),
  \eprint[http://arXiv.org/abs]{hep-ex/0011037}.

\bibitem[{\citenamefont{Benslama et~al.}(2002)}]{Benslama:2002pa}
\bibinfo{author}{\bibfnamefont{K.}~\bibnamefont{Benslama}} \bibnamefont{et~al.}
  (\bibinfo{collaboration}{CLEO}), \bibinfo{journal}{Phys. Rev.}
  \textbf{\bibinfo{volume}{D66}}, \bibinfo{pages}{077101}
  (\bibinfo{year}{2002}), \eprint[http://arXiv.org/abs]{hep-ex/0204019}.

\bibitem[{\citenamefont{Symanzik}(1983{\natexlab{a}})}]{Symanzik:1983dc}
\bibinfo{author}{\bibfnamefont{K.}~\bibnamefont{Symanzik}},
  \bibinfo{journal}{Nucl. Phys.} \textbf{\bibinfo{volume}{B226}},
  \bibinfo{pages}{187} (\bibinfo{year}{1983}{\natexlab{a}}).

\bibitem[{\citenamefont{Symanzik}(1983{\natexlab{b}})}]{Symanzik:1983gh}
\bibinfo{author}{\bibfnamefont{K.}~\bibnamefont{Symanzik}},
  \bibinfo{journal}{Nucl. Phys.} \textbf{\bibinfo{volume}{B226}},
  \bibinfo{pages}{205} (\bibinfo{year}{1983}{\natexlab{b}}).

\bibitem[{\citenamefont{Jegerlehner et~al.}(2000)\citenamefont{Jegerlehner,
  Kenway, Martinelli, Michael, P{\`e}ne, Petersson, Petronzio, Sachrajda, and
  Schilling}}]{Jegerlehner:2000xt}
\bibinfo{author}{\bibfnamefont{F.}~\bibnamefont{Jegerlehner}},
  \bibinfo{author}{\bibfnamefont{R.~D.} \bibnamefont{Kenway}},
  \bibinfo{author}{\bibfnamefont{G.}~\bibnamefont{Martinelli}},
  \bibinfo{author}{\bibfnamefont{C.}~\bibnamefont{Michael}},
  \bibinfo{author}{\bibfnamefont{O.}~\bibnamefont{P{\`e}ne}},
  \bibinfo{author}{\bibfnamefont{B.}~\bibnamefont{Petersson}},
  \bibinfo{author}{\bibfnamefont{R.}~\bibnamefont{Petronzio}},
  \bibinfo{author}{\bibfnamefont{C.~T.} \bibnamefont{Sachrajda}},
  \bibnamefont{and} \bibinfo{author}{\bibfnamefont{K.}~\bibnamefont{Schilling}}
  (\bibinfo{year}{2000}), \bibinfo{note}{{ECFA-99-200}, CERN-2000-002}.

\bibitem[{\citenamefont{Lepage}(1996)}]{Lepage:1996jw}
\bibinfo{author}{\bibfnamefont{G.~P.} \bibnamefont{Lepage}}, in
  \emph{\bibinfo{booktitle}{Perturbative and Nonperturbative Aspects of Quantum
  Field Theory : Proceedings of the 35. Internationale Universit{\"a}tswochen
  f{\"u}r Kern- und Teilchenphysik, Schladming, Austria, March 2-9, 1996}},
  edited by \bibinfo{editor}{\bibfnamefont{H.}~\bibnamefont{Latal}}
  \bibnamefont{and} \bibinfo{editor}{\bibfnamefont{W.}~\bibnamefont{Schweiger}}
  (\bibinfo{year}{1996}), p.~\bibinfo{pages}{1}.

\bibitem[{\citenamefont{Luo et~al.}(1999)\citenamefont{Luo, Guo, Kr{\"o}ger,
  and Sch{\"u}tte}}]{Luo:1998dx}
\bibinfo{author}{\bibfnamefont{X.-Q.} \bibnamefont{Luo}},
  \bibinfo{author}{\bibfnamefont{S.-H.} \bibnamefont{Guo}},
  \bibinfo{author}{\bibfnamefont{H.}~\bibnamefont{Kr{\"o}ger}},
  \bibnamefont{and}
  \bibinfo{author}{\bibfnamefont{D.}~\bibnamefont{Sch{\"u}tte}},
  \bibinfo{journal}{Phys. Rev.} \textbf{\bibinfo{volume}{D59}},
  \bibinfo{pages}{034503} (\bibinfo{year}{1999}),
  \eprint[http://arXiv.org/abs]{hep-lat/9804029}.

\bibitem[{\citenamefont{Wichmann}(2001)}]{WichmannPhD}
\bibinfo{author}{\bibfnamefont{A.}~\bibnamefont{Wichmann}}, Ph.D. thesis,
  \bibinfo{school}{ITKP Bonn} (\bibinfo{year}{2001}).

\bibitem[{\citenamefont{Chin et~al.}(1988)\citenamefont{Chin, Long, and
  Robson}}]{Chin:1988at}
\bibinfo{author}{\bibfnamefont{S.~A.} \bibnamefont{Chin}},
  \bibinfo{author}{\bibfnamefont{C.}~\bibnamefont{Long}}, \bibnamefont{and}
  \bibinfo{author}{\bibfnamefont{D.}~\bibnamefont{Robson}},
  \bibinfo{journal}{Phys. Rev.} \textbf{\bibinfo{volume}{D37}},
  \bibinfo{pages}{3001} (\bibinfo{year}{1988}).

\bibitem[{\citenamefont{Long et~al.}(1988)\citenamefont{Long, Robson, and
  Chin}}]{Long:1988qe}
\bibinfo{author}{\bibfnamefont{C.}~\bibnamefont{Long}},
  \bibinfo{author}{\bibfnamefont{D.}~\bibnamefont{Robson}}, \bibnamefont{and}
  \bibinfo{author}{\bibfnamefont{S.~A.} \bibnamefont{Chin}},
  \bibinfo{journal}{Phys. Rev.} \textbf{\bibinfo{volume}{D37}},
  \bibinfo{pages}{3006} (\bibinfo{year}{1988}).

\bibitem[{\citenamefont{Berg and Billoire}(1983)}]{Berg:1983kp}
\bibinfo{author}{\bibfnamefont{B.}~\bibnamefont{Berg}} \bibnamefont{and}
  \bibinfo{author}{\bibfnamefont{A.}~\bibnamefont{Billoire}},
  \bibinfo{journal}{Nucl. Phys.} \textbf{\bibinfo{volume}{B221}},
  \bibinfo{pages}{109} (\bibinfo{year}{1983}).

\bibitem[{\citenamefont{Teper}(1999)}]{Teper:1998te}
\bibinfo{author}{\bibfnamefont{M.~J.} \bibnamefont{Teper}},
  \bibinfo{journal}{Phys. Rev.} \textbf{\bibinfo{volume}{D59}},
  \bibinfo{pages}{014512} (\bibinfo{year}{1999}),
  \eprint[http://arXiv.org/abs]{hep-lat/9804008}.

\bibitem[{\citenamefont{Johnson and Teper}(1999)}]{Johnson:1998ev}
\bibinfo{author}{\bibfnamefont{R.}~\bibnamefont{Johnson}} \bibnamefont{and}
  \bibinfo{author}{\bibfnamefont{M.}~\bibnamefont{Teper}},
  \bibinfo{journal}{Nucl. Phys. Proc. Suppl.} \textbf{\bibinfo{volume}{73}},
  \bibinfo{pages}{267} (\bibinfo{year}{1999}),
  \eprint[http://arXiv.org/abs]{hep-lat/9808012}.

\bibitem[{\citenamefont{Johnson}(2002)}]{Johnson:2002qt}
\bibinfo{author}{\bibfnamefont{R.~W.} \bibnamefont{Johnson}},
  \bibinfo{journal}{Phys. Rev.} \textbf{\bibinfo{volume}{D66}},
  \bibinfo{pages}{074502} (\bibinfo{year}{2002}),
  \eprint[http://arXiv.org/abs]{hep-lat/0206005}.

\bibitem[{\citenamefont{Liu and Wu}(2002)}]{Liu:2001wq}
\bibinfo{author}{\bibfnamefont{D.~Q.} \bibnamefont{Liu}} \bibnamefont{and}
  \bibinfo{author}{\bibfnamefont{J.~M.} \bibnamefont{Wu}},
  \bibinfo{journal}{Mod. Phys. Lett.} \textbf{\bibinfo{volume}{A17}},
  \bibinfo{pages}{1419} (\bibinfo{year}{2002}),
  \eprint[http://arXiv.org/abs]{hep-lat/0105019}.

\bibitem[{\citenamefont{Caswell}(1974)}]{Caswell:1974gg}
\bibinfo{author}{\bibfnamefont{W.~E.} \bibnamefont{Caswell}},
  \bibinfo{journal}{Phys. Rev. Lett.} \textbf{\bibinfo{volume}{33}},
  \bibinfo{pages}{244} (\bibinfo{year}{1974}).

\bibitem[{\citenamefont{Jones}(1974)}]{Jones:1974mm}
\bibinfo{author}{\bibfnamefont{D.~R.~T.} \bibnamefont{Jones}},
  \bibinfo{journal}{Nucl. Phys.} \textbf{\bibinfo{volume}{B75}},
  \bibinfo{pages}{531} (\bibinfo{year}{1974}).

\bibitem[{\citenamefont{van Ritbergen et~al.}(1997)\citenamefont{van Ritbergen,
  Vermaseren, and Larin}}]{vanRitbergen:1997va}
\bibinfo{author}{\bibfnamefont{T.}~\bibnamefont{van Ritbergen}},
  \bibinfo{author}{\bibfnamefont{J.~A.~M.} \bibnamefont{Vermaseren}},
  \bibnamefont{and} \bibinfo{author}{\bibfnamefont{S.~A.} \bibnamefont{Larin}},
  \bibinfo{journal}{Phys. Lett.} \textbf{\bibinfo{volume}{B400}},
  \bibinfo{pages}{379} (\bibinfo{year}{1997}),
  \eprint[http://arXiv.org/abs]{hep-ph/9701390}.

\bibitem[{\citenamefont{Dashen and Gross}(1981)}]{Dashen:1981vm}
\bibinfo{author}{\bibfnamefont{R.~F.} \bibnamefont{Dashen}} \bibnamefont{and}
  \bibinfo{author}{\bibfnamefont{D.~J.} \bibnamefont{Gross}},
  \bibinfo{journal}{Phys. Rev.} \textbf{\bibinfo{volume}{D23}},
  \bibinfo{pages}{2340} (\bibinfo{year}{1981}).

\bibitem[{\citenamefont{Hasenfratz and Hasenfratz}(1981)}]{Hasenfratz:1981tw}
\bibinfo{author}{\bibfnamefont{A.}~\bibnamefont{Hasenfratz}} \bibnamefont{and}
  \bibinfo{author}{\bibfnamefont{P.}~\bibnamefont{Hasenfratz}},
  \bibinfo{journal}{Nucl. Phys.} \textbf{\bibinfo{volume}{B193}},
  \bibinfo{pages}{210} (\bibinfo{year}{1981}).

\bibitem[{\citenamefont{L{\"u}scher and Weisz}(1995)}]{Luscher:1995np}
\bibinfo{author}{\bibfnamefont{M.}~\bibnamefont{L{\"u}scher}} \bibnamefont{and}
  \bibinfo{author}{\bibfnamefont{P.}~\bibnamefont{Weisz}},
  \bibinfo{journal}{Nucl. Phys.} \textbf{\bibinfo{volume}{B452}},
  \bibinfo{pages}{234} (\bibinfo{year}{1995}),
  \eprint[http://arXiv.org/abs]{hep-lat/9505011}.

\bibitem[{\citenamefont{Hamer}(1996)}]{Hamer:1996ub}
\bibinfo{author}{\bibfnamefont{C.~J.} \bibnamefont{Hamer}},
  \bibinfo{journal}{Phys. Rev.} \textbf{\bibinfo{volume}{D53}},
  \bibinfo{pages}{7316} (\bibinfo{year}{1996}).

\bibitem[{\citenamefont{Teper}(1998)}]{Teper:1997am}
\bibinfo{author}{\bibfnamefont{M.~J.} \bibnamefont{Teper}}, in
  \emph{\bibinfo{booktitle}{Confinement, Duality, and Nonperturbative Aspects
  of QCD}}, edited by \bibinfo{editor}{\bibfnamefont{P.}~\bibnamefont{van
  Baal}} (\bibinfo{year}{1998}),
  \eprint[http://arXiv.org/abs]{hep-lat/9711011}.

\bibitem[{\citenamefont{Creutz}(1983)}]{Creutz:1984m}
\bibinfo{author}{\bibfnamefont{M.}~\bibnamefont{Creutz}},
  \emph{\bibinfo{title}{Quarks, Gluons and Lattices}}
  (\bibinfo{publisher}{Cambridge University Press}, \bibinfo{year}{1983}),
  \bibinfo{note}{(Cambridge Monographs On Mathematical Physics)}.

\bibitem[{\citenamefont{Kogut}(1983)}]{Kogut:1983ds}
\bibinfo{author}{\bibfnamefont{J.~B.} \bibnamefont{Kogut}},
  \bibinfo{journal}{Rev. Mod. Phys.} \textbf{\bibinfo{volume}{55}},
  \bibinfo{pages}{775} (\bibinfo{year}{1983}).

\bibitem[{\citenamefont{L{\"u}scher and Weisz}(1985)}]{Luscher:1985zq}
\bibinfo{author}{\bibfnamefont{M.}~\bibnamefont{L{\"u}scher}} \bibnamefont{and}
  \bibinfo{author}{\bibfnamefont{P.}~\bibnamefont{Weisz}},
  \bibinfo{journal}{Phys. Lett.} \textbf{\bibinfo{volume}{B158}},
  \bibinfo{pages}{250} (\bibinfo{year}{1985}).

\bibitem[{\citenamefont{Fang et~al.}(2000)\citenamefont{Fang, Guo, and
  Liu}}]{Fang:2000vm}
\bibinfo{author}{\bibfnamefont{X.-Y.} \bibnamefont{Fang}},
  \bibinfo{author}{\bibfnamefont{S.-H.} \bibnamefont{Guo}}, \bibnamefont{and}
  \bibinfo{author}{\bibfnamefont{J.-M.} \bibnamefont{Liu}},
  \bibinfo{journal}{Mod. Phys. Lett.} \textbf{\bibinfo{volume}{A15}},
  \bibinfo{pages}{737} (\bibinfo{year}{2000}).

\bibitem[{\citenamefont{Lepage and Mackenzie}(1993)}]{Lepage:1993xa}
\bibinfo{author}{\bibfnamefont{G.~P.} \bibnamefont{Lepage}} \bibnamefont{and}
  \bibinfo{author}{\bibfnamefont{P.~B.} \bibnamefont{Mackenzie}},
  \bibinfo{journal}{Phys. Rev.} \textbf{\bibinfo{volume}{D48}},
  \bibinfo{pages}{2250} (\bibinfo{year}{1993}),
  \eprint[http://arXiv.org/abs]{hep-lat/9209022}.

\bibitem[{\citenamefont{Alford et~al.}(1995{\natexlab{a}})\citenamefont{Alford,
  Dimm, Lepage, Hockney, and Mackenzie}}]{Alford:1995ui}
\bibinfo{author}{\bibfnamefont{M.~G.} \bibnamefont{Alford}},
  \bibinfo{author}{\bibfnamefont{W.}~\bibnamefont{Dimm}},
  \bibinfo{author}{\bibfnamefont{G.~P.} \bibnamefont{Lepage}},
  \bibinfo{author}{\bibfnamefont{G.}~\bibnamefont{Hockney}}, \bibnamefont{and}
  \bibinfo{author}{\bibfnamefont{P.~B.} \bibnamefont{Mackenzie}},
  \bibinfo{journal}{Nucl. Phys. Proc. Suppl.} \textbf{\bibinfo{volume}{42}},
  \bibinfo{pages}{787} (\bibinfo{year}{1995}{\natexlab{a}}),
  \eprint{hep-lat/9412035}.

\bibitem[{\citenamefont{Alford et~al.}(1995{\natexlab{b}})\citenamefont{Alford,
  Dimm, Lepage, Hockney, and Mackenzie}}]{Alford:1995hw}
\bibinfo{author}{\bibfnamefont{M.~G.} \bibnamefont{Alford}},
  \bibinfo{author}{\bibfnamefont{W.}~\bibnamefont{Dimm}},
  \bibinfo{author}{\bibfnamefont{G.~P.} \bibnamefont{Lepage}},
  \bibinfo{author}{\bibfnamefont{G.}~\bibnamefont{Hockney}}, \bibnamefont{and}
  \bibinfo{author}{\bibfnamefont{P.~B.} \bibnamefont{Mackenzie}},
  \bibinfo{journal}{Phys. Lett.} \textbf{\bibinfo{volume}{B361}},
  \bibinfo{pages}{87} (\bibinfo{year}{1995}{\natexlab{b}}),
  \eprint{hep-lat/9507010}.

\bibitem[{\citenamefont{Horn et~al.}(1985)\citenamefont{Horn, Karliner, and
  Weinstein}}]{Horn:1985ax}
\bibinfo{author}{\bibfnamefont{D.}~\bibnamefont{Horn}},
  \bibinfo{author}{\bibfnamefont{M.}~\bibnamefont{Karliner}}, \bibnamefont{and}
  \bibinfo{author}{\bibfnamefont{M.}~\bibnamefont{Weinstein}},
  \bibinfo{journal}{Phys. Rev.} \textbf{\bibinfo{volume}{D31}},
  \bibinfo{pages}{2589} (\bibinfo{year}{1985}).

\bibitem[{\citenamefont{Batrouni and Halpern}(1984)}]{Batrouni:1984rb}
\bibinfo{author}{\bibfnamefont{G.~G.} \bibnamefont{Batrouni}} \bibnamefont{and}
  \bibinfo{author}{\bibfnamefont{M.~B.} \bibnamefont{Halpern}},
  \bibinfo{journal}{Phys. Rev.} \textbf{\bibinfo{volume}{D30}},
  \bibinfo{pages}{1782} (\bibinfo{year}{1984}).

\bibitem[{\citenamefont{Maldacena}(1998)}]{Maldacena:1998re}
\bibinfo{author}{\bibfnamefont{J.~M.} \bibnamefont{Maldacena}},
  \bibinfo{journal}{Adv. Theor. Math. Phys.} \textbf{\bibinfo{volume}{2}},
  \bibinfo{pages}{231} (\bibinfo{year}{1998}),
  \eprint[http://arXiv.org/abs]{hep-th/9711200}.

\bibitem[{\citenamefont{Witten}(1998)}]{Witten:1998qj}
\bibinfo{author}{\bibfnamefont{E.}~\bibnamefont{Witten}},
  \bibinfo{journal}{Adv. Theor. Math. Phys.} \textbf{\bibinfo{volume}{2}},
  \bibinfo{pages}{253} (\bibinfo{year}{1998}),
  \eprint[http://arXiv.org/abs]{hep-th/9802150}.

\bibitem[{\citenamefont{Montvay and M{\"u}nster}(1994)}]{Montvay:1994cy}
\bibinfo{author}{\bibfnamefont{I.}~\bibnamefont{Montvay}} \bibnamefont{and}
  \bibinfo{author}{\bibfnamefont{G.}~\bibnamefont{M{\"u}nster}},
  \emph{\bibinfo{title}{Quantum Fields on a Lattice}}
  (\bibinfo{publisher}{Cambridge University Press}, \bibinfo{year}{1994}),
  chap. \bibinfo{chapter}{3.2.3}, \bibinfo{note}{(Cambridge Monographs On
  Mathematical Physics)}.

\bibitem[{\citenamefont{Eriksson et~al.}(1981)\citenamefont{Eriksson,
  Svartholm, and Skagerstam}}]{Eriksson:1981rq}
\bibinfo{author}{\bibfnamefont{K.~E.} \bibnamefont{Eriksson}},
  \bibinfo{author}{\bibfnamefont{N.}~\bibnamefont{Svartholm}},
  \bibnamefont{and} \bibinfo{author}{\bibfnamefont{B.~S.}
  \bibnamefont{Skagerstam}}, \bibinfo{journal}{J. Math. Phys.}
  \textbf{\bibinfo{volume}{22}}, \bibinfo{pages}{2276} (\bibinfo{year}{1981}).

\bibitem[{\citenamefont{Leonard}(2001)}]{ConradPhD}
\bibinfo{author}{\bibfnamefont{C.~R.} \bibnamefont{Leonard}}, Ph.D. thesis,
  \bibinfo{school}{The University of Melbourne} (\bibinfo{year}{2001}).

\bibitem[{\citenamefont{Gessel}(1990)}]{Gessel:1990}
\bibinfo{author}{\bibfnamefont{I.~M.} \bibnamefont{Gessel}},
  \bibinfo{journal}{J. Combin. Theory Ser. A} \textbf{\bibinfo{volume}{53}},
  \bibinfo{pages}{257} (\bibinfo{year}{1990}).

\bibitem[{\citenamefont{Stanley}(1999)}]{Stanley:1999}
\bibinfo{author}{\bibfnamefont{R.~P.} \bibnamefont{Stanley}},
  \emph{\bibinfo{title}{Enumerative Combinatorics}}
  (\bibinfo{publisher}{Cambridge University Press}, \bibinfo{year}{1999}),
  vol.~\bibinfo{volume}{2}, p. \bibinfo{pages}{453}.

\bibitem[{\citenamefont{Sagan}(2001)}]{Sagan:2001}
\bibinfo{author}{\bibfnamefont{B.~E.} \bibnamefont{Sagan}},
  \emph{\bibinfo{title}{The Symmetric Group}}
  (\bibinfo{publisher}{Springer-Verlag}, \bibinfo{year}{2001}), chap.
  \bibinfo{chapter}{2.11}, \bibinfo{edition}{2nd} ed.

\bibitem[{\citenamefont{Baik and Rains}(2001)}]{Baik:2001}
\bibinfo{author}{\bibfnamefont{J.}~\bibnamefont{Baik}} \bibnamefont{and}
  \bibinfo{author}{\bibfnamefont{E.~M.} \bibnamefont{Rains}},
  \bibinfo{journal}{Duke Math. J.} \textbf{\bibinfo{volume}{2}},
  \bibinfo{pages}{issue 1:1} (\bibinfo{year}{2001}),
  \eprint[http://arXiv.org/abs]{math.CO/9905083}.

\bibitem[{\citenamefont{Rains}(1998)}]{Rains:1998}
\bibinfo{author}{\bibfnamefont{E.~M.} \bibnamefont{Rains}},
  \bibinfo{journal}{Electron. J. Combin.} \textbf{\bibinfo{volume}{5(1)}},
  \bibinfo{pages}{R12} (\bibinfo{year}{1998}).

\bibitem[{\citenamefont{Widom et~al.}(2001)\citenamefont{Widom, Its, and
  Tracy}}]{Widom:2001}
\bibinfo{author}{\bibfnamefont{H.}~\bibnamefont{Widom}},
  \bibinfo{author}{\bibfnamefont{A.~R.} \bibnamefont{Its}}, \bibnamefont{and}
  \bibinfo{author}{\bibfnamefont{C.~A.} \bibnamefont{Tracy}},
  \bibinfo{journal}{Physica} \textbf{\bibinfo{volume}{D152-153}},
  \bibinfo{pages}{199} (\bibinfo{year}{2001}).

\bibitem[{\citenamefont{Regev}(1981)}]{Regev:1981}
\bibinfo{author}{\bibfnamefont{A.}~\bibnamefont{Regev}}, \bibinfo{journal}{Adv.
  in Math.} \textbf{\bibinfo{volume}{41}}, \bibinfo{pages}{115}
  (\bibinfo{year}{1981}).

\bibitem[{\citenamefont{Diaconis and Shahshahani}(1994)}]{Diaconis:1994}
\bibinfo{author}{\bibfnamefont{P.}~\bibnamefont{Diaconis}} \bibnamefont{and}
  \bibinfo{author}{\bibfnamefont{M.}~\bibnamefont{Shahshahani}},
  \bibinfo{journal}{J. Appl. Probab.} \textbf{\bibinfo{volume}{31A}},
  \bibinfo{pages}{49} (\bibinfo{year}{1994}).

\bibitem[{\citenamefont{Widom and Tracy}(1999)}]{Widom:1999}
\bibinfo{author}{\bibfnamefont{H.}~\bibnamefont{Widom}} \bibnamefont{and}
  \bibinfo{author}{\bibfnamefont{C.~A.} \bibnamefont{Tracy}},
  \bibinfo{journal}{Comm. Math. Phys.} \textbf{\bibinfo{volume}{207}},
  \bibinfo{pages}{665} (\bibinfo{year}{1999}).

\bibitem[{\citenamefont{Creutz}(1978)}]{Creutz:1978ub}
\bibinfo{author}{\bibfnamefont{M.}~\bibnamefont{Creutz}}, \bibinfo{journal}{J.
  Math. Phys.} \textbf{\bibinfo{volume}{19}}, \bibinfo{pages}{2043}
  (\bibinfo{year}{1978}).

\bibitem[{\citenamefont{De~Pietri}(1997)}]{DePietri:1997pj}
\bibinfo{author}{\bibfnamefont{R.}~\bibnamefont{De~Pietri}},
  \bibinfo{journal}{Class. Quant. Grav.} \textbf{\bibinfo{volume}{14}},
  \bibinfo{pages}{53} (\bibinfo{year}{1997}),
  \eprint[http://arXiv.org/abs]{gr-qc/9605064}.

\bibitem[{\citenamefont{Rovelli and Smolin}(1995)}]{Rovelli:1995ac}
\bibinfo{author}{\bibfnamefont{C.}~\bibnamefont{Rovelli}} \bibnamefont{and}
  \bibinfo{author}{\bibfnamefont{L.}~\bibnamefont{Smolin}},
  \bibinfo{journal}{Phys. Rev.} \textbf{\bibinfo{volume}{D52}},
  \bibinfo{pages}{5743} (\bibinfo{year}{1995}),
  \eprint[http://arXiv.org/abs]{gr-qc/9505006}.

\bibitem[{\citenamefont{Ezawa}(1997)}]{Ezawa:1997bv}
\bibinfo{author}{\bibfnamefont{K.}~\bibnamefont{Ezawa}},
  \bibinfo{journal}{Phys. Rept.} \textbf{\bibinfo{volume}{286}},
  \bibinfo{pages}{271} (\bibinfo{year}{1997}),
  \eprint[http://arXiv.org/abs]{gr-qc/9601050}.

\bibitem[{\citenamefont{Kogut et~al.}(1982)\citenamefont{Kogut, Snow, and
  Stone}}]{Kogut:1982ez}
\bibinfo{author}{\bibfnamefont{J.~B.} \bibnamefont{Kogut}},
  \bibinfo{author}{\bibfnamefont{M.}~\bibnamefont{Snow}}, \bibnamefont{and}
  \bibinfo{author}{\bibfnamefont{M.}~\bibnamefont{Stone}},
  \bibinfo{journal}{Nucl. Phys.} \textbf{\bibinfo{volume}{B200}},
  \bibinfo{pages}{211} (\bibinfo{year}{1982}).

\bibitem[{\citenamefont{Brower et~al.}(1981)\citenamefont{Brower, Rossi, and
  Tan}}]{Brower:1981vt}
\bibinfo{author}{\bibfnamefont{R.}~\bibnamefont{Brower}},
  \bibinfo{author}{\bibfnamefont{P.}~\bibnamefont{Rossi}}, \bibnamefont{and}
  \bibinfo{author}{\bibfnamefont{C.-I.} \bibnamefont{Tan}},
  \bibinfo{journal}{Nucl. Phys.} \textbf{\bibinfo{volume}{B190}},
  \bibinfo{pages}{699} (\bibinfo{year}{1981}).

\bibitem[{\citenamefont{Weyl}(1946)}]{Weyl:1946}
\bibinfo{author}{\bibfnamefont{H.}~\bibnamefont{Weyl}},
  \emph{\bibinfo{title}{The Classical Groups, Their Invariants and
  Representations}} (\bibinfo{publisher}{Princeton University Press},
  \bibinfo{year}{1946}).

\bibitem[{\citenamefont{Gradshteyn and Ryzhik}(1994)}]{Gradshteyn:1994}
\bibinfo{author}{\bibfnamefont{I.~S.} \bibnamefont{Gradshteyn}}
  \bibnamefont{and} \bibinfo{author}{\bibfnamefont{I.~M.}
  \bibnamefont{Ryzhik}}, \emph{\bibinfo{title}{Table of Integrals, Series, and
  Products}} (\bibinfo{publisher}{Academic Press}, \bibinfo{year}{1994}).

\bibitem[{\citenamefont{van Moerbeke}(2001)}]{vanMoerbeke:2001}
\bibinfo{author}{\bibfnamefont{P.}~\bibnamefont{van Moerbeke}}, in
  \emph{\bibinfo{booktitle}{Random Matrices and Their Applications}}
  (\bibinfo{publisher}{Cambridge University Press}, \bibinfo{year}{2001}),
  \bibinfo{note}{{MSRI-publication} {\#}40},
  \eprint[http://arXiv.org/abs]{math.CO/0010135}.

\bibitem[{\citenamefont{Sloane}()}]{Sloane:OE}
\bibinfo{author}{\bibfnamefont{N.~J.~A.} \bibnamefont{Sloane}},
  \emph{\bibinfo{title}{{T}he {O}n-line {E}ncyclopedia of {I}nteger
  {S}equences}},
  \bibinfo{note}{\texttt{http://www.research.att.com/\~{}njas/sequences/}}.

\bibitem[{\citenamefont{Tonkin}(1987)}]{Tonkin:1987nh}
\bibinfo{author}{\bibfnamefont{S.~P.} \bibnamefont{Tonkin}},
  \bibinfo{journal}{Nucl. Phys.} \textbf{\bibinfo{volume}{B292}},
  \bibinfo{pages}{573} (\bibinfo{year}{1987}).

\bibitem[{\citenamefont{Bars}(1980)}]{Bars:1980yy}
\bibinfo{author}{\bibfnamefont{I.}~\bibnamefont{Bars}}, \bibinfo{journal}{J.
  Math. Phys.} \textbf{\bibinfo{volume}{21}}, \bibinfo{pages}{2678}
  (\bibinfo{year}{1980}).

\bibitem[{\citenamefont{Greensite}(1987)}]{Greensite:1987rg}
\bibinfo{author}{\bibfnamefont{J.}~\bibnamefont{Greensite}},
  \bibinfo{journal}{Phys. Lett.} \textbf{\bibinfo{volume}{B191}},
  \bibinfo{pages}{431} (\bibinfo{year}{1987}).

\bibitem[{\citenamefont{Greensite}(1979)}]{Greensite:1979yn}
\bibinfo{author}{\bibfnamefont{J.~P.} \bibnamefont{Greensite}},
  \bibinfo{journal}{Nucl. Phys.} \textbf{\bibinfo{volume}{B158}},
  \bibinfo{pages}{469} (\bibinfo{year}{1979}).

\bibitem[{\citenamefont{Hamer et~al.}(1992)\citenamefont{Hamer, Oitmaa, and
  Zheng}}]{Hamer:1992ic}
\bibinfo{author}{\bibfnamefont{C.~J.} \bibnamefont{Hamer}},
  \bibinfo{author}{\bibfnamefont{J.}~\bibnamefont{Oitmaa}}, \bibnamefont{and}
  \bibinfo{author}{\bibfnamefont{W.-H.} \bibnamefont{Zheng}},
  \bibinfo{journal}{Phys. Rev.} \textbf{\bibinfo{volume}{D45}},
  \bibinfo{pages}{4652} (\bibinfo{year}{1992}).

\bibitem[{\citenamefont{Llewellyn~Smith and
  Watson}(1993)}]{LlewellynSmith:1993ig}
\bibinfo{author}{\bibfnamefont{C.~H.} \bibnamefont{Llewellyn~Smith}}
  \bibnamefont{and} \bibinfo{author}{\bibfnamefont{N.~J.}
  \bibnamefont{Watson}}, \bibinfo{journal}{Phys. Lett.}
  \textbf{\bibinfo{volume}{B302}}, \bibinfo{pages}{463} (\bibinfo{year}{1993}),
  \eprint{hep-lat/9212025}.

\bibitem[{\citenamefont{Chen et~al.}(1994)\citenamefont{Chen, Guo, Fang, and
  Zheng}}]{Chen:1994ri}
\bibinfo{author}{\bibfnamefont{Q.-Z.} \bibnamefont{Chen}},
  \bibinfo{author}{\bibfnamefont{S.-H.} \bibnamefont{Guo}},
  \bibinfo{author}{\bibfnamefont{X.-Y.} \bibnamefont{Fang}}, \bibnamefont{and}
  \bibinfo{author}{\bibfnamefont{W.-H.} \bibnamefont{Zheng}},
  \bibinfo{journal}{Phys. Rev.} \textbf{\bibinfo{volume}{D50}},
  \bibinfo{pages}{3564} (\bibinfo{year}{1994}).

\bibitem[{\citenamefont{Luo and Chen}(1996)}]{Luo:1996ha}
\bibinfo{author}{\bibfnamefont{X.-Q.} \bibnamefont{Luo}} \bibnamefont{and}
  \bibinfo{author}{\bibfnamefont{Q.-Z.} \bibnamefont{Chen}},
  \bibinfo{journal}{Mod. Phys. Lett.} \textbf{\bibinfo{volume}{A11}},
  \bibinfo{pages}{2435} (\bibinfo{year}{1996}), \eprint{hep-ph/9604395}.

\bibitem[{\citenamefont{Samuel}(1997)}]{Samuel:1997bt}
\bibinfo{author}{\bibfnamefont{S.}~\bibnamefont{Samuel}},
  \bibinfo{journal}{Phys. Rev.} \textbf{\bibinfo{volume}{D55}},
  \bibinfo{pages}{4189} (\bibinfo{year}{1997}), \eprint{hep-ph/9604405}.

\bibitem[{\citenamefont{Li et~al.}(2000)\citenamefont{Li, Guo, and
  Luo}}]{Li:2000bg}
\bibinfo{author}{\bibfnamefont{J.-M.} \bibnamefont{Li}},
  \bibinfo{author}{\bibfnamefont{S.}~\bibnamefont{Guo}}, \bibnamefont{and}
  \bibinfo{author}{\bibfnamefont{X.-Q.} \bibnamefont{Luo}},
  \bibinfo{journal}{Commun. Theor. Phys.} \textbf{\bibinfo{volume}{34}},
  \bibinfo{pages}{301} (\bibinfo{year}{2000}).

\bibitem[{\citenamefont{Lucini and Teper}(2002)}]{Lucini:2002wg}
\bibinfo{author}{\bibfnamefont{B.}~\bibnamefont{Lucini}} \bibnamefont{and}
  \bibinfo{author}{\bibfnamefont{M.}~\bibnamefont{Teper}}
  (\bibinfo{year}{2002}), \eprint[http://arXiv.org/abs]{hep-lat/0206027}.

\bibitem[{\citenamefont{'t~Hooft}(1974)}]{'tHooft:1974jz}
\bibinfo{author}{\bibfnamefont{G.}~\bibnamefont{'t~Hooft}},
  \bibinfo{journal}{Nucl. Phys.} \textbf{\bibinfo{volume}{B72}},
  \bibinfo{pages}{461} (\bibinfo{year}{1974}).

\bibitem[{\citenamefont{Witten}(1979)}]{Witten:1979kh}
\bibinfo{author}{\bibfnamefont{E.}~\bibnamefont{Witten}},
  \bibinfo{journal}{Nucl. Phys.} \textbf{\bibinfo{volume}{B160}},
  \bibinfo{pages}{57} (\bibinfo{year}{1979}).

\bibitem[{\citenamefont{Eguchi and Kawai}(1982)}]{Eguchi:1982nm}
\bibinfo{author}{\bibfnamefont{T.}~\bibnamefont{Eguchi}} \bibnamefont{and}
  \bibinfo{author}{\bibfnamefont{H.}~\bibnamefont{Kawai}},
  \bibinfo{journal}{Phys. Rev. Lett.} \textbf{\bibinfo{volume}{48}},
  \bibinfo{pages}{1063} (\bibinfo{year}{1982}).

\bibitem[{\citenamefont{Csaki et~al.}(1999{\natexlab{a}})\citenamefont{Csaki,
  Ooguri, Oz, and Terning}}]{Csaki:1998qr}
\bibinfo{author}{\bibfnamefont{C.}~\bibnamefont{Csaki}},
  \bibinfo{author}{\bibfnamefont{H.}~\bibnamefont{Ooguri}},
  \bibinfo{author}{\bibfnamefont{Y.}~\bibnamefont{Oz}}, \bibnamefont{and}
  \bibinfo{author}{\bibfnamefont{J.}~\bibnamefont{Terning}},
  \bibinfo{journal}{JHEP} \textbf{\bibinfo{volume}{01}}, \bibinfo{pages}{017}
  (\bibinfo{year}{1999}{\natexlab{a}}),
  \eprint[http://arXiv.org/abs]{hep-th/9806021}.

\bibitem[{\citenamefont{de~Mello~Koch et~al.}(1998)\citenamefont{de~Mello~Koch,
  Jevicki, Mihailescu, and Nunes}}]{deMelloKoch:1998qs}
\bibinfo{author}{\bibfnamefont{R.}~\bibnamefont{de~Mello~Koch}},
  \bibinfo{author}{\bibfnamefont{A.}~\bibnamefont{Jevicki}},
  \bibinfo{author}{\bibfnamefont{M.}~\bibnamefont{Mihailescu}},
  \bibnamefont{and} \bibinfo{author}{\bibfnamefont{J.~P.} \bibnamefont{Nunes}},
  \bibinfo{journal}{Phys. Rev.} \textbf{\bibinfo{volume}{D58}},
  \bibinfo{pages}{105009} (\bibinfo{year}{1998}),
  \eprint[http://arXiv.org/abs]{hep-th/9806125}.

\bibitem[{\citenamefont{Zyskin}(1998)}]{Zyskin:1998tg}
\bibinfo{author}{\bibfnamefont{M.}~\bibnamefont{Zyskin}},
  \bibinfo{journal}{Phys. Lett.} \textbf{\bibinfo{volume}{B439}},
  \bibinfo{pages}{373} (\bibinfo{year}{1998}),
  \eprint[http://arXiv.org/abs]{hep-th/9806128}.

\bibitem[{\citenamefont{Brower et~al.}(2000{\natexlab{a}})\citenamefont{Brower,
  Mathur, and Tan}}]{Brower:1999nj}
\bibinfo{author}{\bibfnamefont{R.~C.} \bibnamefont{Brower}},
  \bibinfo{author}{\bibfnamefont{S.~D.} \bibnamefont{Mathur}},
  \bibnamefont{and} \bibinfo{author}{\bibfnamefont{C.-I.} \bibnamefont{Tan}},
  \bibinfo{journal}{Nucl. Phys.} \textbf{\bibinfo{volume}{B574}},
  \bibinfo{pages}{219} (\bibinfo{year}{2000}{\natexlab{a}}),
  \eprint[http://arXiv.org/abs]{hep-th/9908196}.

\bibitem[{\citenamefont{Brower et~al.}(2000{\natexlab{b}})\citenamefont{Brower,
  Mathur, and Tan}}]{Brower:2000rp}
\bibinfo{author}{\bibfnamefont{R.~C.} \bibnamefont{Brower}},
  \bibinfo{author}{\bibfnamefont{S.~D.} \bibnamefont{Mathur}},
  \bibnamefont{and} \bibinfo{author}{\bibfnamefont{C.-I.} \bibnamefont{Tan}},
  \bibinfo{journal}{Nucl. Phys.} \textbf{\bibinfo{volume}{B587}},
  \bibinfo{pages}{249} (\bibinfo{year}{2000}{\natexlab{b}}),
  \eprint[http://arXiv.org/abs]{hep-th/0003115}.

\bibitem[{\citenamefont{Russo}(1999)}]{Russo:1998mm}
\bibinfo{author}{\bibfnamefont{J.~G.} \bibnamefont{Russo}},
  \bibinfo{journal}{Nucl. Phys.} \textbf{\bibinfo{volume}{B543}},
  \bibinfo{pages}{183} (\bibinfo{year}{1999}),
  \eprint[http://arXiv.org/abs]{hep-th/9808117}.

\bibitem[{\citenamefont{Csaki et~al.}(1999{\natexlab{b}})\citenamefont{Csaki,
  Russo, Sfetsos, and Terning}}]{Csaki:1999vb}
\bibinfo{author}{\bibfnamefont{C.}~\bibnamefont{Csaki}},
  \bibinfo{author}{\bibfnamefont{J.}~\bibnamefont{Russo}},
  \bibinfo{author}{\bibfnamefont{K.}~\bibnamefont{Sfetsos}}, \bibnamefont{and}
  \bibinfo{author}{\bibfnamefont{J.}~\bibnamefont{Terning}},
  \bibinfo{journal}{Phys. Rev.} \textbf{\bibinfo{volume}{D60}},
  \bibinfo{pages}{044001} (\bibinfo{year}{1999}{\natexlab{b}}),
  \eprint[http://arXiv.org/abs]{hep-th/9902067}.

\bibitem[{\citenamefont{Dalley and van~de Sande}(2001)}]{Dalley:2000ye}
\bibinfo{author}{\bibfnamefont{S.}~\bibnamefont{Dalley}} \bibnamefont{and}
  \bibinfo{author}{\bibfnamefont{B.}~\bibnamefont{van~de Sande}},
  \bibinfo{journal}{Phys. Rev.} \textbf{\bibinfo{volume}{D63}},
  \bibinfo{pages}{076004} (\bibinfo{year}{2001}),
  \eprint[http://arXiv.org/abs]{hep-lat/0010082}.

\bibitem[{\citenamefont{Karabali and Nair}(1996)}]{Karabali:1996je}
\bibinfo{author}{\bibfnamefont{D.}~\bibnamefont{Karabali}} \bibnamefont{and}
  \bibinfo{author}{\bibfnamefont{V.~P.} \bibnamefont{Nair}},
  \bibinfo{journal}{Phys. Lett.} \textbf{\bibinfo{volume}{B379}},
  \bibinfo{pages}{141} (\bibinfo{year}{1996}),
  \eprint[http://arXiv.org/abs]{hep-th/9602155}.

\bibitem[{\citenamefont{Karabali
  et~al.}(1998{\natexlab{a}})\citenamefont{Karabali, Kim, and
  Nair}}]{Karabali:1998wk}
\bibinfo{author}{\bibfnamefont{D.}~\bibnamefont{Karabali}},
  \bibinfo{author}{\bibfnamefont{C.-J.} \bibnamefont{Kim}}, \bibnamefont{and}
  \bibinfo{author}{\bibfnamefont{V.~P.} \bibnamefont{Nair}},
  \bibinfo{journal}{Nucl. Phys.} \textbf{\bibinfo{volume}{B524}},
  \bibinfo{pages}{661} (\bibinfo{year}{1998}{\natexlab{a}}),
  \eprint[http://arXiv.org/abs]{hep-th/9705087}.

\bibitem[{\citenamefont{Karabali
  et~al.}(1998{\natexlab{b}})\citenamefont{Karabali, Kim, and
  Nair}}]{Karabali:1998yq}
\bibinfo{author}{\bibfnamefont{D.}~\bibnamefont{Karabali}},
  \bibinfo{author}{\bibfnamefont{C.-J.} \bibnamefont{Kim}}, \bibnamefont{and}
  \bibinfo{author}{\bibfnamefont{V.~P.} \bibnamefont{Nair}},
  \bibinfo{journal}{Phys. Lett.} \textbf{\bibinfo{volume}{B434}},
  \bibinfo{pages}{103} (\bibinfo{year}{1998}{\natexlab{b}}),
  \eprint[http://arXiv.org/abs]{hep-th/9804132}.

\bibitem[{\citenamefont{Nair}(2002)}]{Nair:2002uj}
\bibinfo{author}{\bibfnamefont{V.~P.} \bibnamefont{Nair}},
  \bibinfo{journal}{Nucl. Phys. Proc. Suppl.} \textbf{\bibinfo{volume}{108}},
  \bibinfo{pages}{194} (\bibinfo{year}{2002}).

\bibitem[{\citenamefont{Krattenhaler}(1999)}]{Krattenhaler}
\bibinfo{author}{\bibfnamefont{C.}~\bibnamefont{Krattenhaler}},
  \bibinfo{journal}{S\'eminaire Lotharingien Combin.}
  \textbf{\bibinfo{volume}{42}}, \bibinfo{pages}{B42q} (\bibinfo{year}{1999}),
  \bibinfo{note}{{T}he {A}ndrews {F}estschrift}.

\bibitem[{\citenamefont{Fang et~al.}(2002)\citenamefont{Fang, Hui, Chen, and
  Sch{\"u}tte}}]{Fang:2002ps}
\bibinfo{author}{\bibfnamefont{X.-Y.} \bibnamefont{Fang}},
  \bibinfo{author}{\bibfnamefont{P.}~\bibnamefont{Hui}},
  \bibinfo{author}{\bibfnamefont{Q.-Z.} \bibnamefont{Chen}}, \bibnamefont{and}
  \bibinfo{author}{\bibfnamefont{D.}~\bibnamefont{Sch{\"u}tte}},
  \bibinfo{journal}{Phys. Rev.} \textbf{\bibinfo{volume}{D65}},
  \bibinfo{pages}{114505} (\bibinfo{year}{2002}).

\bibitem[{\citenamefont{Chen et~al.}(1995)\citenamefont{Chen, Luo, Guo, and
  Fang}}]{Chen:1995ca}
\bibinfo{author}{\bibfnamefont{Q.-Z.} \bibnamefont{Chen}},
  \bibinfo{author}{\bibfnamefont{X.-Q.} \bibnamefont{Luo}},
  \bibinfo{author}{\bibfnamefont{S.-H.} \bibnamefont{Guo}}, \bibnamefont{and}
  \bibinfo{author}{\bibfnamefont{X.-Y.} \bibnamefont{Fang}},
  \bibinfo{journal}{Phys. Lett.} \textbf{\bibinfo{volume}{B348}},
  \bibinfo{pages}{560} (\bibinfo{year}{1995}),
  \eprint[http://arXiv.org/abs]{hep-ph/9502235}.

\bibitem[{\citenamefont{Batrouni}(1982{\natexlab{a}})}]{Batrouni:1982bg}
\bibinfo{author}{\bibfnamefont{G.~G.} \bibnamefont{Batrouni}},
  \bibinfo{journal}{Nucl. Phys.} \textbf{\bibinfo{volume}{B208}},
  \bibinfo{pages}{467} (\bibinfo{year}{1982}{\natexlab{a}}).

\bibitem[{\citenamefont{Batrouni}(1983)}]{Batrouni:1983ch}
\bibinfo{author}{\bibfnamefont{G.~G.} \bibnamefont{Batrouni}}, Ph.D. thesis,
  \bibinfo{school}{Lawrence Berkeley Laboratory, University of California}
  (\bibinfo{year}{1983}), \bibinfo{note}{lBL-15993}.

\bibitem[{\citenamefont{Kogut et~al.}(1979)\citenamefont{Kogut, Pearson, and
  Shigemitsu}}]{Kogut:1979vg}
\bibinfo{author}{\bibfnamefont{J.~B.} \bibnamefont{Kogut}},
  \bibinfo{author}{\bibfnamefont{R.~B.} \bibnamefont{Pearson}},
  \bibnamefont{and}
  \bibinfo{author}{\bibfnamefont{J.}~\bibnamefont{Shigemitsu}},
  \bibinfo{journal}{Phys. Rev. Lett.} \textbf{\bibinfo{volume}{43}},
  \bibinfo{pages}{484} (\bibinfo{year}{1979}).

\bibitem[{\citenamefont{Hamer}(1989)}]{Hamer:1989qm}
\bibinfo{author}{\bibfnamefont{C.~J.} \bibnamefont{Hamer}},
  \bibinfo{journal}{Phys. Lett.} \textbf{\bibinfo{volume}{B224}},
  \bibinfo{pages}{339} (\bibinfo{year}{1989}).

\bibitem[{\citenamefont{Horn and Weinstein}(1984)}]{Horn:1984bq}
\bibinfo{author}{\bibfnamefont{D.}~\bibnamefont{Horn}} \bibnamefont{and}
  \bibinfo{author}{\bibfnamefont{M.}~\bibnamefont{Weinstein}},
  \bibinfo{journal}{Phys. Rev.} \textbf{\bibinfo{volume}{D30}},
  \bibinfo{pages}{1256} (\bibinfo{year}{1984}).

\bibitem[{\citenamefont{{v}an~{d}en Doel and Horn}(1986)}]{VanDenDoel:1986bw}
\bibinfo{author}{\bibfnamefont{C.~P.} \bibnamefont{{v}an~{d}en Doel}}
  \bibnamefont{and} \bibinfo{author}{\bibfnamefont{D.}~\bibnamefont{Horn}},
  \bibinfo{journal}{Phys. Rev.} \textbf{\bibinfo{volume}{D33}},
  \bibinfo{pages}{3011} (\bibinfo{year}{1986}).

\bibitem[{\citenamefont{{v}an~{d}en Doel and Horn}(1987)}]{vandenDoel:1987xk}
\bibinfo{author}{\bibfnamefont{C.~P.} \bibnamefont{{v}an~{d}en Doel}}
  \bibnamefont{and} \bibinfo{author}{\bibfnamefont{D.}~\bibnamefont{Horn}},
  \bibinfo{journal}{Phys. Rev.} \textbf{\bibinfo{volume}{D35}},
  \bibinfo{pages}{2824} (\bibinfo{year}{1987}).

\bibitem[{\citenamefont{Horn and Lana}(1991)}]{Horn:1991fz}
\bibinfo{author}{\bibfnamefont{D.}~\bibnamefont{Horn}} \bibnamefont{and}
  \bibinfo{author}{\bibfnamefont{G.}~\bibnamefont{Lana}},
  \bibinfo{journal}{Phys. Rev.} \textbf{\bibinfo{volume}{D44}},
  \bibinfo{pages}{2864} (\bibinfo{year}{1991}).

\bibitem[{\citenamefont{Horn and Schreiber}(1993)}]{Horn:1993wb}
\bibinfo{author}{\bibfnamefont{D.}~\bibnamefont{Horn}} \bibnamefont{and}
  \bibinfo{author}{\bibfnamefont{D.}~\bibnamefont{Schreiber}},
  \bibinfo{journal}{Phys. Rev.} \textbf{\bibinfo{volume}{D47}},
  \bibinfo{pages}{2081} (\bibinfo{year}{1993}).

\bibitem[{\citenamefont{Bishop and K{\"u}mmel}(1987)}]{Bishop:1987}
\bibinfo{author}{\bibfnamefont{R.~F.} \bibnamefont{Bishop}} \bibnamefont{and}
  \bibinfo{author}{\bibfnamefont{H.~G.} \bibnamefont{K{\"u}mmel}},
  \bibinfo{journal}{Physics Today} \textbf{\bibinfo{volume}{40(3)}},
  \bibinfo{pages}{52} (\bibinfo{year}{1987}).

\bibitem[{\citenamefont{Sch{\"u}tte et~al.}(1997)\citenamefont{Sch{\"u}tte,
  Zheng, and Hamer}}]{Schutte:1997du}
\bibinfo{author}{\bibfnamefont{D.}~\bibnamefont{Sch{\"u}tte}},
  \bibinfo{author}{\bibfnamefont{W.-H.} \bibnamefont{Zheng}}, \bibnamefont{and}
  \bibinfo{author}{\bibfnamefont{C.~J.} \bibnamefont{Hamer}},
  \bibinfo{journal}{Phys. Rev.} \textbf{\bibinfo{volume}{D55}},
  \bibinfo{pages}{2974} (\bibinfo{year}{1997}),
  \eprint[http://arXiv.org/abs]{hep-lat/9603026}.

\bibitem[{\citenamefont{Guo et~al.}(1994)\citenamefont{Guo, Chen, and
  Li}}]{Guo:1994vq}
\bibinfo{author}{\bibfnamefont{S.-H.} \bibnamefont{Guo}},
  \bibinfo{author}{\bibfnamefont{Q.-Z.} \bibnamefont{Chen}}, \bibnamefont{and}
  \bibinfo{author}{\bibfnamefont{L.}~\bibnamefont{Li}}, \bibinfo{journal}{Phys.
  Rev.} \textbf{\bibinfo{volume}{D49}}, \bibinfo{pages}{507}
  (\bibinfo{year}{1994}).

\bibitem[{\citenamefont{Hu et~al.}(1997)\citenamefont{Hu, Luo, Chen, Fang, and
  Guo}}]{Hu:1997ys}
\bibinfo{author}{\bibfnamefont{L.}~\bibnamefont{Hu}},
  \bibinfo{author}{\bibfnamefont{X.-Q.} \bibnamefont{Luo}},
  \bibinfo{author}{\bibfnamefont{Q.-Z.} \bibnamefont{Chen}},
  \bibinfo{author}{\bibfnamefont{X.-Y.} \bibnamefont{Fang}}, \bibnamefont{and}
  \bibinfo{author}{\bibfnamefont{S.-H.} \bibnamefont{Guo}},
  \bibinfo{journal}{Commun. Theor. Phys.} \textbf{\bibinfo{volume}{28}},
  \bibinfo{pages}{327} (\bibinfo{year}{1997}),
  \eprint[http://arXiv.org/abs]{hep-ph/9609435}.

\bibitem[{\citenamefont{Chin et~al.}(1986)\citenamefont{Chin, Long, and
  Robson}}]{Chin:1986fe}
\bibinfo{author}{\bibfnamefont{S.~A.} \bibnamefont{Chin}},
  \bibinfo{author}{\bibfnamefont{C.}~\bibnamefont{Long}}, \bibnamefont{and}
  \bibinfo{author}{\bibfnamefont{D.}~\bibnamefont{Robson}},
  \bibinfo{journal}{Phys. Rev. Lett.} \textbf{\bibinfo{volume}{57}},
  \bibinfo{pages}{2779} (\bibinfo{year}{1986}).

\bibitem[{\citenamefont{Halpern}(1979)}]{Halpern:1979ik}
\bibinfo{author}{\bibfnamefont{M.~B.} \bibnamefont{Halpern}},
  \bibinfo{journal}{Phys. Rev.} \textbf{\bibinfo{volume}{D19}},
  \bibinfo{pages}{517} (\bibinfo{year}{1979}).

\bibitem[{\citenamefont{Batrouni}(1982{\natexlab{b}})}]{Batrouni:1982dx}
\bibinfo{author}{\bibfnamefont{G.~G.} \bibnamefont{Batrouni}},
  \bibinfo{journal}{Nucl. Phys.} \textbf{\bibinfo{volume}{B208}},
  \bibinfo{pages}{12} (\bibinfo{year}{1982}{\natexlab{b}}).

\bibitem[{\citenamefont{Fadeev and Slavnov}(1980)}]{Fadeev&Slavnov}
\bibinfo{author}{\bibfnamefont{L.~D.} \bibnamefont{Fadeev}} \bibnamefont{and}
  \bibinfo{author}{\bibfnamefont{A.~A.} \bibnamefont{Slavnov}},
  \emph{\bibinfo{title}{Gauge Fields: Introduction to Quantum Theory}}
  (\bibinfo{publisher}{The Benjamin/Cummings Publishing Company, Inc.},
  \bibinfo{year}{1980}), chap. \bibinfo{chapter}{3.2}.

\bibitem[{\citenamefont{Jackiw}(1997)}]{Jackiw1993}
\bibinfo{author}{\bibfnamefont{R.}~\bibnamefont{Jackiw}}, in
  \emph{\bibinfo{booktitle}{Lectures on QCD}}, edited by
  \bibinfo{editor}{\bibfnamefont{F.}~\bibnamefont{Lenz}},
  \bibinfo{editor}{\bibfnamefont{H.}~\bibnamefont{Griesshammer}},
  \bibnamefont{and} \bibinfo{editor}{\bibfnamefont{D.}~\bibnamefont{Stoll}}
  (\bibinfo{publisher}{Springer-Varlag}, \bibinfo{year}{1997}), pp.
  \bibinfo{pages}{90--127}.

\bibitem[{\citenamefont{Kogut and Shigemitsu}(1980)}]{Kogut:1980pm}
\bibinfo{author}{\bibfnamefont{J.~B.} \bibnamefont{Kogut}} \bibnamefont{and}
  \bibinfo{author}{\bibfnamefont{J.}~\bibnamefont{Shigemitsu}},
  \bibinfo{journal}{Phys. Rev. Lett.} \textbf{\bibinfo{volume}{45}},
  \bibinfo{pages}{410} (\bibinfo{year}{1980}), \bibinfo{note}{erratum-ibid.{\bf
  45}:1217,1980}.

\end{thebibliography}

\appendix
\chapter{Commutation Relations}
\label{commutationrelations}
In improved Hamiltonian LGT calculations, one encounters matrix elements of the form: 
\bea
\langle \phi_0 |\sum_{\boldx,i}\Tr\left[\LE_i(\boldx)\LE_i(\boldx)\right]| \phi_0 
\rangle\quad {\rm and} \quad 
\langle \phi_0 |\sum_{\boldx,i}\Tr \left[\LE_i(\boldx)U_i(\boldx)\LE_i(\boldx+a\boldsymbol{i})U^\dagger_i(\boldx)
\right]| \phi_0
 \rangle.
\label{me}
\eea
The first of these is easily handled. One simply writes the electric field operators as $\LE_i(\boldx)=\LE^\alpha_i(\boldx)\lambda^\alpha $ and uses the normalisation condition, \eqn{lambdanormalisation}, to give
\bea
\langle \phi_0 |\sum_{\boldx,i}\Tr \left[\LE_i(\boldx)\LE_i(\boldx)\right]| \phi_0 \rangle 
&=& \frac{1}{2}\sum_{\boldx,i}\langle \phi_0 |\LE^\alpha_i(\boldx)\LE^\alpha_i(\boldx)| \phi_0 \rangle.
\eea
Let the trial state, $|\phi_0\rangle $, have the form $|\phi_0\rangle = e^S|0\rangle$, where $S$ is a function of link variables and $S^\dagger = S$. Then
\bea
\sum_{\boldx,i}\langle \phi_0| \LE^\alpha_i(\boldx) \LE^\alpha_i(\boldx) |\phi_0\rangle = -\sum_{\boldx,i} \langle 0| \left[\LE^\alpha_i(\boldx), e^S \right]\left[\LE^\alpha_i(\boldx), e^S\right] |0\rangle .
\label{blerk}
\eea
Making use of the Baker-Hausdorff formula, \eqn{cbh}, we can derive the following result,
\bea
e^S \LE^\alpha_i(\boldx) e^{-S} &=& \LE^\alpha_i(\boldx) - [\LE^\alpha_i(\boldx),S] + \frac{1}{2}[[\LE^\alpha_i(\boldx),S],S] 
+\ldots \nn\\
&=& \LE^\alpha_i(\boldx) - [\LE^\alpha_i(\boldx),S].
\label{comcom}
\eea
The last line follows from the fact that $S$ is a function of link variables 
and the commutation relation of \eqn{commutations1}.
Some rearrangement of \eqn{comcom} leads to the useful result,
\bea
\left[\LE^\alpha_i(\boldx), e^S\right] = \left[\LE^\alpha_i(\boldx), S\right] e^S =
e^S \left[\LE^\alpha_i(\boldx), S\right] ,\label{useful}
\eea 
which we apply in \eqn{blerk} to obtain,
\bea
\sum_{\boldx,i}\langle \phi_0| \LE^\alpha_i(\boldx) \LE^\alpha_i(\boldx) |\phi_0\rangle &=& -\sum_{\boldx,i}\langle \phi_0|\left[\LE^\alpha_i(\boldx), S \right]\left[\LE^\alpha_i(\boldx), S\right] |\phi_0\rangle \nn\\
&=&  \frac{1}{2}\sum_{\boldx,i} \langle \phi_0| \left[\LE^\alpha_i(\boldx), 
\left[\LE^\alpha_i(\boldx),S\right] \right]|\phi_0 \rangle.\label{simplify}
\eea
In the last line we have used integration by parts~\cite{Arisue:1983tt}.\\ 

The second of the matrix elements in \eqn{me} is more difficult to handle.
For this case it is convenient to expand $\LE_i(\boldx+a\boldsymbol{i})$ in the basis ${\tilde{\lambda}^\alpha_i(\boldx+a\boldsymbol{i})}$:
\bea
\LE_i(\boldx+a\boldsymbol{i}) = \tilde{\lambda}^\alpha_i(\boldx+a\boldsymbol{i}) \tilde{\LE}^\alpha_i(\boldx+a\boldsymbol{i}),
\label{crud}
\eea
where,
\bea
\tilde{\lambda}^\alpha_i(\boldx+a\boldsymbol{i}) = U^\dagger_i(\boldx) \lambda^\alpha  U_i(\boldx).  
\label{def}
\eea 
We now derive the commutation relations between $\tilde{\LE}^\alpha_i(\boldx)$ and $U_{j}(\boldsymbol{y})$. We start with the analogous relation 
to \eqn{lambdanormalisation}, 
\bea
\Tr \left[\tilde{\lambda}^\alpha_i(\boldx) \tilde{\lambda}^\beta_i(\boldx) \right]
= \frac{1}{2} \delta_{\alpha \beta}, \label{trrel}
\eea 
which follows trivially from \eqn{def}.
This relation allows us to invert \eqn{crud} and write $\tilde{\LE}^\alpha$ in terms of $\LE^\alpha $
\bea
\tilde{\LE}^\alpha_i(\boldx) &=& 2 \Tr\left[\LE_i(\boldx)\tilde {\lambda}^\alpha_i(x)
\right] \nn\\
&=& 2 \Tr\left[\tilde{\lambda}^\alpha_i(\boldx) 
\lambda^\beta \right] \LE^\beta_i(\boldx).
\label{thatone}
\eea
We can form a similar relation between $\tilde{\lambda}^\alpha$ and $\lambda^\alpha$ as follows. Since $\tilde{\lambda}^\alpha_i(\boldx)\in {\rm SU}(N)$, we can expand in the Gell-Mann basis as follows,
\bea
\tilde{\lambda}^\alpha_i(\boldx) = c^{\alpha \gamma}_i(\boldx) \lambda ^\gamma.
\label{thisone}
\eea
Here $c^{\alpha \beta}_i(\boldx)$ are constants which we now determine.
Multiplying \eqn{thisone} throughout by $\lambda^\beta$, tracing and making use of \eqn{lambdanormalisation} gives 
\bea
c^{\alpha \beta}_i(\boldx) = 
2\Tr\left[\tilde{\lambda}^\alpha_i(\boldx) \lambda ^\beta \right].
\eea
From \eqn{thatone} we see that $\tilde{\LE}^\alpha_i(\boldx) =  c^{\alpha \beta}_i(\boldx) \LE^\beta_i(\boldx)$. The commutation relations follow immediately,
\bea
\left[ \tilde{\LE}^\alpha_i(\boldx),U_{j}(\boldsymbol{y})\right] &=& c^{\alpha \beta}_i(\boldx)
\left[ \LE^\beta_i(\boldx),U_{j}(\boldsymbol{y})\right] = c^{\alpha \beta}_i(\boldx) \lambda^\beta
\delta_{ij} \delta_{\boldx\boldsymbol{y}}U_{i}(\boldsymbol{x}) \nn\\
&=& \delta_{ij} \delta_{\boldx \boldsymbol{y}}\tilde{\lambda}^\alpha_i(\boldx)U_{i}(\boldx). 
\label{modcom}
\eea
These commutation relations enable the simplification the second of the matrix elements in \eqn{me} using the basis defined in \eqn{crud}. Making use of \eqns{thatone}{modcom}
leads to the following simplification: 
\bea
\!\!\!\!\!\!\!\!\!\!\!\!\!\!\!
\langle \phi_0 |\sum_{\boldx,i}\Tr \left[\LE_i(\boldx)U_i(\boldx)\LE_i(\boldx\!+\!a\boldsymbol{i})U^\dagger_i(\boldx)
\right]\!| \phi_0 \rangle &\!\!\!=\!\!\!& \frac{1}{4}\sum_{\boldx,i}\langle \phi_0 |\!
\left[\LE^\alpha_i(\boldx),[\tilde{\LE}^\alpha_i(\boldx\!+\!a\boldsymbol{i}),S\right]|\! \phi_0\rangle .
\label{simplify2}
\eea
In order to calculate such matrix elements the following results are useful
\bea
\hspace{-1.5cm}
\left[ \LE^\alpha_i(\boldx),\left[\tilde{\LE}^\alpha_i(\boldx+a\boldsymbol{i}),U_i(\boldx)U_i(\boldx+a\boldsymbol{i})\right]\right] 
&\!\!\!=\!\!\!& \frac{N^2-1}{2N}U_i(\boldx)U_i(\boldx+a\boldsymbol{i}),\label{com1}\\
\hspace{-1.5cm}\left[\LE^\alpha_i(\boldx),\{U_i(\boldx)\}_{AB}\right]\left[\tilde{\LE}^\alpha_i(\boldx+a\boldsymbol{i}),\{U_i(\boldx+a\boldsymbol{i})\}_{CD}\right]&\!\!\!=\!\!\!& \nn\\
 && \hspace{-7.5cm}\frac{1}{2}\delta_{B C}\{U_i(\boldx)\}_{A B'}
\{U_i(\boldx+a\boldsymbol{i})\}_{B'D} -  \frac{1}{2N}\{U_i(\boldx)\}_{AB}
\{U_i(\boldx+a\boldsymbol{i})\}_{CD}, \\
\hspace{-1.5cm}\left[\LE^\alpha_i(\boldx),\{U^\dagger_i(\boldx)\}_{AB}\right]\left[\tilde{\LE}^\alpha_i(\boldx+a\boldsymbol{i}),\{U_i(\boldx+a\boldsymbol{i})\}_{CD}\right]&\!\!\!=\!\!\!& \nn\\
 && \hspace{-7cm}-\frac{1}{2}\{U^\dagger_i(\boldx)\}_{C B}
\{U_i(\boldx+a\boldsymbol{i})\}_{A D} +  \frac{1}{2N}\{U^\dagger_i(\boldx)\}_{AB}
\{U_i(\boldx+a\boldsymbol{i})\}_{CD},\\
\hspace{-1.5cm}\left[\LE^\alpha_i(\boldx),\{U_i(\boldx)\}_{AB}\right]\left[\tilde{\LE}^\alpha_i(\boldx+a\boldsymbol{i}),\{U^\dagger_i(\boldx+a\boldsymbol{i})\}_{CD}\right]&\!\!\!=\!\!\!& \nn\\
 && \hspace{-7cm}-\frac{1}{2}\{U_i(\boldx)\}_{C B}
\{U^\dagger_i(\boldx+a\boldsymbol{i})\}_{AD} +  
\frac{1}{2N}\{U_i(\boldx)\}_{AB}
\{U^\dagger_i(\boldx+a\boldsymbol{i})\}_{CD}.\label{com4}
\eea
Here $\{X\}_{AB}$ denote the colour indices of the matrix $X$ and an implicit sum over repeated colour indices is understood. 
These results follow simply from the commutation relations of 
\eqns{commutations1}{modcom}, and the following SU($N$) formula:
\bea
\lambda_{AB}^\alpha \lambda_{CD}^\alpha = 
\frac{1}{2}\delta_{AD}\delta_{BC} - \frac{1}{2N} \delta_{AB}\delta_{CD}.
\eea
\chapter{Mandelstam Constraints}
\label{mandelstamconstraints}
\bea
\Tr U^n &=& \Tr U^{n-3}-\Tr U^\dagger \Tr U^{n-2} + \Tr U \Tr U^{n-1}
\qquad \forall U\in \rm{SU}(3) \\
\Tr U^n &=& -\Tr U^{n-4}+\Tr U^\dagger \Tr U^{n-3} -\frac{1}{2} (\Tr
U)^2 \Tr U^{n-2} + \frac{1}{2} \Tr
U^2 \Tr U^{n-2} \nn\\
&& + \Tr U \Tr U^{n-1}
\qquad \forall U\in \rm{SU}(4)
\\
\Tr U^n &=&
    \Tr U^{n-5} - \Tr U^\dagger \Tr U^{n-4} + 
      \frac{1}{6}  (\Tr U)^3 \Tr U^{n-3} - 
      \frac{1}{2} \Tr U \Tr U^2 \Tr U^{n-3} \nn\\
&&  + 
      \frac{1}{3} \Tr U^3 \Tr U^{n-3} - \frac{1}{2} (\Tr U)^2 \Tr U^{n-2} + 
      \frac{1}{2} \Tr U^2 \Tr U^{n-2} \nn\\
&& + \Tr U \Tr U^{n-1} \qquad \forall U\in \rm{SU}(5)
\\
\Tr U^n &=& -\Tr U^{n-6} + 
      \Tr U^\dagger\Tr U^{n-5} - 
      \frac{1}{24} (\Tr U)^4\Tr U^{n-4}+ 
      \frac{1}{4} (\Tr U)^2\ \Tr U^2 \Tr U^{n-4}\nn\\
&& - 
      \frac{1}{8} (\Tr U^2)^2 \Tr U^{n-4} - 
      \frac{1}{3} \Tr U \Tr U^3 \Tr U^{n-4}+ 
      \frac{1}{4} \Tr U^4\Tr U^{n-4}+ 
      \frac{1}{6} (\Tr U)^3\Tr U^{n-3} 
\nn\\ 
&&
     - \frac{1}{2} \Tr U \Tr U^2 \Tr U^{n-3}+ 
      \frac{1}{3} \Tr U^3\Tr U^{n-3} - 
      \frac{1}{2} \Tr U^2\Tr U^{n-2} + 
      \frac{1}{2} \Tr U^2 \Tr U^{n-2} \nn\\
&& +\Tr U \Tr U^{n-1}  \qquad \forall U\in \rm{SU}(6)
\\
 \Tr  U^n  &=& 
    \Tr U^{n-7} - \Tr U^\dagger \Tr U^{n-6} + 
      \frac{1}{120}(\Tr U)^5\Tr U^{n-5} - 
      \frac{1}{12}(\Tr U)^3\Tr U^2 \Tr U^{n-5} 
\nn\\&&+ 
      \frac{1}{8}\Tr U (\Tr U^2)^2\Tr U^{n-5} + 
      \frac{1}{6}(\Tr U)^2\Tr U^3 \Tr U^{n-5} - 
      \frac{1}{6}\Tr U^2 \Tr U^3 \Tr U^{n-5}
\nn\\&& - 
      \frac{1}{4}\Tr U \Tr U^4 \Tr U^{n-5}+ 
      \frac{1}{5}\Tr U^5 \Tr U^{n-5} - 
      \frac{1}{24}(\Tr U)^4\Tr U^{n-4} + 
      \frac{1}{4}(\Tr U)^2\Tr  U^2 \Tr U^{n-4}
\nn\\&&  - 
      \frac{1}{8}(\Tr U^2)^2\Tr U^{n-4}  - 
      \frac{1}{3}\Tr U \Tr U^3 \Tr U^{n-4}+ 
      \frac{1}{4}\Tr U^4 \Tr U^{n-4}  + 
      \frac{1}{6}(\Tr U)^3\Tr U^{n-3}
\nn\\&&  - 
      \frac{1}{2}\Tr U \Tr U^2 \Tr U^{n-3}+ 
      \frac{1}{3}\Tr U^3 \Tr U^{n-3}- 
      \frac{1}{2}(\Tr U)^2\Tr U^{n-2}  + 
      \frac{1}{2}\Tr U^2 \Tr U^{n-2}
\nn\\&&  + \Tr U \Tr U^{n-1}\qquad \forall U\in \rm{SU}(7)
 \eea
\bea  
 \Tr U^n  &=& -\Tr U^{n-8}  + 
      \Tr U^\dagger \Tr U^{n-7}  - 
      \frac{1}{720}(\Tr U)^6\Tr U^{n-6}  + 
      \frac{1}{48}(\Tr U)^4\Tr U^2 \Tr U^{n-6} 
\nn\\&& - 
      \frac{1}{16}(\Tr U)^2(\Tr U^2)^2\Tr U^{n-6} + 
      \frac{1}{48}(\Tr U^2)^3\Tr U^{n-6}  - 
      \frac{1}{18}(\Tr U)^3\Tr U^3 \Tr U^{n-6}  
\nn\\&&+ 
      \frac{1}{6}\Tr U \Tr U^2 \Tr U^3 \Tr U^{n-6}  - 
      \frac{1}{18}(\Tr U^3)^2\Tr U^{n-6} + 
      \frac{1}{8}(\Tr U)^2\Tr U^4 \Tr U^{n-6}
\nn\\&&  - 
      \frac{1}{8}\Tr U^2 \Tr U^4 \Tr U^{n-6}  - 
      \frac{1}{5}\Tr U \Tr U^5 \Tr U^{n-6}+ 
      \frac{1}{6}\Tr U^6 \Tr U^{n-6}  + 
      \frac{1}{120}(\Tr U)^5\Tr U^{n-5} 
\nn\\&& - 
      \frac{1}{12}(\Tr U)^3\Tr U^2 \Tr U^{n-5}  + 
      \frac{1}{8}\Tr U (\Tr U^2)^2\Tr U^{n-5} + 
      \frac{1}{6}(\Tr U)^2\Tr U^3 \Tr U^{n-5}
\nn\\&& - 
      \frac{1}{6}\Tr U^2 \Tr U^3 \Tr U^{n-5}  - 
      \frac{1}{4}\Tr U \Tr U^4 \Tr U^{n-5}+ 
      \frac{1}{5}\Tr U^5 \Tr U^{n-5} - 
      \frac{1}{24}(\Tr U)^4\Tr U^{n-4}
\nn\\&& + 
      \frac{1}{4}(\Tr U)^2\Tr U^2 \Tr U^{n-4}  - 
      \frac{1}{8}(\Tr U^2)^2\Tr U^{n-4}  - 
      \frac{1}{3}\Tr U \Tr U^3 \Tr U^{n-4}+ 
      \frac{1}{4}\Tr U^4 \Tr U^{n-4} 
\nn\\&&+ 
      \frac{1}{6}(\Tr U)^3\Tr U^{n-3}  - 
      \frac{1}{2}\Tr U \Tr U^2 \Tr U^{n-3}  + 
      \frac{1}{3}\Tr U^3 \Tr U^{n-3}  - 
      \frac{1}{2}(\Tr U)^2\Tr U^{n-2}
\nn\\&& + 
      \frac{1}{2}\Tr U^2 \Tr U^{n-2}
  + \Tr U \Tr U^{n-1}
\qquad \forall U\in \rm{SU}(8)
\eea

\end{document}